\documentclass[twoside,10pt,openany]{book}
\usepackage{amssymb}
\usepackage{amsmath}

\input{TCILATEX.TEX}
\begin{document}

\author{I.Lubashevsky, V.Gafiychuk}
\title{Mathematical description \\
of heat transfer in living tissue}
\maketitle
\tableofcontents

\newpage

\pagestyle{myheadings}

\vspace{3cm}

{\hspace{1em}\LARGE \textbf{Preface }}

\markboth{\sc{Preface}}{\sc{Preface}}

\vspace{1cm}
How living organisms function and organize themselves is a very attractive
and challenging problem. This problem is extremely complex because living
organisms involve a great number of hierarchy levels from biomacromolecules
up to total organisms functioning as a whole which are related to each other
by energy and mass flow. Nevertheless, one of the possible ways to solve
this problem is to divide the whole hierarchical structure into different
levels that can be considered individually in the framework of a single
branch of science. Living tissue forms a basic level of this hierarchy which
in turn contains own complex hierarchical substructure and, from the stand
point of heat and mass transfer, can be treated as a certain medium.

Traditionally theoretical and mathematical physics deals with continuous
media for investigation of which a large number of various methods have been
developed. It is natural to apply the methods of theoretical and
mathematical physics to analysis of transport phenomena, in particular, heat
and mass transfer in living tissues. In this way one can obtain not only
particular results important for specific problems in biology, medicine and
biophysics, but also penetrate deeply into the main principles of
functioning and organization of living organism as a whole because these
principles are likely to be similar for each of the hierarchy levels.

For the theory of heat and mass transfer in living tissue one of the central
issues is how to create good models that describe these transport phenomena,
at least on the mesoscopic level, in terms of certain physical fields and
the corresponding governing equations accounting for interaction between
different levels of the living tissue hierarchy. In this way it is quite
possible to meet new problems that can be of significant interest from the
stand point of other natural hierarchical systems. The present book states
the bioheat transfer problem, which from our point of view describes the
main properties of transport phenomena peculiar to such media.

Let us make clear the subtitle of the present book which contains three key
characteristics of living tissue, mainly, ``hierarchically organized'',
``active'' and ``heterogeneous'' medium.

Roughly speaking, living tissue consists of two subsystems: the cellular
tissue treated as a uniform medium and a highly branching hierarchical
vascular network involving arterial and venous beds. Blood flow through the
arterial bed supplies the cellular tissue with oxygen, nutritious products,
etc. and controls heat balance in the system. Through the venous bed blood
flow withdraws products resulting from a life activity of the cellular
tissue.

The vascular network is embedded into the cellular tissue and in spite of
its small relative volume the vascular network mainly determines heat and
mass propagation. This is the case due to the fast convective transport%
\index{convective transport} with blood flow in vessels. Such a
characteristic feature makes living tissue as highly heterogeneous media
where low - dimension heterogeneities form fast transport paths controlling
bioheat transfer in living tissue.

Such significant effect of low - dimensional heterogeneities on transport
phenomena%
\index{transport phenomena} is also met in diffusion
\index{diffusion} processes in polycrystals and crystals with dislocations
where grain boundaries or dislocations form the fast diffusion paths.
However, living tissue differs significantly from such media in that vessels
make up a highly branching network of unique architectonics and blood flow
in vessels of one level are directly related to blood flow in all other
vessels. Therefore, in order to describe influence of blood flow on heat and
mass propagation the vascular network should be taken into account as a
whole rather than in terms of individual vessels. This characteristic
feature of living tissue is reflected in the term ``hierarchically organized
media''.

Living tissue is not only a highly heterogeneous but also an active medium%
\index{active medium}. The fact is that the cellular tissue under various
conditions requires different amount of oxygen, nutritious products, etc.
Therefore, the vascular network must respond to variations in the cellular
tissue state in the proper way. Due to expansion of a single vessel leading,
in principle, to blood flow redistribution over the whole vascular actually
all vessels should take part in the vascular networks response to local
variations of the tissue state parameters. In other words, the vascular
network response is the cooperative action of all the vessels. Variations in
vessel parameters lead to alterations of heterogeneity characteristics.
Therefore, for example, oxygen and heat propagation affect the state of
living tissue, leading to blood flow redistribution over the vascular
network due to its response, which in turn alter heterogeneity properties
and affects oxygen and heat propagation. Thus, such transport phenomena%
\index{transport phenomena} should exhibit nonlinear behavior, and living
tissue is an active distributed system with self-regulation.

The three inalienable characteristics of living tissue are the essence of
the bioheat transfer problem in its own right.

In this monograph we do not claim the complete solution of the bioheat
transfer problem that could be used in comparing with particular
experimental data. In fact, in the present monograph we formulate a simple
model for heat transfer in living tissue with self-regulation. The initial
point of the model is the governing equations describing heat transfer in
living tissue at the mesoscopic level, i.e. considering different vessels
individually. Then, basing on the well known equivalence of the diffusion
type process and random walks, we develop a certain regular procedure that
enables us to average these mesoscopic equations practically over all scales
of the hierarchical vascular network. The microscopic governing equations
obtained in this way describe living tissue in terms of an active medium
with continuously distributed self-regulation.

One of the interesting results obtained in the present monograph is that
there can be the phenomena of ideal self-regulation in large active
hierarchical systems. Large hierarchical systems are characterized by such a
great in%
\index{active medium}formation flow that none of its elements can possess
whole information required of governing the system behavior. Nevertheless,
there exists a cooperative mechanism of regulation which involves individual
response of each element to the corresponding hierarchical piece of
information and leads to ideal system response due to self - processing of
information. The particular results are obtained for bioheat transfer.
However, self-regulation in other natural hierarchical systems seems to be
organized in a similar way.

The characteristics of large hierarchical systems occurring in nature are
discussed from the stand point of regulation problems. By way of example,
some ecological and economic systems are considered. An cooperative
mechanism of self-regulation which enables the system to function ideally is
proposed.

The authors are very grateful to Arthur Cadjan, in cooperation with whom
many scientific results of the chapters \ref{ch.4}--\ref{ch.9}, \ref{ch.13},
\ref{ch.14} were obtained.

The results presented in this book were partially supported by research
grants U1I000, U1I200 from the International Science Foundation.

\vspace{0.2in}

Moscow, Russia \qquad \qquad \qquad \qquad I.Lubashevsky,

Lviv, Ukraine \qquad \qquad \qquad \qquad \qquad V.Gafiychuk,

June 1999.

\clearpage

\pagenumbering{arabic}

\part{The basis of the bioheat transfer theory}

\markboth{{\sc \thepart.{ } The basis of the bioheat transfer theory}}{}

\chapter{INTRODUCTION: Essentials of the description of transport phenomena
in highly heterogeneous media. What the bioheat transfer problem is}

\label{ch.0}
\markright
{{\sc \thechapter. INTRODUCTION: Essentials of the description \ldots}
}

Transport phenomena%
\index{transport phenomena} such as diffusion and heat propagation in
solids, liquids as well as in condensed materials of complex structure can
be usually described in the framework of the phenomenological approach
where, for example, gradients in concentrations of diffusing species and
temperature give rise to local mass and heat flow in the medium. In the
simplest case diffusion of a scalar field $c$ is characterized by its flux $%
\mathbf{J}_{c}$ proportional to the gradient \textbf{$\nabla $}$c$. When the
medium itself can move in space, at least locally, the flux $\mathbf{J}_{c}$
contains the term proportional also to the local velocity $\mathbf{v}$ of
the medium motion. In other words,
\begin{equation}
\mathbf{J}_{c}=-D\mathbf{\nabla }c+\mathbf{v}c  \label{i1}
\end{equation}
and the governing equation of evolution of the field $c$ follows from the
conservation law
\begin{equation}
\frac{\partial c}{\partial t}=-\mathbf{\nabla J}_{c}+q.  \label{i2}
\end{equation}
Here $D$ is the diffusion coefficient and $q$ is the net density of sources
and sinks.

There is a large number of various media where kinetic coefficients (for
example, the diffusion coefficient $D$) or the local velocity $\mathbf{v}$
vary in space dramatically. From a mathematical point of view the
consideration of heat and mass transfer in these media is closely connected
with bioheat transfer in living tissue. For this reason we direct our
attention to the discussion of characteristic features of these media. In
particular, in polycrystals and crystals with dislocation diffusion
coefficients in the regular crystal lattice and in the vicinity of the grain
boundaries and dislocations differ in magnitude by a factor of $%
10^{5}-10^{8} $. In the system formed by materials of different structures
and composition (composite materials) the heat conductivity also vary in
space considerably. In porous media transport phenomena are controlled by
convective flow of gas or liquid through channels of complex branched form.
Therefore, in such media the local velocity $\mathbf{v}(\mathbf{r},t)$ of
convective flux governing mass and heat propagation depends on the spatial
coordinates $\mathbf{r}$ and, may be, the time $t$ significantly. Turbulent
fluid is also a highly heterogeneous medium%
\index{heterogeneous medium} from the standpoint of mass and heat transfer.
Indeed, due to laminar flow instability the velocity $\mathbf{v}(\mathbf{r}%
,t)$ is actually a random vector whose space-time distribution is
characterized by a large number of scales.

It should be noted that transport phenomena%
\index{transport phenomena} in a biological and natural environment system
can be also treated in such terms, at least qualitatively. For example,
contaminant propagation controlled by rivers, winds blowing in a forest, and
mountains, as well as epidemic propagation over regions of non-uniform
population can be described in this approach.

For such media the substantial spatial (and may be temporal) dependence of
kinetic coefficients and the velocity $\mathbf{v}$ on small scales $\ell $
is actually the essence of their highly inhomogeneity. Indeed, equations
similar to (\ref{i2}) which govern transport phenomena form practically
microscopic description%
\index{microscopic description} of these processes because they explicitly
contain spatial inhomogeneities of scales $\ell $. In particular, for the
scalar field $c$ we get

\begin{equation}
\frac{\partial c}{\partial t}=\mathbf{\nabla }\left[ D(\mathbf{r},t,\ell )%
\mathbf{\nabla }c-\mathbf{v}(\mathbf{r},t,\ell )c\right] +q(\mathbf{r},t).
\label{i3}
\end{equation}
The solution $c(\mathbf{r},t,\ell )$ of this equation contains all the
details of the field $c$ distribution on small scales of order $\ell $ as
well as on large scales $\mathcal{L}$ characterizing the medium as a whole.
However, evolution of such systems and transport phenomena in them usually
are of importance only on spatial scales much greater than $\ell $. So, the
theory of mass and heat transfer in these highly heterogeneous media can be
based on the corresponding diffusing field averaged on scales $\ell $. In
other words, we should find the field 
\begin{equation*}
c_{a}(\mathbf{r},t)=\left\langle c(\mathbf{r},t,\ell )\right\rangle _{\ell }
\end{equation*}
where the symbol $\left\langle \ldots \right\rangle $ stands for averaging
on scales $\ell $. In addition, directly\ finding the solution of equation~(%
\ref{i3}) is a stubborn mathematical problem. Therefore, for such
heterogeneous media the main aim of the transport problem is to reduce the
microscopic equations similar to (\ref{i3}) to certain macroscopic equations
describing the system evolution in terms of the averaged diffusing fields ($%
c_{a}$). The generality of this problem for different branches of the
theoretical and mathematical physics, applied mathematics makes the
development of the corresponding averaging technique an interesting
mathematical problem in its own right.

Obviously that an adequate averaging technique cannot be constructed for
entire system in the general case. When all the microscopic spatial scales
as well as the corresponding temporal scales can be treated as small
parameters such technique has practically been developed and there is a
great number of works devoted to this problem for different systems(for a
review see, e.g., \cite{BP84,BPC82,LNP,OYS90,S-P80,G} and references
therein). In this case for media such as composite materials, polycrystals,
porous media, etc. the obtained macroscopic governing equations usually
retain their initial form similar to (\ref{i3}) and contain certain smoothed
effective kinetic coefficients and the mean local velocity.

When the conditions of the microscopic scales are small they are violated
and the problem becomes more complicated. For example, the macroscopic
equation of grain boundary diffusion will contain the partial derivative of
fractal order with respect to the time when certain temporal microscopic
scales are not small \cite{42}.

For turbulent fluids the velocity field $\mathbf{v}(\mathbf{r},t,\{\ell
,\tau _{\ell }\})$ is a result of cooperative interaction between a huge
number of vortexes characterized by a wide range of spatial scales $\{\ell
\} $ from a characteristic dimension $\mathcal{L}$ of the system as a whole
up to an extremely small scale $\ell _{\min }\ll \mathcal{L}$. The
corresponding temporal scales $\{\tau _{\ell }\}$ of the velocity
nonuniformities also vary over a wide range \cite{38,TL72}. Therefore,
although transport phenomena 
\index{transport phenomena}in turbulent fluids have been considered for many
years. The theory of these processes is far from being completed \cite
{MH92,JR92,VGC91,YKK93}. The basic difficulty is that one should take into
account the simultaneous effect of a large number of different vortexes in
diffusion processes and it is impossible to single out beforehand a vortex
controlling transport phenomena. Moreover, diffusing field, for example,
temperature can affect liquid motion. In this case the macroscopic governing
equations of heat transfer should allow for the nonlinear interaction of
liquid motion and heat propagation.

Such a problem of multiscale averaging the microscopic equations of
turbulent transport is also met in describing transport phenomena in other
systems. Contaminant diffusion in a river flowing through a brush of reeds
and the wind blowing in a forest are two typical examples of these processes 
\cite{PO94}.

Heat transfer along with mass transport in living tissue comprises all the
characteristic features inherent in the aforementioned systems. In fact,
living tissue (Fig.~\ref{F.i1}) is a heterogeneous medium involving blood
vessels embedded into cellular tissue and heat propagation in cellular
tissue and inside vessels with blood flow differs significantly in
properties. So, like composite materials, polycrystals and crystals with
dislocations living tissue contains certain regions with various kinetic
coefficients and heat transfer is governed by an equation similar to (\ref
{i3}). The main difference between heat propagation in the cellular tissue
and vessels is that heat slowly diffuses inside the former region and blood
flow in vessels forms paths of its fast convective transport
\index{convective transport} \cite{30}.

\FRAME{ftbpFU}{2.5763in}{1.5748in}{0pt}{\Qcb{Vascular network of real living
tissue.}}{\Qlb{F.i1}}{SUD.GIF}{\special{language "Scientific Word";type
"GRAPHIC";maintain-aspect-ratio TRUE;display "USEDEF";valid_file "F";width
2.5763in;height 1.5748in;depth 0pt;original-width 14.5003in;original-height
8.8332in;cropleft "0";croptop "1";cropright "1";cropbottom "0";filename
'SUD.GIF';file-properties "XNPEU";}}

However, in contrast to such physical systems the heterogeneities of living
tissue due to vascular network are characterized by hierarchical
architectonics: the vascular network 
\index{vascular network} involves vessels of different lengths, from large
arteries and veins of length $\mathcal{L}$ up to small capillaries of length 
$\ell _{%
\text{cap}}$. Since smaller vessels and larger ones are connected through
the branching points the velocity field $\mathbf{v}(\mathbf{r},t)$ of blood
in vessels of one hierarchy level is correlated with that of other levels
rather than independent of each other. From this point of view the problem
of heat and mass transfer in living tissue is closely connected with the
turbulent transport problem in hydrodynamics because heat and mass
propagation in living tissue is also governed by cooperative influence of
blood velocity field $\mathbf{v}(\mathbf{r},t,\ell )$ in vessels of all the
lengths from $\mathcal{L}$ to $\ell _{\text{cap}}$.

Living tissue also possesses a peculiarity that makes it distinct from
physical and mechanical media. This difference is that the vascular network 
\index{vascular network} is active and responds to variations in the
cellular tissue state. In cellular tissue the temperature, oxygen
concentration, etc. vary in time the vessels will expand or contract,
increasing or decreasing blood flow%
\index{blood flow} in order to supply cellular tissue with blood amount
required. Therefore, in development of the theory of transport phenomena%
\index{transport phenomena} in living tissue one should also take into
account the physical parameters of vessels belonging to all the hierarchy
levels vary in such a self--consistent way it enables the vascular network%
\index{vascular network}\ to respond properly and, so, living tissue to
adapt to new conditions.

Thus, any theory that claims to describe adequately real transport phenomena
in living tissue should account for these basic properties. One of the first
steps in this direction is the development of an averaging technique%
\index{averaging technique} converting the microscopic equations similar to (%
\ref{i3}) into macroscopic governing equations describing evolution of
certain smoothed fields. Obtaining such macroscopic governing equations for
heat propagation in living tissue is the essence of the bioheat transfer
problem. It should be noted that, in fact, this problem involves basic
parts, the former is constructing the averaging technique in its own right,
the latter is the description of the living tissue active behavior.
Correspondingly, the present book considers these questions successively.

The book is organized as follows. The theory developed previously treats
heat transfer in living tissue actually at the phenomenological level, based
mainly on the conservation of blood and energy and practically does not
account details of heat interaction between vessels of different levels.
These models and their background are briefly reviewed in Chapter~\ref{ch.1}%
. Here we also represent the well known rough classification of all blood
vessels according to their influence on heat propagation which is the first
physiological background for any model of living tissue. Besides, we discuss
what macroscopic physical variables (continuous fields) bioheat transfer
theory should deal with. In particular, in addition to the smoothed
temperature the blood flow rate $j(\mathbf{r},t)$ is such a state variable.

Real living tissues are extremely complex systems and there is a large
number of processes where heat transfer occurs. So, the model proposed in
this book is certain not to be able to describe real processes of bioheat
transfer in full measure. It solely takes into account the main
characteristic features of living tissue and can be the basis for analysis
of temperature distribution under extreme conditions (e.g., during
hyperthermia treatment) when the temperature is a leading parameter the
state. Certain physiological properties of real living tissues, including
architectonics of vascular networks, what determines a region of living
tissue that can be treated as a distributed medium, mechanisms of tissue
response to temperature variations are considered in Chapter~\ref{ch.2}. In
no case this Chapter can be regarded as an introduction to physiology of
living tissues in its own right. We understand that a large number of
important physiological problems is beyond the scope of our discussion. We
consider only those forming the starting point of the proposed model and
motivating the properties to be ascribed to blood vessels.

In Chapter~\ref{ch.3} we specify microscopic equations, governing
temperature distribution in cellular tissue and vessels individually, as
well as the vascular network architectonics. A particular form of vascular
network, on one hand, must meet certain conditions (Chapter~\ref{ch.2}) and,
on the other hand, may be chosen for convenience. The latter is possible
because, as will be shown in Chapters~\ref{ch.5} and \ref{ch.6}, the heat
propagation exhibits the self - averaging properties and solely
characteristic properties of vessel branching have remarkable effect on heat
transfer. In this Chapter we also formulate the specific model for the
vessel response to temperature variations in cellular tissue.

Then, in Chapters~\ref{ch.4}--\ref{ch.6} instead of solving the microscopic
temperature evolution equations directly we describe bioheat transfer in
terms of random walks in living tissue. This is possible due to the well
known equivalence of diffusive type processes and random motion of certain
Brownian particles. The characteristic path of walker motion in living
tissue involves an alternating sequence of walker motion inside the vessels
with blood flow%
\index{blood flow} and in the cellular tissue. Finding the mean displacement
of a typical walker at a given time we gain capability to trace the typical
way of walker motion through the hierarchical vessel system and to propose
the desired averaging procedure. In this way we will be able to obtain
specific form of the macroscopic bioheat equation under various conditions.
The developed procedure enables us to classify all the vessels of the given
vascular network according to their influence on heat transfer. The latter
forms the basis of thermoregulation theory dealing with temperature response
of individual vessels.

The result of averaging the microscopic description of the hierarchical
model is presented in Chapter~\ref{ch.7} by the generalized bioheat
equation. This equation, first, contains terms, treating living tissue as an
effective homogeneous medium. This medium, however, has additional effective
heat sinks caused by blood flow. In other words, averaging initial
microscopic equations of the divergence form leads to the appearance of
sinks whose density is proportional to the blood flow rate. The fact is that
large vessels form traps of the Brownian particles rather than paths of
their fast transport from the standpoint of their motion in cellular tissue.
In addition, it turns out that under certain conditions the renormalization
coefficient of the temperature diffusivity practically does not depend on
the physical parameters of the system. This is a direct consequence of the
vascular network being hierarchically organized. The generalized bioheat
equation contains other terms whose appearance is caused by the discrete
distribution of small vessels in the space and which are regarded as random
spatial inhomogeneities. The corresponding characteristic properties of
random spatial nonuniformities and fluctuations of the tissue temperature
are analyzed in detail in Chapters~\ref{ch.13}, \ref{ch.14}.

When blood flow distribution over the vascular network becomes substantially
non-uniform the bioheat equation should be modified which is the subject of
Chapters~\ref{ch.8}--\ref{ch.9}. In this case not only the temperature, but
also the 
\index{blood flow} rate must be smoothed and the 
\index{macroscopic description} contains two equations: averaged temperature
and the averaged blood flow rate equations.

By Chapter~\ref{ch.9} we complete, the development of the averaging
technique. Then, we consider the active behavior of living tissue, namely,
the vascular network response to variations of temperature in cellular
tissue. At this point we meet a certain fundamental problem that can be
stated in the general case and is typical not only for living tissues but
also for a large number of hierarchically organized living systems in
nature. All of them need permanent supply of external products for life
activities and inside these systems the products are delivered to different
elements through supplying networks organized hierarchically. Their peculiar
property is the capacity for responding and adapting to changes in the
environment. The latter requires redistribution of the product flow inside a
system over the supplying network. Since, as a rule, the products for life
activity enter a living system centrally there must be a certain mechanism
that governs the proper response of the supplying network at all the
hierarchy levels. Such control of the supplying network dynamics requires
processing a great amount of information characterizing the system behavior
at the all hierarchy levels. However, none of its elements can possess all
the information required of the governing system. Therefore, how a natural
large hierarchically organized systems can respond properly to changes in
the environment is a challenging problem. In the present book we deal with
the bioheat transfer problem by showing that there can be a cooperative
mechanism of self-regulation which involves individual response of each
element to the corresponding hierarchical piece of information and leads to
the ideal system functioning due to self-processing of information. It is
believed that such a cooperative mechanism of self-regulation is inherent
practically to all natural large hierarchical systems.

In living tissue blood flows through the vascular network involving arterial
and venous beds supplies cellular tissue with oxygen, nutritious products,
etc. At the same time blood withdraws carbon dioxide and products resulting
from life activities of the cellular tissue. Both the arterial and venous
beds are of the tree form contain a large number of hierarchy levels and are
similar in structure. The living tissue responses to disturbances in life
activity, for instance, through vessel response to variations of the blood
temperature, the carbon dioxide concentration, which gives rise to expansion
or contraction of arteries. These aspects form the base of the model for
vascular network response developed in Chapters~\ref{ch.10}--\ref{ch.11}.
The general equations governing the active living tissue behavior are
reduced to the local relation of the blood flow rate $j(\mathbf{r},t)$ and
the tissue temperature $T(\mathbf{r},t)$ in the spatial case which we call t%
\index{blood flow}he ideal self--regulation. The existence of the local
relationship between $j(\mathbf{r},t)$ and $T(\mathbf{r},t)$ is a surprise
because it is the consequence of a complicated mutual compensation between
blood flows at all the hierarchy levels.

In Chapter~\ref{ch.12} we apply the developed theory to analysis of heat
transfer in living tissue containing a tumor. We do not pretend to total
description of temperature distribution in this case, we only try to grasp
the main rough characteristics of temperature distribution near small tumors
which can occur in hyperthermia treatment. The same concerns the problems of
cryosurgery treatment.

As it has been mentioned above the cooperative mechanism of self--regulation 
\index{self-regulation} is inherent not only in biological organisms, but
also in large ecological and economic systems. So, in Appendix we, firstly,
generalize the model for ideal self--regulation proposed for living tissue.
Then, we show that the market with perfect competition can possess ideal
self--regulation too. From this point of view we also consider some problems
that occur in ecological models of the Lotka--Volterra type.

\chapter{Mean field approach to the bioheat transfer problem}

\label{ch.1}
\markright
{{\sc  \thechapter. Mean field approach to the bioheat transfer\ldots }
}

\section{Modern models for heat transfer in living tissue}

\label{s1.1}

The theory of heat transfer in living tissue that has been developed in the
last years is mainly aimed at promoting a better understanding of real
processes that take place in living tissue during its strong heating or
cooling.

Mathematical analysis of temperature distribution in living tissue on scales
of a single organ is, on one hand, of considerable interest for
understanding fundamental problems of human physiology as well as for
treatment of specific diseases. In fact, for example, in hyperthermia
treatment of a small tumor a tissue region containing the tumor is locally
heated to high temperature by external power sources. In this case
mathematical modelling%
\index{mathematical modelling} of temperature distribution is required to
optimize the treatment (for a review see e.g. \cite{2,30,49,56} and
references therein).

On the other hand, description of transport phenomena, in particular, heat
transfer in living tissues, is a challenging problem of mathematical
biophysics in its own right. The matter is that blood flow in vessels forms
a branching network of fast heat transport and, from the standpoint of heat
transfer, living tissue is a highly inhomogeneous and hierarchically
organized medium. Besides, due to vessel response to temperature variations
this medium is characterized by nonlinear phenomena responsible for
thermoregulation.

By now a number of models for heat transfer in living tissue have been
proposed. Reviews on bioheat transfer in living tissue can be found in \cite
{6,11,19,37,50}. Below we shall outline some of these models and their
physical background.

The simplest approach to description of bioheat transfer is to consider
living tissue in terms of an effective homogeneous 
\index{continuum}continuum where thermal interaction between cellular tissue
and blood is treated as a distributed heat sink. In physical sense, the heat
sink is caused, for example, by heating of blood in its passage through the
tissue. Within the framework of this approach we may write the following
equation for the tissue temperature

\begin{equation}
c_{t}\rho _{t}%
\frac{\partial T}{\partial t}=\mathbf{\nabla }(\kappa \mathbf{\nabla }%
T)-c_{b}\rho _{b}j(T-T_{a})+q_{h},  \label{1.1}
\end{equation}
where $c_{t},\rho _{t}$ are the density and heat capacity of the tissue, $%
c_{b},\rho _{b}$ are the same values for blood, $\kappa $ is the tissue
thermal conductivity, $q_{h}$ is the heat generation rate caused by
metabolic processes and external power sources, $T_{a}$ is the temperature
of blood in large arteries of a systemic circulation, and $j$ is the blood
flow rate, i.e. the volume of blood flowing through unit tissue volume per
unit time.

This equation has been firstly introduced by Pennes \cite{47} and now is
widely known as the conventional bioheat transfer equation. He assumes that
arterial blood enters capillaries of a tissue domain under consideration
without heat exchange with the surrounding cellular tissue, then attains
thermal equilibrium practically instantaneously , and leaves this tissue
domain through a venous bed without heat exchange again.

Since blood flow in vessels gives rise to convective heat transport it has
been proposed a model \cite{63} where living tissue is treated as a continuum%
\index{continuum} with effective convective flux $\mathbf{v}_{\mathrm{eff}%
}(r)$, and the governing bioheat equation is of the form

\begin{equation}
c_{t}\rho _{t}%
\frac{\partial T}{\partial t}=\mathbf{\nabla }(\kappa \mathbf{\nabla }%
T)-c_{b}\rho _{b}\mathbf{v}_{\mathrm{eff}}\mathbf{\nabla }T+q_{h}.
\label{1.2}
\end{equation}

Chen and Holmes \cite{13} have considered heat transfer in living tissue
containing hierarchical system of vessels and analyzed the main properties
of heat exchange between blood in different vessels and the surrounding
cellular tissue. This allowed them to justify adequately the continuum
approach to bioheat description and to propose more adequate equation for
the tissue temperature evolution in small-scale living tissue domain \cite
{13,14}

\begin{equation}
c_{t}\rho _{t}\frac{\partial T}{\partial t}=\mathbf{\nabla }(\kappa _{%
\mathrm{eff}}\mathbf{\nabla }T)-c_{b}\rho _{b}j_{v}^{\ast }(T-T_{a}^{\ast
})-c_{b}\rho _{b}\mathbf{v}_{\mathrm{eff}}\mathbf{\nabla }T+q_{h}.
\label{1.3}
\end{equation}
Here $\kappa _{\mathrm{eff}}$ is the effective thermal conductivity, $%
j_{v}^{\ast }$ -- the blood flow rate determined by arteries where blood
practically attains thermal equilibrium with the cellular tissue for the
first time and $T_{a}^{\ast }$ -- the initial blood temperature in these
arteries.

In real living tissues the arterial bed is typically located in the
immediate vicinity of the corresponding venous bed. The blood temperature in
arteries can differ significantly from the blood temperature in veins, so,
in principle, there must be an essential heat exchange between an artery and
the nearest vein of the same level. This effect is phenomenologically taken
into account in the effective conductivity model \cite{34,35,57}

\begin{equation}
c_{t}\rho _{t}\frac{\partial T}{\partial t}=\mathbf{\nabla }(\kappa _{%
\mathrm{eff}}\mathbf{\nabla }T)+q_{h}  \label{1.4}
\end{equation}
and in \cite{57} the particular relationship between the effective and
intrinsic thermal conductivity $\kappa _{\mathrm{eff}}$ and the blood flow
parameters has been found. It should be noted that effective conductivity
model has both the strong and weak sides (for more details see \cite
{6,7,9,10,59,60,63}.

All these models allow for various features of the bioheat transfer process.
So, each of them may be valid, at least at the qualitative level, under
certain conditions. Therefore, taking into account the present state of the
bioheat transfer theory it has been suggested to use for application the
following generalized bioheat equation which combines the main models
mentioned above \cite{18}:

\begin{equation}
c_{t}\rho _{t}\frac{\partial T}{\partial t}=\mathbf{\nabla }(\kappa _{%
\mathrm{eff}}\mathbf{\nabla }T)-fc_{b}\rho _{b}j(T-T_{a})+q_{h}.  \label{1.5}
\end{equation}
Here the effective thermal conductivity $\kappa _{\mathrm{eff}}$ and the
factor $f$ ranging from $0\;$to$\;1$ are phenomenological parameters.

Concluding this section we would like to point out that in order to find the
valid limits of the given collection of bioheat equations, including the
generalized equation (\ref{1.5}), as well as to obtain the specific
expressions for the corresponding kinetic coefficients one needs a
successive procedure that would enable to get a macroscopic equation by
averaging directly the corresponding microscopic governing equations. This
procedure should take into account the hierarchical structure of vascular
network, correlations in mutual arrangement of vessels belonging to
different levels, vessel response to temperature variations, etc. In the
present work we intend to develop such a procedure.

\section{Rough classification of blood vessels according to their influence
on heat propagation}

\label{s1.2}

From the standpoint of heat transfer living tissue may be represented as a
homogeneous continuum (cellular tissue) in which a hierarchical vascular
network%
\index{vascular network} is embedded. Heat propagation in the cellular
tissue and in blood flowing through the vessels is different in properties,
viz., in the cellular tissue heat propagation is controlled by thermal
conduction and inside vessels the convective heat transport can play a main
role. The number of vessel levels in the vascular network is typically large 
$N\approx 10-20$ \cite{44,55}, so blood flow in vessels of different
hierarchy levels affects variously heat transfer.

In order to characterize the effect of blood flow in a single vessel on heat
propagation it is usually used the quantity $l_{\parallel }$ defined as the
length after which the blood temperature in the vessel has practically
approached the temperature of the surrounding tissue in the corresponding
tissue cylinder, i.e. the tissue domain falling on one vessel of the given
level. If for a given vessel the value $l_{\parallel }$ is much larger than
the vessel length $l\,(l_{\parallel }\gg l)$ blood flow in it will affect
heat transfer significantly. Otherwise, $l_{\parallel }\ll l$, the blood
flow effect is ignorable \cite{12,13}. In order to imagine the extent to
which blood flow in vessels of different levels can affect heat transfer we
represent \thinspace \thinspace Table 1 where characteristic properties of
blood flow in various vessels as

\vspace{0.2in} \textbf{Table 1}\newline
\vspace{0.2in} {\footnotesize \vspace{0.1in} \noindent 
\begin{tabular}{|l|l|l|l|l|l|l|}
\hline
vessel type & diameter & length $l$ & flow & number & $l_{\parallel }$ & $%
l_{\parallel }/l$ \\ \hline
& (mm) & (cm) & (cm/s) &  & (cm) &  \\ \hline
&  &  &  &  &  &  \\ \hline
aorta & 10 & 40 & 50 & 1 & 12500 & 310 \\ \hline
large arteries & 3 & 20 & 13 & 40 & 290 & 15 \\ \hline
main branches & 1 & 10 & 8 & 600 & 20 & 2.0 \\ \hline
secondary branches & 0.6 & 4 & 8 & 1800 & 7.2 & 1.8 \\ \hline
tertiary branches & 0.14 & 1.4 & 3.4 & 7.6 $\cdot 10^{4}$ & 0.17 & 0.1 \\
\hline
terminal branches & 0.05 & 0.1 & 2 & $10^{6}$ & 0.013 & 0.1 \\ \hline
terminal arteries & 0.03 & 0.15 & 0.4 & 1.3 $\cdot 10^{7}$ & 0.0009 & 0.006
\\ \hline
arterioles & 0.02 & 0.2 & 0.3 & 4 $\cdot 10^{7}$ & 0.0003 & 0.002 \\ \hline
capillaries & 0.008 & 0.1 & 0.07 & 1.2 $\cdot 10^{9}$ & 0.00001 & 0.0001 \\ 
\hline
venules & 0.03 & 0.2 & 0.07 & 8 $\cdot 10^{7}$ & 0.00016 & 0.001 \\ \hline
terminal branches & 0.07 & 0.15 & 0.07 & 1.3 $\cdot 10^{7}$ & 0.0009 & 0.006
\\ \hline
terminal vein & 0.13 & 0.1 & 0.3 & $10^{6}$ & 0.013 & 0.1 \\ \hline
tertiary veins & 0.28 & 1.4 & 0.8 & 7.6 $\cdot 10^{4}$ & 0.16 & 0.1 \\ \hline
secondary veins & 1.5 & 4 & 1.3 & 1800 & 7.3 & 1.8 \\ \hline
main veins & 2.4 & 10 & 1.5 & 600 & 22 & 2.2 \\ \hline
large veins & 6 & 20 & 3.6 & 40 & 320 & 16 \\ \hline
vena cava & 12.5 & 40 & 33 & 1 & 12900 & 320 \\ \hline
\end{tabular}
}

\noindent well as vessels themselves are introduced \cite{6,13,19,58}.

As it follows from Table 1 the vessels where arterial blood attains thermal
equilibrium with the surrounding cellular tissue for the first time are
approximately of the length $l_{\parallel }\sim 2$\thinspace cm. Typically a
regional vascular network contains vessels whose length is much larger and
much smaller than $l_{\parallel }$. Thus, the bioheat transfer models should
take into account that the vascular network can contain vessels
significantly different in effect on heat transfer. In particular, blood
flow in vessels of length $l>l_{\parallel }$ forms a complex system of fast
heat transport paths, leading to high heterogeneity of living tissue.

\section{Mean field approach%
\index{Mean field approach}}

\label{s1.3}

In the first section of this chapter we represented various forms of the
macroscopic bioheat equation proposed by different authors. All these
mo\-dels are actually based on the mean field approach, firstly used in the
simplest form by Pennes \cite{47}. This approach principally grasps the
essence of heat transfer in living tissue so we discuss it in more detail.

Let us consider a certain living tissue domain $\mathcal{Q}$ of size $\ell $
that, on one hand, is not too small and vessels in which blood is in thermal
equilibrium with the surrounding cellular tissue are entirely\ located in
this domain (Fig.~\ref{Fig_a2}). The maximal length of such vessels is about
$l_{\parallel }$, so $\ell >l_{\parallel }$. The averaged temperature $T$ of
the cellular tissue is considered to be approximately constant over the
domain $\mathcal{Q}$. Besides, we assume that only one large artery and vein
of length $l>l_{\parallel }$ go into the region $\mathcal{Q}$ where they
join branching into small vessels.

\FRAME{ftFU}{5.3092cm}{4.9907cm}{0pt}{\Qcb{ A schematic representation of
artery and vein trees in living tissue domain $\mathcal{Q}$.}}{\Qlb{Fig_a2}}{%
Fig_a2}{\special{language "Scientific Word";type
"GRAPHIC";maintain-aspect-ratio TRUE;display "USEDEF";valid_file "F";width
5.3092cm;height 4.9907cm;depth 0pt;original-width 8.8704in;original-height
8.3333in;cropleft "0";croptop "1.0003";cropright "1";cropbottom "0";filename
'FIG_A2.GIF';file-properties "XNPEU";}}

For the points of the domain $\mathcal{Q}$ not belonging to a small
neighborhood of large vessels the true tissue temperature $T_{tr}$ obeys the
equation 
\begin{equation}
\frac{\partial T_{tr}}{\partial t}=D\mathbf{\nabla }^{2}T_{tr}-\mathbf{v}(%
\mathbf{r})\mathbf{\nabla }T_{tr}+\frac{q_{h}}{c_{t}\rho _{t}},  \label{a1}
\end{equation}
where $D=\kappa /(c_{t}\rho _{t})$ is the temperature diffusivity and for
the sake of simplicity we have ignored the difference between physical
parameters of the cellular tissue and blood. The value of $\mathbf{v}(%
\mathbf{r})$ is determined by blood flow in small vessels randomly oriented
in space, thus, the velocity $\mathbf{v}(\mathbf{r})$ is also treated as
random field $\mathbf{v}(\mathbf{r})=\left\{ v_{1}(\mathbf{r}),v_{2}(\mathbf{%
r}),v_{3}(\mathbf{r})\right\} $ with a small correlation length $l_{\text{cor%
}}\ll l_{\parallel }$. In other words, we set 
\begin{equation}
\left\langle \mathbf{v}(\mathbf{r})\right\rangle =0  \label{c1}
\end{equation}
and 
\begin{equation}
\left\langle v_{i}(\mathbf{r})v_{j}(\mathbf{r}^{\prime })\right\rangle
=v_{f}^{2}g_{ij}\left( \frac{\mathbf{r}-\mathbf{r}^{\prime }}{l_{\text{cor}}}%
\right) ,  \label{c2}
\end{equation}
where the symbol $\left\langle \ldots \right\rangle $ stands for averaging
over the small scales, $v_{f}$ is the mean amplitude of the velocity $%
\mathbf{v}(\mathbf{r})$, and $g_{ij}\left( \mathbf{r}\right) $ is a certain
function of order unity for $|\mathbf{r|}\sim 1$ which tends to zero as $|%
\mathbf{r|}\rightarrow \infty $. In addition, due to blood incompressibility
the field $\mathbf{v}(\mathbf{r})$ must obey the equation 
\begin{equation}
\mathbf{\nabla v}(\mathbf{r})=0  \label{a2}
\end{equation}
which allows us to write 
\begin{equation}
\mathbf{v}(\mathbf{r})=\mathbf{\nabla }\times \mathbf{a}(\mathbf{r}),
\label{c3}
\end{equation}
where $\mathbf{a}(\mathbf{r})=[a_{1}(\mathbf{r}),a_{2}(\mathbf{r}),a_{3}(%
\mathbf{r})]$ is a certain random field determined practically by the
concentration of the small vessels and the mean velocity of blood flow in
them. The latter allows us to set 
\begin{equation}
\left\langle \mathbf{a}(\mathbf{r})\right\rangle =0  \label{c4}
\end{equation}
and 
\begin{equation}
\left\langle a_{i}(\mathbf{r})a_{j}(\mathbf{r}^{\prime })\right\rangle
=a_{f}^{2}\delta _{ij}g_{a}\left( \frac{|\mathbf{r}-\mathbf{r}^{\prime }|}{%
l_{\text{cor}}}\right) .  \label{c5}
\end{equation}
Here $a_{f}=v_{f}l_{\text{cor}}$ is the mean amplitude of the value $\mathbf{%
a}(\mathbf{r})$, $\delta _{ij}$ is the Kronecker delta, and the function $%
g_{a}(r)$ is such that $g_{a}(r)>0$, for $r\sim 1$ the value of $%
g_{a}(r)\sim 1$, and $g_{a}(r)\rightarrow 0$ as $r\rightarrow \infty $. In
these terms the expression~(\ref{c2}) may be rewritten as 
\begin{equation}
\left\langle v_{i}(\mathbf{r})v_{j}(\mathbf{r}^{\prime })\right\rangle
=-(v_{f}l_{\text{cor}})^{2}\left[ \delta _{ij}\mathbf{\nabla }^{2}-\mathbf{%
\nabla }_{i}\mathbf{\nabla }_{j}\right] g_{a}\left( \frac{|\mathbf{r}-%
\mathbf{r}^{\prime }|}{l_{\text{cor}}}\right) .  \label{c6}
\end{equation}

The true tissue temperature $T_{tr}$ involves two parts: one is the averaged
temperature $T$, the other, $\tilde{T}$, characterizes random
nonuniformities in temperature distribution caused by the field $\mathbf{v}(%
\mathbf{r})$. The value of $\tilde{T}$ is considered to be small and the
averaged temperature distribution $T(\mathbf{r},t)$ is regarded as a smooth
field varying slowly in time. This allows us to separate equation~(\ref{a1})
into the following two equations governing the fields $\tilde{T}$, $T$
individually 
\begin{equation}
\frac{\partial T}{\partial t}=D\mathbf{\nabla }^{2}T-\mathbf{\nabla }%
\left\langle \mathbf{v}(\mathbf{r})\tilde{T}\right\rangle +\frac{q_{h}}{%
c_{t}\rho _{t}},  \label{a3}
\end{equation}
\begin{equation}
D\mathbf{\nabla }^{2}\tilde{T}-\mathbf{v}(\mathbf{r})\mathbf{\nabla }T=0.
\label{a4}
\end{equation}
The solution of equation~(\ref{a4}) is of the form 
\begin{equation}
\tilde{T}(\mathbf{r})=-\frac{1}{4\pi D}\int d\mathbf{r}^{\prime }\frac{1}{%
\left| \mathbf{r-r}^{\prime }\right| }\mathbf{v}(\mathbf{r}^{\prime })%
\mathbf{\nabla }T(\mathbf{r}^{\prime }).  \label{a5}
\end{equation}
Substituting (\ref{a5}) into (\ref{a3}) we get 
\begin{equation}
\frac{\partial T}{\partial t}=D\mathbf{\nabla }^{2}T+\frac{1}{4\pi D}%
\sum_{i,j=1}^{3}\left[ \int d\mathbf{r}^{\prime }\frac{1}{\left| \mathbf{r-r}%
^{\prime }\right| }\left\langle v_{i}(\mathbf{r})v_{j}(\mathbf{r}^{\prime
})\right\rangle \right] \mathbf{\nabla }_{i}T\mathbf{\nabla }_{j}T+\frac{%
q_{h}}{c_{t}\rho _{t}},  \label{a6}
\end{equation}
where we have also taken into account that the value $\mathbf{\nabla }T\,$
is practically constant on the scale $l_{\text{cor}}$. Substituting (\ref{c6}%
) into (\ref{a6}) we find that evolution of the averaged temperature at such
points of living tissue can be described in terms of the effective medium
model, namely 
\begin{equation}
\frac{\partial T}{\partial t}=D_{\mathrm{eff}}\mathbf{\nabla }^{2}T+\frac{%
q_{h}}{c_{t}\rho _{t}},  \label{a7}
\end{equation}
where the effective temperature diffusivity is 
\begin{equation}
D_{\mathrm{eff}}=D\left[ 1+g_{a}\left( 0\right) \left( \frac{v_{f}l_{\text{%
cor}}}{D}\right) ^{2}\right] .  \label{a8}
\end{equation}

In the vicinity of large vessels equation~(\ref{a7}) does not hold which is
due to heat interaction between the cellular tissue and blood in such
arteries and veins. In order to allow for this interaction let us write the
heat conservation equation for the domain $\mathcal{Q}$ as a whole 
\begin{equation}
\frac{d}{dt}\int\limits_{\mathcal{Q}}d\mathbf{r}T=D_{\mathrm{eff}%
}\oint\limits_{\partial \mathcal{Q}}ds\mathbf{\nabla }_{n}T+J(T_{a}-T_{v})+%
\frac{1}{c_{t}\rho _{t}}\int\limits_{\mathcal{Q}}d\mathbf{r}q_{h}.
\label{a9}
\end{equation}
Here the first term on the left--hand side of equation~(\ref{a9}) describes
heat flow through the boundary $\partial \mathcal{Q}$ of the domain $%
\mathcal{Q}$, the second one is caused by heat flow going into and out of
the domain $\mathcal{Q}$ with blood through large artery and vein, $T_{a}$
and $T_{v}$ are the temperatures of blood in these vessels at the boundary $%
\partial \mathcal{Q}$, and $J$ is the total blood flow going through the
given domain. For the value of $J$ we may write 
\begin{equation}
J=\int\limits_{\mathcal{Q}}d\mathbf{r}j(\mathbf{r}).  \label{a10}
\end{equation}
Depending on the vessel architectonics, the temperature $T_{v}$ of blood in
the large vein going out of the domain $\mathcal{Q}$ is approximately equal
to the averaged tissue temperature $T$ or there can be a substantial
difference between these temperatures. If the venous and arterial beds are
not located in a closed vicinity of each other then any vein is far enough
from the arteries of the same length and thereby $T_{v}\approx T$ because
blood in small vessels is in thermal equilibrium with the surrounding
cellular tissue. When the vascular network is organized in such way that
arteries and veins are located in the vicinity of one another there is a
strong heat exchange between blood flows in a vein and in the corresponding
artery (counter current vessels). So, in this case $|T_{v}-T_{a}|<|T-T_{a}|$%
. In order to allow for the given heat interaction we may introduce a
cofactor $f<1$ into the relation 
\begin{equation}
(T_{v}-T_{a})=(T-T_{a})f.  \label{aa1}
\end{equation}
Then assuming the fields $T(\mathbf{r})$ and $j(\mathbf{r})$ to be
approximately constant over the domain $\mathcal{Q}$ and taking into account
expressions~(\ref{a10}) and (\ref{aa1}) we can convert equation~(\ref{a9})
into the following partial differential equation 
\begin{equation}
\frac{\partial T}{\partial t}=D_{\mathrm{eff}}\mathbf{\nabla }%
^{2}T-fj(T-T_{a})+\frac{q_{h}}{c_{t}\rho _{t}},  \label{a12}
\end{equation}
which practically exactly coincides with the generalized bio\-he\-at
equa\-ti\-on~(\ref{1.5}).

Equation (~\ref{a12}) actually is phenomenological rather than the reliable
result of averaging the microscopic bioheat equations. It contains at least
two parameters, the effective diffusivity $D_{\mathrm{eff}}$ and the
cofactor $f$ which cannot be found in the framework of the mean field
approach. In addition, the question of whether the averaged tissue
temperature can be practically constant on spatial scales of order $%
l_{\parallel }$ is beyond the mean field theory of bioheat transfer. The
same concerns the form of the microscopic bioheat equation when the blood
flow rate is extremely non-uniform distributed over living tissue.

Due to the vessel system being hierarchically organized blood flow
distribution over the vascular network as well as over the tissue domain has
to be characterized by strong correlations between different hierarchy
le\-vels and also by spatial correlations. Therefore, in order to describe
the blood flow effect on heat transfer one should take into account the
vascular network as a whole rather than consider vessels of different levels
individually. The vascular network models (see, e.g., \cite{6,7,62}) dealing
with living tissue phantoms containing infinitely long vessels or models
where the effect of blood flow in different vessels on heat transfer are
treated in the same terms \cite{31,57,58} cannot form the basis of the
successive procedure of averaging the microscopic equations. The next
characteristic property of living tissue is its active response to
temperature variations. Living tissue tries to remain it temperature within
a certain vital interval $[T_{-},T_{+}]$. Therefore, if a certain tissue
domain is, for example, heated, the vessels supplying this domain with blood
will expand and the blood flow rate will increase. In order to find specific
relationship between the blood flow rate $j(\mathbf{r})$ and the tissue
temperature field $T(\mathbf{r})$ one should, in principle, account for the
temperature response of the vascular network as a whole. Thus, the bioheat
equations discussed in section~\ref{s1.1} can use only phenomenological
models for the relation $j(T)$ (see, e.g., \cite{13,30}). It should be noted
that blood flow rate can increase locally by tenfold \cite{54}.

In order to go out of the framework of the mean field approach we will
describe bioheat transfer in terms of random walks%
\index{random walk} in living tissue, instead of solving the microscopic
temperature evolution equations directly. This is possible due to the well
known equivalence of diffusive type processes and random motion of certain
Brownian particles. For example, in the theory of grain boundary diffusion
and diffusion in crystals with dislocations such an approach has enabled to
obtain rigorous equations for anomalous diffusion \cite{1,39,40,41,42}.

\chapter{Physiological background}

\label{ch.2} \markright
{ {\sc \thechapter.  Physiological background}
}

\section{Microcirculatory region as a basic fundamental domain of bioheat
transfer theory}

\label{s2.1}

In the theory of bioheat transfer living tissue is regarded as a certain
part of a living organism. The models mentioned in Introduction treat living
tissue as a continuum containing vessels where blood flow is predetermined.
So, these models consider vessels of different levels practically
independently of each other \cite{6}. In the same time blood flow
distribution over vessels belonging to different levels must be self -
consistent due to the hierarchical organization of vascular networks.
Therefore, on one hand, in order to develop the desired successive averaging
procedure we need to take into account blood flow distribution over all
hierarchy levels that can affect the blood flow rate at a point under
consideration. On the other hand, it is impossible to describe heat transfer
in a living organism as a whole, including blood flow distribution over its
systemic circulation, in the context of continuous theory. The latter also
is of a little consequence when the tissue region affected directly, for
example, heated by external power sources is not sufficiently large.

Therefore, first of all, we should specify a minimal region of a living
organism for which a complete theory of heat transfer can be developed. In
other words, such a theory has to describe in the self-consistent way the
distribution of the tissue temperature as well as the blood flow rate over
the tissue domain under consideration. This minimal region of living
organism will be called the basic fundamental domain of living tissue.

In living organisms a microcirculatory bed%
\index{microcirculatory bed} region can be treated as a basic fundamental
domain. In fact, the main aim of systemic circulation is to maintain the
arterial-venous pressure drop $P$ at a given constant. A regional
circulation, i.e. the vascular network of a single organ, varying its
resistance to blood flow supplies different points of the organ with such
amount of blood that is needed for the organ activities \cite{44,Pop1,Pop2}.
For a relatively simple organ its whole regional circulation is a
microcirculatory bed%
\index{microcirculatory bed}. In other organs a microcirculatory bed%
\index{microcirculatory bed} is a certain part of the organ regional
circulation, for example, brain pial arterial network forms a single
microcirculatory bed. In this case blood flow in different microcirculatory
beds, in principle, can be analyzed independently of each other \cite{44}.

In the present work we consider heat transfer in a living tissue domain that
contains a complete vessel system forming a single microcirculatory bed%
\index{microcirculatory bed}.

We would like to point out that analysis of temperature distribution in
living tissue under strong local heating touches one of the fundamental
problems in mathematical biophysics, viz. mathematical description of heat
and mass transfer, and associated self-regulation processes in living
organisms on scales of a single microcirculatory bed.

The matter is that, on one hand, a tissue domain containing a single
microcirculatory bed is ordinarily large enough so, that first, transport of
oxygen, possibly other nutrients as well as heat over the domain is mainly
caused by blood flow in the vascular bed. Second, due to self-regulations
processes the distribution of $O_{2},CO_{2}$, etc. as well as the
temperature field in their turn control blood flow redistribution over the
vascular network. Thus, already on such scales mass and heat transfer
possesses properties peculiar to living organisms. On the other hand,
different parts of the same microcirculatory bed seem to be similar in
physiological function and structure. So, in the context of field theory
there can be a suitable mathematical model%
\index{mathematical model} for heat and mass transfer in living tissue on
scales of a single microcirculatory bed. In addition, due to the relative
volume of vascular network being typically small, from the viewpoint of mass
and heat transfer living tissue is an active extremely heterogeneous medium
which is characterized by a peculiar geometry of fast transport paths whose
properties depend on the temperature field and concentrations of diffusing
elements. Therefore, description of mass and heat transfer also forms a
mathematical problem in its own right.

Let us now discuss some fundamental properties of real microcirculatory beds
and heat transfer in living tissue that form the ground for the following
constructions.

When a living tissue domain containing a microcirculatory bed is in the
normal state, i.e. the temperature as well as the concentration of $%
O_{2},\,CO_{2}$, etc. are constant over the domain, the blood flow rate is
uniformly distributed over this domain too. If the latter takes place, every
small part of the domain (in comparison with the domain itself) is bound to
contain, on the average, an equal number of vessels, whose lengths are
smaller than the size of this part. In addition, in the normal state the
parameters describing blood flow in vessels of the same level must be equal
for these vessels because, otherwise, it would give rise to nonuniform
distribution of the blood flow rate $j(\mathbf{r}\,)$. Therefore,
architectonics of such a microcirculatory bed have to satisfy the condition
that any path along the vascular network from the host artery to a small
arteriole and then from the corresponding venule to the host vein (or, at
least, along the arteries and veins determining the resistance of the
vascular network to blood flow) must be of an equal length.

Typically, a vascular network involves arterial and venous parts in the tree
form as well as the system of artery-artery, vein-vein and artery-vein
anastomoses \cite{44,52}. However, there are organs containing a few number
of anastomoses or practically no one at all. Moreover, it seems that the
basic role of anastomoses is to compensate, for the blood flow
redistribution over the vascular network when self-regulation%
\index{self-regulation} processes cause, for example, substantial expansion
of some vessels \cite{44} and this subject deserves an individual
investigation. Therefore, in the present work we shall ignore anastomoses
and assume that arteries and veins make up the arterial and venous bed of
the tree form.

The real capillary system connecting arterioles and venules with each other
involves host and minor capillaries \cite{44}. As a rule, the former join
arterioles to the nearest venules, whereas, the latter are transformed into
a capillary network by a large number of capillary anastomoses \cite{44}.
This capillary network can connect not only the nearest venules and
arterioles but also distant ones. It should be noted that such a capillary
bed has been previously considered in terms of a porous medium in modelling
heat transfer \cite{51,63}. Under certain conditions rheological properties
of blood give rise to switching on or switching off the minor capillary bed 
\cite{44}. Therefore, as it follows from the percolation theory, (see. e.g., 
\cite{4,5,21}) connection between distant arterioles and venules can play a
significant role in heat transport, at least in the vicinity of the
switching points. In addition, in accordance with the results obtained below
(see Section~\ref{s5.5}), capillary influence on heat transport is
collective, i.e. only the mean properties of the capillary system geometry
are the factor. We may use any model for a capillary system which is
equivalent to the real one with respect to the main characteristic details.

From the standpoint of heat transfer the specific geometry of vascular
network is not a factor (see Chapters~\ref{ch.4}-\ref{ch.6}); thus, solely
the characteristic details of vessel branching (for example, the mean number
of arteries formed by branching of one artery whose length is twice as
large) should be taken into account. The latter allows us to choose the
specific vascular network architectonics for convenience. Typically the
resistance of a microcirculatory bed to blood flow is mainly determined by
an artery collection involving vessels of different lengths. Therefore, we
may assume that, at least in normal living tissue venules, capillaries, and
arterioles have no significant direct effect on the blood flow
redistribution. For real microcirculatory beds the resistance to blood flow
in venous parts is not a factor. Arterial and venous parts of the same
microcirculatory bed, on the average, are approximately similar in geometry
\cite{44,52,Pop1,Pop2}. Veins have wider diameters than arteries of the same
length, thus, the blood pressure drop across a microcirculatory bed is
mainly caused by the resistance of its arterial bed. However, as it follows
from the results obtained in Chapter~\ref{ch.4}, heat transfer in living
tissue actually depends on blood currents in vessels rather than on the
velocity field of blood flow in the vessels and their radii individually.
So, due to the arterial and venous beds being of the tree form the blood
current patterns on the arterial and venous parts of the same vascular
network are approximately identical. There is a certain self-averaging
property of heat transfer in living tissue (see Chapter~\ref{ch.4}) owing to
which specific features of the arterial and venous beds are not the factor.
So, for simplicity we may consider the symmetrical model for the vascular
network where the arterial and venous parts are the mirror images of each
other and the total pressure drop is twice as large.%
\vspace{0.2in} \noindent

\section{Characteristics of temperature distribution in living tissue}

\label{s2.2}

Temperature distribution in living tissue is characterized by a number of
spatial scales. One of them is the distance $l_D$ on which the tissue
temperature variations are directly controlled by heat conduction in the
cellular tissue. According to the conventional bioheat equation (\ref{1.1})

\begin{equation}
l_{D}\sim \left( \frac{D}{j}\right) ^{1/2}  \label{2.1}
\end{equation}
where $D=\kappa /(c_{t}\rho _{t})$ is the tissue thermal diffusivity and we
have taken into account that $\rho _{b}\sim \rho _{t}\,;c_{b}\sim c_{t}$. It
will be shown below in the present work, the estimate (\ref{2.1}) holds also
true for living tissue with countercurrent vascular networks. The spatial
scale $l_{D}$ is associated with the temporal scale $\tau _{D}\sim
l_{D}^{2}/D\sim 1/j$. There is another characteristic spatial scale $l_{v}$
that is the mean length of a single vessel where arterial blood attains
thermal equilibrium with the surrounding cellular tissue for the first time
in its motion through the vessel system from large arteries to arterioles.
The value of $l_{v}$ practically coincides with the thermal equilibrium
length $l_{\parallel }$ after which the blood temperature in a vessel has
approached the temperature of the surrounding tissue: $l_{v}\sim
l_{\parallel }$. According to \cite{12,58}

\begin{equation}
l_{\parallel} \sim \frac{a^2 v}{D} \ln \left ( \frac{d}{a} \right )
\label{2.2}
\end{equation}
where $a$ is the radius of a vessel under consideration, $d$ is the mean
distance between vessels of the same length $l$ and $v$ is the blood
velocity averaged over the vessel cross section. Typically $l \sim d$ \cite
{61}, then from the condition $l_v \sim l_{\parallel}$, and expression (\ref
{2.2}) we get

\begin{equation}
l_{v}\sim \left( \frac{D}{j}\frac{1}{\ln (d/a)}\right)  \label{2.3}
\end{equation}
where we have also taken into account the relation $j\sim (a^{2}v)/l^{3}$
because $\pi ^{2}a^{2}v$ is the total blood current flowing through the
tissue domain falling on one vessel of this type whose volume is about $%
d^{2}l_{v}\sim l_{v}^{3}$.

The quantity $l_{v}$ classifies vessels by their influence on heat transfer.
For typical values of the tissue thermal diffusivity $D\sim 2\cdot
10^{-3}(cm^{2}/s)$, the blood flow rate $j\sim 6\cdot 10^{-3}\cdot s^{-1}$,
and the ratio $d/a\sim 40$ we get $l_{D}\sim 0,6cm$ ; $l_{v}\sim 0,3cm$ and $%
\tau _{D}\sim 3\min $. For real microcirculatory beds ordinarily the mean
length of the shortest vessels (capillaries, venules, arterioles) is well
below $l_{v}$ , whereas that of host arteries and veins is substantially
larger that $l_{v}$ . In this case, as it will be shown in Chapter~\ref{ch.5}%
, the value $l_{v}$ divides all vessels of a microcirculatory bed into two
classes according to length. The first class involves the vessels whose
length is larger than $l_{v}$ , and the second class consists of the vessels
whose length is smaller than $l_{v}$.

Blood flowing through the first class vessels has actually no time to attain
thermodynamic equilibrium with the surrounding cellular tissue and, thereby,
may be considered to take no part, on the average, in heat exchange with the
cellular tissue \cite{13,14}. The latter allows us to suppose that at
branching points of the first class vessels not only the conservation law of
blood current but also the conservation law of heat current are true. For
this reason in the first class arteries the blood temperature is practically
equal to the temperature $T_{a}$ of blood in large arteries of systemic
circulation, which is assumed to be a predetermined quantity. In the first
class veins blood must be characterized by its own temperature $T^{\ast }$.
Indeed, into a given vein of the first class through smaller veins, also
belonging to the first class, blood comes without loss in heat from
different point of a tissue domain whose size is about the vein length. So,
when, for example, the tissue temperature $T(\mathbf{r},t)$ is nonuniform on
scales of order $l>l_{v}$ the temperature of blood in a vein, whose length
is larger than $l$, will not coincide with the mean tissue temperature in
the vein neighborhood of radius $l$.

In the second class vessels blood is in thermodynamic equilibrium with the
cellular tissue and has no significant direct effect on heat transfer in
living tissue. Below the vessels of the first and second classes will be
also called heat conservation (or thermoimpermeable) and heat dissipation
(or thermopermeable), respectively.

Keeping in mind microcirculatory beds such as those of kidney, muscles etc. 
\cite{32} we consider a three-dimensional vascular network embedded in a
tissue domain $Q_{0}$ three spatial sizes of which $(L_{x},L_{y},L_{z})$ are
of the same order. Moreover, the given domain $Q_{0}$ may be assumed to be
of the cube form because in this case its specific geometry practically is
not a factor. When one of the spatial sizes, e.g. $L_{x}$, is substantially
smaller than the others but well above $l_{D}$ (i.e. $l_{D}\ll L_{x}\ll
L_{y},L_{z}$) the resistance to blood flow in vessels that are responsible
for blood redistribution over the domain $Q_{0}$ on scales being larger than 
$L_{x}$ seems to be ignorable. Thus, in this case we may divide the domain $%
Q_{0}$ into cubes of the volume $L_{x}^{3}$ and consider heat transfer in
the obtained subdomains individually. When $L_{x}\ll l_{D}$ heat transfer in
such a domain should be analyzed within the framework of a two-dimen\-sional
model which is the subject of an individual investigation. Besides, for real
living tissues it is quite natural to assume that if both the tissue
temperature $T$ and the concentration of $O_{2},CO_{2}$, etc. are constant
over a domain, which contains a single microcirculatory bed, then the blood
flow rate $j$ will be also constant over this domain. Within the framework
of the model under consideration such an assumption gives rise to the
requirement that the vascular network is uniformly distributed over the
domain $Q_{0}$. The term $^{\text{``}}$uniformly'' means that each subdomain
of diameter $l$ contains approximately an equal number of vessels, whose
lengths are smaller than $l$ , as it is observed in real living tissues \cite
{44,52}.\vspace{0.2in} \noindent

\section{The main properties of temperature self-re\-gu\-la\-tion in living
tissue}

\label{s2.3}

The main objective of the thermoregulation theory is to obtain the equation
governing evolution of the blood flow rate $j(r)$ at every point of the
microcirculatory bed domain as the tissue temperature varies in this domain.
Previously, in mathematical analysis of the tissue temperature distribution
the blood flow rate $j(\mathbf{r},t)$ is usually taken into consideration in
terms of a predetermined function of the coordinates $\mathbf{r}$ and the
time $t$. However, when the tissue temperature $T$ attains values about $%
42-44^{0}C$ self-regulation processes in the living organ lead to a strong
dependence of the blood flow rate $j$ on the temperature field $T(\mathbf{r}%
,t)$. According to the available experimental data \cite{55} self-regulation
processes can locally increase the blood flow rate by tenfold. It should be
pointed out that in the general case the relation $j=j\{T\}$ is of a
functional form. In other words a value of $j$ at a point $\mathbf{r}$ must
depend on characteristics of temperature distribution over a certain domain
rather than on the temperature $T(\mathbf{r}\,)$ at the given point $\mathbf{%
r}$ only. Indeed, let a certain artery directly supplies a domain $Q^{\prime
}$ with blood. Then, for example, increase in its diameter induced by
temperature variation at some point causes increase in the blood flow rate $%
j $ at least in the whole domain $Q^{\prime }$. Therefore, the proposed
local models for the $j\{T\}$ dependence, based on experimental data, i.e.
the models where the functional $j\{T\}$ is given in terms of a local
function $j(T)$, seems to be justified for relatively uniform temperature
distribution only.

Now let us discuss some characteristics of the real vascular network
response to temperature variations. In principle, strong heating of a living
tissue domain can cause a response in the organism as a whole system.
However, when the domain affected directly is small enough increase in blood
flow inside the given tissue domain will be mainly controlled by the
corresponding microcirculatory bed. In this case large vessels of systemic
circulation, which supply different organs with blood, can be treated as an
infinitely large blood reservoir where the pressure drop is maintained at a
constant value $P$ \cite{44}. Due to the conservation law of heat current in
the first class veins the temperature $T^{\ast }$ of blood in these veins is
a natural parameter for control of the temperature field $T(\mathbf{r},t)$
in the cellular tissue. Indeed, since heat exchange between blood in the
first class veins and the cellular tissue is ignorable., the temperature $%
T_{i}^{\ast }$ of blood, for example, in a vein $i$ of the first class
should be approximately equal to the mean tissue temperature $T_{i}$ in the
subdomain $Q_{i}$ from which blood is carried away through the given vein.
So, the value of $T_{i}^{\ast }$ can immediately specify the response of the
corresponding artery directly supplying the given tissue subdomain $Q_{i}$
with blood.

The precise details of temperature self-regulation processes in living
tissue, in particular, the position of temperature receptors and the
mechanism of their response are still the subject of investigations. So, let
us discuss some speculations on this problem. It is quite natural to assume
that such temperature receptors should be located in veins. Indeed, in this
case their ``readings'' immediately enable the organism to get information
on the temperature distribution over the tissue on all scales. If they were
uniformly distributed in the cellular tissue, then to respond to temperature
variations in the proper way, for each artery the individual transformation
of their ``readings'' would be required. This would make the system of the
organism response to temperature variations more complicated and, therefore,
less reliable. Nowadays such receptors are usually assumed to be located in
veins, in arteries as well as in the cellular tissue \cite{44}. However, the
position of temperature receptors is practically a factor only for
sufficiently large vessels of a microcirculatory bed, viz., for the first
class vessels. So, in the sequel we shall assume that all temperature
receptors are in the veins. As to the mechanism of the vascular network
response to temperature variations, the available experimental date indicate
that there are certain receptors located in veins which by means of the
nervous system determine response of the corresponding arteries \cite{44}.
These receptors are sensitive to the concentration of such components as $%
H^{+},K^{+},CO_{2}$ etc. However, variations in the tissue temperature give
rise to change in the metabolic process and vice versa. So, there must be a
local relation between the temperature field $T(\mathbf{r},t)$ and the
concentration of such components because their diffusion coefficients are
much less than the thermal diffusivity. Besides, for the first class veins
propagation of these components and heat propagation possess the same
properties. Therefore, it is quite justified to assume that there is a
similar local relation between the blood temperature $T^{\ast }$ and the
concentration of these components in blood. So, such receptors can be
regarded as ones responding to the blood temperature $T^{\ast }$. Taking
into account the aforementioned characteristics of tissue thermoregulation
we, first, assume that the response of an artery $i$ is directly controlled
by the temperature $T_{i}^{\ast }$ of blood in the corresponding vein $i$
and, second, describe the temperature self-regulation process in terms of
time variations in vessel resistances to blood flow, which are caused by
variations in the blood temperature.

By definition, the vessel resistance $R$ to blood flow is the quantity
appearing in the relationship $\Delta P=RJ$ between the pressure drop $%
\Delta P$ across a given vessel and the blood current $J$ in it. Due to
blood being complex in structure, nonlinear effects in its rheology give
rise, in particular, to dependence of the vessel resistance on the mean
blood velocity. In other words, in the general case the resistance $R$ is a
function of $J$. However, under normal conditions such nonlinear effects
exert appreciable influence on blood flow only in vessels whose radius $a$
is less than $50\div 100\mu m$ \cite{44}. For real microcirculatory beds a
typical value of the ratio between the length $l$ and radius $a$ of a given
artery is about $(l/a)\sim 30-40$. Therefore, the possible dependence $R(J)$
has to be taken into account only for vessels whose length is smaller than $%
0.15\div 0.3cm$. As it has been obtained above, the value $l_v\sim 0.3cm$,
thus, the nonlinear effects in blood rheology can play an important role in
blood redistribution over the second class vessels only, at least when the
blood flow rate is not extremely high. For this reason in the present
manuscript we shall ignore the blood current dependence of the vessel
resistances. The influence of these nonlinear effects on thermoregulation
and heat transfer in living tissue may be the subject of individual
investigations. In addition, in the quasistationary case the resistance $R_i$
of the artery $i$ as well as the vein $i$ is supposed to be a given explicit
function of the temperature $T_i^{*}$ of blood in the vein $%
i:R_i=R_n(T_i^{*})$ where $n$ is the level number of this vessel pair.

Finally, we discuss the properties that the function $R_{n}(T^{\ast })$ must
possess. Let $T_{\text{nor}}$ be the temperature required for the normal
functioning of the organ and $q_{n}$ be the corresponding value of the heat
generation rate. For $T_{\text{nor}}>T_{a}$ increase or decrease in $q$
gives rise to increase or decrease in the tissue temperature $T_{a}$
respectively. To compensate these temperature variations the response of the
arteries should cause an increase or appropriate decrease in the blood flow
rate, because for $T_{\text{nor}}>T_{a}$ arterial blood can be regarded as a
cooling source. This effect takes place when $R_{n}(T^{\ast })$ is a
decreasing function in the region $T^{\ast }>T_{a}$. If $T_{\text{nor}%
}<T_{a} $, then, as it can be shown similarly, $R_{n}(T^{\ast })$ must be an
increasing function for $T^{\ast }<T_{a}$. Such a behavior of the function $%
R_{n}(T^{\ast })$ in the general form is displayed in Fig.~\ref{Fig1}a by
the line ``$r$'', and Fig.~\ref{Fig1}b shows the corresponding \FRAME{ftbpFU%
}{6.48cm}{3.1719cm}{0pt}{\Qcb{The vessels resistance $R$ as a function of
the blood temperature $T^{\ast }$ (a) and the corresponding quasistationary
dependence of the tissue temperature $T$ on the uniform heat generation rate 
$q$ (b) (thecurves $r$ and $i$ display the real and ideal dependencies).}}{%
\Qlb{Fig1}}{Fig1}{\special{language "Scientific Word";type "GRAPHIC";display
"USEDEF";valid_file "F";width 6.48cm;height 3.1719cm;depth
0pt;original-width 13.3337in;original-height 14.1665in;cropleft "0";croptop
"0.9991";cropright "1.0009";cropbottom "0";filename
'FIG1.GIF';file-properties "XNPEU";}} quasistationary dependence of the
tissue temperature $T^{\ast }$ on the heat generation rate $q$ for heat
sources uniformly distributed over the tissue. When the heat generation rate
becomes large enough and the vessels exhaust their possibility of responding
the temperature self-regulation process is depressed. In this case the
resistance $R_{n}=R_{\min }$ ceases to depend on $T^{\ast }$ and, as the
value of the heat generation rate $q$ increases, the temperature $T$ goes
beyond the interval $[T_{-},T_{+}]$ where the organ tissue can survive. For $%
T>T_{+}$ or $T<T_{-}$ after a certain time the organ capacity for
functioning is irreversibly depressed. However, description of the latter
process is an individual problem and is not considered here.

In the present work special attention will be focused on a certain idealized
model for the $R_{n}(T^{\ast })$ dependence that is shown in Fig.~\ref{Fig1}%
a by the line ``$i$'' and can be represented as

\begin{equation}
R_{n}^{id}(T^{\ast })=\left\{ 
\begin{array}{cccc}
R_{n}^{0}\left[ 1-\frac{\mid T^{\ast }-T_{a}\mid }{\Delta }\right] & \!; & 
\text{if}\; & \mid T^{\ast }-T_{a}\mid \leqslant \Delta \\ 
0 & \!; & \text{if}\; & \mid T^{\ast }-T_{a}\mid >\Delta
\end{array}
\right. \;,  \label{2.4}
\end{equation}
where $R_{n}^{0}$ is a certain constant for a given vessel pair, $\Delta $
is the half width of the vital temperature interval, and for simplicity we
have set $T_{+}=T_{a}+\Delta $ and $T_{-}=T_{a}-\Delta $. The corresponding
ideal dependence of $T$ on $q$ is displayed in Fig.~\ref{Fig1}b by the curve
`$i$'. It should be pointed out that within the framework of the given model
the resistance $R_{n}^{id}(T_{i}^{\ast })$ goes to zero at the same
temperature $T_{i}^{\ast }=T_{+}$ or $T_{i}^{\ast }=T_{-}$ for each vessel
pair and, thus, in the quasistationary case the tissue temperature $T(%
\mathbf{r},t)$ cannot go beyond the interval [$T_{-};T_{+}$]. For this
reason, when the vessel response can be described by the function $%
R_{n}^{id}(T^{\ast })$ the temperature self-regulation process will be also
referred to as ideal.

Characteristics of heat transfer and the temperature self-regulation process
that we have discussed above are the main ground essentials for the
following model proposed in the present work.

\chapter{Hierarchical model for living tissue and the governing equations}

\label{ch.3}
\markright
{ {\sc \thechapter.  Hierarchical model for living\ldots}
}

\section{Evolution equations for the temperature of cellular tissue and
blood in vessels}

\label{s3.1}

Due to heat conduction of living tissue practically all spatial scales of
the temperature field $T(\mathbf{r},t)$ are well above the single cell size.
This allows us to consider the cellular tissue in terms of an isotropic
continuum in which the temperature obeys the equation:

\begin{equation}
\rho _{t}c_{t}\frac{\partial T}{\partial t}=\kappa \mathbf{\nabla }%
^{2}T+q_{h}.  \label{3.1}
\end{equation}
Here as well as in the Introduction $q_{h}$ is the heat generation rate
caused by metabolic processes and electromagnetic or ultrasonic radiation
absorption, $\rho _{t},\,c_{t},\,\kappa $ are the density, specific heat
capacity, and thermal conductivity of the tissue which are regarded as
constant quantities.

Let us describe each vessel as a pipe of length $l$ and radius $a$. The
temperature field $T^*$ of blood inside vessels satisfies the equation:

\begin{equation}
\rho _{t}c_{t}\left( \frac{\partial T^{\ast }}{\partial t}+\mathbf{v}(%
\mathbf{r}\,)\mathbf{\nabla }T^{\ast }\right) =\kappa \mathbf{\nabla }%
^{2}T^{\ast }+q_{h}  \label{3.2}
\end{equation}
subject to the boundary conditions at the vessel interface $\sigma $

\begin{equation}
T|_{\sigma }=T^{\ast }|_{\sigma },  \label{3.3}
\end{equation}
\begin{equation}
\mathbf{\nabla }_{n}T\left| _{\sigma }\right. =\mathbf{\nabla }_{n}T^{\ast
}\left| _{\sigma }\right. .  \label{3.4}
\end{equation}
Here for the sake of simplicity the density, heat capacity and thermal
conductivity of blood are assumed to be the same as for the cellular tissue, 
$\mathbf{v}(\mathbf{r}\,)$ is the velocity field of blood flow inside the
vessel system. The velocity $\mathbf{v}(\mathbf{r}\,)$ substantially varies
over the cross section of any vessel and, in particular, attains its maximum
at the center of a vessel and is equal to zero at its boundary. However, as
it follows from the results obtained below (see Chapter~\ref{ch.4}), to
describe the influence of blood flow in a given vessel on heat transfer we
may take into account only the blood flow velocity $v$ averaged over its
cross section. Boundary conditions (\ref{3.3}), (\ref{3.4}) represent
equality of the blood and tissue temperature $T,\,T^{\ast }$ and continuity
of the normal component of the heat flux at the vessel interface. In
addition, the total relative volume of the vessels in the tissue domain
under consideration is assumed to be small, which allows us to ignore the
term $q_{h}$ in equation (\ref{3.2}).

\section{Hierarchical model for the vascular network}

\label{s3.2}

\index{hierarchical model}The vascular network involves vessels of all
lengths from capillaries to the host artery and vein, however, as it will be
shown in Part.2, just only vessels of a certain length $l_{v}$ directly
control the tissue temperature. Moreover, the effect of the latter vessels
on heat transfer depends solely on the characteristic properties of their
spatial distribution. This results from the fact that the tissue temperature
at a given point $\mathbf{r}$ is practically determined by the mean heat
generation rate in the neighborhood $Q$ of the point $\mathbf{r}$, whose
size is above $l_{D}$, and by the collective influence of $l_{v}$ length
vessels which are within this neighborhood.

In heat transfer the role of vessels whose lengths are substantially larger
than $l_v$ is to transport blood practically without heat exchange with the
cellular tissue. Arteries and veins whose lengths are smaller than $l_v$
have practically no effect on heat transfer at all. Since the mean distance
between capillaries can be much less than their characteristic lengths the
capillary network is able, in principle, to influence on heat transport,
however, their effect is also collective.

Thus, heat transfer actually depends only on characteristic properties of
the vascular network architectonics and blood flow distribution over the
vascular network. This means, for example, that the tissue temperature at
the point $\mathbf{r}$ depends on the total number of $l_{v}$ length vessels
be inside $Q$ rather than on details of their interconnections. This
characteristic of heat transfer in living tissue, which will be called the
self-averaging property%
\index{self-averaging property} of heat transfer, allows us to choose the
following model for the microcirculatory bed.

In accordance with Chapter~\ref{ch.2} we assume that the vascular network
(the microcirculatory bed) under consideration is embedded in a cube $Q_{0}$
of the volume $(2l_{0}/%
\sqrt{3})^{3}$. The host artery and the host vein of length $l_{0}$ and
radius $a_{0}$ go into and out of the cube $Q_{0}$ through one of its
corners (Fig.\ref{Fig2}). \FRAME{ftbpFU}{4.5184cm}{4.3889cm}{0pt}{\Qcb{
Fragment of the vascular network architectonics under consideration
including interconnection between vessels of different levels. (the solid
and point lines represent arteries and veins respectively).}}{\Qlb{Fig2}}{%
Fig2}{\special{language "Scientific Word";type
"GRAPHIC";maintain-aspect-ratio TRUE;display "USEDEF";valid_file "F";width
4.5184cm;height 4.3889cm;depth 0pt;original-width 6.9263in;original-height
6.7317in;cropleft "0";croptop "1.0006";cropright "1.0007";cropbottom
"0";filename 'FIG2.GIF';file-properties "XNPEU";}}They form the initial
(zeros) level of the vascular network. The host artery reaches the cube
center $O_{0}$ where it branches out into eight arteries of the first level.
Each artery of the first level reaches a center $O_{1}$ of one of the eight
cubes $\{Q_{1}\}$ (called the fundamental domains of the first level) which
compose together the cube $Q_{0}$ . In the centers $\{O_{1}\}$ each of the
first level arteries in its turn branches out into eight second level
arteries. Then the artery branching is continued in a similar way up to
level $N\gg 1$. The geometry of the venous bed is identical to the arterial
one at all the levels.

\FRAME{ftFU}{4.4108cm}{4.4108cm}{0pt}{\Qcb{Dichotomously branching vascular
network model.}}{\Qlb{Fig3}}{Fig3}{\special{language "Scientific Word";type
"GRAPHIC";maintain-aspect-ratio TRUE;display "USEDEF";valid_file "F";width
4.4108cm;height 4.4108cm;depth 0pt;original-width 7.4815in;original-height
7.472in;cropleft "0";croptop "1.0006";cropright "0.9993";cropbottom
"0";filename 'FIG3.GIF';file-properties "XNPEU";}}

There is an alternative to the proposed model for the vascular network
architectonics that is based on the dichotomously branching tree%
\index{dichotomously branching tree}. The characteristic fragment of this
vascular network is shown in Fig.~\ref{Fig3}. As for the first model the
dichotomously branching vascular network is uniformly distributed over the
basic fundamental domain $Q_{0}$, i.e. each of the fundamental domain of
level $n$ contains identical collection of vessels whose lengths are smaller
than its size and the path on the vascular network from any of the smallest
arteries (or veins) to the tree stem is the same.

Therefore, from the stand point of heat transfer these two models for
vascular network are practically equivalent because the particular form of
vessel branching points is of little consequence. The similarity between the
two vascular network models as well as a real vascular network can be
established by identifying vascular network fragments that connect the
center points $\{O_{n},O_{n+1}\}$ of fundamental domains of neighboring
hierarchy levels. This identification for a real vascular network and the
model represented in Fig.~\ref{Fig2} is illustrated in Fig.~\ref{Fig4}. In
Fig.~\ref{Fig3} the arterial bed of the dichotomously branching vascular
network is represented by the thick solid line and venous one is shown by
the pointed line. We assume that the arteries and veins whose level number $%
n $ is less or equal to $n_{cc}(n\mathbf{\leqslant }n_{cc})$ are located in
the immediate vicinity of each other whereas the arteries and veins of
higher levels, $n>n_{cc}$, do not correlate in orientation. The possibility
of such behavior is demonstrated in Fig.~\ref{Fig3} and stems from the fact
that the vessels can go out of branching points in different directions. The
value $n_{cc}$ is a parameter of the vascular network model and may be equal
to any number of the collection $\{0,1,...,N\}$. If $n_{cc}=0$ there is no
counter current vessels pair, whereas when $n_{cc}=N$ all the arteries and
the corresponding veins are parallel to each other.

In the eight - fold branching point model (Fig.~\ref{Fig2}) we also
introduce the parameter $n_{cc}$ which divides all vessels into two groups.
The arteries and veins of the first group whose level number $n\leqslant
n_{cc}$ form countercurrent pairs and the arteries and veins of the second
group for which $n>n_{cc}$ are formally not considered to correlate in
orientation.

The main attention in the present work will be focused on the first vascular
network model because dealing with this model we avoid awkward mathematical
calculations.

\FRAME{ftbpFU}{3.7031in}{1.6137in}{0pt}{\Qcb{Representation of a real
fragment of vascular network (a) as the eightfold branching point (b).}}{%
\Qlb{Fig4}}{Fig4}{\special{language "Scientific Word";type
"GRAPHIC";maintain-aspect-ratio TRUE;display "USEDEF";valid_file "F";width
3.7031in;height 1.6137in;depth 0pt;original-width 17.9163in;original-height
7.76in;cropleft "0";croptop "1";cropright "1";cropbottom "0";filename
'FIG4.GIF';file-properties "XNPEU";}}

When the blood temperature is equal to the arterial temperature of systemic
circulation $T_{a}$, all vessels of one level (for example of level $n$) and
blood flows in them are assumed to be equivalent and are described by the
same set of parameters, such as their length $l_{n}$, radius $a_{n}$ and the
mean velocity $v_{n}$ of blood in these vessels, i.e.

\begin{equation}
\{ l_n = l_0 2^{-n} ; \,\, a_n; \,\, v_n \} .  \label{3.5}
\end{equation}

\noindent Due to the blood current conservation law at the branching points
for the given vascular network architectonics these parameters satisfy the
relation:

\begin{equation}
\pi a^2_n v_n M_n = J_0 ,  \label{3.6}
\end{equation}
where $n$ is the level number, $M_n = 2^{3n}$ is the total number of
arteries or veins belonging to level $n$, and $J_0$ is the total blood
current going into the cube $Q_0$ or, what is the same, is the blood current
in the host artery or vein. For real vascular networks the ratio of the
length to radius of a single vessel weakly depends on its length and is
about $30-40$ \cite{44}. Therefore, in addition, we assume that

\begin{equation}
\frac{l_n}{a_n}=w(n)\frac{l_0}{a_0},  \label{3.7}
\end{equation}
where $w(n)$ is such a smooth function of the level number $n$ that $w(0)=1$
and $w(n)$ is of order unity. In the following numerical estimations we also
shall set $l_n/a_n\sim 30-40$. Each countercurrent pair, for example, that
of level $n$ is also characterized by the distance $b_n$ between its
vessels. In the given model we assume that the ratio $b_n/a_n$ is a constant

\begin{equation}
\frac{b_{n}}{a_{n}}=\mu
\end{equation}
and $\mu \geqslant 2$.

The last level vessels of the arterial and venous beds (called below
arterioles and venules, respectively) are interconnected by the capillary
system in the following way. Each arteriole branches out into $m\;\;(m\gg 1)$
capillaries forming a ``chimney brush''%
\index{chimney brush} structure (CB structure) (Fig.~\ref{Fig5}). Then
without branching capillaries connect with venules in a similar way. Each
capillary is described by the set of the parameters $l_{c},a_{c},v_{c}$
obeying a relation similar to (\ref{3.6}), viz%
\FRAME{fhFU}{7.8046cm}{11.1984cm}{0pt}{\Qcb{Geometry of capillary system.}}{%
\Qlb{Fig5}}{Fig5}{\special{language "Scientific Word";type
"GRAPHIC";maintain-aspect-ratio TRUE;display "PICT";valid_file "F";width
7.8046cm;height 11.1984cm;depth 0pt;original-width 11.4259in;original-height
16.4263in;cropleft "0";croptop "1.0009";cropright "0.9997";cropbottom
"0";filename 'FIG5.GIF';file-properties "XNPEU";}}

\begin{equation}
\pi a_{c}^{2}v_{c}m=\pi a_{N}^{2}v_{N}=\frac{1}{M}J_{0}  \label{3.8}
\end{equation}
where $M_{N}=M=2^{3N}$ is the total number of arterioles or venules in the
microcirculatory bed, $l_{c}$ and $a_{c}$ are the length and radius of
capillary and $v_{c}$ is the mean velocity of blood in it.

Below we shall consider individually two different structures of the
capillary system that correspond to one of the following two inequalities $%
l_c \geq l_N$ and $l_c \gg l_N$.

In the former case each arteriole can be connected with the first nearest
neighbor venule only and capillaries are practically straight pipes. In the
latter case we assume that each arteriole is joined to every venule being

\noindent inside its neighborhood of size of order $\mathcal{R}$ by an
approximately equal number of capillaries. In addition, we also assume that
each capillary keeps its spatial direction within the scale $\lambda \geq
l_{N}$. So, $\lambda $ is also the mean curvature radius%
\index{curvature radius} of a capillary regarded as a line, and on scales
above $\lambda $ different parts of a capillary are oriented independently
from one another. Therefore, in this case we may represent each capillary as
a set of randomly oriented in space rectilinear portions of length $\lambda $
and assume that every arteriole and venule spaced at distance $r$ are joined
to each other, on the average, by $m(r\,)$ capillaries where

\begin{equation}
m(r\,)=%
\frac{V_{N}m}{(2\pi \mathcal{R}^{2})^{3/2}}\exp \left\{ -\frac{r^{2}}{2%
\mathcal{R}^{2}}\right\} ,  \label{3.9}
\end{equation}
$V_{N}=(2l_{N}/\sqrt{3})^{3}$ is the volume of fundamental domain of the
last level, and

\begin{equation}
\mathcal{R}^{2}=\frac{1}{3}(l_{c}\lambda )  \label{3.10}
\end{equation}
Expressions (\ref{3.9}) and (\ref{3.10}) result from the fact that the
terminal point of a broken line made up of $\mathcal{N}=l_{c}/\lambda $ such
randomly oriented portions, is characterized by the spatial distribution $%
P_{bl}(r\,)$of the form \cite{27}

\begin{equation}
P_{bl}(r\,)=\frac{1}{\left( \frac{2}{3}\pi \mathcal{N}\lambda ^{2}\right)
^{3/2}}\exp \left( -\frac{r^{2}}{\frac{2}{3}\mathcal{N}\lambda ^{2}}\right) ,
\end{equation}
where $\mathcal{N}\lambda ^{2}=\lambda l_{c}=3\mathcal{R}^{2}$ can be
treated as the mean squared distance between the initial and terminal points
of this broken line. Therefore, the probability%
\index{probability} for a capillary generated by an arteriole to reach a
venule located at a distance $r$ is $P_{bl}(r\,)V_{N}$ because all the
capillaries terminating inside an elementary domain must go into the venule
in this domain. Multiplying the last value by $m$ we obtain the mean number
of capillaries joining an arteriole and a venule spaced at the distance $r$.
Immediately follows expression (\ref{3.9}).

Equations (\ref{3.1}),(\ref{3.2}) and boundary conditions (\ref{3.3}),(\ref
{3.4}) with the vascular network model described above are the mathematical
formulation of the heat transfer problem. It should be mentioned that the
temperature $T_a$ of blood going into the tissue domain $Q_0$ through the
host artery is the given parameter of the system.

For the purpose of the following investigations we also estimate the mean
distance between vessels of certain types. By definition, the mean distance $%
d$ between vessels of a given type is a quantity appearing in the expression 
$V=d^2l$ for the mean volume $V$ falling on each vessel of this type. Due to
every fundamental domain of level $n$ containing just one artery of level $n$
the mean distance between these arteries is

\begin{equation}
d_{n}=\left( 
\frac{2}{\sqrt{3}}\right) ^{3/2}l_{n}.  \label{3.11}
\end{equation}
Obviously, $d_{n}$ is also the mean distance between the veins of level $n$.
The tissue domain $Q_{0}$ contains $(l_{0}/l_{N})^{3}$ arterioles (because $%
M=2^{3N}=(l_{0}/l_{N})^{3}$) and, thus, $m(l_{c}/\lambda )(l_{0}/l_{N})^{3}$
a practically rectilinear portions of the capillaries. Therefore, the
quantity $d_{c}^{2}\lambda m(l_{c}/\lambda )(l_{0}/l_{N})^{3}$, where $%
(d_{c}^{2}\lambda )$ is the mean volume falling on one capillary rectilinear
portion, must be equal to the volume $V_{0}=(2l_{0}/\sqrt{3})^{3}$ of the
domain $Q_{0}$. We obtain the desired expression for the mean distance
between the capillary rectilinear portions or, what is the same, for the
mean distance $d_{c}$ between the capillaries themselves

\begin{equation}
d_c = \left ( \frac{2}{\sqrt{3}} \right )^{3/2} \left ( \frac{l_N^3}{ml_c}
\right )^{1/2}.  \label{3.12}
\end{equation}

Another quantity required for the further analysis is the blood flow rate $j(%
\mathbf{r}\,)$ regarded as a continuous field defined at every point of the
tissue domain%
\index{tissue domain} $Q_{0}$. In the general case, including also
nonuniform blood flow distribution over a vascular network, this quantity
can be defined in the following way. Let us divide the total tissue domain
into domains $\{Q_{i}\}$ of volume $V$ and consider solely the arterial part
of the vascular network. Then, for each domain $Q_{i}$ we can find the total
blood current $J_{i}$ going into this domain through the arterial bed. By
definition, the blood flow rate averaged over the domain $Q_{i}$ is

\begin{equation}
{\langle j\rangle }_{Q_{i}}=%
\frac{J_{i}}{V}.  \label{3.13}
\end{equation}
In this way we obtain the system of quantities $\{\langle j\rangle
_{Q_{i}}\} $ associated with the given partition $\{Q_{i}\}$ of the total
tissue domain. Let there exist such a partition $\{Q_{i}\}$ that, on one
hand, each domain $Q_{i}$ contains a large number of arteries, thereby, the
quantities $\{J_{i}\}$ cannot change significantly at least for the nearest
domains and, on the other hand, their characteristic spatial size $(V)^{1/3}$
is small enough. The latter means that all physical processes under
consideration are controlled by collective influence of blood flow in
arteries which belong to different domains rather than to the same. In this
case we may interpolate the system $\{\langle j\rangle _{Q_{i}}\}$ by a
certain smooth field $j(\mathbf{r}\,)$ called the blood flow rate field or
the blood flow rate. In particular, when for each branch of the arterial bed
after a certain level blood flow is practically uniformly distributed over
it, the obtained field $j(\mathbf{r}\,)$ obviously does not depend on a
partition of the total tissue domain provided its domains are small enough.
So, at least in the latter case, the definition of the blood flow rate is
worthwhile.

For the model of the living tissue described in this section it is natural
to consider a partition of the total tissue domain $Q_0$ into fundamental
domains $\{Q_n\}$ of level $n$. Since each of these fundamental domains is
supplied with blood by an artery of level $n$, which is contained in it, the
total blood current $J_i$ going into a given domain $Q_i$ coincides with the
blood current in the corresponding artery of level $n$. When blood flow is
uniformly distributed over the vascular network, thus, according to (\ref
{3.6}) the blood current in each artery of level $n$ is equal to $J_i = J_0
2^{-3n}$ where $J_0$ can be regarded as the total blood current going into
the tissue domain $Q_0$. Expression (\ref{3.5}) enables us to represent the
volume $V_n$ of a fundamental domain belonging to level $n$ in terms of $V_n
=(2l_n/\sqrt{3})^3 = V_0 2^{-3n}$, where $V_0$ is the volume of the domain $%
Q_0$. Thereby, from (\ref{3.13}) we find that the quantities $\langle
j\rangle _{Q_n}$ are equal to the same value

\begin{equation}
j_{c}=\frac{J_{0}}{V_{0}}\equiv \left( \frac{\sqrt{3}}{2}\right) ^{3}\frac{%
J_{0}}{l_{0}^{3}}  \label{3.14}
\end{equation}
but independent of the level number. Therefore, in the given case the blood
flow rate is the uniform field $j(\mathbf{r}\,)=j_{0}.$

\section{Governing equations for blood flow distribution over the vascular
network}

\label{s3.3}

Now we formulate the governing equations for blood flow distribution over
the vascular network.

Due to the self-averaging property the effect of blood flow on heat transfer
is mainly governed by the blood current pattern on the vascular network
rather than by particular details of vessel geometry and the blood velocity
field $\mathbf{v}$ in the vessels individually. Therefore, in order to
describe the vessel response to temperature variations we may deal with the
vascular network considering every vessel as a whole, i.e. we need not to
take into account particular details of the blood flux inside a given vessel
and may allow for the mean properties of the blood flux in this vessel. So,
in accordance with Chapters~\ref{ch.2},\ref{ch.3}, each vessel is
characterized by the resistance $R$ to blood current $J$ in it, which is
assumed to be independent of $J$ and governed by the blood temperature $%
T^{\ast }$ in the corresponding vein.

The blood current distribution $\{J_{i}\}$ over the vascular network, obeys
the conservation law%
\index{conservation law} of blood at branching points%
\index{branching point}. In particular, for a given branching point $B_{a}$
of the arterial bed we have

\begin{equation}
\sum\limits_{B_{a}}J_{%
\text{out}}=J_{\text{in}},  \label{3.15}
\end{equation}
where $J_{\text{in}}$ and $J_{\text{out}}$ are the blood currents in the
arteries going in and out of the branching point $B_{a}$ and the sum runs
over all the arteries going out of the branching point $B_{a}$. For a
branching point $B_{v}$ of the venous bed we get a similar expression, viz.:

\begin{equation}
\sum\limits_{B_{v}}J_{\text{in}}=J_{\text{out}},  \label{3.16}
\end{equation}
where $J_{\text{in}}$ and $J_{\text{out}}$ are the same quantities for the
veins forming the branching point $B_{v}$, but the sum runs over all the
veins going in the branching point $B_{v}$. Secondly, the blood current
pattern%
\index{current pattern} $\{J_{i}\}$ is related with the pressure distribution%
\index{pressure distribution} $\{P_{i}\}$ over the branching points of the
vascular network by the expression

\begin{equation}
J_{i}R_{i}=\Delta P_{i}\;,  \label{3.17}
\end{equation}
where $\Delta P_{i}$ is the pressure drop%
\index{pressure drop} across the vessel $i$ and $R_{i}$ -- the resistance
\index{resistance} of the vessel $i$. The additional condition, that the
total pressure drop across the vascular network is equal to $2P$, and the
collection of equations (\ref{3.15}),(\ref{3.16}) and (\ref{3.17}) for
different branching points and vessels, respectively, forms the complete
system of Kirchhoff's equations%
\index{Kirchhoff's equations} governing the blood current pattern on the
vascular network. It should be noted that within the framework of the
symmetrical model for the microcirculatory bed (see Chapters~\ref{ch.2},\ref
{ch.3}) we may confine ourselves to analysis of the blood current pattern on
the venous part only.

In the framework of the classical approach to description of heat transfer
in living tissue blood flow is conventionally characterized by the blood
flow rate distribution $j(\mathbf{r},t)$ determined at every point of the
tissue domain. So, for the purpose of the following analysis we need to
specify the relationship between the blood current pattern $\{J_{i}\}$ and
the blood flow rate $j(\mathbf{r},t)$. Since the last level number $N\gg 1$
and, thus, the length $l_{N}=l_{0}2^{-N}$ of the last level arteries and
veins may be treated as a small spatial scale, it is natural to assume that
the blood flow distribution over the vascular network is uniform on scales
of order $l_{N}$. In this case, according to the definition of blood flow
rate, we can write

\begin{equation}
J_{i_{\mathbf{r}}}=V_{N}j(\mathbf{r},t),  \label{3.18}
\end{equation}
where $J_{i_{\mathbf{r}}}$ is the blood current in the last level artery
(arteriole) $i_{\mathbf{r}}$ (or, what is the same, in the last level vein
(venule) $i_{\mathbf{r}}$ that is located in the elementary domain%
\index{elementary domain} $Q_{N\mathbf{r}}$ containing the point $\mathbf{r}$%
. By virtue of expressions (\ref{3.16}),(\ref{3.17}) and the adopted model
for the vascular network embedding from (\ref{3.18}) it immediately follows
that the blood current $J_{i}$ in the artery $i$ or vein $i$ contained in
the fundamental domain $Q_{i}$ of the same level $n$ and the blood flow rate 
$j(\mathbf{r},t)$ are related as

\begin{equation}
J_{i}=\int\limits_{Q_{i}}d\mathbf{r}j(\mathbf{r},t),  \label{3.19}
\end{equation}
because the artery $i$ directly supplies the whole domain $Q_{i}$ with blood
and the vein $i$ drains it. Expression (\ref{3.19}) is the desired
relationship between the blood current pattern and the blood flow rate
distribution.

\section{Model for vessel response to temperature variations. Mechanism of
temperature self-re\-gu\-la\-tion}

\label{s3.4}

Thermoregulation%
\index{thermoregulation} in living tissue gives rise to vessel temperature
variations which are described as variations in the vessel resistances%
\index{vessel resistance} $\{R_{i}\}$ to blood flow. Therefore, in order to
complete the bioheat transfer theory we need to specify the equations
governing the vessel resistance evolution.

When the tissue temperature is constant over the domain $Q_{0}$ and
coincides with the systemic arterial blood temperature%
\index{blood temperature} $T_{a}\,(T=T_{a})$, the temperature $T^{\ast }$ of
blood in all the vessels will coincide with $T_{a}$ too $(T^{\ast }=T_{a})$.
In this case all vessels of the same level are assumed to be equivalent and
are characterized by the resistance

\begin{equation}
R_{n}^{0}=R_{0}2^{3n}\rho (n),  \label{3.20}
\end{equation}
where $n$ is the number of the given level and $\rho (n)$ is a smooth
function%
\index{smooth function} of $n$ such as $\rho (0)=1$ and $\rho (n)\rightarrow
0$ as $n\rightarrow \infty $. It should be pointed out that from (\ref{3.20}%
) of the $R_{n}^{0}$ dependence follows from the requirement of the blood
flow redistribution%
\index{flow redistribution} over the vascular network controlled by an
artery group comprising vessels of different lengths. This property is
typical for real microcirculatory beds and its relation with form (\ref{3.20}%
) of the $R_{n}$ dependence can be illustrated as follows.

When the resistances of vessels belonging to one level are equal at each
branching point the blood current splits into eight equal parts. Therefore,
for any path on the vascular network leading from the arterial bed stem to
the capillaries and then to the venous bed stem we can write

\begin{equation}
2P = 2 \left \{ J_0 R_0 + \frac 18 J_0 R_1 + 
\frac{1}{8^2} J_0 R_2 + \ldots \right \} = 2 J_0 \sum_{n=0}^{N} 2^{-3n}
R^0_n,
\end{equation}
where we have ignored the capillary bed resistance. From the latter
expression and formula (\ref{3.20}) we find the total resistance $R^*$ of
the vascular network

\begin{equation}
R^{\ast }=R_{0}\sum_{n=0}^{N}\rho (n).  \label{3.21}
\end{equation}
It follows that the total resistance of the microcirculatory bed is
determined by an artery group comprising vessels of different levels if $%
\rho (n)$ is a smooth function of $n$. Otherwise, the bed resistance and,
consequently, the blood flow rate redistribution due to thermoregulation
will be controlled by vessels practically of one level.

Besides, form (\ref{3.20}) of the $R_{n}$ dependence can be partly justified
in physical terms. It is natural to assume that the vessel radius $a_{n}$
changes in the same way as the length $l_{n}=l_{0}2^{-n}$ during vessel
branching. In other words, we may suppose that $a_{n}\simeq a_{0}2^{-\gamma
n}$ where $a_{0}$ is the host vessel radius and the constant $\gamma \approx
1$. Characterizing rheological properties of blood by the effective
coefficient $\mu _{\mathrm{eff}}$ of viscosity the vessel resistance $R$ to
blood flow is given by the formula \cite{38}

\begin{equation}
R=\frac{8}{\pi }\mu _{\mathrm{eff}}\frac{l}{a^{4}}.
\end{equation}
Thus, when, for example, $\mu _{\mathrm{eff}}(a)\sim a^{\alpha }$, where $%
\alpha $ is a constant \cite{44}, this expression leads to formula (\ref
{3.20}) for

\begin{equation}
\gamma =\frac{4}{4-\alpha }.
\end{equation}
In particular, for $\alpha \leqslant 1$ \cite{44} the value $\gamma \approx
1 $.

When the tissue temperature is nonuniform distributed over the domain $Q_{0}$
the blood temperature is also nonuniform and due to self-regulation
processes the resistances of all the vessels can become different. In this
case, in accordance with the remarks on the temperature self-regulation
discussed in Section~\ref{s2.3}, the response of any artery to temperature
variations is assumed to coincide with that of the corresponding vein and to
be governed by the temperature of blood in this vein. The term
``corresponding'' means that the given artery and vein belong to the same
level and are connected with each other through vessels of the higher
levels. In mathematical terms we specify the response of such two vessels,
for instance, of an artery $i$ or vein $i$ by the following model.

According to Section~\ref{s2.3} the response of a vessel to temperature
variations can be described in terms of the dependence of the vessel radius%
\index{vessel radius} $a$ on the temperature $T^{\ast }$ of blood in this
vessel if it is a vein or in the corresponding vein for an artery. The
vessel response is assumed to be specified by the following simple
phenomenological equation

\begin{equation}
\tau ^{\ast }(a)%
\frac{da}{dt}+a_{0}f^{\ast }\left( \frac{a}{a_{0}}\right) =a_{0}\frac{\left| 
{\ T^{\ast }-T_{a}}\right| }{\Delta }.  \label{3.22}
\end{equation}
Here the first term on the left-hand side describes delay of the vessel
response to the dimensionless ``signal'' $\left| T^{\ast }-T_{a}\right|
/\Delta $ generated by ``tempe\-rature'' receptors, $\tau ^{\ast }(a)$ is
the characteristic time delay of this process, $a_{0}$ is the vessel radius
for $T^{\ast }=T_{a}$, and the second term represents the vessel
counteraction to its expansion. The qualitative behavior of the function $%
f^{\ast }(a/a_{0})$ is shown in Fig.~\ref{Fig6}. \FRAME{fthFU}{5.184cm}{%
4.5514cm}{0pt}{\Qcb{The qualitative shape of the function $f^{\ast
}(a/a_{0}) $}}{\Qlb{Fig6}}{Fig6}{\special{language "Scientific Word";type
"GRAPHIC";display "USEDEF";valid_file "F";width 5.184cm;height
4.5514cm;depth 0pt;original-width 6.0001in;original-height 6.7775in;cropleft
"0";croptop "1.0003";cropright "0.9986";cropbottom "0";filename
'FIG6.GIF';file-properties "XNPEU";}}Let the resistance $R$ of the given
vessel depend on its radius $a$ as:

\begin{equation}
R=\frac{\mathrm{constant}}{a^{\beta }}\;,  \label{3.23}
\end{equation}
where the constant $\beta >0$. Then taking into account (\ref{3.22}) and (%
\ref{3.23}) we find 
\begin{equation}
\tau \frac{dR}{dt}+(R-R^{0})f\left( \frac{R}{R_{0}}\right) =-R_{0}\frac{%
\left| T^{\ast }-T_{a}\right| }{\Delta }\;.  \label{3.24}
\end{equation}
Here we have introduced the quantities 
\begin{equation}
R^{0}=R\mid _{a=a_{0}};\tau =\tau ^{\ast }(a)\frac{1}{\beta }\left( \frac{a}{%
a_{0}}\right) ^{\beta +1}  \label{3.25}
\end{equation}
and the function 
\begin{equation}
f(x)=\frac{1}{(1-x)}f^{\ast }\left[ \left( \frac{1}{x}\right) ^{1/\beta }%
\right] \;.  \label{3.26}
\end{equation}

Within the framework of the proposed model in equation (\ref{3.24}) the
value of $\tau $ is determined by the time delay in the response of both the
vessel muscles and the nervous system. There are also other additional
mechanisms of the time delay. One of them is associated with the time
required for an appropriate heated portion of blood to reach a given vein.
Another can be realized when the response of vessels to temperature
variations occurs through their response to variations in the concentrations
of $K^{+},H^{+}$, etc. (see Introduction and Chapter~\ref{ch.2}). Indeed, in
this case in order to cause variations in the $K^{+},H^{+}$, etc.
concentrations a certain time $\tau $ is required for change in the
metabolic process is due to the tissue temperature alteration. At least
within the framework of a qualitative analysis we may combine all the
mechanisms of the time delay%
\index{time delay} in one mathematical term $\tau 
\frac{\partial R}{\partial t}$ and describe dynamics of the vessel response
by equation (\ref{3.24}). Setting for simplicity $\tau =$\textrm{constant},
where $\tau $ is a certain phenomenological time delay, we immediately
obtain the desired equation for the resistance $R_{i}$:

\begin{equation}
\tau _{n}\frac{dR_{i}}{dt}+(R_{i}-R_{n}^{0})f\left( \frac{R_{i}}{R_{n}^{0}}%
\right) =-R_{n}^{0}\frac{\left| T_{i}^{\ast }-T_{a}\right| }{\Delta }\;,
\label{3.27}
\end{equation}
where $n$ is the level number of the given vessel pair, $\tau _{n}$ is the
characteristic time of the n-th level vessel response; $\Delta $ is the
halfwidth of the temperature vital interval, and the function $f(x)$
describes the vessel capacity for responding. In addition, following Section~%
\ref{s2.3} we assume that in the quasistationary case the resistance $%
R_{i}^{q}$ of the vein $i$ or the artery $i$ is an explicit function of $%
T_{i}^{\ast }$, viz.

\begin{equation}
R_{i}^{q}=R_{n}^{0}\varphi \left( \frac{\left| T_{i}^{\ast }-T_{a}\right| }{%
\Delta }\right) \;,  \label{3.28}
\end{equation}
where $\varphi (x)$ is a certain given function universal for all the
vessels. The properties of this function have been actually discussed in
Section~\ref{s2.3}. Besides, as it follows from (\ref{3.27}) and (\ref{3.28}%
) the functions $f(x)$ and $\varphi (x)$ are related by the expression

\begin{equation}
[1-\varphi (x)]\cdot f[\varphi (x)]=x\;.  \label{3.29}
\end{equation}

In particular, according to the definition given in Section~\ref{s2.3} for
the ideal temperature self-regulation process

\begin{equation}
\varphi ^{id}(x)=\left\{ 
\begin{array}{cccc}
1-x & , & \;\text{if}\; & 0\mathbf{\leqslant }x\mathbf{\leqslant }1, \\ 
0 & , & \;\text{if}\; & 1<x
\end{array}
\right.  \label{3.30}
\end{equation}
and, by virtue of (\ref{3.29}), $f^{id}(x)=1$ for $0<x\mathbf{\leqslant }1$
and $f^{id}(x)$ is undefined at $x=0$. The behavior of the functions $f(x)$
and $\varphi (x)$ for the real and ideal thermoregulation is displayed in
Fig.~\ref{Fig7}.\FRAME{fthFU}{8.6371cm}{4.1516cm}{0pt}{\Qcb{Behavior of the
function $\protect\varphi (x)$ (a) and the function $f(x)$ (b) for the real
(curves $r$) and ideal (curves $i$) thermoregulation.}}{\Qlb{Fig7}}{Fig7}{%
\special{language "Scientific Word";type "GRAPHIC";maintain-aspect-ratio
TRUE;display "USEDEF";valid_file "F";width 8.6371cm;height 4.1516cm;depth
0pt;original-width 12.9263in;original-height 6.1851in;cropleft "0";croptop
"0.9989";cropright "0.9996";cropbottom "0";filename
'FIG7.GIF';file-properties "XNPEU";}}The blood temperature $T_{i}^{\ast }$
appearing in equation (\ref{3.27}) and expression (\ref{3.28}) is actually
the mean temperature of blood in the vessel $i$. However, due to the blood
temperature field $T^{\ast }(\mathbf{r},t)$ being practically constant over
a single vessel we may not distinguish between the quantity $T_{i}^{\ast }$
and the true temperature of blood in the given vessel.

\chapter{Random walk description of heat transfer}

\label{ch.4} 
\markright
{ {\sc \thechapter.  Random walk description \ldots}
}

\section[The Fok\-ker-Planck equa\-ti\-on and ran\-dom walk
des\-cri\-pti\-on of he\-at pro\-pa\-ga\-ti\-on]{The Fok\-ker-Planck
equa\-ti\-on and ran\-dom \newline
walk des\-cri\-pti\-on of he\-at pro\-pa\-ga\-ti\-on}

\label{s4.1}

The bioheat transfer problem stated in Sections~\ref{s3.1}, \ref{s3.2}
describes the tissue temperature evolution at the microscopic level, i.e.
considers each vessel individually. In order to develop an averaging
procedure%
\index{averaging procedure} of these microscopic equations and to obtain a
macroscopic model for bioheat transfer we shall make use of the random walk%
\index{random walk} description of heat transfer. For this purpose we should
write equations (\ref{3.1}),(\ref{3.2}) in the form of the Fokker - Planck
equation%
\index{Fokker - Planck equation}. Its solution can be represented in terms
of a path integral%
\index{path integral}, which immediately leads to the desired random walk%
\index{random walk} description.

Due to blood being an incompressible liquid%
\index{incompressible liquid} and, thus, $\mathbf{\nabla v}=0$, the system
of equations (\ref{3.1}), (\ref{3.2}) subject to boundary conditions%
\index{boundary conditions} (\ref{3.3}), (\ref{3.4}) can be rewritten in the
form of the Fokker - Planck equation%
\index{Fokker - Planck equation} with sources, i.e.

\begin{equation}
\frac{\partial C}{\partial t}=\mathbf{\nabla }^{2}(DC)-\mathbf{\nabla }(%
\mathbf{v}(\mathbf{r}\,)C)+q.  \label{4.1}
\end{equation}
Here $C=(T-T_{a})/(T_{a}V_{N})$ and $C=(T^{\ast }-T_{a})/(T_{a}V_{N})$ in
the cellular tissue and inside the vessels, respectively, $q=q_{h}/(\rho
_{t}c_{t}T_{a}V_{N}),\,\,D=\kappa /(\rho _{t}c_{t})$ is the thermal
diffusivity%
\index{thermal diffusivity} of the tissue, and $V_{N}$ is the volume of
fundamental domain of the last level. We note that in the cellular tissue $%
\mathbf{v}\,(\mathbf{r}\,)=0$ and the value $C$ has the dimension of
concentration, and its variation from zero can be caused by heat generation
only when the size of the tissue domain $Q_{0}$ is large enough, viz, $%
l_{0}\gg l_{D}$.

If we ignore effect of the domain $Q_{0}$ boundaries on heat transfer, then
the solution of equation (\ref{4.1}) can be written as 
\begin{equation}
C(\mathbf{r},t)=\int\limits_{-\infty }^{t}dt^{\prime }\int\limits_{Q_{0}}d%
\mathbf{r}^{\prime }G(\mathbf{r},t\mid \mathbf{r}^{\prime },t^{\prime })q(%
\mathbf{r}^{\prime },t^{\prime }).  \label{4.2}
\end{equation}
where the Green function%
\index{Green function} $G(\mathbf{r},t\mid \mathbf{r}^{\prime },t^{\prime })$
admits the path integral 
\index{path integral} representation

\begin{equation}
G(\mathbf{r},t\mid \mathbf{r}^{\prime },t^{\prime })=\int \mathcal{D}%
\{r[t^{\prime \prime }]\}\exp \left\{ -%
\frac{1}{4D}\int\limits_{t^{\prime }}^{t}dt^{\prime \prime }[\mathbf{r}%
\,(t^{\prime \prime })-\mathbf{v}\,(\mathbf{r}\,[t^{\prime \prime
}],t^{\prime \prime })]^{2}\right\} ,  \label{4.3}
\end{equation}
where $\mathcal{D}\{r[t^{\prime \prime }]\}$ is the Wiener integral measure
of the paths and $\{\mathbf{r}\,[t^{\prime \prime }]\}$ is a given
realization of paths connecting the initial and terminal points: $\mathbf{r}%
\,(t^{\prime })=\mathbf{r}^{\prime },\mathbf{r}\,(t)=\mathbf{r}$. The well
known relationship between the Fokker - Planck equation%
\index{Fokker - Planck equation} and random motion of certain Brownian
particles%
\index{Brownian particle} enables us to mimic heat transfer in living tissue
described by equation (\ref{4.1}) or path integral%
\index{path integral} (\ref{4.3}) as random motion%
\index{random motion} of certain walkers governed by the stochastic equation 
\index{stochastic equation}

\begin{equation}
\mathbf{r}=\mathbf{v}(r,t)+\mathbf{f}(t),  \label{4.4}
\end{equation}
where $\mathbf{f}(t)=\left\{ f_{x}(t),f_{y}(t),f_{z}(t)\right\} $ is a
random force such that

\begin{equation}
\langle \mathbf{f}(t)\rangle =0,  \label{4.5}
\end{equation}
\begin{equation}
\left\langle \mathbf{f}_{\alpha }(t)\mathbf{f}_{\alpha ^{\prime }}(t^{\prime
})\right\rangle =2D\delta (t-t^{\prime })\delta _{\alpha \alpha ^{\prime }}.
\label{4.6}
\end{equation}

Here $\langle (\ldots )\rangle $ is the mean value of $(\ldots )$, $\alpha
,\alpha ^{\prime }=x,y,z;\delta _{\alpha \alpha ^{\prime }}$ is the
Kronecker delta 
\index{Kronecker delta} and $\delta (t)$ is the Dirac delta function%
\index{Dirac delta function}.

Equation (\ref{4.4}) determines the path of a walker in living tissue. The
walker created in the cellular tissue randomly moves in the cellular tissue
until it reaches the vessel boundary which is permeable for it due to
conditions (\ref{3.3}) and (\ref{3.4}). Then, the walker is transported with
blood flow until it either goes out of the vessels into the cellular tissue
again or leaves the domain $Q_{0}$, containing the microcirculatory bed,
through the host vein with blood. Therefore, each path of a walker involves
a sequence of portions associated with its migration either inside the
cellular tissue or in the vessels.

In this description of heat transfer $q$ is the generation rate of the
walks, $C$ is the walker concentration in the tissue and in blood, and $D$
is their diffusion coefficient%
\index{diffusion coefficient}.

It should be noted that the value of $q$ can be negative. So, to describe
heat transfer in terms of random walker we have to introduce two types of
walkers:``positive" and ``negative" ones. However, due to equations (\ref
{3.1}), (\ref{3.2}) and boundary conditions (\ref{3.3}),(\ref{3.4}) being
linear, the motion of a walker can be considered without regard to the
arrangement of other walkers. Since from the walker motion standpoint the
two types of the walkers are identical we may not distinguish them.

To simplify the following analysis, in this Chapter let us, first, consider
characteristic properties of random walks in the tissue phantom containing
the vessel system in the form of hexagonal array of straight identical pipes
parallel, for example, to the $z$-axis (Fig.~\ref{Fig8}). We assume that the
array spacing $d_{p}$ is well above the pipe radius%
\index{pipe radius} $a(d_{p}\gg a)$ and blood currents in different pipes
are randomly oriented in space, with the probability of blood flow in the
positive $z$-direction being equal to $1/2(1+\xi )$ and that of the opposite
direction being equal to $1/2(1-\xi )$ where $\mid \xi \mid <1$.

\FRAME{ftbpFU}{8.5075cm}{11.306cm}{0pt}{\Qcb{System of parallel vessels. The
directed wavy line represents characteristics path of the walker motion in
the cellular tissue and in vessels.}}{\Qlb{Fig8}}{Fig8}{\special{language
"Scientific Word";type "GRAPHIC";maintain-aspect-ratio TRUE;display
"USEDEF";valid_file "F";width 8.5075cm;height 11.306cm;depth
0pt;original-width 9.8519in;original-height 13.1209in;cropleft "0";croptop
"1.0009";cropright "1.0007";cropbottom "0";filename
'FIG6.GIF';file-properties "XNPEU";}}

Second, we shall analyse characteristic properties of walker random motion
in the tissue phantom containing the system of counter current%
\index{counter current} pipes pairs arranged in a similar way.

\section{Random walks in the tissue phantom containing the hexagonal array
of straight pa\-rallel pipes}

\label{s4.2}

Here we consider random walks in the tissue phantom that contains the vessel
system involving straight identical pipes parallel to the $z$-axis (Fig.~\ref
{Fig8}). We assume that the pipes make up a hexagonal array%
\index{hexagonal array} of spacing $d_{p}$, which is well above the pipe
radius $a$ $(d_{p}\gg a)$, and in the $xy$-plane normal to the pipes this
array forms a hexagonal lattice%
\index{hexagonal lattice} of discs centered at points $\{\mathbf{\rho }%
_{i}\} $ (Fig.~\ref{Fig8}). The velocity field $\mathbf{v}(\mathbf{r}%
\,)=\{0,0,v(\mathbf{\rho }\,)\}$ of blood flow in the given vessel system
will be described in terms of

\begin{equation}
v(\mathbf{\rho }\,)=\pi a^{2}v\sum_{i}\xi _{i}U(\mathbf{\rho }-\mathbf{\rho }%
_{i})  \label{4.7}
\end{equation}
Here $v$ is the mean velocity of blood in the pipes taken individually, the
sum runs over all the pipes, the quantity $\xi _{i}$ specifying the
direction of blood flow in the pipe $i$ takes the values $+1$ and $-1$ for
the positive and negative $z$-directions, respectively, $\mathbf{\rho }$ is
the projection of the vector $\mathbf{r}$ into the $xy$-plane $(S_{xy})$.
The function $U(\mathbf{\rho }\,)$ normalized to unity describes the blood
velocity distribution over the pipe cross section satisfies the equality, by
definition,

\begin{equation}
\int\limits_{S_{xy}}d\mathbf{\rho }\,U(\mathbf{\rho }-\mathbf{\rho }%
_{i})\equiv \int\limits_{S_{xy}}d\mathbf{\rho }_{i}U(\mathbf{\rho }-\mathbf{%
\rho }_{i})=1.  \label{4.8}
\end{equation}
The quantities $\{\xi _{i}\}$ are regarded as pairwise independent random
variables obeying the conditions

\begin{equation}
\langle \xi _{i}\rangle =\xi ,  \label{4.9}
\end{equation}
\begin{equation}
\langle \xi _{i}\xi _{i^{\prime }}\rangle =\delta _{ii^{\prime }}+\xi
^{2}(1-\delta _{ii^{\prime }}),  \label{4.10}
\end{equation}
where $\xi $ is a constant such that $\mid \xi \mid <1,\,\delta _{ii^{\prime
}}$ is the Kronecker delta. For the given tissue phantom the velocity $%
\mathbf{v}\,(\mathbf{r},t)$ of blood flux solely depends on the vector $%
\mathbf{\rho }$ rather than on the coordinate $z$ and the time $t$.
Therefore, according to equation (\ref{4.4}) random walks in the given
medium can be described in terms of two-dimen\-sional random walks $\{%
\mathbf{\rho }\,[t]\}$ in the $xy$-plane

\noindent and one-dimensional random walks $\{ z[t] \}$ along the $z$-axis,
with the former being independent of the latter.

Firstly, we shall analyse the characteristic properties of the distance $%
l_{\parallel }(t,\mathbf{\rho }_{0})$ that walkers starting from the point $%
\mathbf{r}_{0}=\{\mathbf{\rho }_{0},0\}$ travel along the $z$-axis during
the time $t$. By virtue of (\ref{4.4}) for a given realization $\mathbf{r}%
\,[t]=\{\mathbf{\rho }\,[t],z[t]\}$ of walker paths we may write

\begin{equation}
l_{\parallel }(t,\mathbf{\rho }_{0})=\int\limits_{0}^{t}dt^{\prime }\left\{
f_{z}(t^{\prime })+\int\limits_{S_{xy}}d\mathbf{\rho }\,^{\prime }v(\mathbf{%
\rho }\,^{\prime })\delta (\mathbf{\rho }\,^{\prime }-\mathbf{\rho }%
\,[t^{\prime }])\right\} ,  \label{4.11}
\end{equation}
where $f_{z}(t^{\prime })$ is the random force causing the walker chaotic
motion along the $z$-axis and satisfying conditions (\ref{4.5}),(\ref{4.6}),
and $\delta (\mathbf{\rho }\,)$ are the spatial $\delta $- functions.

Below in this Section we shall consider two different limits that
practically describe characteristic properties of the walker motion in the
vicinity of a single pipe and the collective pipe influence.

\subsection{A single pipe}

\label{s4.2.1}

When $t\ll d_{p}^{2}/(2D)$ we confine ourselves to the case where the
initial point $\mathbf{r}_{0}$ of the walker paths is in the vicinity of a
certain pipe, for example, of the pipe $i_{0}$, i.e.$\;\left| \mathbf{\rho }%
_{i_{0}}-\mathbf{\rho }_{0}\right| \sim a$. In this case blood flow in all
the pipes except for the pipe $i_{0}$ has practically no effect on the
walker motion and expression (\ref{4.6}) can be rewritten in the form

\begin{equation}
l_{\parallel }(t,\mathbf{\rho }_{0})=\int\limits_{0}^{t}dt^{\prime }\left\{
f_{z}(t^{\prime })+\pi a^{2}v\xi _{i_{0}}\int\limits_{S_{xy}}d\mathbf{\rho }%
\,^{\prime }U(\mathbf{\rho }\,^{\prime }-\mathbf{\rho }_{i_{0}})\delta (%
\mathbf{\rho }\,^{\prime }-\mathbf{\rho }\,[t^{\prime }])\right\} .
\label{4.12}
\end{equation}
Substituting (\ref{4.2}) into (\ref{4.6}) we may take into account the term $%
i_{0}$ only. Let us calculate the mathematical expectations of $l_{\parallel
}(t,\mathbf{\rho }_{0})$ and $l_{\parallel }^{2}(t,\mathbf{\rho }_{0})$
regarded as functionals of $\mathbf{r}\,[t]$. To do this we make use of the
following relations:

\begin{equation}
\left\langle \delta (\mathbf{\rho }-\mathbf{\rho }\,[t])\right\rangle _{w}=G(%
\mathbf{\rho },\mathbf{\rho }_{0},t),  \label{4.13}
\end{equation}
\begin{equation*}
\left\langle \delta (\mathbf{\rho }-\mathbf{\rho }\,[t])\delta (\mathbf{\rho 
}\,^{\prime }-\mathbf{\rho }\,[t^{\prime }])\right\rangle _{w}=
\end{equation*}
\begin{equation}
=\left\{ 
\begin{array}{ccc}
G(\mathbf{\rho },\mathbf{\rho }\,^{\prime },t-t^{\prime })G(\mathbf{\rho }%
\,^{\prime },\mathbf{\rho }_{0},t^{\prime }), & \;%
\text{if}\; & t^{\prime }<t, \\ 
G(\mathbf{\rho }\,^{\prime },\mathbf{\rho },t^{\prime }-t)G(\mathbf{\rho },%
\mathbf{\rho }_{0},t), & \;\text{if}\; & t^{\prime }>t,
\end{array}
\right.  \label{4.14}
\end{equation}
where the symbol ${\langle (\dots )\rangle }_{w}$ designates the value of $%
(\ldots )$ averaged over all the possible walker paths under consideration
and $G(\mathbf{\rho },\mathbf{\rho }_{0},t)$ is the probability for the
two-dimensional random walker starting at the point $\mathbf{\rho }_{0}$ to
reach the point $\mathbf{\rho }$ in the time $t$. We note that formula (\ref
{4.9}) directly results from the definition of the averaging procedure.
Indeed, for an arbitrary function $\mathcal{F}(\mathbf{\rho }\,)$

\begin{equation}
\left\langle \mathcal{F}(\mathbf{\rho }\,[t])\right\rangle
_{w}=\int\limits_{S_{xy}}d\mathbf{\rho }\,^{\prime }\mathcal{F}(\mathbf{\rho 
}\,^{\prime })G(\mathbf{\rho }\,^{\prime },\mathbf{\rho }_{0},t),
\end{equation}
it immediately follows expression (\ref{4.9}). Formula (\ref{4.10}) can be
proved in a similar way by representation of the probability for walkers
starting at the point $\mathbf{\rho }_{0}$ to reach the point $\mathbf{\rho }
$ in the time $t$ provided they have visited the point $\mathbf{\rho }%
\,^{\prime }$ at time $t^{\prime }$ in terms of

\begin{equation}
G(\mathbf{\rho },\mathbf{\rho }\,^{\prime },t-t^{\prime })G(\mathbf{\rho }%
\,^{\prime },\mathbf{\rho }_{0},t^{\prime }).
\end{equation}
The possibility of such a representation follows from the fact that the
given two - dimensional random walks are the Markov process \cite{32}. The
function $G(\mathbf{\rho },\mathbf{\rho }_{0},t)$ obeys the diffusion
equation \cite{22}

\begin{equation}
\frac{\partial G}{\partial t}=D\mathbf{\nabla }^{2}G  \label{4.15}
\end{equation}
and meets the initial condition 
\begin{equation}
\left. G\right| _{t=0}=\delta (\mathbf{\rho }-\mathbf{\rho }_{0}).
\label{4.16}
\end{equation}
For the two-dimensional random walks subject to no additional constraints
from (\ref{4.15}) and (\ref{4.16}) we find

\begin{equation}
G(\mathbf{\rho },\mathbf{\rho }_{0},t)=\frac{1}{4\pi Dt}\exp \left[ -\frac{(%
\mathbf{\rho }-\mathbf{\rho }_{0})^{2}}{4Dt}\right] .  \label{4.17}
\end{equation}

Now let us obtain the specific expressions for the desired quantities $%
\left\langle l_{\parallel }(t,\mathbf{\rho }_{0})\right\rangle _{w}$ and $%
\left\langle l_{\parallel }^{2}(t,\mathbf{\rho }_{0})\right\rangle _{w}$.
Averaging formula (\ref{4.12})

\begin{equation}
\left\langle l_{\parallel }(t,\mathbf{\rho }_{0})\right\rangle _{w}=\pi
a^{2}v\xi _{i_{0}}\int\limits_{0}^{t}dt^{\prime }\int\limits_{S_{xy}}d%
\mathbf{\rho }\,^{\prime }U(\mathbf{\rho }\,^{\prime }-\mathbf{\rho }%
_{i_{0}})<\delta (\mathbf{\rho }\,^{\prime }-\mathbf{\rho }\,[t^{\prime
}])>_{w}  \label{4.18}
\end{equation}
Substituting (\ref{4.13}) into (\ref{4.18}) we get

\begin{equation}
\left\langle l_{\parallel }(t,\mathbf{\rho }_{0})\right\rangle _{w}=\pi
a^{2}v\xi _{i_{0}}\int\limits_{0}^{t}dt^{\prime }\int\limits_{S_{xy}}d%
\mathbf{\rho }\,^{\prime }U(\mathbf{\rho }\,^{\prime }-\mathbf{\rho }%
_{i_{0}})G(\mathbf{\rho }\,^{\prime }-\mathbf{\rho }_{0},t^{\prime }),
\label{4.19}
\end{equation}

Squaring expression (\ref{4.12}) and taking into account that the random
force $f_{z}(t)$ and the random walkers in the $xy$ - plane are independent
of each other, thus, $\left\langle f_{z}(t^{\prime })\delta (\mathbf{\rho }%
\,^{\prime }-\mathbf{\rho }\,[t^{\prime \prime }])\right\rangle _{w}=0$ we
obtain

\begin{equation*}
\left\langle l_{\parallel }^{2}(t,\mathbf{\rho }_{0})\right\rangle
_{w}=\int\limits_{0}^{t}\int\limits_{0}^{t}dt^{\prime }dt^{\prime \prime
}\left\langle f_{z}(t^{\prime })f_{z}(t^{\prime \prime })\right\rangle _{w}+
\end{equation*}
\begin{equation*}
+(\pi a^{2}v)^{2}\xi _{i_{0}}^{2}\int\limits_{S_{xy}}\int\limits_{S_{xy}}d%
\mathbf{\rho }\,^{\prime }d\mathbf{\rho }\,^{\prime \prime }U(\mathbf{\rho }%
\,^{\prime }-\mathbf{\rho }_{i_{0}})U(\mathbf{\rho }\,^{\prime \prime }-%
\mathbf{\rho }_{i_{0}})\cdot
\end{equation*}
\begin{equation}
\cdot \left\langle \delta (\mathbf{\rho }\,^{\prime }-\mathbf{\rho }%
\,[t^{\prime }])\delta (\mathbf{\rho }\,^{\prime \prime }-\mathbf{\rho }%
\,[t^{\prime \prime }])\right\rangle _{w}.  \label{4.20}
\end{equation}
According to (\ref{4.6}) the first term on the right-hand side of expression
(\ref{4.20}) is equal to $2Dt$. In order to transform the second term we
make use of the identity

\begin{equation}
\int\limits_{0}^{t}\int\limits_{0}^{t}dt^{\prime }dt^{\prime \prime }(\ldots
)=\int\limits_{0}^{t}dt^{\prime }\int\limits_{0}^{t^{\prime }}dt^{\prime
\prime }(\dots )+\int\limits_{0}^{t}dt^{\prime \prime
}\int\limits_{0}^{t^{\prime \prime }}dt^{\prime }(\ldots ).  \label{4.21}
\end{equation}
Then the symmetry of the second term of expression (\ref{4.20}) with respect
to the replacement $\mathbf{\rho }\,^{\prime }\rightarrow \mathbf{\rho }%
\,^{\prime \prime }$ and $\mathbf{\rho }\,^{\prime \prime }\rightarrow 
\mathbf{\rho }\,^{\prime }$ enables us to represent it as

\begin{equation*}
2(\pi a^{2}v)^{2}\int\limits_{0}^{t}dt^{\prime }\int\limits_{0}^{t^{\prime
}}dt^{\prime \prime }\int\limits_{S_{xy}}\int\limits_{S_{xy}}d\mathbf{\rho }%
\,^{\prime }d\mathbf{\rho }\,^{\prime \prime }U(\mathbf{\rho }\,^{\prime }-%
\mathbf{\rho }_{i_{0}})U(\mathbf{\rho }\,^{\prime \prime }-\mathbf{\rho }%
_{i_{0}})\cdot
\end{equation*}
\begin{equation*}
\cdot \left\langle \delta (\mathbf{\rho }\,^{\prime }-\mathbf{\rho }%
\,[t^{\prime }])\delta (\mathbf{\rho }\,^{\prime \prime }-\mathbf{\rho }%
\,[t^{\prime \prime }])\right\rangle _{w}.
\end{equation*}
where we have taken into account that $\xi _{i_{0}}^{2}\equiv 1$.
Substituting (\ref{4.14}) into the last term we obtain the following
expression for quantity $\left\langle l_{\parallel }^{2}(t,\mathbf{\rho }%
_{0})\right\rangle _{w}$:

\begin{equation*}
\left\langle (l_{\parallel }(t,\mathbf{\rho }_{0}))^{2}\right\rangle
_{w}=2Dt+(\pi a^{2}v)^{2}2\int\limits_{0}^{t}dt^{\prime
}\int\limits_{0}^{t^{\prime }}dt^{\prime \prime }\int\limits_{S_{xy}}d%
\mathbf{\rho }\,^{\prime }\int\limits_{S_{xy}}d\mathbf{\rho }\,^{\prime
\prime }\cdot
\end{equation*}
\begin{equation}
\cdot U(\mathbf{\rho }\,^{\prime }-\mathbf{\rho }_{i_{0}})U(\mathbf{\rho }%
\,^{\prime \prime }-\mathbf{\rho }_{i_{0}})G(\mathbf{\rho }\,^{\prime },%
\mathbf{\rho }\,^{\prime \prime },t^{\prime }-t^{\prime \prime })G(\mathbf{%
\rho }\,^{\prime \prime },\mathbf{\rho }_{0},t^{\prime \prime }).
\label{4.22}
\end{equation}

The function $U(\mathbf{\rho }\,)$ differs from zero in the domain $\mid 
\mathbf{\rho }\mid <a$ only, whereas, for $t^{\prime }\gg \tau
_{a}=a^{2}/(2D)$ on scales of order $a$ variations of the function $G(%
\mathbf{\rho }\,^{\prime },\mathbf{\rho }_{0},t^{\prime })$ are ignorable.
The latter, together with (\ref{4.8}), (\ref{4.17}) and the condition $\mid 
\mathbf{\rho }_{0}-\mathbf{\rho }_{i_{0}}\mid \sim a$ allow us to set

\begin{equation}
\int\limits_{S_{xy}}d\mathbf{\rho }\,^{\prime }U(\mathbf{\rho }\,^{\prime }-%
\mathbf{\rho }_{i_{0}})G(\mathbf{\rho }\,^{\prime },\mathbf{\rho }%
_{0},t^{\prime })\simeq \frac{1}{4\pi Dt^{\prime }}.  \label{4.23}
\end{equation}
If we formally substitute (\ref{4.23}) into (\ref{4.19}) then the integral
over $t^{\prime }$ becomes divergent one with a logarithmic singularity%
\index{logarithmic singularity} at $t^{\prime }=+0$. We meet a similar
situation when dealing with expression (\ref{4.22}). Thus, for $t\gg \tau
_{a}$ particular details of the function $U(\mathbf{\rho }\,)$ are of little
consequence and in this case we may choose, for example, the following form
of the given function

\begin{equation}
U(\mathbf{\rho }\,)=%
\frac{1}{\pi a^{2}}\exp \left[ -\frac{(\mathbf{\rho }\,)^{2}}{a^{2}}\right] .
\label{4.24}
\end{equation}
The choice of function (\ref{4.24}) is connected with the useful identity
that will be widely used in the following calculations:

\begin{equation}
\int\limits_{S_{xy}}d\mathbf{\rho }\,^{\prime \prime }\mathcal{G}(\mathbf{%
\rho }-\mathbf{\rho }\,^{\prime \prime };A_{1})\mathcal{G}(\mathbf{\rho }%
\,^{\prime \prime }-\mathbf{\rho }\,^{\prime };A_{2})=\mathcal{G}(\mathbf{%
\rho }-\mathbf{\rho }\,^{\prime };A_{1}+A_{2}),  \label{4.25}
\end{equation}
where $A_{1},A_{2}>0$ and

\begin{equation}
\mathcal{G}(\mathbf{\rho },A)=\frac{1}{\pi A}\exp \left\{ -\frac{\rho ^{2}}{A%
}\right\} .  \label{4.25a}
\end{equation}

Then, substituting (\ref{4.17}) and (\ref{4.24}) into (\ref{4.19}) and (\ref
{4.22}) using (\ref{4.25}), and directly integrating with respect to $%
\mathbf{\rho }\,^{\prime },\mathbf{\rho }\,^{\prime \prime }$ and $t^{\prime
},t^{\prime \prime }$ for $t\gg \tau _{a}$ and $\mid \mathbf{\rho }_{i_{0}}-%
\mathbf{\rho }_{0}\mid \sim a$ we obtain

\begin{equation}
\left\langle l_{\parallel }(t,\mathbf{\rho }_{0})\right\rangle _{w}\approx 
\frac{a^{2}v}{2D}\xi _{i_{0}}\ln \left( \frac{(2Dt)^{1/2}}{a}\right) ,
\label{4.26}
\end{equation}
\begin{equation}
\left\langle (l_{\parallel }(t,\mathbf{\rho }_{0}))^{2}\right\rangle
_{w}\approx 2Dt+2\left\langle l_{\parallel }(t,\mathbf{\rho }%
_{0})\right\rangle ^{2}.  \label{4.27}
\end{equation}
In particular, comparing (\ref{4.26}), (\ref{4.27}) with one another we find
that the mean distance $\left\langle l_{\parallel }(t)\right\rangle $
travelled by a walker along a pipe with blood flow in it for the time $t$
such that $\tau _{a}\ll t\ll d_{p}^{2}/(2D)$ can be estimated as

\begin{equation}
\left\langle l_{\parallel }(t)\right\rangle \approx \frac{a^{2}v}{2D}\ln
\left( \frac{(2Dt)^{1/2}}{a}\right) .  \label{4.28}
\end{equation}
It should be pointed out that expression (\ref{4.28}) can be also obtained
in the following way. Clearly, $\left\langle l_{\parallel }(t)\right\rangle
\approx v\left\langle t_{p}\right\rangle $, where $\left\langle
t_{p}\right\rangle $is the mean total time during which a walker being
initially near a certain pipe, for example, the pipe $i_{0}$, is inside this
pipe. For a given realization $\mathbf{\rho }\,[t]$ of the two-dimensional
random walks the total residence time%
\index{residence time} of walkers inside the pipe $i_{0}$ is

\begin{equation}
t_{p}=\int\limits_{0}^{t}dt^{\prime }\Theta _{i_{0}}(\mathbf{\rho }%
\,[t^{\prime }]),  \label{4.29}
\end{equation}
where $\Theta _{i_{0}}(\mathbf{\rho }\,)$ is the characteristic function of
the pipe $i_{0}$, i.e.

\begin{equation}
\Theta _{i_{0}}(\mathbf{\rho }\,)=\left\{
\begin{array}{cccc}
1 & , & \;%
\text{if}\; & \mid \mathbf{\rho }-\mathbf{\rho }_{i_{0}}\mid <a \\ 
0 & , & \;\text{if}\; & \mid \mathbf{\rho }-\mathbf{\rho }_{i_{0}}\mid >a
\end{array}
\right. .  \label{4.30}
\end{equation}
The identity

\begin{equation}
\Theta _{i_{0}}(\mathbf{\rho }\,)=\int\limits_{\left| \mathbf{\rho }%
\,^{\prime }-\mathbf{\rho }_{i_{0}}\right| <a}d\mathbf{\rho }\,^{\prime
}\delta (\mathbf{\rho }\,^{\prime }-\mathbf{\rho }\,)  \label{4.31}
\end{equation}
enables us to rewrite (\ref{4.29}) as

\begin{equation}
t_{p}=\int\limits_{0}^{t}dt^{\prime }\int\limits_{\left| \mathbf{\rho }%
\,^{\prime }-\mathbf{\rho }_{i_{0}}\right| <a}d\mathbf{\rho }\delta (\mathbf{%
\rho }-\mathbf{\rho }\,[t^{\prime }]).  \label{4.32}
\end{equation}
By virtue of (\ref{4.13}) and (\ref{4.17}) from (\ref{4.32}) we get

\begin{equation}
\left\langle t_{p}\right\rangle \approx \frac{a^{2}}{2D}\ln \left( \frac{%
(2Dt)^{1/2}}{a}\right)  \label{4.33}
\end{equation}
and the estimate $\left\langle l_{\parallel }(t)\right\rangle \approx
v\left\langle t_{p}\right\rangle $ leads to expression (\ref{4.28}).

\subsection{The cooperative effect of the pipe system on wa\-l\-ker motion}

\label{s4.2.2}

When $t\gg d_{p}^{2}/(2D)$ and, thus, practically each walker in its motion
visits a large number of pipes we determine the mathematical expectations of 
$l_{\parallel }(t,\mathbf{\rho }_{0})$ and $(l_{\parallel }(t,\mathbf{\rho }%
_{0}))^{2}$ regarded as functions of the independent random variables $%
\mathbf{r}\,[t],\{\xi _{i}\}$ and $\mathbf{\rho }_{0}$, with the latter
being characterized by uniform distribution over the $xy$-plane.

Substituting (\ref{4.7}) into (\ref{4.11}) and averaging the latter over
these random variables we get

\begin{equation*}
\left\langle l_{\parallel }(t,\mathbf{\rho }_{0})\right\rangle _{w,\rho
_{0},\xi }=\pi a^{2}v<\xi _{i}>\sum_{i}\int\limits_{0}^{t}dt^{\prime
}\int\limits_{S_{xy}}d\mathbf{\rho }\,^{\prime }\cdot
\end{equation*}
\begin{equation}
\cdot U(\mathbf{\rho }\,^{\prime }-\mathbf{\rho }_{i})\left\langle
\left\langle \delta (\mathbf{\rho }\,^{\prime }-\mathbf{\rho }\,[t^{\prime
}])\right\rangle _{w}\right\rangle _{\rho _{0}}.  \label{4.34}
\end{equation}
Here the sum runs over all the pipes and $\left\langle (\ldots
)\right\rangle _{\rho _{0}}$ stands for the value of $(\ldots )$ averaged
over the random variable $\mathbf{\rho }_{0}$ and determined by the formal
expression

\begin{equation}
\left\langle (\ldots )\right\rangle _{\rho _{0}}=\frac{1}{S_{xy}}%
\int\limits_{S_{xy}}d\mathbf{\rho }_{0}(\ldots )  \label{4.35}
\end{equation}
where $S_{xy}$ denotes the area of the $xy$-plane treated as some large
value and $1/S_{xy}$ is the probability density of the variable $\rho _{0}$
distribution. Formulas (\ref{4.9}), (\ref{4.13}) and (\ref{4.35}) allow us
to represent expression (4.34) in the form

\begin{equation*}
\left\langle l_{\parallel }(t,\mathbf{\rho }_{0})\right\rangle _{w,\rho
_{0},\xi }=
\end{equation*}
\begin{equation}
\pi a^{2}v\xi \frac{1}{S_{xy}}\sum_{i}\int\limits_{0}^{t}dt^{\prime
}\int\limits_{S_{xy}}d\mathbf{\rho }_{0}\int\limits_{S_{xy}}d\mathbf{\rho }%
\,^{\prime }U(\mathbf{\rho }\,^{\prime }-\mathbf{\rho }_{i})G(\mathbf{\rho }%
\,^{\prime },\mathbf{\rho }_{0},t^{\prime }).  \label{4.36}
\end{equation}

By virtue of (\ref{4.17})

\begin{equation}
\int\limits_{S_{xy}}d\mathbf{\rho }_{0}G(\mathbf{\rho }\,^{\prime \prime },%
\mathbf{\rho }_{0},t^{\prime \prime })=1.  \label{4.37}
\end{equation}
The latter identity and the chosen form (\ref{4.24}) of the function $U(%
\mathbf{\rho }\,)$ enable us to integrate over $\mathbf{\rho }\,^{\prime }$
and in expressions (\ref{4.36})

\begin{equation}
\left\langle (l_{\parallel }(t,\mathbf{\rho }_{0}))^{2}\right\rangle
_{w,\rho _{0},\xi }=\pi a^{2}v\xi +\frac{1}{S_{xy}}\sum_{i}l_{i},
\label{4.38}
\end{equation}
where $l_{i}\equiv 1$ for each site of the lattice $\{\mathbf{\rho }_{2}\}$.
Following the definition of mean distance between vessels (see Section~\ref
{s3.2}) we may set

\begin{equation}
\frac{1}{S_{xy}}\sum_{i}l_{i}=\frac{1}{d^{2}},  \label{4.39}
\end{equation}
where $d$ is the mean distance%
\index{mean distance} between the pipes or, what is the same, between the
discs of the lattice $\{\mathbf{\rho }_{i}\}$ in the $xy$ - plane. Thus,
from (\ref{4.38}) we find

\begin{equation}
\left\langle l_{\parallel }(t,\mathbf{\rho }_{0})\right\rangle _{w,\rho
_{0},\xi }=%
\frac{\pi a^{2}v}{d^{2}}\xi t.  \label{4.40}
\end{equation}
In order to find the relationship between $d$ and $d_{p}$ we note that, by
definition, $d^{2}$ is the mean area of the $xy$-plane per each disk and for
the given disk lattice the value of $d^{2}$ coincides with the area of the
Wigner - Seitz cell%
\index{Wigner - Seitz cell} \cite{29,33,49,64} shown in Fig.~\ref{Fig8}.
Thus,

\begin{equation}
d^2 =
\frac{\sqrt{3}}{2} d^2_p.  \label{4.41}
\end{equation}

Now let us calculate the averaged value of $l_{\parallel }^{2}(t,\mathbf{%
\rho }_{0})$. Substituting (\ref{4.7}) into (\ref{4.12}), squaring the
obtained result we find

\begin{equation*}
\left\langle l_{\parallel }^{2}(t,\mathbf{\rho }_{0})\right\rangle _{w,\rho
_{0},\xi }=\int\limits_{0}^{t}\int\limits_{0}^{t}dt^{\prime }dt^{\prime
\prime }\left\{ \left\langle f_{z}(t^{\prime })f_{z}(t^{\prime \prime
})\right\rangle +\right.
\end{equation*}
\begin{equation*}
+(\pi a^{2}v)^{2}\sum_{i^{\prime },i^{\prime \prime }}\left\langle \xi
_{i^{\prime }}\xi _{i^{\prime \prime }}\right\rangle _{\xi
}\int\limits_{S_{xy}}\int\limits_{S_{xy}}d\mathbf{\rho }\,^{\prime }d\mathbf{%
\rho }\,^{\prime \prime }U(\mathbf{\rho }\,^{\prime }-\mathbf{\rho }%
_{i^{\prime }})U(\mathbf{\rho }\,^{\prime \prime }-\mathbf{\rho }%
\,_{i^{\prime \prime }}^{\prime })\cdot
\end{equation*}
\begin{equation}
\left. \cdot \left\langle \left\langle \delta (\mathbf{\rho }\,^{\prime }-%
\mathbf{\rho }\,[t^{\prime }])\delta (\mathbf{\rho }\,^{\prime \prime }-%
\mathbf{\rho }\,[t^{\prime \prime }])\right\rangle _{w}\right\rangle _{\rho
_{0}}\right\} .  \label{4.42}
\end{equation}
By virtue of (\ref{4.6}), (\ref{4.10}), (\ref{4.14}), (\ref{4.35}) and
identity (\ref{4.21}) formula (\ref{4.42}) can be rewritten in the form

\begin{equation*}
\left\langle (l_{\parallel }(t,\mathbf{\rho }_{0}))^{2}\right\rangle
_{w,\rho _{0},\xi }=2Dt+2(\pi a^{2}v)^{2}\frac{1}{S_{xy}}\sum_{i^{\prime
},i^{\prime \prime }}[(1-\xi ^{2})\delta _{i^{\prime },i^{\prime \prime
}}+\xi ^{2}]\cdot
\end{equation*}
\begin{equation*}
\cdot \int\limits_{0}^{t}dt^{\prime }\int\limits_{0}^{t^{\prime }}dt^{\prime
\prime }\int\limits_{S_{xy}}d\mathbf{\rho }\,^{\prime }\int\limits_{S_{xy}}d%
\mathbf{\rho }\,^{\prime \prime }\int\limits_{S_{xy}}d\mathbf{\rho }_{0}U(%
\mathbf{\rho }\,^{\prime }-\mathbf{\rho }_{i}^{\prime })U(\mathbf{\rho }%
\,^{\prime \prime }-\mathbf{\rho }\,_{i^{\prime \prime }}^{\prime })\cdot
\end{equation*}
\begin{equation}
\cdot G(\mathbf{\rho }\,^{\prime },\mathbf{\rho }\,^{\prime \prime
},t^{\prime }-t^{\prime \prime })G(\mathbf{\rho }\,^{\prime \prime },\mathbf{%
\rho }_{0},t^{\prime \prime }).  \label{4.43}
\end{equation}

Taking into account identity (\ref{4.25}) and (\ref{4.37}) from (\ref{4.43})
we find

\begin{equation*}
\left\langle (l_{\parallel }(t,\mathbf{\rho }_{0}))^{2}\right\rangle
_{w,\rho _{0},\xi }=2Dt+2(\pi a^{2}v)^{2}\frac{1}{S_{xy}}\sum_{i^{\prime
},i^{\prime \prime }}[(1-\xi ^{2})\delta _{i^{\prime },i^{\prime \prime
}}+\xi ^{2}]\cdot
\end{equation*}
\begin{equation}
\cdot \int\limits_{0}^{t}dt^{\prime }\int\limits_{0}^{t^{\prime }}dt^{\prime
\prime }\frac{1}{4\pi D(t^{\prime }-t^{\prime \prime }+\tau _{a})}\exp
\left\{ -\frac{(\mathbf{\rho }_{i^{\prime }}-\mathbf{\rho }_{i^{\prime
\prime }})^{2}}{4D(t^{\prime }-t^{\prime \prime }+\tau _{a})}\right\} ,
\label{4.44}
\end{equation}
where as before

\begin{equation}
\tau _a = \frac{a^2}{2D}.  \label{4.45}
\end{equation}

Replacing the variable $t^{\prime \prime }$ by the new variable $t^{\prime
}-t^{\prime \prime }$ under integral (\ref{4.44}), introducing the new
vector $\mathbf{\rho }_{i}=\mathbf{\rho }_{i^{\prime }}-\mathbf{\rho }%
_{i^{\prime \prime }}$ of the lattice $\{\rho _{i}\}$ and using (\ref{4.39})
from (\ref{4.44}) we have

\begin{equation*}
\left\langle (l_{\parallel }(t,\mathbf{\rho }_{0}))^{2}\right\rangle =2Dt+2%
\frac{(\pi a^{2}v)^{2}}{d^{2}}\int\limits_{0}^{t}dt^{\prime
}\int\limits_{0}^{t^{\prime }}dt^{\prime \prime }\frac{1}{4\pi D(t^{\prime
\prime }+\tau _{a})}\cdot
\end{equation*}
\begin{equation}
\cdot \left\{ (1-\xi ^{2})+\xi ^{2}\sum_{i}\exp \left[ -\frac{(\mathbf{\rho }%
_{i})^{2}}{4D(t^{\prime \prime }+\tau _{a})}\right] \right\} .  \label{4.46}
\end{equation}

In order to calculate the sum in formula (\ref{4.46}) let us consider the
lattice $\{\mathbf{g}_{i}\}$ reciprocal to the lattice $\{\mathbf{\rho }%
_{i}\}$. Drawing on the definition of reciprocal lattice\cite{29,33,64}we
can show that the lattice $\{\mathbf{g}_{i}\}$ is also hexagonal one, its
spacing

\begin{equation}
d_{p}^{\ast }=\frac{4\pi }{\sqrt{3}}\frac{1}{d_{p}},  \label{4.47}
\end{equation}
and all the vectors $\{\mathbf{g}_{i}\}$ can be represented in terms of

\begin{equation}
\mathbf{g}_{i}=n_{i}d_{p}^{\ast }\left\{ \frac{\sqrt{3}}{2};\,-\frac{1}{2}%
\right\} +m_{i}d_{p}^{\ast }\left\{ 0,1\right\} ,  \label{4.48}
\end{equation}
where $n_{i},m_{i}$ are integers (Fig.~\ref{Fig9}). Expanding the delta
function $\delta (\mathbf{\rho }\,)$ into the Fourier series%
\index{Fourier series} in the Wigner-Seitz cell%
\index{Wigner - Seitz cell} of the lattice $\{\mathbf{\rho }_{i}\}$ we get
the well known formula \cite{33,64} to be used in the following analysis

\FRAME{ftbpFU}{3.1393in}{1.4684in}{0pt}{\Qcb{The lattice $\protect\rho (a)$
and its reciprocal lattice $g_{i}\;$($Q_{w},Q_{w}^{\ast }$ are the
Wig\-ner-Seitz cells).}}{\Qlb{Fig9}}{Fig9}{\special{language "Scientific
Word";type "GRAPHIC";maintain-aspect-ratio TRUE;display "USEDEF";valid_file
"F";width 3.1393in;height 1.4684in;depth 0pt;original-width
16.2775in;original-height 7.574in;cropleft "0";croptop "1";cropright
"1";cropbottom "0";filename 'FIG9.GIF';file-properties "XNPEU";}}

\begin{equation}
\sum_{i}\delta (\mathbf{\rho }-\mathbf{\rho }_{i})=\frac{1}{d^{2}}%
\sum_{i}\exp [i\mathbf{\rho }\,\mathbf{g}_{i}].  \label{4.49}
\end{equation}
because $d^{2}$ coincides with the area of the Wigner - Seitz cell%
\index{Wigner - Seitz cell}. Identity (\ref{4.49}) allows us to rewrite the
last term on the ring-hand side of (\ref{4.46}) in the form

\begin{equation*}
\sum_{i}%
\frac{1}{4\pi D(t^{\prime \prime }+\tau _{a})}\exp \left[ -\frac{(\mathbf{%
\rho }_{i})^{2}}{4\pi D(t^{\prime \prime }+\tau _{a})}\right]
=\int\limits_{S_{xy}}d\mathbf{\rho }\frac{1}{4\pi D(t^{\prime \prime }+\tau
_{a})}\cdot
\end{equation*}
\begin{equation*}
\cdot \exp \left[ -\frac{(\mathbf{\rho }\,)^{2}}{4\pi D(t^{\prime \prime
}+\tau _{a})}\right] \sum_{i}\delta (\mathbf{\rho }-\mathbf{\rho }_{i})=%
\frac{1}{d^{2}}\sum_{i}\int\limits_{S_{xy}}d\mathbf{\rho }\frac{1}{4\pi
D(t^{\prime \prime }+\tau _{a})}\cdot
\end{equation*}
\begin{equation*}
\cdot \exp \left[ i\mathbf{\rho g}_{i}-\frac{(\mathbf{\rho }\,)^{2}}{4\pi
D(t^{\prime \prime }+\tau _{a})}\right] =\frac{1}{d^{2}}\sum_{i}\exp \left[
-(\mathbf{g}_{i})^{2}D(t^{\prime \prime }+\tau _{a})\right] .
\end{equation*}
Substituting (\ref{4.49}) into (\ref{4.46}) and integrating with respect to $%
t^{\prime }$ and $t^{\prime \prime }$ for $t\gg d_{p}^{2}/(2D)$ we find

\begin{equation*}
\left\langle (l_{\parallel }(t,\mathbf{\rho }_{0}))^{2}\right\rangle \approx
\left\langle l_{\parallel }(t,\mathbf{\rho }_{0})\right\rangle
^{2}+2Dt\left\{ 1+\frac{\pi }{4}\left( \frac{a^{2}v}{Dd}\right) ^{2}\left(
(1-\right. \right.
\end{equation*}
\begin{equation}
\left. \left. -\xi ^{2})\ln \left( \frac{t}{\tau _{a}e}\right) +\xi ^{2}%
\frac{4\pi }{d^{2}}\sum^{\prime }\nolimits_{i}\frac{1}{(\mathbf{g}_{i})^{2}}%
\exp [-(\mathbf{g}_{i})^{2}D\tau _{a}]\right) \right\} ,  \label{4.50}
\end{equation}
where $e$ is the base of the natural logarithm and the prime on the sum over 
$i$ indicates that the term corresponding to $\mathbf{g}_{i}=0$ is omitted.
Due to $a\ll d_{p}$ the main contribution to the sum in (\ref{4.50}) is
determined by a large number of sites of lattice $\{\mathbf{g}_{i}\}$.
Therefore, the value of this sum can be found from the following approximate
equality

\begin{equation}
\sum^{\prime }\nolimits_{i}\frac{1}{(\mathbf{g}_{i})^{2}}\exp [-(\mathbf{g}%
_{i})^{2}D\tau _{a}]\approx \frac{2}{\sqrt{3}(d_{p}^{\ast })^{2}}2\pi
\int\limits_{d_{p}^{\ast }}^{\infty }dg\frac{1}{g}\exp [-g^{2}D\tau _{a}],
\label{4.51}
\end{equation}
where $(\sqrt{3}/2)(d_{p}^{\ast })^{2}$ is the area of the Wigner-Seitz cell%
\index{Wigner - Seitz cell} of the reciprocal lattice $\{\mathbf{g}_{i}\}$
and the cut-off parameter%
\index{cut-off parameter} $d_{p}^{\ast }$ for the logarithmic divergence%
\index{logarithmic divergence} at small values of $g$ arises from the
absence of $g_{i}^{2}$ less than $(d_{p}^{\ast })^{2}$ in (\ref{4.51}).
Expressions (\ref{4.41}), (\ref{4.44}) and (\ref{4.51}) enable us to
represent (\ref{4.50}) in the form 
\begin{equation*}
\left\langle (l_{\parallel }(t,\mathbf{\rho }_{0}))^{2}\right\rangle \approx
\left\langle l_{\parallel }(t,\mathbf{\rho }_{0})\right\rangle ^{2}+2Dt%
\Biggl \{1+
\end{equation*}

\begin{equation}
+%
\frac{\pi }{4}\left( \frac{a^{2}v}{Dd}\right) ^{2}\left[ (1-\xi ^{2})\ln
\left( \frac{t}{\tau _{a}e}\right) +\xi ^{2}\ln \left( \frac{\sqrt{3}d^{2}}{%
8\gamma \pi ^{2}D\tau _{a}}\right) \right] \Biggr \},  \label{4.52}
\end{equation}
where $\gamma $ is the 
\index{Euler constant}Euler constant. Expressions (\ref{4.40}) and (\ref
{4.52}) are the desired results of the given Section.

It should be pointed out that in the case under consideration the walker
random motion along the pipes does not obey the regular diffusion law
because, as it follows from (\ref{4.52}), for $\mid \xi \mid <1$, the
dispersion of the random variable $l_{\parallel }(t,\mathbf{\rho }_{0})$
depends on the time $t$ as $t\ln (t/\tau _{a})$ rather than $t$. Thus, from
the standpoint of random walks the tissue phantom containing straight
parallel pipes cannot be described in terms of a medium with an effective
diffusion coefficient 
\index{effective diffusion coefficient}. However, in the general case, i.e.
for the tissue that contains vessels variously oriented in space, this
conclusion does not hold true. This question is the subject of the next
Subsection.

\subsection{The effective medium description of the tissue phantom with
pipes differently oriented in space}

\label{s4.2.3}

In real living tissues vessels are variously oriented in space. In the
present Section we analyse characteristic properties of random walks in a
tissue phantom%
\index{tissue phantom} similar to one considered before, which, however
contains, in addition, pipes parallel to the $xy$ - plane. In particular, we
shall show that the influence of pipes parallel to the $xy$ - plane allows
one to introduce the effective diffusion coefficient%
\index{effective diffusion coefficient|textbf} and estimate its value, based
on the results obtained in the previous Subsection.

The tissue phantom is assumed to contain the pipe system specified in
Subsection~\ref{s4.2.1} and pipes oriented randomly in the $xy$ - plane,
with the mean distance%
\index{mean distance} between these pipes being equal to $d$.

In the previous Subsections we represented three - dimensional random walks%
\index{three - dimensional random walks} $\mathbf{r}\,[t]=(\rho \lbrack
t],z[t])$ as two - dimensional random walks $\mathbf{\rho }\,[t]$ and random
walks $z[t]$ along the $z$ - axis. In the tissue phantom containing solely
pipes parallel to the $z$ - axis the two - dimensional random walks are
independent of the one - dimensional ones. When the tissue phantom contains
pipes variously oriented in space random walks in the $xy$ - plane depend on
the walker motion along the $z$ - axis because during motion along the $z$ -
axis the walker can meet a pipe parallel to the $xy$ - plane, which gives
rise to its fast transport along the $xy$ - plane with blood in these pipes.

Since, the characteristics of walker motion in the $xy$ - plane determine
properties of their motion along the $z$ - axis walker motion in the tissue
phantom with undirected pipes and pipes variously oriented in space can
differ in properties.

Therefore, keeping in mind walker motion%
\index{walker motion} in the $z$-direction, first\ we, analyse the effect of
blood flow on the two - dimensional random walks in the $xy$-plane. When the
time $t$ of the walker motion satisfies the condition $t\ll d^{2}/(2D)$ we
may ignore the influence of blood flow in vessels on the two-dimensional
random walks. This is justified because walkers at initial time near a
vessel parallel to $z$ - axis, practically cannot reach other vessels during
this time. When $t\gg d^{2}/(2D)$ actually each walker in its motion visits
a large number of vessels including the vessels parallel to the $xy$-plane.
In this case we can, at least roughly, describe the walker motion along the $%
xy$-plane in terms of two - dimensional random walks in a medium with an
effective diffusion coefficient $D_{\mathrm{eff}}$. We note that the effect
of blood flow on the walker motion should be taken into account only when $%
D_{\mathrm{eff}}\gg D$. Therefore, for the sake of simplicity, the latter
inequality is assumed to be true beforehand. Within the framework of the
adopted assumptions expression (\ref{4.43}) holds true provided the function 
$G(\mathbf{\rho },\mathbf{\rho }\,^{\prime },t)$ is specified by formula (%
\ref{4.17}) for $t\ll d^{2}/(2D)$ only, whereas for $t\gg d^{2}/(2D)$ it is
given by the formula

\begin{equation}
G(\mathbf{\rho },\mathbf{\rho }\,^{\prime },t)\approx 
\frac{1}{4\pi D_{\mathrm{eff}}t}\exp \left[ -\frac{(\mathbf{\rho }-\mathbf{%
\rho }\,^{\prime })^{2}}{4D_{\mathrm{eff}}t}\right] .  \label{4.53}
\end{equation}
According to (\ref{4.43}) and (\ref{4.46}) the time dependence $t\ln (t/\tau
_{a})$ practically arises from the term

\begin{equation}
\int\limits_{0}^{t}dt^{\prime }\int\limits_{\tau _{a}}^{t^{\prime
}}dt^{\prime \prime }G(0,0,t^{\prime }-t^{\prime \prime }).  \label{4.54}
\end{equation}
Here the function $G(0,0,t^{\prime }-t^{\prime \prime })$ determines the
probability for the two-dimen\-sio\-nal random walks originating at the
point $\mathbf{\rho }=0$ at time $t^{\prime \prime }$ to return to the
initial point at time $t^{\prime }$. As it follows from (\ref{4.17}) and (%
\ref{4.53}) for $t\gg \tau _{a}=t^{2}/(2D)$ term (\ref{4.54}) is
approximately equal to

\begin{equation}
\frac{1}{4\pi }\left[ \frac{1}{D}\ln \frac{\tau _{d}}{\tau _{a}}+\frac{1}{D_{%
\mathrm{eff}}}\ln \frac{t}{\tau _{d}}\right] .
\end{equation}
When the value of the time $t$ is not extremely large, i.e.

\begin{equation}
\frac{d^{2}}{2D}\ll t\ll \frac{d^{2}}{2D}\left( \frac{d^{2}}{a^{2}}\right) ^{%
\frac{D_{\mathrm{eff}}}{D}}  \label{4.55}
\end{equation}
this term can be estimated as

\begin{equation}
\frac{t}{4\pi D}\ln \left( \frac{d^{2}}{2D\tau _{a}}\right) .  \label{4.56}
\end{equation}
Expression (\ref{4.56}) shows that on temporal scales (\ref{4.55}) the
effect which blood flow in vessels parallel to the $xy$-plane has on the
two-dimensional random walks is actually reduced to inhibiting these random
walks from returning to their initial point on temporal scales $t\gg
d^{2}/(2D)$. So, in the general case the time dependence of$\;\left\langle
(l_{\parallel }(t,\mathbf{\rho }_{0}))^{2}\right\rangle $ obeys the regular
diffusion law, viz., according to (\ref{4.40}), (\ref{4.50}), (\ref{4.52})
and (\ref{4.56})

\begin{equation}
\left\langle (l_{\parallel }(t,\mathbf{\rho }_{0}))^{2}\right\rangle \approx
\left\langle l_{\parallel }(t,\mathbf{\rho }_{0})\right\rangle
^{2}+2Dt\left\{ 1+\frac{\pi }{4}\left( \frac{a^{2}v}{Dd}\right) ^{2}\ln
\left( \frac{d^{2}}{2D\tau _{a}}\right) \right\} .  \label{4.57}
\end{equation}

Therefore, by virtue of (\ref{4.40}) and (\ref{4.57}), in the tissue phantom
containing straight parallel pipes where, however, pipes of other directions
are\ also present walker during a time $t\gg d^{2}/(2D)$ travels a distance $%
l_{\parallel }(t)$ along the pipes which is characterized by the mean values

\begin{equation}
\left\langle l_{\parallel }(t)\right\rangle =v_{\mathrm{eff}%
}t,\;\;\;\left\langle (l_{\parallel }(t))^{2}\right\rangle =\left\langle
l_{\parallel }(t)\right\rangle ^{2}+D_{\mathrm{eff}}t,  \label{4.58}
\end{equation}
where the coefficients

\begin{equation}
v_{\mathrm{eff}}=\frac{\pi a^{2}v}{d^{2}}\xi  \label{4.59}
\end{equation}
and 
\begin{equation}
D_{\mathrm{eff}}\simeq D\left[ 1+\frac{\pi }{2}\left( \frac{a^{2}v}{Dd}%
\right) ^{2}\ln \left( \frac{d}{(2D\tau _{a})^{1/2}}\right) \right] .
\label{4.60}
\end{equation}

In considering walker motion in the $z$-direction on a time scale $t\gg
d^{2}/(2D)$ we can choose such a time scale $\tau $ that $d/(2D)\ll \tau \ll
t$ and represent the probability%
\index{probability} $\mathcal{P}(z,z^{\prime }\mid t,t^{\prime })$ for the
walker to reach the point $z$ at time $t$ provided it has been at the point $%
z^{\prime }$ at time $t^{\prime }$ in the form

\begin{equation*}
\mathcal{P}(z,z^{\prime }\left| t,t^{\prime })=\right. \int\limits_{-\infty
}^{+\infty }\dots \int\limits_{-\infty }^{+\infty }dz_{1}\ldots dz_{N}%
\mathcal{P}(z,z_{N}\mid t,t-\tau )...
\end{equation*}

\begin{eqnarray}
\mathcal{P}(z_{N},z_{N}-1 &\mid &t-\tau ,t-2\tau )  \label{4.61} \\
&&\ldots \mathcal{P}(z_{2},z_{1}\left| t^{\prime }+2\tau ,t^{\prime }+\tau )%
\mathcal{P}(z_{1},z^{\prime }\right| t^{\prime }+\tau ,t^{\prime }),  \notag
\end{eqnarray}
where $t=t^{\prime }+(N+1)\tau $. For $N\gg 1$ the specific form of the
function $\mathcal{P}(z,z^{\prime }\mid t+\tau ,t)$ is not a factor and only
must lead to relations (\ref{4.58}) (see, e.g., \cite{26}). The latter
allows us to write

\begin{equation}
\mathcal{P}(z_{1},z^{\prime }\mid t+\tau ,t)=%
\frac{1}{4(\pi D_{\mathrm{eff}}\tau )^{1/2}}\exp \left\{ -\frac{(z-z^{\prime
}-v_{\mathrm{eff}}\tau )^{2}}{4D_{\mathrm{eff}}\tau }\right\} .  \label{4.62}
\end{equation}
Substituting (\ref{4.62}) into (\ref{4.61}) and going to continuum
description we get

\begin{equation}
\mathcal{P}(z,z^{\prime }\mid t,t^{\prime })=\int \mathcal{D}{z[t^{\prime
\prime }]}\exp \left\{ -\frac{1}{4}\int\limits_{t^{\prime }}^{t}dt^{\prime
\prime }\frac{(\dot{z}-v_{\mathrm{eff}})^{2}}{D_{\mathrm{eff}}}\right\} .
\label{4.63}
\end{equation}
Therefore, distribution of walkers in such a tissue phantom evolves
according to the equation

\begin{equation}
\frac{\partial C}{\partial t}=\mathbf{\nabla }[D_{\mathrm{eff}}\mathbf{%
\nabla }C-v_{\mathrm{eff}}C].  \label{4.64}
\end{equation}
In other words, from the standpoint of walker motion this tissue phantom can
be described in term of a medium with the effective diffusion coefficient $%
D_{\mathrm{eff}}$ for the direction parallel to the pipes and with
convective flux whose velocity

\begin{equation}
\mathbf{v}_{\mathrm{eff}}=v_{\mathrm{eff}}\mathbf{n}_{z},  \label{4.65}
\end{equation}
where $\mathbf{n}_{z}$ is the unit vector in the positive $z$-direction.

\section{Random walks in the tis\-sue phan\-tom wi\-th\-out tou\-ching the
pipes}

\label{s4.3}

As will be shown below in the following Chapter large veins can be treated
as walker traps%
\index{walker traps} because if a walker reaches such a vessel it will leave
the fundamental tissue domain with blood for a short time. Therefore,
another characteristic parameter of heat transfer in living tissue is the
mean time during which a walker goes in the cellular tissue without touching
the vessels.

In the present Section we consider this problem for the tissue phantom
described in Section~\ref{s4.2}.

Let us calculate the mean time $\tau _l$ during which walkers created in the
tissue containing the straight parallel pipes (Fig.~\ref{Fig8}) travel in it
without touching the pipes. For the given geometry of the pipe system this
time can be regarded as the mean time during which walkers created in the $%
xy $-plane travel in this plane without touching the disk system shown in
Fig.~\ref{Fig8}.

To calculate the value of $\tau _{l}$ we need to find the probability $F(%
\mathbf{\rho }_{0},t)$ for a walker initially created at point $\mathbf{\rho 
}_{0}$ to reach for the first time the boundary $\sigma =\cup _{i}\{\mid 
\mathbf{\rho }-\mathbf{\rho }_{i}\mid =a\}$ of the disk system in the time $%
t $. Then, the mean time $\tau _{l}(\mathbf{\rho }_{0})$ during which the
walker travels in the $xy$ - plane without touching the boundary $\sigma $ is

\begin{equation}
\tau _{l}(\mathbf{\rho }_{0})=\int\limits_{0}^{\infty }dttF(\mathbf{\rho }%
_{0},t)  \label{4.66}
\end{equation}
and 
\begin{equation}
\tau _{l}=\left\langle \tau _{l}(\mathbf{\rho }_{0})\right\rangle _{\rho
_{0}}.  \label{4.67}
\end{equation}
The function $F(\mathbf{\rho }_{0},t)$ is directly related to the
probability $G(\mathbf{\rho },\mathbf{\rho }_{0},t)$ for such a walker
starting from the point $\mathbf{\rho }_{0}$ at initial time to reach the
point $\mathbf{\rho }$ in the time $t$ without touching the disks. This
probability as a function of $\mathbf{\rho }$ obeys equation (\ref{4.15}) at
points external to the disks, meets initial condition (\ref{4.16}) and the
boundary condition

\begin{equation}
\left. G\right| _{\mathbf{\rho }\in \sigma }=0.  \label{4.68}
\end{equation}
It is well known (see, e.g., \cite{26}), $G(\mathbf{\rho },\mathbf{\rho }%
_{0},t)$ as a function of $\mathbf{\rho }_{0}$ must also satisfy the inverse
Fokker - Planck equation

\begin{equation}
\frac{\partial G}{\partial t}=D\mathbf{\nabla }_{\rho _{0}}^{2}G.
\label{4.69}
\end{equation}
The probability of finding the walker in the tissue at time $t$ can be
equivalently represented as

\begin{equation}
\int\limits_{S_{xy}\setminus \cup _{i}\mid \mathbf{\rho }-\mathbf{\rho }%
_{i}\mid \mathbf{\leqslant }a}d\mathbf{\rho }G(\mathbf{\rho },\mathbf{\rho }%
_{0},t)=\int\limits_{t}^{\infty }dt^{\prime }F(\mathbf{\rho }_{0},t^{\prime
}),  \label{4.70}
\end{equation}
Taking into account expression (\ref{4.70}), integrating equation (\ref{4.69}%
) over $\rho $ in the region external to the disk system and finding the
derivatives of both the parts of the obtained equation we get

\begin{equation}
\frac{\partial F}{\partial t}=D\mathbf{\nabla }_{\rho _{0}}^{2}F.
\label{4.71}
\end{equation}
Obviously, for $\rho _{0}\not\in \sigma $

\begin{equation}
F\mid _{t=0}=0  \label{4.72}
\end{equation}
because a certain time is needed for the walker to reach the disk boundary.
From (\ref{4.71}) and (\ref{4.72}) we obtain that the Laplace transform 
\index{Laplace transform}$F_{L}(\mathbf{\rho }_{0},s)$ of the function $F(%
\mathbf{\rho }_{0},t)$:

\begin{equation}
F_{L}(\mathbf{\rho }_{0},s)=\int\limits_{0}^{\infty }dte^{-st}F(\mathbf{\rho 
}_{0},t)  \label{4.73}
\end{equation}
satisfies the equation

\begin{equation}
sF_{L}=D\mathbf{\nabla }_{\mathbf{\rho }_{0}}^{2}F_{L}  \label{4.74}
\end{equation}
subject to the boundary condition%
\index{boundary condition}

\begin{equation}
\left. F_{L}\right| _{\mathbf{\rho }\in \sigma }=1.  \label{4.75}
\end{equation}
Boundary condition (\ref{4.75}) implies that a walker created in the
immediate vicinity of the boundary $\sigma $ reaches this boundary in no
time.

Obviously, $F_{L}(\mathbf{\rho }_{0},s)$ is a periodic function, thus, the
normal gradient of $F_{L}(\mathbf{\rho }_{0},s)$ at the boundary $\sigma
_{w} $ of any Wigner-Seitz cell of the lattice $\{\mathbf{\rho }_{i}\}$
(Fig.~\ref{Fig8}) must be equal to zero. Thus, in order to find the function
$F_{L}(\mathbf{\rho }_{0},s)$ we may consider the solution of equation (\ref
{4.74}) subject to the boundary conditions

\begin{equation}
\left. F_{L}\right| _{\mid \mathbf{\rho }_{0}\mid =a}=1,  \label{4.76}
\end{equation}
\begin{equation}
\left. \mathbf{\nabla }_{n}F_{L}\right| _{\mathbf{\rho }\in \sigma
_{w}^{0}}=0,  \label{4.77}
\end{equation}
where $\sigma _{w}^{0}$ is the boundary of the Wigner-Seitz cell centered at
the point $\mathbf{\rho }_{i}=0$. To a good approximation we may replace
this Wigner-Seitz cell by a disk of the same area, whose radius $r=(%
\sqrt{3}/(2\pi ))^{1/2}d_{p}$ (Fig.~\ref{Fig8}), and require that the
function $F_{L}(\mathbf{\rho }_{0},s)$ satisfies the condition

\begin{equation}
\left. \mathbf{\nabla }_{n}F_{L}\right| _{\mid \mathbf{\rho }_{0}\mid =r}=0.
\label{4.78}
\end{equation}
The solution of (\ref{4.74}) meeting conditions (\ref{4.76}) and (\ref{4.78}%
) is of the form

\begin{equation}
F_{L}(\mathbf{\rho }_{0},s)=AI_{0}\left[ \rho _{0}\left( \frac{s}{D}\right)
^{1/2}\right] +BK_{0}\left[ \rho _{0}\left( \frac{s}{D}\right) ^{1/2}\right]
,  \label{4.79}
\end{equation}
where 
\begin{equation*}
A=K_{1}\left[ r\left( \frac{s}{D}\right) ^{1/2}\right] \left\{ I_{0}\left[
a\left( \frac{s}{D}\right) ^{1/2}\right] K_{1}\left[ r\left( \frac{s}{D}%
\right) ^{1/2}\right] +\right.
\end{equation*}
\begin{equation}
\left. +K_{0}\left[ a\left( \frac{s}{D}\right) ^{1/2}\right] I_{1}\left[
r\left( \frac{s}{D}\right) ^{1/2}\right] \right\} ^{-1}  \label{4.80}
\end{equation}
and 
\begin{equation*}
B=I_{1}\left[ r\left( \frac{s}{D}\right) ^{1/2}\right] \left\{ I_{0}\left[
a\left( \frac{s}{D}\right) ^{1/2}\right] K_{1}\left[ r\left( \frac{s}{D}%
\right) ^{1/2}\right] +\right.
\end{equation*}
\begin{equation}
\left. +K_{0}\left[ a\left( \frac{s}{D}\right) ^{1/2}\right] I_{1}\left[
r\left( \frac{s}{D}\right) ^{1/2}\right] \right\} ^{-1}.  \label{4.81}
\end{equation}
Here $I_{0},I_{1},K_{0},K_{1}$ are the modified Bessel functions%
\index{modified Bessel function} of order zero and one, of the first and
second kind, respectively. Averaging the function (\ref{4.79}) over the
domain $a\mathbf{\leqslant }\mid \mathbf{\rho }_{0}\mid \mathbf{\leqslant }r$
we get

\begin{equation*}
<F_{L}(\mathbf{\rho }_{0},s)>=%
\frac{2a}{(r^{2}-a^{2})}\left( \frac{D}{s}\right) ^{1/2}\left\{ K_{1}\left[
a\left( \frac{s}{D}\right) ^{1/2}\right] I_{1}\left[ r\left( \frac{s}{D}%
\right) ^{1/2}\right] -\right.
\end{equation*}
\begin{equation*}
\left. -I_{1}\left[ a\left( \frac{s}{D}\right) ^{1/2}\right] K_{1}\left[
r\left( \frac{s}{D}\right) ^{1/2}\right] \right\} \left\{ I_{0}\left[
a\left( \frac{s}{D}\right) ^{1/2}\right] K_{1}\left[ r\left( \frac{s}{D}%
\right) ^{1/2}\right] +\right.
\end{equation*}
\begin{equation}
\left. +K_{0}\left[ a\left( \frac{s}{D}\right) ^{1/2}\right] I_{1}\left[
r\left( \frac{s}{D}\right) ^{1/2}\right] \right\} ^{-1}.  \label{4.82}
\end{equation}

Let us consider the function $\left\langle F_{L}(\mathbf{\rho }%
_{0},s)\right\rangle $ for $s\ll D/r^{2}$. Expanding each function appearing
in (\ref{4.82}) into a power series%
\index{power series} of $s$ and $a/d_{p}$ from (\ref{4.82}) we get

\begin{equation}
\left\langle F_{L}(\mathbf{\rho }_{0},s)\right\rangle \approx 
\frac{1}{1+s\frac{r^{2}}{2D}\ln \left( \frac{r}{a}\right) }.  \label{4.83}
\end{equation}
If $D/r^{2}\ll s\ll D/a^{2}$, then, according to (\ref{4.82}) the function $%
\left\langle F_{L}(\mathbf{\rho }_{0},s)\right\rangle \ll 1$. Therefore, by
virtue of (\ref{4.83}) we may represent the inverse transform of (\ref{4.83}%
) in the form

\begin{equation}
\left\langle F(\mathbf{\rho }_{0},s)\right\rangle =\frac{1}{\tau _{l}}\exp %
\left[ -\frac{t}{\tau _{l}}\right] ,  \label{4.84}
\end{equation}
where 
\begin{equation}
\tau _{l}=\frac{r^{2}}{2D}\ln \left( \frac{r}{a}\right) .  \label{4.85}
\end{equation}

Therefore, according to definition (\ref{4.66}), (\ref{4.67}) we obtain that 
$\tau _{l}$ is the desired mean time of walker motion without touching the
pipes.

Since $d_{p}\gg a$ and, as follows from the definition of the mean distance%
\index{mean distance} between the pipes, $\pi r^{2}=d^{2}$, we may rewrite
formula (\ref{4.85}) in the form

\begin{equation}
\tau _l \approx 
\frac{d^2}{2\pi D} \ln \left ( \frac{d}{a} \right ),  \label{4.86}
\end{equation}
where we have ignored the term $1/2 \ln \pi$ in the cofactor $\ln (r/a) =
\ln (d/a) - 1/2 \ln \pi$. In addition, it should be pointed out that for $%
\rho _0 \sim a$ and $D/d^2 \ll s \ll d/a^2$ according to (\ref{4.79})

\begin{equation}
F_{L}(\mathbf{\rho }_{0},s)\simeq \frac{K_{0}\left[ \rho _{0}\left( \frac{s}{%
D}\right) ^{1/2}\right] }{K_{0}\left[ a\left( \frac{s}{D}\right) ^{1/2}%
\right] }.  \label{4.87}
\end{equation}
Thereby, for the time $t\mathbf{\leqslant }d^{2}/(2D)$ practically each
walker created near a pipe (i.e. for which $\rho _{0}\sim a$) inevitably
crosses this pipe at least one time.

\section{Random walks in the tissue phan\-tom con\-tai\-ning the
h\-exa\-gonal array of countercurrent pairs}

\label{s4.4}

In the present Section we analyse characteristic properties of random walks
in the tissue phantom that contains the vessel system involving straight
identical pipes of radius $a$ parallel to the $z$-axis and grouped in pairs
where blood currents flow in the opposite directions (Fig.~\ref{Fig10}). We
assume that the pipes pairs make up a hexagonal array of spacing $d_{p}\gg a$%
. In other words, all the countercurrent pairs are assumed to be located in
the vicinity of certain straight lines crossing the $xy$-plane at points $\{%
\mathbf{\rho }_{i}\}$ forming the hexagonal array.

\FRAME{ftbpFU}{8.9996cm}{5.0698cm}{0pt}{\Qcb{The $xy$ crossection of the
hexagonal array of counter current pipes}}{\Qlb{Fig10}}{Fig10}{\special%
{language "Scientific Word";type "GRAPHIC";maintain-aspect-ratio
TRUE;display "USEDEF";valid_file "F";width 8.9996cm;height 5.0698cm;depth
0pt;original-width 11.5556in;original-height 6.4818in;cropleft "0";croptop
"1";cropright "0.9993";cropbottom "0";filename 'FIG10.GIF';file-properties
"XNPEU";}}

Blood flow in this vessel system is characterized by the following blood
velocity field $v(\mathbf{\rho }\,)$ in the $z$-direction:

\begin{equation}
v(\mathbf{\rho }\,)=\pi a^{2}v\sum_{i}[U(\mathbf{\rho }-\mathbf{\rho }%
\,_{i}^{+})-U(\mathbf{\rho }-\mathbf{\rho }\,_{i}^{-})],  \label{4.88}
\end{equation}
where as in Section~\ref{s4.2} $v$ is the mean velocity of blood in single
pipes, the sum runs over all the pairs and $\mathbf{\rho }\,_{i}^{+},\mathbf{%
\rho }\,_{i}^{-}$ are the points at which the centre lines of pipes with
blood flow in the positive and negative directions cross the $xy$-plane. The
function $U(\mathbf{\rho }\,)$ normalized to unity describes the blood flow
distribution inside a pipe and, for the sake of simplicity, will be given in
the form (\ref{4.24}). The pipe center line coordinates $\{\mathbf{\rho }%
\,_{i}^{+},\mathbf{\rho }\,_{i}^{-}\}$ are considered to be pairwise
independent random variables characterized by the distribution functions

\begin{equation}
\Phi _{i}(\mathbf{\rho }\,)=\frac{1}{\pi b^{2}}\exp \left\{ -\frac{(\mathbf{%
\rho }-\mathbf{\rho }_{i})^{2}}{b^{2}}\right\} .  \label{4.89}
\end{equation}
Here $b=\mu a$ is the mean distance between pipes of a single pair and $\mu
\geq 2$ is a system parameter.

Following Section~\ref{s4.2} in order to analyse characteristic properties
of random walks in this tissue phantom we calculate the mean values of
distance

\begin{equation}
l_{\parallel }(\vec{t},\mathbf{\rho }_{0})=\int\limits_{0}^{t}dt^{\prime
}\left\{ f_{z}(t^{\prime })+\int\limits_{S_{xy}}d\mathbf{\rho }\,^{\prime }v(%
\mathbf{\rho }\,^{\prime })\delta (\mathbf{\rho }\,^{\prime }-\mathbf{\rho }%
\,[t^{\prime }])\right\}  \label{4.90}
\end{equation}
which a walker created at the point $\{\mathbf{\rho }_{0},t\}$ travel in the 
$z$-direction during the time $t$ providing it moves along the path $\mathbf{%
r}\,[t^{\prime }]=\{\mathbf{\rho }\,[t^{\prime }],z[t^{\prime }]\}$.

Below we shall consider two different limits that characterize influence of
a single counter current pair on walker motion and their cooperative effect.

\subsection{A single counter current pair}

\label{s4.4.1}

In this Subsection we assume that the time $t$ meets the condition

\begin{equation}
\tau _{a}=\frac{a^{2}}{2D}\ll t\ll \frac{d_{p}^{2}}{2D}  \label{4.91}
\end{equation}
and the walker has been created near the pair $i_{0}$ at the point $\mathbf{%
\rho }_{0}\approx \mathbf{\rho }_{i_{0}}$, i.e. $\mid \mathbf{\rho }_{i_{0}}-%
\mathbf{\rho }_{0}\mid \sim a$. In this case we may take into account only
the pipe pair $i_{0}$.

Replicating practically one-to-one calculations of Section~\ref{s4.2} we get

\begin{equation}
\left\langle l_{\parallel }(t,\mathbf{\rho }_{0})\right\rangle _{w}=\pi
a^{2}v\int\limits_{0}^{t}dt^{\prime }\int\limits_{S_{xy}}d\mathbf{\rho }%
\,^{\prime }U_{i_{0}}^{p}(\mathbf{\rho }\,^{\prime })G(\mathbf{\rho }%
\,^{\prime }-\mathbf{\rho }_{0},t^{\prime })  \label{4.92}
\end{equation}
and 
\begin{equation*}
\left\langle (l_{\parallel }(t,\mathbf{\rho }_{0}))^{2}\right\rangle
_{w}=2Dt+(\pi a^{2}v)^{2}2\int\limits_{0}^{t}dt^{\prime
}\int\limits_{0}^{t^{\prime }}dt^{\prime \prime
}\int\limits_{S_{xy}}\int\limits_{S_{xy}}d\mathbf{\rho }\,^{\prime }d\mathbf{%
\rho }\,^{\prime \prime }\cdot
\end{equation*}
\begin{equation}
\cdot U_{i_{0}}^{p}(\mathbf{\rho }\,^{\prime })U_{i_{0}}^{p}(\mathbf{\rho }%
\,^{\prime \prime })G(\mathbf{\rho }\,^{\prime }-\mathbf{\rho }\,^{\prime
\prime },t^{\prime }-t^{\prime })G(\mathbf{\rho }\,^{\prime \prime }-\mathbf{%
\rho }_{0},t^{\prime \prime }).  \label{4.93}
\end{equation}
Here $G(\mathbf{\rho },t)$ is the probability for a walker created at the
point $\mathbf{\rho }_{0}=0$ in the $xy$ - plane to reach the point $\mathbf{%
\rho }$ in the time $t$ and is determined by expression (\ref{4.17}), and

\begin{equation}
U_{i}^{p}(\mathbf{\rho }\,)=U_{i}(\mathbf{\rho }-\mathbf{\rho }_{i}\,^{+})-U(%
\mathbf{\rho }-\mathbf{\rho }_{i}\,^{-}).  \label{4.94}
\end{equation}

Let us, first, estimate integral (\ref{4.92}). Formulas (\ref{4.17}), (\ref
{4.24}) and identity (\ref{4.25}) allow us to integrate directly over $%
\mathbf{\rho }\,^{\prime \prime }$ in expression (\ref{4.92}). In this way
we get

\begin{equation*}
<l_{\parallel }(t,\mathbf{\rho }_{0})>_{w}=\pi
a^{2}v2\int\limits_{0}^{t}dt^{\prime }\frac{1}{\pi \lbrack 4Dt^{\prime
}+a^{2}]}\cdot
\end{equation*}
\begin{equation}
\cdot \left\{ \exp \left[ -\frac{(\mathbf{\rho }_{0}-\mathbf{\rho }%
\,_{i_{0}}^{+})^{2}}{4Dt^{\prime }+a^{2}}\right] -\exp \left[ -\frac{(%
\mathbf{\rho }_{0}-\mathbf{\rho }\,_{i_{0}}^{-})^{2}}{4Dt^{\prime }+a^{2}}%
\right] \right\}  \label{4.95}
\end{equation}

For $\left| \mathbf{\rho }_{0}-\mathbf{\rho }_{i_{0}}\,^{+}\right| \sim
a;\,\left| \mathbf{\rho }_{0}-\mathbf{\rho }_{i_{0}}\,^{-}\right| \sim a$
and $t\gg a^{2}/(2D)$ the integral (\ref{4.95}) is about

\begin{equation}
\frac{a^{2}}{\pi \lbrack 4Dt]^{2}}.
\end{equation}
Since the integral 
\begin{equation}
\int\limits_{T}^{\infty }\frac{dt}{t^{2}}
\end{equation}
converges at the upper limits, the main contribution to integral (\ref{4.95}%
) is associated with values of the time $t^{\prime }$ being about $a^{2}/D$.
The latter enables us to estimate the mean value $\left\langle l_{\parallel
}(t,\mathbf{\rho }_{0})\right\rangle $ as

\begin{equation}
\left| \left\langle l_{\parallel }(t,\mathbf{\rho }_{0})\right\rangle
\right| \sim \frac{a^{2}v}{2D}=v\tau _{a}.  \label{4.96}
\end{equation}

For the sake of simplicity we calculate the value $\left\langle
(l_{\parallel }(t,\mathbf{\rho }_{0}))^{2}\right\rangle $ averaged over all
the possible arrangement of the pipe centre lines $\{\mathbf{\rho }%
_{i_{0}}\,^{+},\mathbf{\rho }_{i_{0}}\,^{-}\}$. By definition

\begin{equation*}
\left\langle U_{i_{0}}^{p}(\mathbf{\rho }\,^{\prime })U_{i_{0}}^{p}(\mathbf{%
\rho }\,^{\prime \prime })\right\rangle _{\pm }=
\end{equation*}
\begin{equation}
=\int\limits_{S_{xy}}\int\limits_{S_{xy}}d\mathbf{\rho }_{i_{0}}\,^{+}d%
\mathbf{\rho }_{i_{0}}{}^{-}\Phi _{i_{0}}(\mathbf{\rho }_{i_{0}}{}^{+})\Phi
_{i_{0}}(\mathbf{\rho }_{i_{0}}{}^{-})U_{i_{0}}^{p}(\mathbf{\rho }^{\prime
})U_{i_{0}}^{p}(\mathbf{\rho }^{\prime \prime }).  \label{4.97}
\end{equation}
Substituting (\ref{4.94}) into (\ref{4.93}), then, averaging according to
rule (\ref{4.97}) and taking into account that the superscripts ``$+$'' and
``$-$'' become the dummy indexes, we get

\begin{align}
\left\langle (l_{\parallel }(t,\mathbf{\rho }_{0}))^{2}\right\rangle _{w,\pm
}& =2Dt+4(\pi a^{2}v)^{2}\int\limits_{0}^{t}dt^{\prime
}\int\limits_{0}^{t^{\prime }}dt^{\prime \prime }\iint_{S_{xy}}d\mathbf{\rho 
}\,^{\prime }d\mathbf{\rho }\,^{\prime \prime }\cdot  \notag \\
& \cdot \Biggl \{\int\limits_{S_{xy}}d\mathbf{\rho }\,^{+}\Phi _{i_{0}}(%
\mathbf{\rho }_{i_{0}}\,^{+})U(\mathbf{\rho }\,^{\prime }-\mathbf{\rho }%
\,^{+})U(\mathbf{\rho }\,^{\prime \prime }-\mathbf{\rho }\,^{+})\cdot  \notag
\\
& \cdot G(\mathbf{\rho }\,^{\prime }-\mathbf{\rho }\,^{\prime \prime
},t^{\prime }-t^{\prime \prime })G(\mathbf{\rho }\,^{\prime \prime }-\mathbf{%
\rho }_{0},t^{\prime \prime })-  \notag \\
& -\iint_{S_{xy}}d\mathbf{\rho }\,^{+}d\mathbf{\rho }\,^{-}\Phi _{i_{0}}(%
\mathbf{\rho }\,^{+})\Phi _{i_{0}}(\mathbf{\rho }\,^{-})U(\mathbf{\rho }%
\,^{\prime }-\mathbf{\rho }\,^{+})U(\mathbf{\rho }\,^{\prime \prime }-%
\mathbf{\rho }\,^{-})\cdot  \notag \\
& \cdot G(\mathbf{\rho }\,^{\prime }-\mathbf{\rho }\,^{\prime \prime
},t^{\prime }-t^{\prime \prime })G(\mathbf{\rho }\,^{\prime \prime }-\mathbf{%
\rho }_{0},t^{\prime \prime })\Biggr \},  \label{4.98}
\end{align}
where we also have integrated over $\rho ^{-}$ in the first term in the
braces.

All the cofactors of the integrals in (\ref{4.98}) are of the form given by
formula (\ref{4.25a}). So, it is convenient to represent the integral term
of (\ref{4.98}) as the diagram shown in Fig.~\ref{Fig11}. The line between
two points (Fig.~\ref{Fig11}a) represents the function $\mathcal{G}(\mathbf{%
\rho }_{0}-\mathbf{\rho }\,^{\prime },A)$ given by (\ref{4.25a}), the solid
circuits show the fixed coordinates and the empty circuits correspond to
variables of integration. Some of the rules of diagram transformation%
\index{diagram transformation} to be used are given in Fig.~\ref{Fig11}b.
Following these rules we reduce formula (\ref{4.98}) whose integral term is
shown in Fig.~\ref{Fig11}c to the expression

\begin{equation*}
\left\langle (l_{\parallel }(t,\mathbf{\rho }_{0}))^{2}\right\rangle _{w,\pm
}=2Dt+%
\frac{4}{\pi ^{2}}(\pi a^{2}v)^{2}\int\limits_{0}^{t}dt^{\prime
}\int\limits_{0}^{t^{\prime }}dt^{\prime \prime }\cdot
\end{equation*}
\begin{equation*}
\cdot \left\{ \frac{1}{4D(t^{\prime }-t^{\prime \prime })+2a^{2}}\frac{1}{%
(4Dt^{\prime \prime }+b^{2}+\sum\nolimits_{-})}\exp \left[ -\frac{(\mathbf{%
\rho }_{0}-\mathbf{\rho }_{i_{0}})^{2}}{4Dt^{\prime \prime
}+b^{2}+\sum\nolimits_{+}}\right] -\right.
\end{equation*}
\begin{equation}
\left. -\frac{1}{4D(t^{\prime }-t^{\prime \prime })+2(a^{2}+b^{2})}\frac{1}{%
(4Dt^{\prime \prime }+\sum\nolimits_{-})}\exp \left[ -\frac{(\mathbf{\rho }%
_{0}-\mathbf{\rho }_{i_{0}})^{2}}{4Dt^{\prime \prime }+\sum\nolimits_{-}}%
\right] \right\} ,  \label{4.99}
\end{equation}
where 
\begin{equation}
\sum\nolimits_{+}=\frac{a^{2}[4D(t^{\prime }-t^{\prime \prime })+a^{2}]}{%
4D(t^{\prime }-t^{\prime \prime })+2a^{2}},  \label{4.100}
\end{equation}
\begin{equation}
\sum\nolimits_{-}=\frac{a^{2}[4D(t^{\prime }-t^{\prime \prime
})+(a^{2}+b^{2})]}{4D(t^{\prime }-t^{\prime \prime })+2(a^{2}+b^{2})}.
\label{4.101}
\end{equation}
For $t^{\prime }-t^{\prime \prime }\gg \tau _{a}\sim a^{2}/D$; $b^{2}/D$
according to (\ref{4.100}), (\ref{4.101}) $\sum\nolimits_{+}\simeq a^{2}$; $%
\sum\nolimits_{-}\simeq a^{2}+b^{2}$. Therefore, as seen from (\ref{4.99})
the main contribution to the integral (\ref{4.99}) is associated with the
region $t^{\prime }-t^{\prime \prime }\sim \tau _{a}$. For $\mid \mathbf{%
\rho }_{0}-\mathbf{\rho }_{i_{0}}\mid \sim a$ and $t^{\prime \prime }\gg
a^{2}/D$ we may rewrite the term in the braces as

\begin{equation}
\frac{1}{4Dt^{\prime \prime }}\frac{2b^{2}}{[4D(t^{\prime }-t^{\prime \prime
})+2a^{2}][4D(t^{\prime }-t^{\prime \prime })+2(a^{2}+b^{2})]}.
\end{equation}
If we substitute this term into integral (\ref{4.99}) then, it will diverge
logarithmically, which may be cut-off at $t^{\prime \prime }\sim \tau _{a}$.
In this way integrating over $t^{\prime }$ and $t^{\prime \prime }$ we obtain

\begin{equation}
\left\langle (l_{\parallel }(t,\mathbf{\rho }_{0}))^{2}\right\rangle _{w,\pm
}=2Dt+\left( \frac{a^{2}v}{2D}\right) ^{2}\ln (1+\mu ^{2})\ln \frac{t}{\tau
_{a}}.  \label{4.102}
\end{equation}

\FRAME{ftbpFU}{8.9029cm}{8.2835cm}{0pt}{\Qcb{Diagrams representation of
integral terms.}}{\Qlb{Fig11}}{Fig11}{\special{language "Scientific
Word";type "GRAPHIC";maintain-aspect-ratio TRUE;display "USEDEF";valid_file
"F";width 8.9029cm;height 8.2835cm;depth 0pt;original-width
8.9543in;original-height 8.3333in;cropleft "0";croptop "1.0006";cropright
"1.0013";cropbottom "0";filename 'FIG11.GIF';file-properties "XNPEU";}}

Formulas (\ref{4.96}) and (\ref{4.102}) give the desired mean values of the
distance travelled by a walker along the pipes during the time $t$.
Concluding the present Subsection we note that blood flow in a single pipe
and a single countercurrent pair affects walker motion in different way.
Indeed, according to (\ref{4.40}), (\ref{4.52}) for a single pipe the blood
flow effect may be treated in terms of convective transport and the mean
displacement $l_{\parallel }$ along the pipe with blood flow is $%
l_{\parallel }\approx vt_{p}$, where $t_{p}$ is the mean residence time of
walkers inside the pipe (see Subsection~\ref{s4.2.1}). For a single counter
current pair, according to (\ref{4.96}) and (\ref{4.102}) the blood flow
effect is diffusive in nature. The matter is that a walker during its motion
visits alternately vessels with the opposite blood flow directions. During
each visit of a pipe the walker with blood flow travels, on the average, a
disk line $v\tau _{a}$. The mean number of such visits can be estimated as $%
t_{p}/\tau _{a}$. Due to the walker visiting each of the pair pipes randomly
the mean squared distance $l_{\parallel }$ travelled by the walker along the
pipes is about

\begin{equation}
l_{\parallel }^{2}\sim (v\tau _{a})^{2}\frac{t_{p}}{\tau _{a}}
\end{equation}
and substituting (\ref{4.33}) into this expression we immediately get
formula (\ref{4.102}).

\subsection{The cooperative effect of countercurrent pairs}

\label{s4.4.2}

When $t \gg d^2_p / (2D)$ walkers during their motion visit a large number
of counter current pairs. Therefore, in this case the total effect of blood
flow in the counter current pairs must be cooperative and the tissue phantom
containing this pipe pairs can be described in terms of an effective medium.

In the present Subsection we calculate the mean squared distance $%
l_{\parallel }(t,\mathbf{\rho }_{0})$ travelled by a worker during the time $%
t$ along the $z$ - axis (see (\ref{4.90})). We note that for the tissue
phantom under consideration $\left\langle l_{\parallel }(t,\mathbf{\rho }%
_{0})\right\rangle _{w,\pm }=0$. Replicating practically one - to - one the
calculations of Subsection ~\ref{s4.2.2} from (\ref{4.88}), (\ref{4.90}) we
obtain an expression similar to (\ref{4.43})

\begin{equation*}
\left\langle (l_{\parallel }(t,\mathbf{\rho }_{0}))^{2}\right\rangle _{%
\mathbf{\rho }_{0},w,\pm }=2Dt+2(\pi a^{2}v)^{2}\frac{1}{S_{xy}}%
\sum_{i^{\prime },i^{\prime \prime }}\int\limits_{0}^{t}dt^{\prime
}\int\limits_{0}^{t^{\prime }}dt^{\prime \prime }\cdot
\end{equation*}
\begin{equation*}
\cdot \int \int\limits_{S_{xy}}\int d\mathbf{\rho }\,^{\prime }d\mathbf{\rho 
}\,^{\prime \prime }d\mathbf{\rho }_{0}<U_{i^{\prime }}^{p}(\mathbf{\rho }%
\,^{\prime })U_{i^{\prime \prime }}^{p}(\mathbf{\rho }\,^{\prime \prime
})>_{\pm }\cdot
\end{equation*}
\begin{equation}
\cdot G(\mathbf{\rho }\,^{\prime }-\mathbf{\rho }\,^{\prime \prime
},t^{\prime }-t^{\prime \prime })G(\mathbf{\rho }\,^{\prime \prime }-\mathbf{%
\rho }_{0},t^{\prime \prime }).  \label{4.103}
\end{equation}
Noting that according to (\ref{4.94}) and (\ref{4.97})

\begin{equation}
\left\langle U_{i}^{p}(\mathbf{\rho }\,)\right\rangle _{\pm }=0
\end{equation}
and the arrangement of pipe center lines for different countercurrent pairs
are pairwise independent we may write

\begin{equation}
\left\langle U_{i^{\prime }}^{p}(\mathbf{\rho }\,^{\prime })U_{i^{\prime
\prime }}^{p}(\mathbf{\rho }\,^{\prime \prime })\right\rangle =\delta
_{i^{\prime }i^{\prime \prime }}\left\langle U_{i^{\prime }}^{p}(\mathbf{%
\rho }\,^{\prime })U_{i^{\prime }}^{p}(\mathbf{\rho }\,^{\prime \prime
})\right\rangle .  \label{4.104}
\end{equation}
The substitution of (\ref{4.104}) into (\ref{4.103}), formula (\ref{4.37})
and (\ref{4.39}) yield

\begin{equation*}
\left\langle (l_{\parallel }(t,\mathbf{\rho }_{0}))^{2}\right\rangle _{%
\mathbf{\rho }_{0},w,\pm }=2Dt+4\frac{(\pi a^{2}v)^{2}}{d^{2}}%
\int\limits_{0}^{t}dt^{\prime }\int\limits_{0}^{t^{\prime }}dt^{\prime
\prime }\int\limits_{S_{xy}}\int\limits_{S_{xy}}d\mathbf{\rho }\,^{\prime }d%
\mathbf{\rho }\,^{\prime \prime }\cdot
\end{equation*}
\begin{equation*}
\cdot \left\{ \int\limits_{S_{xy}}d\mathbf{\rho }\,^{+}\Phi _{i}(\mathbf{%
\rho }^{+})U(\mathbf{\rho }\,^{\prime }-\mathbf{\rho }\,^{+})U(\mathbf{\rho }%
\,^{\prime \prime }-\mathbf{\rho }\,^{+})G(\mathbf{\rho }\,^{\prime }-%
\mathbf{\rho }\,^{\prime \prime },t^{\prime }-t^{\prime \prime })-\right.
\end{equation*}
\begin{equation}
\left. \int\limits_{S_{xy}}\int d\mathbf{\rho }\,^{+}d\mathbf{\rho }%
\,^{-}\Phi _{i}(\mathbf{\rho }\,^{+})\Phi _{i}(\mathbf{\rho }\,^{-})U(%
\mathbf{\rho }\,^{\prime }-\mathbf{\rho }\,^{+})U(\mathbf{\rho }\,^{\prime }-%
\mathbf{\rho }\,^{-})G(\mathbf{\rho }\,^{\prime }-\mathbf{\rho }\,^{\prime
\prime },t^{\prime }-t^{\prime \prime })\right\}  \label{4.105}
\end{equation}
where $d$ is the mean distance between the counter current pair given by
expression (\ref{4.41}) and $i$ denotes any fixed pair, for example, the
pair with $\mathbf{\rho }_{i}=0$. The diagram representing the integral term
in (\ref{4.105}) is shown in Fig.~\ref{Fig12}. Following the rules indicated
in Fig.~\ref{Fig11}b, we get

\begin{equation*}
\left\langle (l_{\parallel }(t,\mathbf{\rho }_{0}))^{2}\right\rangle _{%
\mathbf{\rho }_{0},w,\pm }=2Dt+4\frac{(\pi a^{2}v)^{2}}{d^{2}}\frac{1}{\pi }%
\int\limits_{0}^{t}dt^{\prime }\int\limits_{0}^{t^{\prime }}dt^{\prime
\prime }\cdot
\end{equation*}
\begin{equation}
\cdot \left[ \frac{1}{4D(t^{\prime }-t^{\prime \prime })+2a^{2}}-\frac{1}{%
4D(t^{\prime }-t^{\prime \prime })+2(a^{2}+b^{2})}\right]  \label{4.106}
\end{equation}

\FRAME{ftpFU}{10.3966cm}{5.5135cm}{0pt}{\Qcb{Diagram representation of
integral term in $<(l_{\parallel }(t,\mathbf{\protect\rho }_{0}))^{2}>_{%
\mathbf{\protect\rho }_{0},w,\pm }$.}}{\Qlb{Fig12}}{Fig12}{\special{language
"Scientific Word";type "GRAPHIC";maintain-aspect-ratio TRUE;display
"USEDEF";valid_file "F";width 10.3966cm;height 5.5135cm;depth
0pt;original-width 12.4818in;original-height 6.5925in;cropleft "0";croptop
"1.0005";cropright "1.0003";cropbottom "0";filename
'FIG12.GIF';file-properties "XNPEU";}}It immediately follows that

\begin{equation}
\left\langle (l_{\parallel }(t,\mathbf{\rho }_{0}))^{2}\right\rangle _{%
\mathbf{\rho }_{0},w,\pm }=2Dt\left\{ 1+\frac{\pi }{2}\left( \frac{a^{2}v}{Dd%
}\right) \ln (1+\mu ^{2})\right\} .  \label{4.107}
\end{equation}
Since the mean distance $\left\langle (l_{\parallel }(t,\mathbf{\rho }%
_{0}))^{2}\right\rangle $ travelled by a walker along the pipes during the
time $t$ is directly proportional to the time $t,$ this tissue phantom may
be described in terms of an effective medium with the diffusion coefficient

\begin{equation}
D_{\mathrm{eff}}=\left\{ 1+\frac{\pi }{2}\left( \frac{a^{2}v}{Dd}\right)
^{2}\ln (1+\mu ^{2})\right\} D.  \label{4.108}
\end{equation}

Concluding the present Subsection we point out that for the given tissue
phantom in contrast to the tissue phantom considered in Section~\ref{s4.2}
formula (\ref{4.107}) does not contain a term proportional to $t \ln t$.
This is the case because a walker returning to the same countercurrent pair
for the next time it is travelled with blood flow in arbitrary direction.

\section{Main characteristics of walker motion in li\-ving tissue with
nonhierarchical vascular net\-work}

\label{s4.5}

In previous Sections we have considered heat transfer in living tissue
phantom containing vessels of the same parameters. This model does not
account for the hierarchical nature of vascular networks in real living
tissue. However, the results obtained in the framework of this model form
the base for the following analysis of heat transfer in living tissue with
hierarchical network. Therefore, in the present Section to gain a better
comprehension we outline the main characteristics of random walkers in these
tissue phantoms and discuss their physical meaning.

Random walks in living tissue can be characterized by two time scales
playing important roles in analysis of bioheat transfer.

One of them is the mean residence time $\left\langle t_{p}\right\rangle $ of
a walker inside a given pipe during its motion for the time $t\gg a^{2}/(2D)$
provided it was initially near this pipe. According to (\ref{4.33})

\begin{equation}
\left\langle t_{p}\right\rangle \simeq \frac{a^{2}}{2D}\ln \left[ \frac{%
(2Dt)^{1/2}}{a}\right] .  \label{4.109}
\end{equation}
Obviously, the total mean time $\left\langle t_{p}\right\rangle $ during
which a walker created near a counter current pair is inside the pipes of
this pair is also given by expression (\ref{4.109}). The other is the mean
time $\tau _{l}$ during which a walker goes in the tissue without touching
the pipes. This time scale is the same for both \ of the tissue phantoms
considered above where individual pipes or countercurrent pairs form
hexagonal areas are of equal spacing. Therefore, according to (\ref{4.88})

\begin{equation}
\tau _{l}\simeq \frac{d^{2}}{2\pi D}\ln \left( \frac{d}{a}\right) ,
\label{4.110}
\end{equation}
where $d=(\sqrt{3}/2)^{1/2}d_{p}$ is the mean distance between the pipes or
the counter current pairs (see (\ref{4.41})). However, if the walker is
initially\ near a certain pipe it practically inevitably will cross the
boundary of the pipe within a time $t\ll \tau _{l}$.

Depending on the value of $2Dt/d^{2}$ the walker motion in the direction
along the pipes differs in properties. In particular, for tissue with unit
pipes when $a^{2}/(2D)\ll t\ll d^{2}/(2D)$ a walker, which is initially near
a certain pipe, travels with blood flow in this pipe, the mean distance (see
(\ref{4.26}) and (\ref{4.28}))

\begin{equation}
\left\langle l_{\parallel }(t)\right\rangle \simeq \frac{a^{2}v}{2D}\ln %
\left[ \frac{(2Dt)^{1/2}}{a}\right] ,  \label{4.111}
\end{equation}
where $v$ is the mean velocity of blood in the given pipe. In this case the
mean square of a distance $l_{\parallel }(t)$ travelled by the walker during
the time $t$, including its motion in the cellular tissue as well as in the
blood stream, is (see (\ref{4.27}))

\begin{equation}
\left\langle (l_{\parallel }(t))^{2}\right\rangle \simeq 2Dt+2\left\langle
l_{\parallel }(t)\right\rangle ^{2}.  \label{4.112}
\end{equation}
We note that expression (\ref{4.111}) can be also rewritten in the form $%
\left\langle l_{\parallel }(t)\right\rangle =v\left\langle
t_{p}\right\rangle $. Therefore, in this case blood flow effect may be
treated in terms of convective transport.

For the tissue with counter current pairs the characteristic distance $%
l_{\parallel} (t)$ travelled by a walker created near a pipe pair with blood
along these pipes is about (see (\ref{4.102})).

\begin{equation}
l_{\parallel }(t)\approx \frac{a^{2}v}{2D}\left[ 2\ln (1+\mu ^{2})\ln \frac{%
(2Dt)^{1/2}}{a}\right] ^{1/2}  \label{4.113}
\end{equation}
where $\mu $ is the mean ration of the distance between pipes to the pipes
radius of a single pair. In this case direction of the walker displacement
is random and 
\begin{equation}
l_{\parallel }(t)\sim v[\tau _{a}\left\langle t_{p}\right\rangle ]^{1/2}
\end{equation}
where $\tau _{a}=a^{2}/(2D)$.

When $t \gg d^2/(2D)$, or, more precisely, $t \gg \tau_l$, practically each
walker in its motion visits a large number of pipes and we may describe the
walker motion along the pipes in terms of random walks in homogeneous medium
with the effective diffusion coefficient (see (\ref{4.60}) and (\ref{4.108}))

\begin{equation}
D_{\mathrm{eff}}\approx D\left\{ 1+\frac{\pi }{2}\left( \frac{a^{2}v}{Dd}%
\right) ^{2}\left[ \ln \left( \frac{d}{a}\right) \right] ^{\beta }\right\}
\label{4.114}
\end{equation}
under an uniform convective flux with the velocity (see (\ref{4.59}))

\begin{equation}
\mathbf{v}_{\mathrm{eff}}\approx \frac{\pi a^{2}v}{d^{2}}\xi \mathbf{n}_{Z},
\label{4.115}
\end{equation}
where $\mathbf{n}_{z}$ is the unit vector in the positive $z$ - direction
and $\beta =1$ for tissue with unit vessels and $\beta =0$ for tissue with
counter current pairs.

In the framework of qualitative analysis expressions (\ref{4.114}) and (\ref
{4.115}) can be also obtained in the following way. First, we consider
tissue with unit vessels. Let us assume that at a certain time a walker is
near a given pipe. During a time $t<d^2/(2D)$ blood flow in the other pipes
has practically no effect on its motion. Thereby, according to (\ref{4.111})
and (\ref{4.113}), the mean value of the distance $l_{\parallel}$ travelled
by the walker along the pipe with blood flow during the time $t \sim
d^2/(2D) $ can be estimated as

\begin{equation}
\left\langle l_{\parallel }\right\rangle \approx \frac{a^{2}v}{2D}\ln \left( 
\frac{d}{a}\right)  \label{4.116}
\end{equation}
and, in addition,

\begin{equation}
\left\langle l_{\parallel }^{2}\right\rangle \approx 2\left\langle
l_{\parallel }\right\rangle ^{2}.  \label{4.117}
\end{equation}
then, we shall assume that in the time $t\sim d^{2}/(2D)$ the walker escapes
from the neighborhood of this pipe whose radius is $d_{p}/2$ and in the time 
$\tau _{l}$ it reaches one of the pipes again. Due to $d\gg a$ the time $%
\tau _{l}$ can be considered to be substantially greater than $d^{2}/(2D)$
and, thus, we may assume that the walker reaches another pipe.

In this way we represent the walker path as a sequence of steps on the pipe
array which practically does not contain returning to pipes that have been
visited before. When $t\gg \tau _{l}$ such a sequence involves $t/\tau _{l}$
pairwise independent steps, with the length of one step obeying conditions (%
\ref{4.116}) and (\ref{4.117}). Taking into account that blood flow in a
vessel is oriented in the positive $z$ - direction with probability $%
1/2(1+\xi )$ and in the opposite one with probability $1/2(1-\xi )$ we find
the following expressions for the total length $\tilde{l}_{\parallel }(t)$
of this step sequence along the pipes (cf. e.g., of. \cite{22})

\begin{equation}
\left\langle \tilde{l}_{\parallel }(t)\right\rangle \approx \xi \left\langle
l_{\parallel }\right\rangle \frac{t}{\tau _{l}}=\frac{\pi a^{2}v}{d^{2}}\xi t
\label{4.118}
\end{equation}
and 
\begin{equation*}
\left\langle (\tilde{l}_{\parallel }(t))^{2}\right\rangle =\left(
\left\langle l_{\parallel }^{2}\right\rangle -\xi ^{2}\left\langle
l_{\parallel }\right\rangle ^{2}\right) \frac{t}{\tau _{l}}+\left( \xi
\left\langle l_{\parallel }\right\rangle \frac{t}{\tau _{l}}\right) ^{2}=
\end{equation*}
\begin{equation}
=2D\left[ \frac{(2-\xi ^{2})\pi }{4}\left( \frac{a^{2}v}{Dd}\right) ^{2}\ln
\left( \frac{d}{a}\right) \right] t+\left\langle \tilde{l}_{\parallel
}(t)\right\rangle ^{2}.  \label{4.119}
\end{equation}
Expression (\ref{4.115}) immediately results from (\ref{4.118}) and
expression (\ref{4.119}) leads to formula (\ref{4.114}) if, in addition, we
allow for a random motion of walkers in the tissue itself and do not
distinguish between $(2-\xi ^{2})/4$ and $1/2$.

Besides, we note that for $\xi =\pm 1$, i.e. for the system of pipes with
blood flow in the same direction, we may consider the blood flow effect on
walker motion in terms of convective transport only. Indeed, for example,
according to (\ref{4.118}) and (\ref{4.119}) for $\mid \xi \mid \sim 1$ and $%
t\gg \tau _{l}$ we get $[\left\langle (\tilde{l}_{\parallel
}(t))^{2}\right\rangle -\left\langle \tilde{l}_{\parallel }(t)\right\rangle
^{2}]/\left\langle \tilde{l}_{\parallel }(t)\right\rangle ^{2}\sim (t/\tau
_{l})^{-1}$ and, thereby, in this case the random component of the walker
displacement is not a factor.

For tissue with counter current pairs these speculations also hold true.
Indeed, within replacement the term pipe by the term countercurrent pair,
the random walk representation will be the same. Here, however, we should
set $\xi =0$ and for $\left\langle l_{\parallel }^{2}\right\rangle $ use
formula (\ref{4.112}) with $(2Dt)^{2}\sim d^{2}$. then, formula (\ref{4.119}%
) immediately leads to expression (\ref{4.114}) with $\beta =0$.

\section{Vessel classification}

\label{s4.6}

In the present Section we classify vessels of various lengths and different
levels according to their ``individual'' influence on walker motion. The
term ``individual'' implies that we take into account at the same time only
vessels of one level. In other words, we do not allow for collective effect
of vessels belonging to different levels on walker motion in living tissue.
Dealing with living tissue containing unit vessels we consider vessels
individually. For living tissue with countercurrent vascular network a
single pair of countercurrent vessels should be regarded as a unit basic
structure of the vascular network architectonics.

Description of individual influence of a single unit vessel and a single
countercurrent pair is practically the same. Therefore, we will (in this
Section) focus the main attention on unit vessels. For counter current pair
solely the final results are formulated.

Let at a certain time a walker reaches a vessel of level $n$. then,within
the time $t_{n}\approx d_{n}^{2}/(2D)$ (where $d_{n}$ is the mean distance
between the $n$-th level vessels) the walker is practically located inside
the fundamental domain%
\index{fundamental domain} $Q_{n}$ containing this vessel. Since there are
practically no other vessels of level $n$ in the domain $Q_{n}$ the
individual effect of blood flow in the given vessel on the walker motion in
the domain $Q_{n}$ can be measured in terms of the mean distance $%
l_{\parallel n}$ which the walker would travel along the vessel with blood
flow in it during the time $t_{n}$, if the vessel were infinitely long. In
accordance with (\ref{4.111})

\begin{equation}
l_{\parallel i} = 
\frac{a^2_i v_i}{2D} \ln \left ( \frac{d_n}{a_i} \right ).  \label{4.120}
\end{equation}
where $a_i$ and $v_i$ are the radius of the given vessel and $v$ is the mean
velocity of blood in it.

Due to walker random motion in the cellular tissue itself, the walker
passes, in addition, a distance of order $(2Dt_n)^{1/2} \sim d_n \sim l_n$
during this time. Thus, blood flow in the given vessel will have a
substantial individual effect on the walker motion if $l_{\parallel n} \gg
l_n$. Otherwise, such an effect will be ignorable.

For each vessel of the vascular network let us introduce the parameter

\begin{equation}
\zeta _i = \frac{a^2_i v_i}{2Dl_n} \ln \left ( \frac{d_n}{a_i} \right ),
\label{4.121}
\end{equation}
which, by virtue of (\ref{4.120}), is the ratio $l_{\parallel n}/l_n$. For a
counter current pair, for example, pair $i$ of level $n$

\begin{equation}
l_{\parallel i} = \frac{a^2_i v_i}{2D} \left [\ln \left ( \frac{d_n}{a_i}
\right ) \right ]^{1/2}  \label{4.122}
\end{equation}
and 
\begin{equation}
\zeta _i = \frac{a^2_i v_i}{2Dl_n} \left [ \ln \left ( \frac{d_n}{a_i}
\right ) \right ]^{1/2}  \label{4.123}
\end{equation}

These parameters enable us to divide all the vessels in two classes
according to \ the value of $\zeta _{i}$. By definition, the first class
(Class 1) consists of vessels (counter current pairs) corresponding to the
value $\zeta _{i}>1$ and the second class (Class 2) involves vessels
(counter current pairs) for which $\zeta _{i}<1$. If a given vessel of level 
$n$ meets the condition $\zeta _{i}\ll 1$, i.e. $l_{\parallel i}\ll l_{n}$,
and there is no one vessel of the first class in the fundamental domain%
\index{fundamental domain} $Q_{n}$ containing this vessel, then,in the time $%
t_{n}$ walkers which are near the given vessel will be distributed
practically uniformly over the whole domain $Q_{n}$. In terms of heat
transfer this means that blood flowing through the given vessel has enough
time to attain thermal equilibrium with the tissue in the domain $Q_{n}$. In
other words, the temperature of blood portion during its motion through this
vessel will coincide with the tissue temperature averaged over the domain $%
Q_{n}$. Therefore, vessels of Class 2 will be called heat - dissipation
vessels. If walkers are initially near a vessel of level $n$ corresponding
to the value $\zeta _{i}\gg 1$, i.e. for which $l_{\parallel i}\gg l_{n}$,
they will go out of the domain $Q_{n}$, that contains this vessel, remaining
inside a neighborhood of this vessel whose radius is much smaller than $%
d_{n} $. In other words, due to high velocity of blood in the given vessel
its temperature has practically no time to become equal to the mean tissue
temperature in the domain $Q_{n}$. Therefore, vessels of Class 1 will be
called heat conservation vessels, because, as it will be shown in Section~%
\ref{s5.1}, we may ignore heat exchange between blood in these vessels and
the surrounding cellular tissue.

Vessels, for which $\zeta _i \sim 1$, exhibit the properties of both heat -
dissipation and heat - conservation vessels and to describe in detail their
influence on heat transfer more complicated analysis is required.
Nevertheless, as it will be shown in Section~\ref{s5.1}, we may divide all
the vessels of the vascular network in two classes according to their
influence on heat transfer by the rigorous inequalities $\zeta _i <1$ and $%
\zeta _i >1$.

The same comments regarding the vessel classification refer equally to the
counter current pairs.

In a similar way we can consider the capillary system and introduce the
corresponding parameter

\begin{equation}
\zeta _c = 
\frac{a^2_c v_c}{2Dl_c} \ln \left ( \frac{d_c}{a_c} \right ),  \label{4.124}
\end{equation}
which classifies capillaries as heat-conservation or heat-dissipation
vessels. \clearpage

\part{Transport phenomena caused by blood flow through hierarchical vascular
network}

\markboth{
{\sc \thepart.{ } Transport phenomena caused by blood flow\ldots}}{}

\chapter{The effect of different group vessels on heat transfer in living
tissue}

\label{ch.5} 
\markright
{ {\sc  \thechapter. The effect of different group\ldots}
}

In the previous Chapter we analyzed characteristics of heat transfer in
living tissue that are caused by individual effect of vessels belonging to
different levels. The present Chapter deals with the main properties that
bioheat transfer exhibits due to blood flow distribution over the whole
vessel tree. Therefore, we shall confine our consideration to uniform blood
flow distribution over vascular network.

\section{Dependence of the vessel classification parameter on hierarchy level%
}

\label{s5.1}

Since the vessel classification parameter $\zeta _{i}$ plays an important
role in description of the blood flow effect on heat transfer in this
Section we analyse in detail the classification parameter as a function of
the level number $n$ and the total blood current $J_{0}$ through the
vascular network. In the case under consideration blood current $\pi
a_{n}^{2}v_{n}$ in each vessel of one level is the same and depends on the
level number $n$ only. Thus, by virtue of (\ref{3.5}), (\ref{3.6}) and (\ref
{4.120}), (\ref{4.121}) the classification parameter $\zeta _{n}$ for unit
vessels and countercurrent pairs can be represented as

\begin{equation}
\zeta _n = G 2^{-2n} \left (\ln \frac{l_0}{a_0} \right )^{-\frac 12 [1-
\beta (n)]}  \label{5.1}
\end{equation}
where the function

\begin{equation}
\beta (n)=\left\{ 
\begin{array}{ccc}
0 & ; & n\mathbf{\leqslant }n_{cc} \\ 
1 & ; & n>n_{cc}
\end{array}
\right. \;,  \label{5.2}
\end{equation}
the dimensionless total blood current

\begin{equation}
G = \frac{J_0}{2\pi D l_0} \ln \left ( \frac{l_0}{a_0} \right ) \equiv \frac{%
4jl_0^2}{3\sqrt{3} \pi D} \ln \left ( \frac{l_0}{a_0} \right )  \label{5.3}
\end{equation}
and we have ignored the difference between $\ln (d_n/a_n)$ and $\ln
(l_0/a_0) $. The latter is justified because $l_n/a_n$ is a large parameter
whereas $l_n/d_n$ and $w(n)$ (see (\ref{3.7})) are of order unity. The
function $\beta (n)$ reflects the fact that the vascular network contains
the countercurrent pairs up to level $n_{cc}$.

Let us analyse the solution $n=n_{t}$ of the equation

\begin{equation}
\left. \zeta _{n}\right| _{n=n_{t}}=1.  \label{5.4}
\end{equation}
for various values of $G$. The classification parameter $\zeta _{n}$ as a
function of $n$ for different values of $G$ is shown in Fig.~\ref{Fig13}
where $n$ is treated as a continuous variable. As seen in Fig.~\ref{Fig13}
there can exist one $n_{t}$ or two roots $n_{t},n_{t}^{\ast }$ of equation (%
\ref{5.4}) depending on the value of $G$. The jump on the curve $\zeta (n)$
is due to countercurrent pairs vanishing at level $n_{cc}$. The dependence
of the roots on the total blood current $(G)$ is shown in Fig.~\ref{Fig13}b
and in the region $G_{1}<G<G_{2}$ involves two single-valued branches.
However, as it will become clear from the following analysis and is
discussed in detail in Chapter~\ref{ch.8} we should choose the lower root $%
n_{t}$ of equation (\ref{5.4}) because it is a connected part of the
vascular network involving the first class vessels among with the tree stem
that determine the collective effect of vessels of different levels on heat
transfer. The solid line in Fig.~\ref{Fig13}b shows the dependence of this
root $n_{t}$ on the dimensionless total blood current $G$ which can be
specified in terms of

\begin{equation}
n_{t}=\left\{ 
\begin{array}{ccc}
\frac{1}{2\ln 2}\ln \frac{G}{G_{0}}\,, & \,\text{{for}} & G_{0}\mathbf{%
\leqslant }G\mathbf{\leqslant }G_{cc}, \\ 
\frac{1}{2\ln 2}\ln G\,, & \,\text{{for}}\, & G_{cc}\mathbf{\leqslant }G,
\end{array}
\right.  \label{5.5}
\end{equation}
where $G_{0}=(\ln (l_{0}/a_{0}))^{1/2}$ and $G_{cc}=2^{2n_{cc}}(\ln
(l_{0}/a_{0}))^{1/2}$.

As it will be shown below, heat exchange between blood and the cellular
tissue is directly controlled by vessels of level $n_{t}$. Therefore, when $%
G<G_{0}$ blood flow has practically no effect on heat transfer and for this
reason we shall examine only the case

\begin{equation}
G\gg G_{0}.  \label{5.6}
\end{equation}
In particular , for typical values of $l_{0}\sim 5cm,\,c_{t}\sim 3.5J/g\cdot
K,\,\rho _{t}\sim g/cm^{3},\,\kappa \sim 7\cdot 10^{-3}J/s\cdot cm\cdot K$,
the mean blood flow rate $j\sim 10^{-3}s^{-1}$, and setting $l_{0}/a_{0}\sim
40$ from (\ref{5.3}) we find $G\sim 10$. For kidney, where blood flow is of
the highest level \cite{46} setting $j\sim 10^{-2}s^{-1}$ we get $G\sim 100$.

\FRAME{ftbpFU}{4.1425in}{2.1508in}{0pt}{\Qcb{ The classification parameter $%
\protect\zeta _{n}$ as a function of $n$ for different values $%
G_{1}<G_{2}<G_{3}$ (a) and the solution $n_{t}$ as a function of the
dimensionless total blood current $G$ (b).($n_{st}=n_{cc}+\frac{1}{2}\ln \ln 
\frac{l_{0}}{a_{0}};$ {$G_{0}=\left( \ln \frac{l_{0}}{a_{0}}\right) ^{1/2};$ 
$\ G_{cc}^{\ast }=2^{2n_{cc}};\;\;G_{cc}=2^{2n_{cc}}(\ln
(l_{0}/a_{0}))^{1/2})$;.}$G_{0}=\left( \ln \frac{l_{0}}{a_{0}}\right)
^{1/2}; $\ \ $G_{cc}^{\ast }=2^{2n_{cc}};$\ \ $G_{cc}=2^{2n_{cc}}(\ln
(l_{0}/a_{0}))^{1/2}$)..}}{\Qlb{Fig13}}{Fig13}{\special{language "Scientific
Word";type "GRAPHIC";maintain-aspect-ratio TRUE;display "USEDEF";valid_file
"F";width 4.1425in;height 2.1508in;depth 0pt;original-width
15.0278in;original-height 7.7686in;cropleft "0";croptop "1";cropright
"1";cropbottom "0";filename 'FIG13.GIF';file-properties "XNPEU";}}

In this Chapter we shall show that the vessels of the vascular network
described in Chapter~\ref{ch.2} can be divided, in principle, into four
possible groups in accordance with their influence on heat transfer. These
groups are the countercurrent pairs of Class 1 and the veins of Class 1 for
which level number $n\mathbf{\leqslant }n_{t}$ and countercurrent pairs of
Class 2 and, may be, the first class veins for which $n_{cc}\mathbf{%
\leqslant }n<n_{t}$; the arteries of Class 1; the arteries,veins; and the
capillary system. Below their influence on heat transfer will be considered
individually.

\section{The effect of the heat conservation countercurrent pairs and veins
on heat transfer}

\label{s5.2}

Let us, first, consider the characteristic properties of the walker motion
that are caused by the first class veins. If at a certain time a walker
reaches a vein of level $n$ for which $\zeta _{n}\gg 1$ then,it will be
transported by blood flow in this vein into a small (in comparison with $%
Q_{n}$) neighborhood of a vein belonging to level ($n-1$) and connected with
the given one through the corresponding branching point. This is the case
because the mean time $\tau _{b_{n}}$ of this process satisfies the
inequality $\tau _{b_{n}}\ll d_{n}^{2}/(2D)$. Indeed, when $1\ll \zeta
_{n}\ll \ln (d_{n}/a_{n})$ and, thereby, the time $\tau _{b_{n}}\gg
d_{n}^{2}/(2D)$ (see (\ref{5.7})) its value can be estimated by setting $%
t=\tau _{b_{n}}$ and $<l_{\parallel }(t)>=l_{n}$ in expression (\ref{4.111}%
). In this way taking into account (\ref{4.121}) we obtain

\begin{equation}
\tau _{bn} \sim \frac{a^2_n}{2D} \exp \left \{ \frac{2\ln ( d_n/a_n )}{\zeta
_n} \right \} \equiv \frac{d^2_n}{2D} \exp \left \{ - \frac{(\zeta _n -1)}{%
\zeta _n} 2 \ln \left ( \frac{d_n}{a_n} \right ) \right \} .  \label{5.7}
\end{equation}
For $\zeta _n \gg \ln (d_n/a_n)$ there are two possible cases differing in
way by which the walker goes into the given vein for the first time. When
the walker goes into this vein with blood flow from a vein of level $(n+1)$
through the corresponding branching point it immediately arrives at central
points of the given vein. Clearly, in this case the value of $\tau _{bn}$
can be estimated as

\begin{equation}
\tau _{bn} \sim \frac{l_n}{v_n} \equiv \frac{a^2_n}{2D\zeta _n} \ln \left ( 
\frac{d_n}{a_n} \right )  \label{5.8}
\end{equation}
because during the time $\tau _{bn} \ll a^2_n /(2D)$ the walker practically
cannot reach the boundary of this vein. If the walker goes into the vein
through its boundary, the walker during its motion along this vein will be
located inside the boundary neighborhood whose thickness is about $(2D\tau
_{bn})^{1/2} \ll a_n$, alternately going inside the vein and in the cellular
tissue. Assuming that at vessel boundaries the blood velocity is equal to
zero and inside vein and arteries of microcirculatory beds the velocity
field of blood flow is approximately specified by parabolic distribution
over the vessel cross section, we can estimate the blood velocity averaged
over the given vein boundary neighborhood as $<v> = v_n (2D \tau
_{bn})^{1/2} /a_n$. Thus, in this case

\begin{equation}
\tau _{bn} \sim \frac{l_n}{<v>} = \frac{a^2_n}{2D} \left [ \frac{1}{\zeta _n}
\ln \left (\frac{d_n}{a_n} \right ) \right ]^{2/3}  \label{5.9}
\end{equation}
So, in the three cases the time $\tau _{bn} \ll d^2_n /(2D)$.

By virtue of (\ref{5.1}) for $n>n_{cc}$ the value $(\ln (d_{n}/a_{n}))/\zeta
_{n}=(2\pi Dl_{0}/J_{0})2^{2n}$. Thereby, according to (\ref{5.7}) - (\ref
{5.9}), the ratio $a_{n}^{2}/(2D\tau _{bn})$ increases as $n$ decreases.
Thus, for all the first class veins of the given group $\tau _{bn}\ll
a_{n}^{2}/(2D)$ except the veins whose level number $n$ satisfies the
inequality $n_{\ast }<n<n_{t}$ for which $\tau _{bn}\gg a_{n}^{2}/(2D)$.
Here the value $n_{\ast }$ is specified by the condition $(2D\tau
_{bn})/a_{n}^{2}\sim 1$ or, what is practically the same according to (\ref
{5.7}) - (\ref{5.9}), by the equality $\zeta _{n_{\ast }}=\ln (d_{n_{\ast
}}/a_{n_{\ast }})$. From the latter equality and expressions (\ref{5.1}), (%
\ref{5.5}) we get

\begin{equation}
n_* \approx n_t - \frac{1}{2 \ln 2} \ln \left [ \ln \left (\frac{l_0}{a_0}
\right ) \right ]  \label{5.10}
\end{equation}
In particular, for $l_0/a_0 \sim 40$ the value $(n_t - n_*) \sim 1$.

Let us show that a walker, being initially near a first class vein of level $%
n$ for which $\zeta _{n}\gg 1$, can be found in the cellular tissue only
within a time of order $\tau _{bn}$. To do this we consider the two
following possible cases.

\FRAME{ftbpFU}{5.8167cm}{5.485cm}{0pt}{\Qcb{Schematic representation of
walker trapping by the first class veins. On intersecting the dashed region
a walker gets the central points of the next level veins and never returns
to the cellular tissue.}}{\Qlb{Fig6_2}}{Fig6_2}{\special{language
"Scientific Word";type "GRAPHIC";maintain-aspect-ratio TRUE;display
"USEDEF";valid_file "F";width 5.8167cm;height 5.485cm;depth
0pt;original-width 8.8513in;original-height 8.3333in;cropleft "0";croptop
"1.0006";cropright "0.9992";cropbottom "0";filename
'FIG6_2.GIF';file-properties "XNPEU";}}

A walker, being initially close to the boundary of a vein whose level number 
$n<n_{\ast }$ is bound to arrive at central points of the vein system
including several nearest levels $(n-1),(n-2)$, etc. To justify this
statement we consider the walker motion in living tissue with a real vessel
system made of two-fold branching points (Fig.~\ref{Fig6_2}). For the eight
- fold branching point model of the vascular network this statement is borne
out to greater accuracy. In this case $\tau _{bn}\ll a_{n}^{2}/(2D)$,
therefore, if the walker starts in the vicinity of the dashed region of the
vein boundary, then,the walker will remain in it during its motion along
this vein and after passing through the branching point it will get the
central points. The probability of this event is about $1/2$. If the walker
starts near the other side of the vein boundary, then,it will get in the
neighborhood of the dashed side of the $(n-1)$-th level vein with
probability $1/2$. In this case after passing through the corresponding
branching point the walker gets the central points again. The total
probability for the walker to reach the central points after passing through
these two branching points is approximately $3/4$. It is obvious that after
passing $m^{\prime }$ branching points the probability of getting the
central points is about $(1-(2)^{-m^{\prime }})$.

So, within the framework of the eight-fold branching point model for the
vascular network it is natural to assume that the walker after passing $%
m^{\prime}$ branching points arrives at the central points with probability $%
(1-(2)^{-3m^{\prime}})$ because the eight-fold branching points represent
practically the fragment of real vascular network containing three two-fold
branching points. Therefore, the walker practically gets the central points
after passing several branching points. Besides, due to $\zeta _n$ and $a_n$
depending on $n$ approximately as $2^{2n}$ and $2^{-n}$, respectively,
according to (\ref{5.9}) in this case $\tau _{bn}$ varies with $n$ as $%
2^{-(2n)/3}$ and the mean total time of reaching the central points by the
walker is about $\tau _{bn}$. By virtue of (\ref{5.7}) for walkers moving
inside veins the ratio $a^2_n / (2D \tau _{bn})$ varies with $n$ practically
as $2^{-2n}$. Therefore, after getting the central points the walker never
returns to the cellular tissue again and leaves the microcirculatory bed
domain $Q_0$ through the host vein with blood flow, because it has
practically no time to reach vein boundaries during its motion with blood
flow through the vein system.

As will be seen from the results obtained below in this Section, vessels of
other types have no significant effect on the walker motion in the vicinity
of a vein whose level number $n^{\ast }<n<n_{t}$. For example, according to (%
\ref{4.111}) and (\ref{4.113}), on temporal scales $t\gg a_{n}^{2}/(2D)$ the
effect of blood flow in such a vein on the walker motion is practically
determined by the blood current $\pi a_{n}^{2}v_{n}$ in it rather than the
mean blood velocity $v_{n}$ and the vessel radius $a_{n}$ individually.
Indeed, by virtue of (\ref{4.36}), in this case the mean distance $%
<l_{\parallel }(t)>$ travelled by a walker along the vein depends on $n$
through the term $a_{n}^{2}v_{n}$ as $2^{-3n}$, whereas the logarithmic
cofactor leads to a linear dependence only. Therefore, to show that a
walker, which is initially near a vein whose level number $n_{\ast }<n<n_{t}$%
, reaches a vein of level $n_{\ast }$ in a time $\tau _{bn}^{\prime }$ of
order $\tau _{bn}$ we may consider its motion in the vicinity of the
following inhomogeneous pipe. We assume that this imaginary pipe of radius $%
a_{n}$ involves different parts of lengths $l_{n},\,l_{n-1},\,l_{n-2},\ldots
,l_{0}$, where the mean blood velocity is equal to $u_{n}=v_{n},%
\,u_{n-1}=v_{n-1}(a_{n-1}/a_{n})^{2},\,u_{n-2}=v_{n-2}(a_{n-2}/a_{n})^{2},%
\ldots $ The time required for the walker to pass the whole length of this
pipe in its motion with blood in the pipe is about

\begin{equation}
\sum_{n^{\prime }=0}^{n}\frac{l_{n^{\prime }}}{u_{n^{\prime }}}%
=a_{n}^{2}\sum_{n^{\prime }=0}^{n}\frac{l_{n^{\prime }}}{v_{n^{\prime
}}a_{n^{\prime }}^{2}}\sim \frac{l_{n}}{v_{n}},  \label{5.11}
\end{equation}
where we have taken into account (\ref{3.5}) and (\ref{3.6}). According to (%
\ref{4.109}) the mean residence time$\;\left\langle t_{p}\right\rangle $ of
the walker inside the pipe during its motion in the vicinity of this pipe
for the time $t\gg a_{n}^{2}/(2D)$ can be estimated as $a_{n}^{2}/(2D)\ln
[(2Dt)^{1/2}/a_{n}]$. In addition, as it follows from comparison between (%
\ref{4.111}) and (\ref{4.113}) the dispersion of the walker residence time $%
t_{p}$ is about the square of its mean value. Therefore, by virtue of (\ref
{5.11}), the desired time $\tau _{bn}^{\prime }$ in which the walker reaches
a vein of level $n_{\ast }$ can be found from the expression

\begin{equation}
\frac{l_{n}}{v_{n}}\sim \frac{a_{n}^{2}}{2D}\ln \left[ \frac{(2D\tau
_{bn}^{\prime })^{1/2}}{a_{n}}\right] .  \label{5.12}
\end{equation}
We get formula (\ref{5.7}) again, and, thus, $\tau _{bn}^{\prime }\sim \tau
_{bn}$.

After reaching a vein of level $n_*$ the walker will arrive at central
points of veins whose level number $n<n_*$ in a time of order $\tau
_{bn_{*}} $, which, by virtue of (\ref{5.9}), is well below the time $\tau
_{bn} : \tau _{bn_{*}} \ll \tau _{bn}$. So, in the given case the time
required for the walker to get central points of the vessel system involving
veins of levels $n<n_*$, where upon it hardly ever returns to the cellular
tissue, is about $\tau ^{\prime}_{bn} + \tau _{bn^{*}} \sim \tau _{bn}$.

Another time scale characterizing the influence of blood flow in the first
class veins on the walker motion is the mean time $\tau _{ln}$ during which
a walker can go in the cellular tissue without touching the veins of level $%
n $ provided all other vessels have no effect on its motion. From the
viewpoint of walker motion in the cellular tissue near a vein of level $n$,
this vein can be regarded as an infinitely long pipe. In addition, every
fundamental domain%
\index{fundamental domain} $Q_{n}$ of level $n$ contains just one vein of
the same level, with its characteristic spatial size being about the vein
length. Therefore, to estimate the value of the time $\tau _{l_{n}}$ we may
conceive the veins of level $n$ as a collection of pipes forming a hexagonal
array for which the mean distance between the pipes coincides with the mean
distance $d_{n}$ between these veins. So, according to (\ref{4.110}), we get

\begin{equation}
\tau _{l_{n}}\sim 
\frac{d_{n}^{2}}{2\pi D}\ln \left( \frac{d_{n}}{a_{n}}\right) .  \label{5.13}
\end{equation}
If we had taken into account veins of Class 2 and arteries then, their
influence on such walker motion could not significantly modify expression (%
\ref{5.13}) (see the next Section).

The results obtained above allow us to regard the first class veins for
which $\zeta \gg 1$ as walker traps. Indeed, on one hand, if wandering
through the cellular tissue a walker crosses one of these veins, for
example, a vein of a level $n$ then,practically after the time $\tau _{bn}$
it will arrive at central points of veins whose level number $n^{\prime
}<n_{\ast }$ whereupon the walker hardly ever returns to the cellular
tissue. According to (\ref{5.7})-(\ref{5.9}) the time $\tau _{bn}\ll
d_{n}^{2}/(2D)$. On the other hand, as it follows from (\ref{5.13}), the
mean time, after which a walker created in the cellular tissue can reach a
vein of a level $n$, satisfies the inequality $\tau _{l_{n}}>d_{n}^{2}/(2D)$%
. Therefore, a walker may be thought of as arriving at the central points in
no time ( in comparison with $d_{n}^{2}/(2D)$) after crossing one of the
first class veins. In other words, these veins may be considered to be
walker traps.

Capillaries can exert some effect on the walker motion in the tissue.
However, when their effect is significant and thereby expression (\ref{5.13}%
) should be modified, capillaries themselves transport walkers practically
inside veins for which $n<n_*$. Thus, also in this case the first class
veins may be regarded as walker traps.

In accordance with (\ref{5.7}) veins of level $n\sim n_{t}$, i.e. for which $%
\zeta _{n}>1$, meet the inequality $\tau _{bn}<d_{n}^{2}/(2D)$. However,
owing to expression (\ref{5.13}) containing the logarithmic cofactor $\ln
(d_{n}/a_{n})$ treated in the given model as a large value, we may assume
that $\tau _{bn}\ll \tau _{l_{n}}$ for such veins too. Therefore, veins of
levels $n\sim n_{t}$ also can be regarded as walker traps. The latter allows
us to treat all the veins for which $\zeta _{n}>1$ as walker traps and to
divide all the veins of the venous bed in Class 1 and Class 2 by the
rigorous inequalities $\zeta _{n}>1$ and $\zeta _{n}<1$ respectively.

Below in this Section we shall ignore the effect of blood flow in the second
class countercurrent pair veins, arteries and capillaries on the walker
motion in the tissue. Due to the relative volume of the vascular network
being small value the mean time during which a walker can go in the cellular
tissue without touching the first class veins is determined by the expression

\begin{equation}
\frac{1}{\tau }\sim \sum_{n=0}^{n_{t}}\frac{1}{\tau _{l_{n}}}  \label{5.14}
\end{equation}
where $n_{t}$ is regarded as a cutoff parameter. The value $d_{n}$ depends
on $n$ as $2^{-n}$ whereas $\ln (d_{n}/a_{n})$ is a smooth function of $n$.
So, by virtue of (\ref{5.13}), the main contribution to sum (\ref{5.14}) is
associated with the last term, i.e. $\tau \sim \tau _{ln_{t}}$. Therefore,
although the walker can be trapped by any vein of Class 1 the veins of the
smallest length in this group, i.e. that of level $n_{t}$ play the main role
in walker trapping. The role of the larger veins is practically to transport
trapped walkers with blood flow to the host vein. The $n_{t}$-th level veins
possess the properties of traps as well as fast migration paths in the
cellular tissue with the latter being the characteristic property of the
second class vessels. Nevertheless, the veins of level $n_{t}$ trap only and
assuming that walker trapping is controlled by these veins set $\tau =\tau
_{ln_{t}}$.

Keeping in mind these assumptions and the definition of the blood flow rate $%
j$ described in Chapter~\ref{ch.3}, from (\ref{5.13}) we get

\begin{equation}
\tau =\frac{d_{n_{t}}^{2}}{2\pi D}\ln \frac{d_{n_{t}}}{a_{n_{t}}}=\frac{1}{j}%
,  \label{5.15a}
\end{equation}
where we have also taken into account the relationship $%
V_{n_{t}}=l_{n_{t}}d_{n_{t}}^{2}$ and represented the blood flow rate, i.e.
the volume of blood flowing through the unit tissue volume per unit time, in
terms of $j=\pi a_{n_{t}}^{2}v_{n_{t}}/V_{n_{t}}$.

In a similar way we can analyse influence of the first class countercurrent
pairs. In particular, we find that these countercurrent pairs may be treated
as walker traps and the life time of walker migration in the cellular tissue
is

\begin{equation}
\tau _{l}=\frac{d_{n_{t}}^{2}}{2\pi D}\ln \frac{d_{n_{t}}}{a_{n_{t}}}=\frac{1%
}{j}\sqrt{\ln \frac{l_{0}}{a_{0}}}  \label{5.15b}
\end{equation}
where we have taken into account formula (\ref{5.1}). Depending on the
relation between the values of $n_{t}$ and $n_{cc}$ either expression (\ref
{5.15a}) or expression (\ref{5.15b}) describes the total effect of blood
flow in these first class vessels on heat transfer.

In conclusion of this Section we point out that according to (\ref{5.15a}),(%
\ref{5.15b}) during the lifetime $\tau $ a walker can pass in the tissue a
distance of order $\sqrt{6D\tau }\sim d_{n_{t}}\sqrt{\ln
(d_{n_{t}}/a_{n_{t}})}$. Since in the given model $\ln (d_{n}/a_{n})$ is
treated as a large parameter this distance is considered to be well about
the mean distance between the veins of level $n_{t}$ which control walker
trapping. So, a walker in its motion in the cellular tissue has possibility
of crossing not only the first nearest veins but any vein in the tissue
domain whose radius is about $(6D\tau )^{1/2}$. Due to $\ln
(d_{n_{t}}/a_{n_{t}})$ being a large parameter such a domain contains a
large number of veins belonging to level $n_{t}$. Thereby, specific details
of their spatial arrangement in the given domain are of little consequence
and solely the mean properties of their distribution in this domain, in
particular, the mean distance between the $n_{t}$-th level veins, have the
main effect on heat transfer in living tissue. The given characteristics of
the walker motion is the ground of the self-averaging property conservation
vessels. The effect of the specific details of the vein arrangement on heat
transfer can be described in terms of random spatial nonuniformities of the
tissue temperature, which is the subject of Chapter~\ref{ch.14}.

It is also should be noted that according to Chapter~\ref{ch.4} all the
first class vessels are called heat conservation vessels, although this is
rigorously justified in reference, for example, to veins whose level number $%
n<n_{\ast }$. Indeed, for such veins $2D\tau _{bn}\ll a_{n}^{2}$ and,
thereby, a trapped walker cannot return to the cellular tissue during its
motion with blood through these veins. In terms of heat transfer blood
flowing through such a vein has no time to attain thermal equilibrium with
cellular tissue surrounding it. In this case we may ignore heat exchange
between blood and the surrounding cellular tissue when analyzing
distribution of heat currents over the vessel system involving such a first
class vein. For a vein whose level number $n_{\ast }<n<n_{t}$ the value $%
2D\tau _{bn}\gg a_{n}^{2}$ and blood flowing through it has enough time to
attain thermal equilibrium with the cellular tissue in a small neighborhood
of this vein. Nevertheless, along the vein transport of walkers located in
this neighborhood is mainly caused by blood flow in the given vein rather
than by random motion of walkers in the tissue. Thus, for such veins the
total current of the walkers along a given vein, i.e. the total heat current
associated with this vein, can be estimated as $\pi a_{n}^{2}v_{n}C_{n}$ (or 
$\pi a_{n}^{2}v_{n}\rho _{t}c_{t}(T_{n}^{\ast }-T_{a})$), where $%
C_{n}(T_{n}^{\ast })$ is the mean walker concentration (the mean blood
temperature) in the given vein. Therefore, distribution of heat flow over
such veins is also controlled by blood flow in the vessels. The latter
allows us to refer to these veins as heat conservation vessels too.

These comments with respect to the heat conservation countercurrent pair
also hold true.

It should be pointed out that from the standpoint of heat transport with
blood through vessels the division of vessels into the countercurrent pairs
and unit arteries and veins is justified solely by vessels whose level
number $n\sim n_{t}$. Indeed, when $\zeta _{n}>\ln (l_{0}/a_{0})$ and, thus, 
$a_{n}^{2}v_{n}/(2D)>l_{n}$ a walker inside this vessel has no time to reach
the vessel boundary during its motion with blood through this vessel. In
other words, blood flow through this vessel practically does not lose its
energy due to heat exchange with the surrounding cellular tissue. Solving
the equation $\zeta _{n}\mid _{n^{\ast }}=1$ we find that $n_{t}-n_{\ast
}\sim 1$ (see \ref{5.10}).

Therefore, the countercurrent pairs whose level number $n<n_{\ast }$
practically consist of arteries and veins for which heat exchange is not
essential.

\section{The effect of heat conservation arteries on the walker motion}

\label{s5.3}

The next group which we shall consider is the arteries of Class 1. In the
given model due to both the arterial and venous beds of the same geometry,
this group involves all the arteries, whose level number $n<n_{t}$. As in
the case analyzed in Section~\ref{s5.2}, if a walker crosses a boundary of
one of these arteries, for example, an artery of level $n$, it will be
transported by blood flow to an artery of the next level within the time $%
\tau _{bn}\ll d_{n}^{2}/(2D)$. However, because of the opposite direction of
blood flow in the arterial system in comparison with the venous one, the
walker will be transported to arteries whose level numbers are $%
(n+1),(n+2),(n+3),\ldots $ until it reaches an artery of level $n_{t}$.

After reaching such an artery the walker leaves the first class arteries and
begins to migrate randomly in the tissue again because arteries whose level
number $n> n_t$ cannot substantially affect heat transfer.

Therefore, if the walker crosses a boundary of an $n$-th level artery,
then,for a time of order $\tau _{bn}\ll \tau $ it will be transported by
blood flow over a distance being approximately equal to $l_{n}$. Indeed, as
it can be seen from Fig.~\ref{Fig15} travelling along a possible path in
blood stream (the directed line) from the given artery to arteries of level $%
n_{t}$ the walker can merely arrive at points which are inside the
fundamental domain $Q_{n}$ of volume $(2\l _{n}/\sqrt{3})^{3}$ that contains
the given $n$-th level artery. In addition, due to the geometric structure%
\index{geometric structure} of the given vascular network, walkers initially
crossing this artery will be uniformly distributed over the domain $Q_{n}$
in a time of order $\tau _{bn}$.

\FRAME{ftbpFU}{6.1989cm}{5.6321cm}{0pt}{\Qcb{Geometry of a possible walker
path (directed line) in the system of the first class arteries.}}{\Qlb{Fig15}%
}{Fig15}{\special{language "Scientific Word";type
"GRAPHIC";maintain-aspect-ratio TRUE;display "USEDEF";valid_file "F";width
6.1989cm;height 5.6321cm;depth 0pt;original-width 7.3336in;original-height
6.6573in;cropleft "0";croptop "1.0003";cropright "1.0003";cropbottom
"0";filename 'FIG15.GIF';file-properties "XNPEU";}}

When the capillary system has no substantial effect on heat transfer, we may
assume that after reaching the $n_{t}$ -th level arteries the walker
randomly goes in the cellular tissue until it crosses a boundary of an
artery or a vein whose level number $n\sim n_{t}$. In this case the walker
meets an artery of level $n_{t}$ as well as a vein of the same level with
the probability about 1/2. The latter follows from the fact that, on one
hand, the probability for a walker to meet larger arteries or veins is small
enough because the mean distances between the arteries whose level number $%
n\ll n_{t}$ are well above $d_{n_{t}}$ and, on the other hand, the influence
of the second class vessels on the random walker is not significant. A
possible effect of the capillary system on the random walker can solely
decrease the probability of meeting arteries (see Section~\ref{s5.5}). So,
after reaching an artery of level $n<n_{t}$ for the first time and before
being trapped by one of the $n_{t}$ -th level veins and then,being carried
away by blood flow from the microcirculatory bed domain, the walker has the
possibility of meeting a few arteries whose level number is about $n_{t}$.
Each meeting with such an artery gives rise to walker displacement with
blood flow over the distance of order $l_{n_{t}}$. Thus, after meeting a
vein of level $n<n_{t}$ and before being trapped in a time of order $\tau $
the walker can pass in the tissue a distance of order $(l_{n}^{2}+\tilde{g}%
l_{n_{t}}^{2}+6D\tau )^{1/2}$, where the constant $\tilde{g}\sim 1$.
Therefore, the first class arteries merely cause fast migration of walkers,
and their influence on the walker motion can be described in terms of an
effective diffusion coefficient $D_{\mathrm{eff}}$. Indeed, on the average,
the probability for a walker to meet an artery of level $n<n_{t}$ before
being trapped by the first class veins is small enough because for such
arteries $d_{n}\gg d_{n_{t}}$. So, when describing the effect of the first
class arteries on the walker motion solely arteries whose level number $%
n\sim n_{t}$ should be taken into account within the framework of the
countercurrent model. In this case the effective diffusion coefficient can
be estimated as $D_{\mathrm{eff}}\sim (\tilde{g}l_{n_{t}}^{2}+6D\tau )/6\tau
\sim D[1+\tilde{g}/\ln (d_{n_{t}}/a_{n_{t}})]$, where the constant $\tilde{g}
$ is also about one and expressions (\ref{5.15a}), (\ref{5.15b}) have been
taken into account. The renormalization coefficient $(1+\tilde{g}/\ln
(d_{n_{t}}/a_{n_{t}}))$ is of order unity, therefore, on the average, the
first class arteries have no substantial direct effect on heat transfer.
However, on scales smaller than $l_{n_{t}}$ these arteries can give rise to
appearance of marked variations in the temperature field.

Effect of blood flow in large vessels on bioheat transfer requires an
individual investigation and cannot be rigorously described within the
framework of the continuum model. Large vessels should be taken into account
separately as much as possible \cite{16,19,45}.

\section{Influence of the heat dissipation vessels on the walker motion}

\label{s5.4}

The following group, which we shall consider, is the vessels of Class 2. By
virtue of (\ref{5.3}) and (\ref{5.5}) this group involves all vessels whose
level number $n>n_{t}$. To investigate the influence of this vessel group on
heat transfer we analyse random walks in the vicinity of these vessels, for
example, in the vicinity of a vessel belonging to level $n$.

If a walker is near this vessel at initial time, it will be inside its
neighborhood of radius $d_{n}$ within the time $d_{n}^{2}/(2D)$. During this
time (as it has been discussed in Section~\ref{s5.3}) the walker can be
transported by blood flow over the distance $l_{\parallel n}$ (see formula (%
\ref{4.73})) which is substantially smaller than the vessel length $%
l_{n}:l_{\parallel n}\ll l_{n}$. The latter allows us to ignore the
influence of such a vessel on the random walks in its neighborhood because
the distance $l_{\parallel n}$ which the walker passes inside the vessel is
small in comparison with the distance that it passes in the tissue for the
time $d_{n}^{2}/(2D)$.

In a time of order $d_n^2/(2D)$ the walker reaches points being at a
distance of order $l_n$ from the vessel and practically never comes back to
this vessel. Therefore, the main amount of the second class vessels has no
significant effect on heat transfer.

The exception is the arteries, veins, and may be the countercurrent pairs
whose level number $n\approx n_{t}$ for which $l_{\parallel n}\mathbf{%
\leqslant }l_{n_{t}}$. Such vessels may have the cooperative effect on heat
transfer only, which may be described in terms of a continuum medium with
the effective diffusion coefficient%
\index{effective diffusion coefficient} $D_{\mathrm{eff}}$. Due to vessels
being oriented randomly in space on scale of order $l_{n_{t}}$ the mean
blood velocity$\;\left\langle \mathbf{v}(\mathbf{r},t)\right\rangle $ is
equal to zero: $\left\langle \mathbf{v}(\mathbf{r},t)\right\rangle =0$.

In order to estimate the value of $D_{\mathrm{eff}}$ we can make use of the
results obtained in Sections~\ref{s4.2} and \ref{s4.3}, because $%
l_{\parallel n}\mathbf{\leqslant }l_{n}$ and the mean distance between
vessels of one level is about their length. Expressing $a_{n}^{2}v_{n}/(2D)$
as a function of $\zeta _{n}$ from (\ref{4.121}) and (\ref{4.123})
substituting into (\ref{4.114}), and summing up influence of all the vessels
whose level number $n\geq n_{t}$, we get

\begin{equation}
D_{\mathrm{eff}}=D\left\{ 1+%
\frac{\pi 3\sqrt{3}}{4}\frac{1}{\ln (l_{0}/a_{0})}\sum_{n\geq
n_{t}}^{N}\zeta _{n}^{2}\right\} .
\end{equation}
where expression (\ref{3.11}) has been also taken into account. Assuming
that $(N-n_{t}\gg 1)$ and accounting for number (\ref{5.1}) we obtain

\begin{equation}
D_{\mathrm{eff}}=D\left\{ 1+\pi \sqrt{3}\frac{1}{\ln (l_{0}/a_{0})}\right\}
\label{5.16a}
\end{equation}
It should be pointed out that this value of $D_{\mathrm{eff}}$ is
independent of the total blood current through the vascular network which is
actually determined by its architectonics only due to the self - similarity
of the vascular network at various levels. However, there is an exception to
the latter conclusion when the dimensionless total blood current $G$ belongs
to the interval $[G_{cc}^{\ast },G_{cc}]$ where $G_{cc}^{\ast }=2^{2n_{cc}}$
and $G_{cc}=2^{2n_{cc}}(\ln l_{0}/a_{0})^{1/2}$ (see Fig.~\ref{Fig13}). For
these values of $G$ Class 2 contains heat - conservation veins and arteries
whose level number $n$ meets the condition $n_{cc}\mathbf{\leqslant }n%
\mathbf{\leqslant }n_{t}^{\ast }$ where $n_{t}^{\ast }$ is the second root
of equation (\ref{5.4}). These vessels are separated from the tree stem by
heat dissipation countercurrent pairs, so, their influence on heat transfer
is reduced to renormalization of the diffusion coefficient only. Then,
replicating practically one-to-one the latter calculations we obtain

\begin{equation}
D_{\mathrm{eff}}=D\left\{ 1+\pi \sqrt{3}\frac{1}{\ln (l_{0}/a_{0})}\left( 
\frac{G}{G_{cc}^{\ast }}\right) ^{2}\right\}  \label{5.16b}
\end{equation}
for $G_{cc}^{\ast }\mathbf{\leqslant }G\mathbf{\leqslant }%
G_{cc}=G_{cc}^{\ast }[\ln (l_{0}/a_{0})]^{1/2}$.

\section{Influence of the capillary on the walker motion}

\label{s5.5}

Finally, let us analyse the effect of the capillary system on heat transfer.
By virtue of (\ref{3.8}), (\ref{4.121}), and (\ref{4.124}), the
classification parameters $\zeta _{N}$ and $\zeta _{c}$ for the last level
vessels and capillaries are related as

\begin{equation}
\zeta _{c}=\frac{l_{N}}{ml_{c}}\frac{\ln (d_{c}/a_{c})}{\ln (d_{N}/a_{N})}%
\zeta _{N}.
\end{equation}
In the adopted model for the vascular network $l_{N}<l_{c}$ and $m\gg 1$,
thus, according to the latter expression we may assume that $\zeta
_{c}<\zeta _{N}$. So, if capillaries belonged to Class 1, i.e. $\zeta _{c}>1$%
, then,all the vessels would be heat-impermeable, because in this case the
inequalities $\{\zeta _{n}>1\}$ would be true for all the levels. In other
words, in this case blood flowing through the given microcirculatory bed
could not attain thermal equilibrium with the cellular tissue and from the
standpoint of thermoregulation such a high blood flow rate would not be
justified. For this reason we shall only examine the case $\zeta _{c}\ll 1$,
i.e. due to (\ref{3.8}) and (\ref{4.124})

\begin{equation}
\zeta _{c}=\frac{J_{0}}{2\pi mMDl_{c}}\ln \left( \frac{d_{c}}{a_{c}}\right)
\ll 1.  \label{5.17}
\end{equation}
Thereby, the capillaries are assumed to belong to Class 2. However, their
effect on heat transfer can be significantly in contrast to the arteries and
veins of Class 2. Indeed, the mean distance $d_{c}$ between the capillaries
is well below their length $l_{c}$. Therefore, the characteristic distance $%
l_{\parallel n}$, over which a walker is transported along a capillary by
blood flow before going out of its neighborhood of radius $d_{c}$, can be
substantially larger than $d_{c}(l_{\parallel c}\gg d_{c})$ although $%
l_{\parallel c}\ll l_{c}$.

First, we shall estimate $l_{\parallel c}$. Taking into account expressions (%
\ref{3.8}), (\ref{4.111}), (\ref{4.121}) and setting $t \sim d^2_c /(2D)$ we
get the desired value

\begin{equation}
l_{\parallel c} \sim \frac{a^2_cv_c}{2D} \ln \left ( \frac{d_c}{a_c} \right
) = l_N \frac{\zeta _N}{m} \frac{\ln (d_c/a_c)}{\ln (d_N/a_N)} .
\label{5.18}
\end{equation}
Now let us consider a certain vessel of level $N$, for example, a venule $i$
and its neighborhood $Q_i(d_N)$ whose diameter is about $d_N$. The system of
capillaries, or more correctly, of their portions being in this neighborhood
can be divided into two subgroups.

The first subgroup consists of all capillary portions connected directly
with the given venule, the second subgroup involves the rest. Blood currents
in the first subgroup capillaries is believed to be directed along the
radius $\mathbf{r}$ to the center of the neighborhood, i.e. to the venule.
Blood currents in the second subgroup capillaries as well as the
orientations of these capillaries in space are assumed to be random.
Therefore, the effects of the two subgroups on heat transfer are different
and we shall analyse them individually. In particular the influence of the
second subgroup capillaries can be described in terms of the effective
diffusion coefficient $D_{\mathrm{eff}}$.

Since each capillary keeps its spatial direction within the scale $\lambda >
l_N$ (see Chapter~\ref{ch.2}) the first subgroup comprises $m$ rectilinear
capillary portions of the length $d_N$ and is of the CB structure form (Fig.~%
\ref{Fig5}).

The quantity to be used in the following analysis is the mean distance $%
d_{I}(r)$ between the first subgroup capillaries at distance $r$ from the
venule $i$. The value $d_{I}(r)$ is defined as follows. Let us consider a
cylindrical neighborhood $Q_{i}(r)$ of the given venule whose radius is $%
r<d_{N}$. The first subgroup capillaries must intersect its boundary at $m$
points. By definition, $d_{I}^{2}(r)$ is the mean area of the neighborhood
boundary that falls on one intersection point. Therefore, the mean distance $%
d_{I}(r)$ satisfies the relation $2\pi rl_{N}\sim m[d_{I}(r)]^{2}$, where $%
2\pi rl_{N}$ is the boundary area of the given venule neighborhood. Whence,
we find

\begin{equation}
d_I(r) \sim \left ( \frac{2\pi rl_N}{m} \right )^{1/2}.  \label{5.19}
\end{equation}

The first subgroup of capillaries and the system of corresponding pipes with
equally directed blood currents (considered in Chapter~\ref{ch.4}) have
approximately the same effect on heat transfer. Therefore, the effect of the
first capillary subgroup on the random walks in the neighborhood $%
Q_{i}(d_{N})$ of the venule $i$ can be described in terms of walker motion
in an effective convective stream whose velocity field, by virtue of (\ref
{4.115}), is specified by the expression

\begin{equation}
\mathbf{v}_{\mathrm{eff}}^{\vee }(\mathbf{r}\,)\sim \frac{\pi a_{c}^{2}}{%
d_{I}^{2}(r)}\mathbf{v}_{c}  \label{5.20}
\end{equation}
and is directed to the neighborhood center, i.e. to the venule. Substituting
(\ref{5.19}) into (\ref{5.20}) we get

\begin{equation}
\mathbf{v}_{\mathrm{eff}}^{\vee }(\mathbf{r}\,)\sim -\frac{1}{2\pi }\frac{%
\pi a_{c}^{2}v_{c}m}{l_{N}}\frac{\mathbf{r}}{r^{2}}.  \label{5.21}
\end{equation}
In addition, taking into account (\ref{3.8}) and the definition of the blood
flow rate, according to which we may set $j=\pi
a_{N}^{2}v_{N}/(l_{N}d_{N}^{2})$, the (\ref{5.21}) can be rewritten as

\begin{equation}
\mathbf{v}_{\mathrm{eff}}^{\vee }(\mathbf{r}\,)\sim -\frac{1}{2\pi }j\frac{%
d_{N}^{2}}{r}\frac{\mathbf{r}}{r}.  \label{5.22}
\end{equation}
Expressions (\ref{5.21}) and (\ref{5.22}) hold true until the radius $r$
becomes less than the distance $l_{\parallel c}$ over which a walker is
transported by the blood flow in a capillary during its location in the
vicinity of this capillary: $r>l_{\parallel c}$. Otherwise $(r<l_{\parallel
c})$, the walker will go into the venule actually with blood flow in a
single capillary and the representation of the walker motion in terms of
random walks in a homogeneous medium does not hold.

Description of the first subgroup effect on random walks in the neighborhood 
$Q_{i}(d_{N})$ containing an arteriole in its center is the same but the
direction of velocity field (\ref{5.22}) must be changed to the opposite
one, i.e.

\begin{equation}
\mathbf{v}_{\mathrm{eff}}^{a}(\mathbf{r}\,)\sim \frac{1}{2\pi }j\frac{%
d_{N}^{2}}{r}\frac{\mathbf{r}}{r}.  \label{5.23}
\end{equation}

In particular, expression (\ref{5.22}) allows us to find the time $\tau _{v}$
required for a walker to reach a venule, provided at initial time it was at
a distance $r$ of order $d_{N}$ from the venule and its motion towards the
venule is mainly caused by blood flow in capillaries of the first subgroup.
In this case the characteristic time required for the walker to go from one
of the nearest arterioles to the given venule is also about $\tau _{v}$.
Solving the equation

\begin{equation}
\frac{d\mathbf{r}}{dt}=\mathbf{v}_{\mathrm{eff}}^{\vee }(\mathbf{r}\,)
\label{5.24}
\end{equation}
under the initial condition $r|_{t=0}=r_{N}=d_{N}/\sqrt{\pi }$ we get $%
r(t)=d_{N}/\left[ \sqrt{\pi }(1-jt)\right] $ and, therefore, the desired
time is

\begin{equation}
\tau _v \sim \frac{1}{j} ,  \label{5.25}
\end{equation}
We point out that the given initial condition corresponds to the estimate of
the volume of the venule neighborhood, $Q_i(d_N)$ as the mean tissue volume
per one venule, i.e. $\pi r^2_N l_N =d^2_N l_N$.

Resulting walker motion in the domain $Q_{i}(r_{N})$ is affected not only by
the blood flow in the capillaries of the first subgroup but also by the
blood flow in the second subgroup capillaries as well as walker motion in
the cellular tissue itself. Thus, in addition, we shall obtain the condition
when the influence of the first subgroup capillaries on walker motion in the
neighborhood $Q_{i}(r_{N})$ should be taken into account. In the general
case stationary distribution $C_{\text{st}}(r)$ of the walkers in $%
Q_{i}(r_{N})$ obeys the equation.

\begin{equation}
D_{\mathrm{eff}}\mathbf{\nabla }C_{\text{st}}-\mathbf{v}_{\mathrm{eff}%
}^{\vee }C_{\text{st}}=0
\end{equation}
Whence, taking into account that expression (\ref{5.22}) can be rewritten in
the form

\begin{equation}
\mathbf{v}_{\mathrm{eff}}^{\vee }=-\mathbf{\nabla }\left[ \frac{1}{2\pi }%
jd_{N}^{2}l_{n}r\right] ,  \label{5.26}
\end{equation}
we find

\begin{equation}
C_{\text{st}}(r_{N})\approx C_{\text{st}}(a_{N})\exp \left\{ -\frac{%
d_{N}^{2}j}{2\pi D_{\mathrm{eff}}}\ln \left( \frac{r_{N}}{a_{N}}\right)
\right\} ,  \label{5.27}
\end{equation}
where $C_{\text{st}}(a_{N})$ and $C_{\text{st}}(r_{N})$ are the walker
concentrations near the venule $i$ and at the boundary of $Q_{i}(r_{N})$.

The capillaries of the first subgroup have a significant effect on the
walker motion when $C_{\text{st}}(r_{N})\gg C_{\text{st}}(a_{N})$. Then,
representing the blood flow rate as $j=\pi a_{N}^{2}v_{N}/(l_{N}d_{N}^{2})$
from (\ref{4.121}) and (\ref{5.27}), we obtain the desired condition

\begin{equation}
\frac{d_{N}^{2}j}{2\pi D_{\mathrm{eff}}}\ln \left( \frac{d_{N}}{a_{N}}%
\right) \equiv \frac{D}{D_{\mathrm{eff}}}\zeta _{N}\gg 1,  \label{5.28}
\end{equation}
where we have ignored $\ln \sqrt{\pi }$ in comparison with $\ln
(d_{N}/a_{N}) $ which plays an important role in heat transfer.

The second subgroup of capillaries plays an important role in heat transfer
in the case $l_c \gg l_N$ only, when the number of the capillary portions
which belong to this subgroup is substantially larger than that of the first
subgroup. Due to both spatial orientation of the capillary portions and the
direction of the blood currents in them being random, the second group
affects the random walks in a similar way as does the system of the
corresponding pipes considered in Section~\ref{s4.2} for which $\zeta =0$.
Therefore, according to (\ref{4.114}), on scales well above $l_{\parallel c}$
we can describe the effect of blood flow in the second subgroup capillaries
on heat transfer in terms of an effective isotropic medium with the
diffusion coefficient

\begin{equation}
D_{\mathrm{eff}}\sim D\left[ 1+\frac{\pi }{2}\left( \frac{a_{c}^{2}v_{c}}{%
Dd_{c}}\right) ^{2}\ln \left( \frac{d_{c}}{a_{c}}\right) \right] .
\label{5.29}
\end{equation}
Taking into account (\ref{3.8}), (\ref{3.12}), (\ref{4.121}) we rewrite (\ref
{5.29}) in the form

\begin{equation}
D_{\mathrm{eff}}\sim D\left[ 1+\frac{3\sqrt{3}\pi }{4}\frac{\ln (d_{c}/a_{c})%
}{[\ln (d_{N}/a_{N})]^{2}}\frac{l_{c}}{l_{N}m}\zeta _{N}^{2}\right] .
\label{5.30}
\end{equation}

On scales smaller than $l_{\parallel c}$ the motion of a walker in
capillaries with blood flow and its migration in the cellular tissue should
be considered individually.

As it follows from expression (\ref{5.30}) it is convenient to characterize
the renormalization of the diffusion coefficient due to blood flow in the
capillaries by the parameter

\begin{equation}
\gamma = \frac{3\sqrt{3} \pi}{4} \frac{\ln (d_c/a_c)}{[ \ln (d_N/a_N)]^2} 
\frac{l_c}{m l_N}  \label{5.31}
\end{equation}
If $\gamma \gg 1$, i.e. the capillary length $l_c$ is large enough, the
influence of the capillary system can become essential at not-too-high blood
flow rates when $n_t < N$. When $\gamma \ll 1$ the blood flow rate must
attain large values for the capillary system to be able to change the
diffusion coefficient. Therefore, capillaries forming the network with $%
\gamma \ll 1$ will be called short, and other wise, $\gamma \gg 1$ will be
referred to as long.

Concluding Chapter~\ref{ch.5} we would like to outline which vessels
contribute to what kind of heat transport. The countercurrent pairs of Class
1 and the first class veins whose level number $n\mathbf{\leqslant }n_{t}$
form the walker traps with the smallest vessels of this class playing a
dominant role in walker trapping. With blood flow through the large first
class veins walkers go out of the tissue domain without returning into the
cellular tissue.

The arteries of Class 1 also form paths of the fast walker migration in
tissue, however, due to the opposite direction of blood flow in them in
comparison with that of veins the role of the first class arteries in the
walker propagation is not significant. From the heat transfer standpoint the
large first class veins and arteries form the vessel system through which
blood goes into and out of the tissue practically without heat exchange with
the cellular tissue. Large vessels, however, are responsible for spatial
nonuniformities in the tissue temperature and should be analyzed
individually.

The effect of the arteries, veins, and countercurrent pairs of Class 2 is
reduced to the renormalization of the diffusion coefficient. Capillaries,
which are assumed to belong to Class 2, can affect the walker motion in
tissue because of $d_{c}\ll l_{c}$. However, their influence causes more
fast migration of walkers in tissue rather than gives rise to the walker
escape from the tissue domain. The effect of the first subgroup capillaries
in the walker motion may be described in terms of the walker motion in an
effective convective stream and the effect of the second subgroup
capillaries is also reduced to renormalization of the diffusion coefficient.

\chapter{Form of the bioheat equation in different limits, depending on the
blood flow rate}

\label{ch.6} 
\markright
{ {\sc  \thechapter. Form of the bioheat equation in different\ldots}
}

In the previous Chapter we have described bioheat transfer in terms of
random walks in living tissue containing the hierarchical vascular network
and analyzed characteristic properties of walker motion. In the present
Chapter based on the obtained results we develop a continuum description of
heat transfer in such living tissue.

\section{Continuum model for walker trapping}

\label{s6.1}

When a walker during its motion reaches one of the countercurrent pairs of
Class 1 or a first class vein it will be transported with blood flow to
large veins for a short time. Whereupon it never returns to the cellular
tissue and leaves the microcirculatory bed domain $Q_{0}$ with blood through
the host vein. From the standpoint of the cellular tissue this event can be
treated as walker disappearance and such a vessel plays the role of a region
where walkers die at a certain rate. In this way the total rate $\Gamma
_{i}\{C\}$ at which walkers are destroyed in the region of first class
vessels, for example, the vessel of level $n$ may be written as

\begin{equation}
\Gamma _{i}\{C\}=\Gamma _{n}\int\limits_{Q_{n}}d\mathbf{r}%
\int\limits_{0}^{l_{n}}ds\delta \lbrack \mathbf{r}-\mathbf{\xi }_{i}(s)]C(%
\mathbf{r}\,)  \label{6.1}
\end{equation}
where $Q_{n}$ is the fundamental domain containing this vessel, $\mathbf{\xi 
}_{i}(s)$ specifies the position of vein or countercurrent pair treated as a
line, $s$ is its natural parameter and $\Gamma _{n}$ is a certain constant.
At the same time the total walker dissipation rate in the domain $Q_{n}$,
according to Chapter~\ref{ch.5} must be equal to

\begin{equation}
\Gamma _{n}\{C\}=\int\limits_{Q_{n}}d\mathbf{r}C(\mathbf{r}\,)\frac{1}{\tau
_{l}}.  \label{6.2}
\end{equation}

For vessels of level $n_t$, which are the smallest vessels capable of
trapping walkers, $l_{\parallel n} \sim l_{n_t}$, so the walker
concentration is approximately uniform on scales of order $l_{n_t}$. The
latter allows us to regard the concentration $C$ in expressions (\ref{6.1})
and (\ref{6.2}) as a constant. The equating the two expressions we find

\begin{equation}
\Gamma _{n_{t}}=d_{n_{t}}^{2}\frac{1}{\tau _{l}}.  \label{6.3}
\end{equation}
Taking into account (\ref{5.15a}), (\ref{5.15b}) from (\ref{6.3}) we obtain

\begin{equation}
\Gamma _{n_{t}}=j\left( \ln \frac{l_{0}}{a_{0}}\right) ^{(\beta
(n_{t})-1)/2}d_{n_{t}}^{2}.  \label{6.4}
\end{equation}

The vessels of level $n_t$ practically control the walker trapping because
the number of the first class vessels per unit volume is practically
determined by vessels of this level. The latter enables us to take into
account solely the vessels $\{i_t\}$ of level $n_t$ in description of walker
disappearance. In this way the living tissue is represented as a medium with
distributed traps where the rate of walker disappearance is given by the
expression

\begin{equation}
\Gamma \{C\}=j\left( \ln \frac{l_{0}}{a_{0}}\right) ^{(\beta
(n_{t})-1)/2}\int\limits_{Q_{0}}d\mathbf{r}C(\mathbf{r}\,)\chi (\mathbf{r}\,)
\label{6.5}
\end{equation}
where

\begin{equation}
\chi (\mathbf{r}\,)=d_{n_{t}}^{2}\sum_{i_{t}}\int\limits_{0}^{l_{n_{t}}}ds%
\delta \lbrack \mathbf{r}-\mathbf{\xi }_{i_{t}}(s)]  \label{6.6}
\end{equation}
is the dimensionless density of traps and the sum runs over all the vessels
of level $n_{t}$.

In what follows the trap density $\chi (\mathbf{r}\,)$ will be treated as a
field with random nonuniformities. This approach enables us to describe both
the mean characteristics of the tissue temperature and nonuniformities in
the tissue temperature due to the discreteness of the $n_{t}$ - th level
distribution.

The mean value of $\chi (\mathbf{r}\,)$ is equal to one because every
fundamental domain of level $n_{t}$ contains just one vein or countercurrent
pair of the same level. Therefore, bearing in mind values averaged on scales
of order $l_{n_{t}}$, we may replace quantity (\ref{6.64}) by a certain
nonuniform field characterized by spatial scales larger than $l_{n_{t}}$ or
of the same order, viz

\begin{equation}
d_{n_{t}}^{2}\sum_{i_{t}}\int\limits_{0}^{l_{n_{t}}}ds\delta \lbrack \mathbf{%
r}-\mathbf{\xi }_{i_{t}}(s)]=1+\chi _{t}(\mathbf{r}\,),  \label{6.7}
\end{equation}
where $\chi _{t}(\mathbf{r}\,)$ is a nonuniform field whose mean value is
equal to zero.

The field $\chi _{t}(\mathbf{r}\,)$ is specified by particular details of
the vessel arrangement in space on the scale $l_{n_{t}}$. However, when we
consider solely characteristic features of the walker concentration $C(%
\mathbf{r},t)$ on these scales, the particular details of the vessel
arrangement is of little consequence. Thereby, to analyse the characteristic
properties of spatial nonuniformities in the 
\index{temperature distribution} temperature distribution%
\index{temperature distribution}, i.e. in the walker concentration, the
vessel arrangement may be described as a system of $n_{t}$-th level vessels
randomly distributed in the cellular tissue. For real microcirculatory beds
small arteries and veins must be uniformly distributed in the tissue
because, other wise, for example, lack of the tissue oxygen supply can
occur. So, in the given analysis we assume that the arrangement of the $%
n_{t} $-th level veins is random on the scale $l_{n_{t}}$ only and does not
exhibit spatial random nonuniformities of scales larger than $l_{n_{t}}$.
Under such conditions the field $\chi _{t}(\mathbf{r}\,)$ can be considered
to be random one obeying the conditions

\begin{equation}
\left\langle \chi _{t}(\mathbf{r}\,)\right\rangle =0  \label{6.8}
\end{equation}
and 
\begin{equation}
\left\langle \chi _{t}(\mathbf{r}\,)\chi _{t}(\mathbf{r}^{\prime
})\right\rangle =g\left( 
\frac{\left| \mathbf{r}-\mathbf{r}^{\prime }\right| }{l_{n_{t}}}\right)
\label{6.9}
\end{equation}
where 
\begin{equation}
g(x)=\frac{1}{2^{3/2}}\exp \left[ -\frac{3\pi }{8}x^{2}\right] -\frac{1}{%
2^{3/2}}\exp \left[ -\frac{\pi }{4}x^{2}\right] .  \label{6.10}
\end{equation}

In order to derive the form (\ref{6.10}) of the correlation function (\ref
{6.9}) let us consider properties of spatial nonuniformities in the vessel
distribution. For this purpose we transform the quantity

\begin{equation}
d_{n}^{2}\sum_{i}\int\limits_{0}^{l_{n}}ds\delta \lbrack \mathbf{r}-\mathbf{%
\xi }_{i}(s)].  \label{6.11}
\end{equation}

Let $\mathbf{r}_{i}$ be the middle point of the curve $\mathbf{\xi }_{i}(s)$%
. Then at a fixed value of $s$ we define the smoothing ($\left\langle
...\right\rangle _{d_{n}}$) of the function $\delta \lbrack \mathbf{r}-%
\mathbf{\xi }_{i}(s)]$ on the scale $d_{n}$ as follows,

\begin{equation}
\left\langle \delta \lbrack \mathbf{r}-\mathbf{\xi }_{i}(s)]\right\rangle
_{d_{n}}=\int d\mathbf{\xi }\delta \lbrack \mathbf{r}-\mathbf{\xi }]\mathcal{%
P}_{i}(\mathbf{\xi })  \label{6.12}
\end{equation}
where the function

\begin{equation}
\mathcal{P}_{i}(\mathbf{\xi })=\frac{1}{V_{n}}\exp \left\{ -\frac{(\mathbf{%
\xi }-\mathbf{r}_{i})^{2}}{2\tilde{l}^{2}}\right\}  \label{6.13}
\end{equation}
where $V_{n}=d_{n}^{2}l_{n}$ is the volume of a domain $Q_{n}$ and the
length $\tilde{l}$ is determined by the relation $(2\pi \tilde{l}%
^{2})^{3/2}=V_{n}$, i.e. $\tilde{l}=V_{n}^{1/3}(2\pi )^{-1/2}$,
characterizes the disposition of the given vessel treated as a random curve.
Whence, it follows that

\begin{equation*}
\left\langle \delta \lbrack \mathbf{r}-\mathbf{\xi }_{i}(s)]\right\rangle
_{d_{n}}=
\end{equation*}
\begin{equation}
=\frac{1}{V_{n}}\int\limits_{Q_{0}}d\mathbf{\xi }\delta \lbrack \mathbf{r}-%
\mathbf{\xi }]\exp \left[ -\frac{(\mathbf{\xi }-\mathbf{r}_{i})^{2}}{2\tilde{%
l}^{2}}\right] =\frac{1}{V_{n}}\exp \left[ -\frac{(\mathbf{r}-\mathbf{r}%
_{i})^{2}}{2\tilde{l}^{2}}\right] .  \label{6.14}
\end{equation}
Following (\ref{6.7}) for level $n$ we represent quantity (\ref{6.11}) as
then smoothing quantity (\ref{6.14}) we get

\begin{equation}
d_{n}^{2}\sum_{i}\int\limits_{0}^{l_{n}}ds\left\langle \delta \lbrack 
\mathbf{r}-\mathbf{\xi }_{i}(s)]\right\rangle _{d_{n}}=1+\chi _{n}(\mathbf{r}%
\,),  \label{6.15}
\end{equation}
where the random field $\chi _{n}(\mathbf{r}\,)$ is specified by the
expression

\begin{equation}
\chi _{n}(\mathbf{r})=\sum_{i}\exp \left[ -\frac{(\mathbf{r}-\mathbf{r}%
_{i})^{2}}{2\tilde{l}^{2}}\right] -1.  \label{6.16}
\end{equation}

Distribution of the points $\{\mathbf{r}_{i}\}$ is assumed to be random on
scales of order $l_{n}$ only. In other words we shall suppose that each
vessel of level $n$ is randomly and practically uniformly distributed in a
domain whose volume is about $V_{n}$ and intersection of such domains
containing different vessels of level $n$ is not considerable. These
conditions can be described in terms of the following one- and two-point
distribution functions. The one-point distribution function%
\index{distribution function} $g_{1}(r)$, i.e. the probability density of
finding a given point $i$ at the point $\mathbf{r}$ averaged over all
possible realizations of the other point arrangement, is supposed to be
equal to $1/V_{0}$:

\begin{equation}
\left\langle \delta \lbrack \mathbf{r}-\mathbf{r}_{i}]\right\rangle =g_{1}(%
\mathbf{r}\,)=%
\frac{1}{V_{0}},  \label{6.17}
\end{equation}
where $\delta (\mathbf{r}\,)$ is the spatial $\delta $-function, the symbol $%
\left\langle ...\right\rangle $ means averaging over arrangement of all the
points $\{r_{i}\}$ and $V_{0}$ is the volume of the domain $Q_{0}$. The
two-point distribution function $g(\mathbf{r},\mathbf{r}^{\prime })$, i.e.
the probability density of finding a given pair of the points $i,i^{\prime }$
at $\mathbf{r}$ and $\mathbf{r}^{\prime }$, respectively, averaged over all
possible realizations of the other point arrangement, is specified, for
simplicity, in the form

\begin{equation}
\left\langle \delta \lbrack \mathbf{r}-\mathbf{r}_{i}]\delta \lbrack \mathbf{%
r}^{\prime }-\mathbf{r}_{i^{\prime }}]\right\rangle =g_{2}(\mathbf{r},%
\mathbf{r}^{\prime })=\frac{1}{V_{0}^{2}}\left\{ 1-\exp \left[ -\frac{(%
\mathbf{r}-\mathbf{r}^{\prime })^{2}}{2\tilde{l}^{2}}\right] \right\} .
\label{6.18}
\end{equation}
Formula (\ref{6.18}) means that, on one hand, two points $i$ and $i^{\prime
} $ are practically independent of each other when the distance $\left| 
\mathbf{r}-\mathbf{r}^{\prime }\right| $ between them substantial large than 
$\tilde{l}$. On the other hand, point pairs characterized by small distance $%
\left| \mathbf{r}-\mathbf{r}^{\prime }\right| \ll \tilde{l}$ are absent. We
note that function $g_{2}(\mathbf{r},\mathbf{r}^{\prime })$ satisfies the
condition $\left\langle \tilde{N}^{2}\right\rangle =\left\langle \tilde{N}%
\right\rangle ^{2}$, where $\tilde{N}$ is the total number of points $\{%
\mathbf{r}_{i}\}$ contained in arbitrary domain $Q$ whose volume $V_{Q}$ is
well above $V_{n}=l_{n}d_{n}^{2}$. This condition implies that on scales
larger than $l_{n}$ the arrangement of the points $\{\mathbf{r}_{i}\}$
exhibits no spatial random nonuniformities. Indeed

\begin{equation}
\tilde{N}=\sum_{i}\int\limits_{Q}d\mathbf{r}\delta \lbrack \mathbf{r}-%
\mathbf{r}_{i}]  \label{6.19}
\end{equation}
and taking into account (\ref{6.17}) and (\ref{6.18}) we get 
\begin{equation}
\left\langle \tilde{N}\right\rangle =\sum_{i}\frac{V_{Q}}{V_{0}}=\frac{V_{Q}%
}{V_{n}},  \label{6.20}
\end{equation}
because $V_{n}$ is equal to the mean volume of the domain $Q_{0}$ which
falls on one vein or one countercurrent pair of level $n$, i.e. falls on one
point of the collection $\{\mathbf{r}_{i}\}$. Then

\begin{align}
\left\langle \tilde{N}^{2}\right\rangle & =\sum_{i}\frac{V_{Q}}{V_{0}}\cdot 
\notag \\
& \left\{ 1+\sum^{\prime }\nolimits_{i^{\prime }}\frac{1}{V_{0}}\left[
V_{Q}-\iint_{Q}d\mathbf{r}d\mathbf{r}^{\prime }\exp \left( -\frac{(\mathbf{r}%
-\mathbf{r}^{\prime })^{2}}{2\tilde{l}^{2}}\right) \right] \right\} ,
\label{6.21}
\end{align}
where the prime on the sum over $i^{\prime }$ indicates that summation is
carried out over all $i^{\prime }\not=i$ and we also have used the relation

\begin{equation}
\left\langle \delta \lbrack \mathbf{r}-\mathbf{r}_{i}]\delta \lbrack \mathbf{%
r}^{\prime }-\mathbf{r}_{i}]\right\rangle =\int\limits_{Q_{0}}d\mathbf{r}%
^{\prime }g_{1}(\mathbf{r}\,^{\prime \prime })\delta \lbrack \mathbf{r}-%
\mathbf{r}\,^{\prime \prime }]\delta \lbrack \mathbf{r}^{\prime }-\mathbf{r}%
\,^{\prime \prime }]=\frac{1}{V_{0}}\delta \lbrack \mathbf{r}-\mathbf{r}%
^{\prime }].  \label{6.22}
\end{equation}

For $V_{Q}\gg V_{n}$ in the limit $V_{0}\rightarrow \infty $ from (\ref{6.20}%
) and (\ref{6.21}) we obtain

\begin{equation}
\left\langle \tilde{N}^{2}\right\rangle \simeq \left\langle \tilde{N}%
\right\rangle ^{2}+\frac{V_{Q}}{V_{n}}\left[ 1-\frac{(2\pi \tilde{l}%
^{2})^{3/2}}{V_{n}}\right] =\left\langle \tilde{N}\right\rangle ^{2}
\label{6.23}
\end{equation}
due to, by definition, $(2\pi \tilde{l}^{2})^{3/2}=V_{n}$.

Within the framework of the adopted assumptions from (\ref{6.16}) we get

\begin{equation}
\left\langle \chi _{n}(\mathbf{r}\,)\right\rangle
=\sum_{i}\int\limits_{Q_{0}}d\mathbf{r}_{i}g_{1}(\mathbf{r}_{i})\exp \left[ -%
\frac{(\mathbf{r}-\mathbf{r}_{i})^{2}}{2\tilde{l}^{2}}\right] -1
\label{6.24}
\end{equation}
and due to (\ref{6.17})

\begin{equation}
\left\langle \chi _{n}(\mathbf{r}\,)\right\rangle =\sum_{i}\frac{1}{V_{0}}%
(2\pi \tilde{l}^{2})^{3/2}-1=\frac{(2\pi \tilde{l}^{2})^{3/2}}{V_{n}}-1=0.
\label{6.25}
\end{equation}
Then taken into account (\ref{6.18}), we find, by definition

\begin{equation*}
g\left( \frac{\mid \mathbf{r}-\mathbf{r}^{\prime }\mid }{l_{n}}\right)
=<\chi _{n}(\mathbf{r}\,)\chi _{n}(\mathbf{r}^{\prime })>=
\end{equation*}
\begin{equation*}
=\sum^{\prime }\nolimits_{i,i^{\prime
}}\int\limits_{Q_{0}}\int\limits_{Q_{0}}d\mathbf{r}_{i}d\mathbf{r}%
_{i^{\prime }}g_{2}(\mathbf{r}_{i},\mathbf{r}_{i^{\prime }})\exp \left[ -%
\frac{(\mathbf{r}-\mathbf{r}_{i})^{2}}{2\tilde{l}^{2}}\right] \cdot
\end{equation*}
\begin{equation*}
\cdot \exp \left[ -\frac{(\mathbf{r}-\mathbf{r}_{i^{\prime }})^{2}}{2\tilde{l%
}^{2}}\right] +\sum_{i}\int\limits_{Q_{0}}d\mathbf{r}_{i}g_{1}(\mathbf{r}%
_{i})\exp \left[ -\frac{(\mathbf{r}-\mathbf{r}_{i})^{2}}{2\tilde{l}^{2}}%
\right] \cdot
\end{equation*}
\begin{equation}
\cdot \exp \left[ -\frac{(\mathbf{r}^{\prime }-\mathbf{r}_{i})^{2}}{2\tilde{l%
}^{2}}\right] -2\sum_{i}\int\limits_{Q_{0}}d\mathbf{r}_{i}g_{1}(\mathbf{r}%
_{i})\exp \left[ -\frac{(\mathbf{r}-\mathbf{r}_{i})^{2}}{2\tilde{l}^{2}}%
\right] +1.  \label{6.26}
\end{equation}
Then substituting (\ref{6.17}), (\ref{6.18}) into (\ref{6.26}) and passing
to the limit $V_{0}\rightarrow \infty $, we can rewrite the latter
expression in the form

\begin{equation*}
g\left( \frac{\mid \mathbf{r}-\mathbf{r}^{\prime }\mid }{l_{n}}\right) =%
\frac{1}{V_{N}}\int\limits_{\mathcal{R}^{3}}d\mathbf{r}\,^{\prime \prime
}\exp \left[ -\frac{(\mathbf{r}-\mathbf{r}\,^{\prime \prime })^{2}}{2\tilde{l%
}^{2}}-\frac{(\mathbf{r}^{\prime }-\mathbf{r}\,^{\prime \prime })^{2}}{2%
\tilde{l}^{2}}\right] -
\end{equation*}
\begin{equation}
-\frac{1}{V_{N}^{2}}\int\limits_{\mathcal{R}^{3}}\int d\mathbf{r}\,^{\prime
\prime }d\mathbf{r}\,^{\prime \prime \prime }\exp \left[ -\frac{(\mathbf{r}-%
\mathbf{r}\,^{\prime \prime })^{2}}{2\tilde{l}^{2}}-\frac{(\mathbf{r}%
\,^{\prime \prime }-\mathbf{r}\,^{\prime \prime \prime })^{2}}{2\tilde{l}^{2}%
}-\frac{(\mathbf{r}\,^{\prime \prime \prime }-\mathbf{r}^{\prime })^{2}}{2%
\tilde{l}^{2}}\right] .  \label{6.27}
\end{equation}
Here, in addition, we have taken into account that the last integral term in
expression (\ref{6.26}) is equal to two, $i$ is a dummy index, and

\begin{equation*}
\frac{1}{V_{0}}\sum_{i}=\frac{\mathcal{N}}{V_{0}}=\frac{1}{V_{n}}
\end{equation*}
\begin{equation*}
\frac{1}{V_{0}^{2}}\sum^{\prime }\nolimits_{i,i^{\prime }}=\frac{\mathcal{N}(%
\mathcal{N}-1)}{V_{0}^{2}}\rightarrow \frac{1}{V_{n}^{2}}
\end{equation*}
where $\mathcal{N}$ is the total number of the points in the collection $\{%
\mathbf{r}_{i}\}$. Using the transformation rules shown in Fig.~\ref{Fig11}b
from (\ref{6.27}) we obtain

\begin{equation*}
g\left( \frac{\mid \mathbf{r}-\mathbf{r}^{\prime }\mid }{l_{n}}\right) =%
\frac{1}{2^{3/2}}\exp \left[ -\frac{(\mathbf{r}-\mathbf{r}^{\prime })^{2}}{4%
\tilde{l}^{2}}\right] -
\end{equation*}
\begin{equation*}
-\frac{1}{3^{3/2}}\exp \left[ -\frac{(\mathbf{r}-\mathbf{r}^{\prime })^{2}}{8%
\tilde{l}^{2}}\right] .
\end{equation*}
Whence it immediately follows expression (\ref{6.10}) where $n=n_{t}$.

\section{Parameters determining the form of bioheat tra\-nsfer equation. The
case of no influence of capillary system}

\label{s6.2}

The specific form of bioheat equation depends on the characteristic
parameters of the vascular network architectonics and the total blood
current flowing through the microcirculatory bed. In order to classify the
form of bioheat equation it is convenient to specify different possible
limits by the parameters $\gamma$ (see (\ref{5.31})) and $G$. The former
characterizes the way in which the capillary system can affect heat transfer
and the latter determines the level number $n_t$ of the vessels directly
controlling heat exchange between blood and the cellular tissue.

The final result of the present Chapter is illustrated in Fig.~\ref{Fig16}
which shows how the form of bioheat equation changes as the dimensionless
total blood current $G$ increases for vascular network with short $(\gamma
\ll 1)$ and long $(\gamma \gg 1)$ capillaries. Below in this Section we
shall consider heat transfer in living tissue with short capillaries when $%
n_t <N$ e.i. there are both heat - conservation and heat - dissipation
vessels viz:

\begin{equation}
\gamma \ll 1;\quad G_{0}\ll G\ll M^{2/3},  \label{6.28}
\end{equation}
where $G_{0}=\left( \ln \frac{l_{0}}{a_{0}}\right) ^{1/2}$ and $M=2^{3N}$ is
the total number of the venules or arterioles (see Chapter~\ref{ch.3}). It
should be noted that the first inequality is the condition on geometry of
the vascular network whereas the last two are actually the condition on the
value of the total blood current $J_{0}$, with the former ($G\gg G_{0}$)
being the overall restriction assumed in the present work (see (\ref{5.6}))
and the latter, according to (\ref{5.1}),(\ref{5.3}), implying that $\zeta
_{N}\ll 1$. In this case the capillary system practically has no effect on
heat transfer. In fact, the renormalization of the diffusion coefficient
caused by the effect of the second subgroup capillaries (see expression (\ref
{5.30})) is not appreciable. Thus solely the venous and arterial trees
should be taken into account.

\FRAME{ftbpFU}{11.3368cm}{13.9815cm}{0pt}{\Qcb{Characteristic types of heat
transfer in living tissue depending on the total blood current (in units of $%
G$) in the microcirculatory bed. Figures ``a'' and ``b'' correspond to the
vascular network with short $(\protect\gamma \ll 1)$ and long $(\protect%
\gamma \gg 1)$ capillaries, respectively. ($M$ is the total number of the
arterioles).}}{\Qlb{Fig16}}{Fig16}{\special{language "Scientific Word";type
"GRAPHIC";maintain-aspect-ratio TRUE;display "USEDEF";valid_file "F";width
11.3368cm;height 13.9815cm;depth 0pt;original-width
14.5185in;original-height 17.9725in;cropleft "-0.001846";croptop
"1.005211";cropright "0.999254";cropbottom "0.006011";filename
'FIG16.GIF';file-properties "XNPEU";}}

Such situation has been practically considered in Section~\ref{s5.2} and
Section~\ref{s5.4} where the effect of blood flow on heat transfer has been
treated in terms of walker trapping and renormalization%
\index{renormalization} of the diffusion coefficient. Therefore, taking into
account continuum description of walker trapping stated in Section~\ref{s6.1}
we can write the following bioheat equation

\begin{equation}
\frac{\partial C}{\partial t}=\mathbf{\nabla }(D_{\mathrm{eff}}\mathbf{%
\nabla }C)-j[1+\chi _{t}(\mathbf{r}\,)]C\left[ \ln \left( \frac{l_{0}}{a_{0}}%
\right) \right] ^{(\beta (n_{t})-1)/2}+q.  \label{6.29}
\end{equation}
Here $D_{\mathrm{eff}}$ is given by formula (\ref{5.16a}) or (\ref{5.16b}), $%
\chi _{t}(\mathbf{r}\,)$ is the random field obeying conditions (\ref{6.8}),
(\ref{6.9}) and the value of $n_{t}$ as function of $G$ is specified by
expression (\ref{5.5}) and shown in Fig.~\ref{Fig13}b.

In particular, if we ignore the spatial nonuniformities in the vessel
distribution (the field $\xi _{t}(\mathbf{r}\,)$) and consider $G>G_{cc}$,
then equation (\ref{6.29}) will become.

\begin{equation}
\frac{\partial C}{\partial t}=\mathbf{\nabla }(D_{\mathrm{eff}}\mathbf{%
\nabla }C)-jC+q.  \label{6.30a}
\end{equation}
The latter equation practically coincides with the conventional bioheat
equation with the replacement $C\rightarrow (T-T_{a})/(T_{a}V_{N})$. For $%
G>G_{cc}$ in a similar way we get

\begin{equation}
\frac{\partial C}{\partial t}=\mathbf{\nabla }(D_{\mathrm{eff}}\mathbf{%
\nabla }C)-j\frac{1}{[\ln (l_{0}/a_{0})]^{1/2}}C+q.  \label{6.30b}
\end{equation}

Concluding this Section we discuss the properties of heat exchange between
blood and the cellular tissue. Let in a fundamental domain $Q_{n}$ of level $%
n<n_{t}$ whose size $l_{n}\gg l_{n_{t}}$ the tissue temperature $T$ be
approximately uniform. Then, when the blood - tissue heat exchange is
directly controlled by unit veins of level $n_{t}$, i.e. $G>G_{cc}$, the
temperature $T^{\ast }$ of blood in the large vein $i_{n}$ of level $n$
drawing this domain must be equal to the tissue temperature $T$. The latter
is the case because blood in veins of level $n_{t}$ is approximately in
thermodynamic equilibrium with the surrounding cellular tissue and arteries
of level $n_{t}$ are, at, on the average, the distance $d_{n_{t}}\sim
l_{n_{t}}$. This future of bioheat transfer is reflected in the form of the
second term $(jC)$ on the right - hand side of equations (\ref{6.30a}), (\ref
{6.30b}). If the blood - tissue heat exchange is controlled by the
countercurrent pairs of level $n_{t}$, i.e. if $G<G_{cc}$, then heat
exchange between arterial and venous blood in the countercurrent vessels of
this pairs is significant. So, venous blood initially flowing through small
vessels where it was in thermodynamic equilibrium with the cellular tissue
and until it reaches large veins of level $n<n_{\ast }$, where heat
conduction does not play a significant role, inevitably loses a certain
portion of heat. Therefore, in this case the temperature $T^{\ast }$ of
blood in the vein $i_{n}$ is not equal to the tissue temperature. The
relationship between $T^{\ast },T$ and $T_{a}$, in general, may be written
as $T^{\ast }=T_{a}+(T-T_{a})\sigma _{av}$ where $\sigma _{av}$ is a certain
coefficient. In the given case the volumetric dissipation rate of heat in
living tissue can be represented in terms of $jc_{t}\rho _{t}(T^{\ast
}-T_{a})=jc_{t}\rho _{t}\sigma _{av}(T-T_{a})$. Comparing the latter with
the second term of (\ref{6.30b}) we find that $\sigma _{av}=[\ln
(l_{0}/a_{0})]^{-1/2}$. Therefore, in the given situation the relationship
between the temperatures $T^{\ast },T_{a}$ and $T$ is of the form

\begin{equation}
T^{\ast }=T_{a}+\frac{1}{[\ln (l_{0}/a_{0})]^{1/2}}(T-T_{a}).  \label{6.31}
\end{equation}
Expression (\ref{6.31}) will be used in the theory of thermoregulation (see
Part 4).

\begin{equation*}
\frac{\partial C}{\partial t}=D_{\mathrm{eff}}\mathbf{\nabla }^{2}C-j[1+\chi
_{t}(\mathbf{r}\,)]C+q.
\end{equation*}
It should be pointed out that (\ref{6.16}) will practically coincide with
bioheat equation (\ref{1.1}) within the replacement $C\rightarrow
(T-T_{a})/(T_{a}V_{N})$, if we ignore the term $\chi _{t}(\mathbf{r}\,)$.

\section{Bioheat equation for living tissue with short capillaries}

\label{s6.3}

As the blood flow rate increases, the value of $n_{t}$ decreases and at
certain $G$ it becomes equal to $N$. For greater values of the blood flow
rate the capillary network affect significantly heat transfer. In this case
there are two limits differ in properties of heat transfer which will be
analyzed individually in the present Section.

\subsection{Convective type heat transport. The porous medium model}

\label{s6.4}

Let us consider the limit

\begin{equation}
\gamma \ll 1; \quad M^{2/3} \ll G \ll \frac{1}{\gamma} M^{2/3}.  \label{6.32}
\end{equation}

The last two inequalities can be also represented as $1\ll \zeta _{N}\ll
1/\gamma $, therefore, in this case there are no arteries and veins of Class
2, and Class 1 involves all the arteries and veins of the vascular network
with the vessels of the shortest length belonging to level $N$ but not to
level $n_{t}$.

According to Section~\ref{s5.2} we can describe the effect of such a vessel
system on heat transfer in terms of the walker motion in a medium containing
walker traps in the form of the venules distributed in the tissue at the
mean distance $d_N$ from each other. However, in this case the walker motion
in the neighborhood $Q_i (d_N)$ of a given venule or arteriole is mainly
controlled by blood flow in capillaries of the first subgroup. Indeed, the
influence of the second subgroup capillaries on the walker motion gives rise
to renormalization of the diffusion coefficient (see (\ref{5.30})). Thereby,
as it follows from (\ref{5.28}) and (\ref{5.30}), the first subgroup
capillaries mainly control the walker motion in $Q_i(d_N)$ if

\begin{equation}
\frac{1}{\zeta _N} + \gamma \zeta _N \ll 1.  \label{6.33}
\end{equation}
The first term is associated with diffusion in the cellular tissue and the
second one describes the effect of the second subgroup capillaries. Due to (%
\ref{6.32}) both these terms are small. Therefore, we may account for solely
convective type transport of the walkers.

Thus, in a cylindrical neighborhood $Q_i(r_N)$ of the radius $r_N = d_N / 
\sqrt{\pi}$ containing a venule $i$ in its center, the walker distribution
is practically of two-dimensional form and can be described by the
two-dimensional equation

\begin{equation}
\frac{\partial C}{\partial t}=-\mathbf{\nabla }[\mathbf{v}_{\mathrm{eff}%
}^{\vee }(\mathbf{r}\,)C]-\left. jd_{N}^{2}C\right| _{r=0}\delta (\mathbf{r}%
\,)+q(\mathbf{r},t).  \label{6.34}
\end{equation}
Here following Section~\ref{s6.1} we have represented the venule as a trap
of the two - dimensional $\delta $ - function form whose parameters satisfy
the condition that the rate of the walker escape from $Q_{i}(r_{N})$ is
determined by blood flow through the venule as it must be if the venules
belong to Class 1. Indeed, in the given case the total rate of the walker
escape with blood flow through the venule is equal to $\pi
a_{N}^{2}v_{N}C|_{r=0}$ where $C|_{r=0}$ is the walker concentration in the
venule. According to the definition of the blood flow rate (see Chapter~\ref
{ch.3}) we may set $j=\pi a_{N}^{2}v_{N}/(l_{N}d_{N}^{2})$. Thereby, the
walker escape rate per unit length of the venule is $jd_{N}^{2}C|_{r=0}$
which is exactly the coefficient of the $\delta $-function in (\ref{6.34}).

As it follows from (\ref{3.11}), (\ref{5.18}), (\ref{5.31}), and (\ref{6.32}%
), in this case the mean distance $l_{\parallel c}$ which a walker passes
with blood in a capillary before going out of its neighborhood of radius $%
d_{N}/\sqrt{\pi }$, satisfies the condition $l_{\parallel c}\ll \frac{\ln
(d_{N}/a_{N})}{2\pi }d_{N}^{2}/l_{c}<d_{N}$ because, at least $l_{c}>d_{N}$
according to the adopted model for the capillary system (see Chapter~\ref
{ch.3}). Thus, on one hand, expression (\ref{5.22}) is true for all values
of $r<d_{N}$ except for a small neighborhood of the point $\mathbf{r}=0$
whose radius is about $l_{\parallel c}$. On the other hand, when $q(\mathbf{r%
},t)$ is approximately constant on the scale $d_{N}$, the formal stationary
solution of equation (\ref{6.34}) with $\mathbf{v}_{\mathrm{eff}}^{\vee }$
specified by expression (\ref{5.22}) for all $r<d_{N}$ cannot vary
substantially on scales of order $l_{\parallel c}$. Therefore, we may assume
that in the given case expression (\ref{5.22}) holds true for all $r<d_{N}$,
and , thereby, represent it in the form

\begin{equation}
\mathbf{v}_{\mathrm{eff}}^{\vee }(\mathbf{r})=\mathbf{\nabla }\mathcal{P}%
^{\vee }(\mathbf{r}\,),  \label{6.35}
\end{equation}
where $\mathcal{P}^{\vee }(\mathbf{r}\,)$ is the velocity potential
satisfying the equation

\begin{equation}
\mathbf{\nabla }^{2}\mathcal{P}^{\vee }(\mathbf{r}\,)=-jd_{N}^{2}\delta (%
\mathbf{r}\,).  \label{6.36}
\end{equation}
Taking into account (\ref{6.36}) we also may rewrite equation (\ref{6.34}) as

\begin{equation}
\frac{\partial C}{\partial t}=-\mathbf{v}_{\mathrm{eff}}^{\vee }(\mathbf{r})%
\mathbf{\nabla }C+q.  \label{6.37}
\end{equation}

It should be pointed out that although when obtaining equation (\ref{6.37})
we have used only approximate values of the corresponding coefficients, for
example, the cofactor $j d^2_N$ in the second term on the right-hand side of
equation (\ref{6.34}), its form complies with the general laws of blood and
heat conservation.

For a similar neighborhood $Q_i(r_N)$ containing an arteriole in its center,
practically in the same way we obtain the following equation for the walker
distribution

\begin{equation}
\frac{\partial C}{\partial t}=-\mathbf{\nabla }[\mathbf{v}_{\mathrm{eff}%
}^{a}(\mathbf{r}\,)C]+q(\mathbf{r},t)  \label{6.38}
\end{equation}
where

\begin{equation}
\mathbf{v}_{\mathrm{eff}}^{a}(\mathbf{r}\,)=\mathbf{\nabla }\mathcal{P}^{a}(%
\mathbf{r}\,)  \label{6.39}
\end{equation}
and 
\begin{equation}
\mathbf{\nabla }^{2}\mathcal{P}^{a}=jd_{N}^{2}\delta (\mathbf{r}\,).
\label{6.40}
\end{equation}
We point out that equation (\ref{6.38}) contains no terms like $%
jd_{N}^{2}C|_{r=0}\delta (\mathbf{r}\,)$ because the effect of the
arterioles on the walker motion is not significant (see Section~\ref{s5.2},%
\ref{s5.3}).

Generalizing these results on the whose microcirculatory bed domain in the
case under consideration we can describe the walker distribution in the
tissue by the equation

\begin{equation}
\frac{\partial C}{\partial t}+\mathbf{v}_{\mathrm{eff}}^{\vee }\mathbf{%
\nabla }C+\mathbf{\nabla }[\mathbf{v}_{\mathrm{eff}}^{a}C]=q.  \label{6.41}
\end{equation}
Here, 
\begin{equation}
\mathbf{v}_{\mathrm{eff}}^{\vee }=\mathbf{\nabla }\mathcal{P}^{\vee };\quad 
\mathbf{v}_{\mathrm{eff}}^{a}=\mathbf{\nabla }\mathcal{P}^{a},  \label{6.42}
\end{equation}
where 
\begin{equation}
\mathbf{\nabla }^{2}\mathcal{P}^{\vee
}=-jd_{N}^{2}\sum_{i_{N}}\int\limits_{0}^{l_{N}}ds\delta \lbrack \mathbf{r}-%
\mathbf{\xi }_{i_{N}}^{\vee }(s)]  \label{6.43}
\end{equation}
\begin{equation}
\mathbf{\nabla }^{2}\mathcal{P}^{a}=jd_{N}^{2}\sum_{i_{N}}\int%
\limits_{0}^{l_{N}}ds\delta \lbrack \mathbf{r}-\mathbf{\xi }_{i_{N}}^{a}(s)],
\label{6.44}
\end{equation}
the functions $\mathbf{\xi }_{i_{N}}^{\vee }(s)$ and $\mathbf{\xi }%
_{i_{N}}^{a}(s)$ specify the spatial position of the center lines of the
arterioles and venules, and $s$ is their natural parameter. We note that the
given model considers the collection of the arterioles and venules as
rectilinear paths $\{\mathbf{\xi }_{i_{N}}^{a}(s),\,\mathbf{\xi }%
_{i_{N}}^{\vee }(s)\}$, respectively, and deals with the capillary system in
terms of a porous medium.

As it follows from the solution of equations (\ref{6.34}) and (\ref{6.38}),
the walker concentration (i.e. the temperature field) practically exhibits
no nonuniformities on scales much smaller than $d_{N}$. Therefore, we may
average the right-hand side of equations (\ref{6.43}) and (\ref{6.44}) over
the coordinates $\{\mathbf{\xi }_{i_{N}}^{a}(s),\,\mathbf{\xi }%
_{i_{N}}^{\vee }(s)\}$ on scales smaller that $d_{N}$. Besides, just as in
the previous case, we regard the venules and arterioles as vessels randomly
distributed in the tissue and their arrangement can exhibit random spatial
fluctuations%
\index{spatial fluctuations} of scale $l_{N}$ only. In addition, due to the
arterioles and venules being separated from each other by capillaries,
fluctuations in their distribution are assumed to be opposite on scales of
order $l_{N}$. Then, as it is shown in Section~\ref{s6.1}, we may replace
the following quantities on the right-hand side of equations (\ref{6.43}), (%
\ref{6.44}) by some random field

\begin{equation}
d_{N}^{2}\sum_{i_{N}}\int\limits_{0}^{l_{N}}ds\delta \lbrack \mathbf{r}-%
\mathbf{\xi }_{i_{N}}^{\vee }(s)]\rightarrow 1+\chi _{v}(\mathbf{r}\,),
\label{6.45}
\end{equation}
\begin{equation}
d_{N}^{2}\sum_{i_{N}}\int\limits_{0}^{l_{N}}ds\delta \lbrack \mathbf{r}-%
\mathbf{\xi }_{i_{N}}^{a}(s)]\rightarrow 1-\chi _{v}(\mathbf{r}\,),
\label{6.46}
\end{equation}
where the second replacement results from the latter assumption. In (\ref
{6.45}), (\ref{6.46}) $\chi _{v}(\mathbf{r}\,)$ is a random function of $%
\mathbf{r}$ satisfying the conditions

\begin{equation}
\left\langle \chi _{v}(\mathbf{r}\,)\right\rangle =0,  \label{6.47}
\end{equation}
\begin{equation}
\left\langle \chi _{v}(\mathbf{r}\,)\chi _{v}(\mathbf{r}^{\prime
})\right\rangle =g\left( 
\frac{\mid \mathbf{r}-\mathbf{r}^{\prime }\mid }{l_{N}}\right) .
\label{6.48}
\end{equation}
The function $g(x)$ is specified by expression (\ref{6.10}).

The identity

\begin{equation}
\mathbf{v}_{\mathrm{eff}}^{\vee }\mathbf{\nabla }C\equiv \mathbf{\nabla }[%
\mathbf{v}_{\mathrm{eff}}^{\vee }C]-C\mathbf{\nabla v}_{\mathrm{eff}}^{\vee
},  \label{6.49}
\end{equation}
expressions (\ref{6.42}), and also transformations (\ref{6.45}),(\ref{6.46})
allow us to rewrite equation (\ref{6.41}) in the form

\begin{equation}
\frac{\partial C}{\partial t}+\mathbf{\nabla }[\mathbf{v}_{\mathrm{eff}%
}C]+j[1+\chi _{v}(\mathbf{r}\,)]C=q.  \label{6.50}
\end{equation}
Here, $\mathbf{v}_{\mathrm{eff}}(\mathbf{r})$ is the velocity of effective
potential blood flow defined by the formula

\begin{equation}
\mathbf{v}_{\mathrm{eff}}=\mathbf{\nabla }\mathcal{P},  \label{6.51}
\end{equation}
where $\mathcal{P}=\mathcal{P}^{a}+\mathcal{P}^{\vee }$ is the velocity
potential which satisfies the equation

\begin{equation}
\mathbf{\nabla }^{2}\mathcal{P}=-2j\chi _{v}(\mathbf{r}\,).  \label{6.52}
\end{equation}

\subsection{The diffusive type heat transport}

\label{s6.4.1}

Here we analyse heat transfer in the limit

\begin{equation}
\gamma \ll 1; \quad l_N \ll l_c ; \quad \frac{M^{2/3}}{\gamma} \ll G \ll
M^{2/3} \frac{ml_c}{l_N} \frac{\ln (d_N/a_N)}{\ln (d_c/a_c)}.  \label{6.53}
\end{equation}

Under these conditions, as in the previous case, venules can be regarded as
walker traps. However, here, until a walker reaches one of the venules, i.e.
until it is trapped, its motion in the tissue is mainly controlled by
capillaries of the second subgroup. Indeed, $\zeta _{N}\approx GM^{-2/3}$
(see (\ref{5.2}), (\ref{5.3})) and, thereby, first, according to (\ref{5.30}%
), (\ref{5.31}), (\ref{6.53}), the value of the effective diffusion
coefficient $D_{\mathrm{eff}}$ is determined by the effect of blood flow in
capillaries. Second, the influence of blood flow in the first subgroup
capillaries on the walker motion in the vicinity of a given venule or
arteriole is ignorable, as it follows from (\ref{6.33}) and (\ref{6.53}). We
also point out that the last inequality of (\ref{6.53}) due to (\ref{5.16a}%
), (\ref{5.16b}) practically coincides with the assumed overall restriction (%
\ref{5.17}) implying that capillaries are heat-permeable vessels.

There is a certain small neighborhood of a given venule or arteriole where
the first subgroup capillaries are dominant in number. At the boundary of
this neighborhood the mean distance $d_{I}(r_{c})$ between the capillaries
of the first subgroup must be equal to the mean distance between the
capillaries of the second subgroup which is about $d_{c}$ due to $l_{c}\gg
l_{N}$. Whence, using (\ref{3.12}), (\ref{5.19}) and (\ref{5.31}) we find
that the radius of this neighborhood is

\begin{equation}
r_{c}\sim \frac{l_{N}}{m}\frac{1}{\gamma }\frac{\ln (d_{c}/a_{c})}{[\ln
(d_{N}/a_{N})]^{2}}.  \label{6.54}
\end{equation}
However, the mean distance $l_{\parallel c}$ over which a walker is
transported along a capillary by blood flow during its location in the
capillary is substantially larger than $r_{c}$. Indeed, taking into account (%
\ref{5.18}),(\ref{6.54}) and the third inequality of (\ref{6.53}) we get

\begin{equation}
\frac{r_c}{l_{\parallel c}} \sim \frac{1}{\gamma \zeta _N} \frac{1}{\ln
(d_N/a_N)} \ll 1.  \label{6.55}
\end{equation}
Therefore, to describe the manner in which a walker can arrive at a given
venule we have to consider in more detail its motion in the vicinity of this
venule.

When $l_{\parallel c}\ll l_{N}$ for a walker that has come into a
neighborhood $Q_{i}(l_{\parallel c})$ of a given venule $i$ whose radius is
about $l_{\parallel c}$, there are practically two ways of going into the
venule. In one way the walker may get into one of the capillaries leading to
the venule, i.e. belonging to the first subgroup, and with blood flow arrive
at the venule. In the other way, the walker, first, gets into a capillary of
the second subgroup, that passes near this venule, then leaves this
capillary in the vicinity of the venule and going through the cellular
tissue reaches the venule. On scales larger than $l_{\parallel c}$ we may
describe the walker motion in terms of random walks in an effective
homogeneous medium with the diffusion coefficient $D_{\mathrm{eff}}\gg D$.

Let us find, first, the probabilities of the walker reaching the venule in
the two ways. By virtue of (\ref{6.55}) in the neighborhood $%
Q_{i}(l_{\parallel c})$ the second subgroup capillaries are dominant in
number and, thus, the mean distance between these capillaries is about $%
d_{c} $. The distance $d_{I}(r)$ between the first subgroup capillaries
depends on the distance $r$ from the venule $i$ (see (\ref{5.19})). However,
due to (\ref{6.55}) its value averaged over the neighborhood $%
Q_{i}(l_{\parallel c})$ is well above $d_{c}$ and can be estimated as $%
d_{I}(l_{\parallel c})\gg d_{c}$. So, in the neighborhood $%
Q_{i}(l_{\parallel c})$ approximately $d_{I}(l_{\parallel c})/d_{c}^{2}$
capillaries of the second subgroup fall at one capillary of the first
subgroup. The latter allows us to estimate the probability that the walker
wandering throughout the cellular tissue in $Q_{i}(l_{\parallel c})$ meets a
capillary of the first rather than the second subgroup, i.e. actually the
probability of the walker arriving at the venule in the first way, as

\begin{equation}
p_{1}\sim \frac{d_{c}^{2}}{d_{I}^{2}(l_{\parallel c})}\sim \left[ \gamma
\zeta _{N}\ln \left( \frac{d_{N}}{a_{N}}\right) \right] ^{-1}\ll 1.
\label{6.56}
\end{equation}
When obtaining (\ref{6.56}) expressions (\ref{3.12}), (\ref{5.18}),(\ref
{5.19}) and (\ref{5.31}) have been also taken into account.

As it follows from (\ref{4.110}) and Section~\ref{s5.4} the mean time during
which a walker can go in the cellular tissue without touching the
capillaries is about $\tau _{lc}\sim d_{c}^{2}/(2\pi D)\ln (d_{c}/a_{c})$.
During this time the walker passes a distance of order $(D\tau
_{lc})^{1/2}\sim d_{c}[\frac{1}{(2\pi )}\ln (d_{c}/a_{c})]^{1/2}$. So, in
cases where the walker motion in the neighborhood $Q_{i}(l_{\parallel c})$
on the scale $l_{\parallel c}$ is determined by blood flow in the second
subgroup capillaries only, the walker can arrive at the venule, if it leaves
the capillaries near the venule at a distance smaller than $d_{c}[\frac{1}{%
(2\pi )}\ln (d_{c}/a_{c})]^{1/2}$.

As it is shown in Chapter~\ref{ch.4} the motion of the walker in tissue with
capillaries can be represented in terms of sequential parts of its motion
near one of the capillaries and in the cellular tissue without touching the
capillaries. In motion near a capillary before going out of its nearest
neighborhood the walker passes, on the average, the distance $l_{\parallel
c} $. So, when moving near a capillary passing through the neighborhood $%
Q_{i}(l_{\parallel c})$ the walker leaves the capillary inside $%
Q_{i}(l_{\parallel c})$ and the probability of leaving the capillary in the
vicinity of a given point is approximately the same for all the points of $%
Q_{i}(l_{\parallel c})$. If, then, the walker meets another capillary it
will go out of the neighborhood $Q_{i}(l_{\parallel c})$ with blood flow in
this capillary.

Thus, for the walker to arrive at the venule it is necessary that $(i)$ the
capillary, blood flow wherein has transported the walker into $%
Q_{i}(l_{\parallel c})$, pass near the venule at a distance of order $d_{c}[%
\frac{1}{(2\pi )}\ln (d_{c}/a_{c})]^{1/2}$ $(ii)$ the walker leave the
capillary near the venule at a distance of the same order and $(iii)$ reach
the venule within the time $\tau _{lc}\sim d_{c}^{2}/(2\pi D)\ln
(d_{c}/a_{c})$. Taking into account the aforesaid the probability of the
first event as well as the second one can be estimated as $%
d_{c}/l_{\parallel c}[\frac{1}{(2\pi )}\ln (d_{c}/a_{c})]^{1/2}$. In Chapter~%
\ref{ch.4} we have obtained expression (\ref{4.87}) practically for the
Laplace transform of the probability for a walker going in the cellular
tissue to reach for the first time the boundary of a pipe of radius $a$ at
time $t$ provided the walker has been at a distance $\rho _{0}$ from the
pipe centerline at initial time. Setting in this expression $\rho _{0}\sim
(2D\tau _{lc})^{1/2}$ and the Laplace transform variable $s=1/\tau _{lc}$ we
may estimate the probability of the third event. In this way by virtue of (%
\ref{4.87}) and the obtained expressions for the probability of the first
and second events, the probability of the walker arriving at the venule in
the second way can be represented as

\begin{equation*}
p_2 \sim \frac{d^2_c}{2\pi l^2_{\parallel c}} \ln \left ( \frac{d_c}{a_c}
\right ) \left [ \ln \left ( \frac{(2D\tau _{lc})^{1/2}}{a_N} \right )
\right ]^{-1} \sim
\end{equation*}
\begin{equation}
\sim \left [ \gamma \zeta ^2_N \ln \left ( \frac{(2D\tau _{lc})^{1/2}}{a_N}
\right ) \right ]^{-1} \ll 1.  \label{6.57}
\end{equation}

When obtaining (\ref{6.57}) we have assumed that $2D\tau _{lc} \gg a^2_N$,
used the asymptotic formula for the Bessel function $k_0 (x)$ for $x \ll 1$,
supposed that $k_0 (x) \sim 1$ for $x\sim 1$, and also taken into account (%
\ref{3.12}),(\ref{5.18}), and (\ref{5.31}).

Now let us analyse the walker motion on scales larger than $l_{\parallel c}$%
. If for each venule we specify the neighborhood $Q_{i}(l_{\parallel c})$ of
radius $l_{\parallel c}$ then, we may assert that a walker, during its
motion in the tissue before being trapped by the venules, visits a large
number of such venule neighborhoods. Indeed, on one hand, as it follows from
the results obtained below (see (\ref{6.63})) and, in addition, as one could
expect on the basis of the general laws of heat transfer, the mean time $%
\tau $ during which a walker is inside the tissue before being trapped by
the venules, is about $1/j(\tau \sim 1/j)$. On the other hand, according to (%
\ref{5.13}) the mean time in which a walker reaches one of the given venules
neighborhoods for the first time can be estimated as

\begin{equation}
\frac{d_{N}^{2}}{2\pi D_{\mathrm{eff}}}\ln \left( \frac{d_{N}}{l_{\parallel
c}}\right) \sim \frac{1}{j\gamma \zeta _{N}}\frac{\ln (d_{N}/l_{\parallel c})%
}{\ln (d_{N}/a_{N})},  \label{6.58}
\end{equation}
where also (\ref{4.121}), (\ref{5.30}), (\ref{5.31}) and the relation $j=\pi
a_{N}^{2}v_{N}/(l_{N}d_{N}^{2})$ have been taken into account. Therefore, a
walker, during its motion in tissue for the time $\tau $, visits, on the
average, $N_{d}$ different venule neighborhoods where due to (\ref{5.30}), (%
\ref{6.53}), (\ref{6.58}), and the relation $M=2^{3N}$

\begin{equation}
N_{d}\sim \frac{2\pi D_{\mathrm{eff}}\tau }{d_{N}^{2}}\left[ \ln \left( 
\frac{d_{N}}{l_{\parallel c}}\right) \right] ^{-1}\sim \tau j\gamma \zeta
_{N}\frac{\ln \left( \frac{d_{N}}{a_{N}}\right) }{\ln \left( \frac{d_{N}}{%
l_{\parallel c}}\right) }>\gamma GM^{-2/3}\gg 1,  \label{6.59}
\end{equation}
provided $\tau j\sim 1$.

For a walker that at initial time is inside a given neighborhood $%
Q_{i}(l_{\parallel c})$ of a venule $i$ the total time $<t_{Q}>$, during
which the walker resides in this neighborhood until it goes out of the
fundamental domain $Q_{N}$ containing the venule $i$, can be estimated by
setting in (4.109) $t\sim d_{N}^{2}/(2D_{\mathrm{eff}})$ and replacing $D$
by $D_{\mathrm{eff}}$ and $a$ by $l_{\parallel c}(D\rightarrow D_{\mathrm{eff%
}},a\rightarrow l_{\parallel c})$. In this way we get

\begin{equation}
\left\langle t_{Q}\right\rangle \sim \frac{l_{\parallel c}^{2}}{2D_{\mathrm{%
eff}}}\ln \left( \frac{d_{N}}{l_{\parallel c}}\right) .  \label{6.60}
\end{equation}

According to the description of the walker motion in tissue with capillaries
which has been developed in Chapter~\ref{ch.4}, a walker can go inside the
cellular tissue without touching the capillaries, on the average, during the
time $\tau _{lc}\sim d_{c}^{2}/(2\pi D)\ln (d_{c}/a_{c})$ (cf.(4.110)).
Then, the walker gets into a capillary and with blood flow travels a
distance of order $l_{\parallel c}$ along the capillary until it leaves the
nearest neighborhood $Q_{i}(d_{c})$ of this capillary in a time of order $%
d_{c}^{2}/(2D)$. The sequence of the two types of the walker motion, i.e.
its motion solely in the cellular tissue for a time of order $\tau _{lc}$
and its subsequent motion in $Q_{i}(d_{c})$, can be regarded as one complex
step of the walker on the capillary network. The latter allows us to
consider the walker motion in the tissue, until the walker is trapped by the
venules, in terms of random walker on the capillary network which are made
up of such steps. In limit (\ref{6.53}) $l_{\parallel c}\gg d_{c}(\ln
(d_{c}/a_{c}))^{1/2}$, i.e. $D_{\mathrm{eff}}\gg D$, thereby, the mean
duration of one step is about $l_{\parallel c}^{2}/(2D_{\mathrm{eff}})$
because in this case the total length of one step is practically equal to $%
l_{\parallel c}$. We note that the mean step duration is bound to be of
order $\tau _{lc}\sim d_{c}^{2}/(2\pi D)\ln (d_{c}/a_{c})$ due to in the
given model $\ln (d_{c}/a_{c})$ being regarded as a large parameter. By
virtue of (\ref{5.18}) and (\ref{5.29}) the two estimates of the same
quantity are in agreement with one other. So for random walks originating in
the vicinity of the venule $i$ and corresponding to the walker motion inside
the fundamental domain $Q_{N}$ expression (\ref{6.60}) and the former
estimate of the step duration allow us to represent the mean total number $%
N_{i}$ of the steps inside the neighborhood $Q_{i}(l_{\parallel c})$ in the
form

\begin{equation}
N_{i}\sim \ln \left( \frac{d_{N}}{l_{\parallel c}}\right) .  \label{6.61}
\end{equation}
Thus, if a walker during its motion in tissue during the time $\tau $ visits 
$N_{d}$ such venule neighborhoods, then for the corresponding random walk
the total number of its steps inside these neighborhoods can be estimated as 
$N_{t}\sim N_{i}N_{d}$. Then, from (\ref{6.59}) and (\ref{6.61}) we obtain

\begin{equation}
N_{t}\sim \frac{2\pi D_{\mathrm{eff}}\tau }{d_{N}^{2}}\sim \tau j\gamma
\zeta _{N}\ln \left( \frac{d_{N}}{a_{N}}\right) .  \label{6.62}
\end{equation}

One step of such random walks actually describes the walker motion on scales
of order $l_{\parallel c}$. So, the walker trapping, for example, by a
venule $i$, can be treated in terms of interruption of the corresponding
random walk when its steps reach the neighborhood $Q_{i}(l_{\parallel c})$
of the given venule. The probability of such interruption is determined by
the probability of the walker arriving at the venule after coming into $%
Q_{i}(l_{\parallel c})$. Thereby, it is equal to $p_{1}+p_{2}$.

The results obtained above allow us to estimate the mean time $\tau $ during
which walkers are inside the tissue before being trapped by the venules,
i.e. their lifetime in the tissue. Indeed, if for the time $\tau $ a random
walk, representing the walker motion in the tissue, visits the system of the
neighborhoods $\{Q_{i}(l_{\parallel c})\}$ $N_{t}$ times, the probability of
its interruption will be about $\exp \{-N_{t}(p_{1}+p_{2})\}$ due to $p_{1}$
and $p_{2}$ being much less than one. Therefore, the random walk will be
practically interrupted, i.e. the walker will be trapped by the venules,
when $N_{t}(p_{1}+p_{2})\sim 1$. Whence, also taking into account (\ref{6.56}%
),(\ref{6.57}), and (\ref{6.62}), we get

\begin{equation}
\tau j\left[ 1+\frac{1}{\zeta _{N}}\frac{\ln (d_{N}/a_{N})}{\ln ((2D\tau
_{lc})^{1/2}/a_{N})}\right] \sim 1.  \label{6.63}
\end{equation}
By virtue of (\ref{6.53}), $\zeta _{N}\sim GM^{-2/3}\gg 1/\gamma $ and $%
\gamma \ll 1$, thereby, the second term on the left-hand side of (\ref{6.63}%
) may be ignored. Keeping the latter in mind from (\ref{6.63}) we obtain
expressions (\ref{5.15a}), (\ref{5.15b}) for $\tau $ again. It should be
pointed out that in the given case according to (\ref{6.63}) the walker
trapping mainly comes about in the first way.

When $l_{\parallel c}\gg l_{N}$ the points of successive intersections of
different capillaries by a walker is believed to be randomly distributed on
the scale $l_{N}$. Therefore, in this case, to reach the venules a walker is
bound to go into any capillary at any point being along the capillary at a
distance smaller than $l_{\parallel c}$ from a point where it connects with
a venule. The probability $p_{c}$ of this event can be estimated as

\begin{equation}
p_{c}\sim \frac{l_{\parallel c}}{l_{c}}.  \label{6.64}
\end{equation}
The mean time of walker migration in tissue between sequential intersections
of different capillaries is $\tau _{lc}\sim d_{c}^{2}/(2\pi D)\ln
(d_{c}/a_{c})$. Thus, in this case the lifetime $\tau $ satisfies the
relation

\begin{equation}
p_{c}\frac{\tau }{\tau _{lc}}\sim \tau \frac{l_{\parallel c}}{l_{c}}\frac{%
2\pi D}{d_{c}^{2}\ln (d_{c}/a_{c})}\sim 1.  \label{6.65}
\end{equation}
Whence, taking into account (\ref{3.11}), (\ref{3.12}), (\ref{4.121}) , (\ref
{5.18}) and the relation $j=\pi a_{N}^{2}v_{N}/(l_{N}d_{N}^{2})$ we get
expressions (\ref{5.15a}), (\ref{5.15b}) again.

As it has been mentioned above in limit (\ref{6.53}) the walker motion in
tissue on scales larger than $l_{\parallel c}$ can be treated in terms of
random motion of the walkers in a homogeneous medium with the effective
diffusion coefficient $D_{\mathrm{eff}}$.

So, in this limit the walker concentration $C$ obeys the equation:

\begin{equation}
\frac{\partial C}{\partial t}=\mathbf{\nabla }(D_{\mathrm{eff}}\mathbf{%
\nabla }C)-jC+q.  \label{6.66}
\end{equation}

It should be pointed out that in (\ref{6.66}) we have ignored possible
spatial nonuniformities in the walker distribution that are caused by
features of the venule arrangement on scales of order $l_N$. The latter is
justified because in the given case for a walker to be trapped it has to
visit a large number of the venule neighborhoods $\{Q_i (l_{\parallel c})\}$
due to $p_1 +p_2 \ll 1$.

The three limits considered below and the corresponding main properties of
heat transfer are displayed in Fig.~\ref{Fig16}a.

\section{Influence of long capillaries on heat transfer}

\label{s6.5}

In this Section we shall show that the effect of long capillaries $(\gamma
\gg 1)$ on heat transfer reduces solely to renormalization of the diffusion
coefficient $D$ and the desired form of bioheat equation coincides with (\ref
{6.66}). For these purposes we analyse individually each of the possible
limits.

The limit:

\begin{equation}
\gamma \gg 1; \quad \frac{M^{2/3}}{\gamma ^{1/2}} \ll G \ll M^{2/3}.
\label{6.67}
\end{equation}

For this geometry of the capillary network there is no range of the
parameter $G$ corresponding to the convective type transport because for any
value of $G$ the capillaries of the first subgroup have no significant
effect on the walker motion. Indeed, the first subgroup capillaries can
affect the walker motion if inequality (\ref{6.33}) is true. However ,$%
1/\zeta _{N}+\gamma \zeta _{N}\geq 2\gamma ^{1/2}$, thus, in limit (\ref
{6.67}) inequality (\ref{6.33}) cannot be true for any $\zeta _{N}$, i.e.
any $G$.

When $\gamma \zeta _{N}^{2}\sim \gamma \lbrack GM^{-2/3}]^{2}\ll 1$ the
influence of the second subgroup capillaries is also ignorable as it follows
from (\ref{5.30}). So, in this case and in limit (\ref{6.23}) the properties
of heat transfer are identical. When $\zeta _{N}\sim GM^{-2/3}\gg 1$ the
venules and arterioles are the smallest vessels of the first class, the
second class veins do not exist at all, and the walker transport in the
tissue is controlled by capillaries of the second group. This case has been
practically considered in limit (\ref{6.53}). For this reason when $\gamma
>1 $ we shall examine limit (\ref{6.67}) only when $1/\gamma ^{1/2}\ll \zeta
_{N}\ll 1$.

In this case the walker transport in the tissue on scales larger than $%
l_{\parallel c}$ is controlled by blood flow in the capillaries and can be
described by random motion of the walkers in a medium with the effective
diffusion coefficient $D_{\mathrm{eff}}\gg D$. The venules and arterioles,
however, are heat-permeable vessels, thus, the walkers can be trapped only
by veins whose level number $n\mathbf{\leqslant }n_{t}<N$. None of the
capillaries is connected with the veins directly (except for the venules).
So in this case, as it has been discussed in limit (\ref{6.53}) regarding to
the walker trapping in the second way, to get into a vein of level $n$ a
walker should, first, reach the vein neighborhood of radius $d_{c}[1/(2\pi
)\ln (d_{c}/a_{c})]^{1/2}$ with blood flow in a capillary passing near the
vein and then leave this capillary and travelling through the cellular
tissue go into the vein.

Keeping the latter in mind and replicating one-to-one the analysis leading
to formulas (\ref{6.57}) and (\ref{6.62}) we obtain the following. First,
for a walker that has come into a neighborhood $Q_i (l_{\parallel c})$ of
such a vein, whose radius is about $l_{\parallel c}$, the probability of
getting into the vein before leaving this neighborhood is approximately
equal to

\begin{equation}
p_{2}^{\prime }\sim \left[ \gamma _{N}^{2}\ln \left( \frac{(2D\tau
_{lc})^{1/2}}{a_{n}}\right) \right] ^{-1}  \label{6.68}
\end{equation}
and due to (\ref{6.67}) $p_{2}^{\prime }\ll 1$. Second, on scales larger
than $l_{\parallel c}$ we may represent a path of the walker motion in the
tissue during time $t\gg \tau _{lc}\sim d_{c}^{2}/(2\pi D)\ln (d_{c}/a_{c})$
as a random walk formed by a sequence of steps, the mean length of which is
about $l_{\parallel c}$. In these terms during a time $t\gg d_{n}^{2}/(2\pi
D)\ln (d_{c}/a_{c})$ the random walk can visit such neighborhoods of the $n$%
-th level veins $N_{t}^{\prime }$ times where, on the average,

\begin{equation}
N^{\prime}_t \sim \frac{d^2_N}{d^2_n}(tj) \gamma \zeta _N \ln \left ( \frac{%
d_N}{a_N} \right )  \label{6.69}
\end{equation}
provided the random walk has not been interrupted, i.e. the walker has not
been trapped.

Expressions (\ref{6.68}),(\ref{6.69}) enable us to find the mean time $\tau
^{\prime}_{ln}$ during which a walker traveling through the tissue will get
into one of the $n$-th level veins. Indeed, this time is bound to satisfy
the relation

\begin{equation}
\left. p^{\prime}_2 N^{\prime}_t \right | _{t=\tau ^{\prime}_{ln}} \sim 1.
\label{6.70}
\end{equation}
Substituting (\ref{6.68}),(\ref{6.69}) into (\ref{6.70}) we obtain

\begin{equation}
\tau ^{\prime}_{ln} \sim \frac{1}{j} \zeta _n \frac{\ln ((2D\tau
_{lc})^{1/2}/a_n)}{\ln (d_N/a_N)},  \label{6.71}
\end{equation}
where we also have taken into account the identity

\begin{equation}
\zeta _n \left [ d^2_n \ln \left (\frac{d_n}{a_n} \right ) \right ]^{-1}
\equiv \zeta _n \left [ d^2_N \ln \left (\frac{d_N}{a_N} \right ) \right %
]^{-1}  \label{6.72}
\end{equation}
resulting from (\ref{3.6}),(\ref{3.11}), and (\ref{4.121}).

In the case under consideration not all veins of the first class can trap
walkers but only those whose parameter $\zeta _n$ meets the condition

\begin{equation}
\zeta _n \geq \frac{\ln (d_n/a_n)}{\ln ((2D\tau _{lc})^{1/2}/a_n)}.
\label{6.73}
\end{equation}
In fact, if a walker arrives at central points of a vein whose level number $%
n<n_*$, i.e. for which $\zeta _n \geq \ln (d_n/a_n)$ the walker will be
trapped because in this case it has no time to return to the cellular
tissue. If a walker reaches a vein whose level number $n>n_*$ it will be in
the vicinity of this vein within the time $\tau _{lc}$ only. In a time of
order $\tau _{lc}$ the walker gets into one of the capillaries blood flow
wherein carries the walker away from the vein. So, practically replicating
the analysis of Section~\ref{s5.2} and taking into account (\ref{3.4}) we
find that the walker will be trapped by this vein if

\begin{equation}
\l_n < \frac{a^2_n v_n}{2D} \ln \left ( \frac{(2D\tau _{lc})^{1/2}}{a_n}
\right ).  \label{6.74}
\end{equation}
The latter condition and expression (\ref{3.13}) directly lead us to
condition (\ref{6.73}). The walker trapping is practically controlled by the
shortest veins whose parameter $\zeta _n$ meets condition (\ref{6.73}), i.e.
by the veins whose level number $n^{\prime}_t$ obeys the equation

\begin{equation}
\zeta _{n^{\prime}_{t}} \approx \frac{\ln (d_{n^{\prime}_t}/a_{n^{\prime}_t})%
} {\ln ((2D\tau _{lc})^{1/2}/a_{n^{\prime}_t})} .  \label{6.75}
\end{equation}
Thus, the mean time $\tau$ during which a walker will be trapped, may be
found from (\ref{6.71}) setting $t=\tau$ and $n=n^{\prime}_t$. By virtue of (%
\ref{6.75}), in this way we obtain the estimate $\tau \sim 1/j$ again.

Therefore, also in the case under consideration the walker distribution in
tissue is governed by equation (\ref{6.66}). Besides, it should be pointed
out that in limit (\ref{6.67}) the class of the heat-impermeable vessels
contains the veins whose level number $n<n_{t}^{\prime }$ rather than $%
n<n_{t}$, where $n_{\ast }<n_{t}^{\prime }<n_{t}$. However, due to $%
l_{n}/a_{n}\sim 30-40$ (see the comments just below expression (\ref{3.7}))
and according to (\ref{5.10}) the value $n_{t}-n_{\ast }\sim 1$. For this
reason we have ignored the latter characteristics in the general analysis of
heat transfer in the previous sections.

The possible types of heat transfer in living tissue where capillaries are
sufficiently long $(\gamma \gg 1)$ are displayed in Fig.~\ref{Fig16}b.

In conclusion of this Chapter we note that when the characteristic total
length of capillaries joining a given arteriole to venules is not too long
(i.e. when $\gamma \ll 1$ within the framework of the present model),
depending on the value of $G$ the effect of the capillary system on heat
transfer can be of different types. In particular, in limit (\ref{6.23}) the
capillary system influence is ignorable and in this case the temperature
distribution%
\index{temperature distribution} (in terms of the walker concentration) is
described by equation (\ref{6.29}).

In limit (\ref{6.32}) the capillary system causes convectional type
transport of heat in the tissue and the tissue temperature obeys equation (%
\ref{6.50}). In this case the capillary system may be considered in terms of
a porous medium.

In limit (\ref{6.53}) heat transport is controlled again by diffusive type
process in a certain effective medium but with the effective diffusion
coefficient $D_{\mathrm{eff}}$ being a function of the blood flow rate.

When the capillary length is long enough (i.e. when $\gamma \gg 1$) the
range of convectional type transport is absent. In this case heat transfer
is controlled by diffusive type transport. In limit (\ref{6.67}) the
cellular tissue containing the capillary system can be treated as an
effective uniform medium with the diffusion coefficient $D_{\mathrm{eff}}$.

\chapter{Generalized bioheat equation}

\label{ch.7} \markright
{ {\sc  \thechapter. Generalized bioheat equation}
}

In Chapter~\ref{ch.6} in the possible limits we have obtained the different
forms of the equation governing evolution of the temperature field $T(%
\mathbf{r},t)$ in tissue (in terms of the walker concentration $C(\mathbf{r}%
,t)$). Comparing these equations with each other we can propose a
generalized bioheat equation which describes heat transfer in living tissue.
It should be pointed out that such a model is bound to comprise equations (%
\ref{6.29}), (\ref{6.50}), and (\ref{6.66}) as particular cases. The present
analysis also holds true for the nonuniform blood flow rate $j(\mathbf{r}\,)$
providing all spatial scales of its nonuniformities are well above $%
l_{n_{t}} $ if $\zeta _{N}<1$ and $l_{N}$ for $\zeta _{N}>1$. Due to (\ref
{3.10}) when $l_{c}\gg l_{N}$ the value $\mathcal{R}\gg l_{N}$ and, thus,
there are cases where the influence of such nonuniformities in $j(r)$ on
heat transfer should be taken into account in the generalized model.

We note that equations (\ref{6.29}) and (\ref{6.66}) are practically of the
same form because for $D_{\mathrm{eff}}\gg D$ the influence of the random
field $\chi _{t}(\mathbf{r}\,)$ on heat transfer is suppressed. To combine
these equations with equation (\ref{6.50}) we may represent the mean walker
flux $\mathbf{J}_{w}$ as the sum of the diffusive type flux and the
convective one:

\begin{equation}
\mathbf{J}_{w}=-\mathbf{\nabla }[D_{\mathrm{eff}}C]+\mathbf{v}_{\mathrm{eff}%
}C,  \label{7.1}
\end{equation}
where $D_{\mathrm{eff}}$ and $\mathbf{v}_{\mathrm{eff}}$ are determined by
expressions (\ref{5.16a}), (\ref{5.16b}), (\ref{5.26}) (or (\ref{5.27})) and
(\ref{6.39}), respectively. In all the limits considered in Chapter~\ref
{ch.6} expression (\ref{7.1}) yields the correct results. Also we point out
that on scales larger than $l_{\parallel c}$ heat transport, i.e. the walker
motion, in the tissue seems to be described by stochastic processes of the
Ito type. So in expression (\ref{7.1}) we have to put the effective
diffusion coefficient $D_{\mathrm{eff}}$ after the operator $\mathbf{\nabla }
$ \cite{26}. Besides it is possible to generalize the bioheat equations (\ref
{6.29}) and (\ref{6.50}) in such a way that the generalized equation allows
for nonuniformities in the trap distribution and the convective treat
transport due to blood flow in the capillary system.

\section{Effective diffusion coefficient}

\label{s7.1}

Due to the effective diffusion coefficient $D_{\mathrm{eff}}$ being the
result of cooperative influence of all the vessels the total value of $D_{%
\mathrm{eff}}$ can be represented as the sum of particular values of
effective diffusion coefficients induced by blood flow in vessels of various
groups. In other words we may write

\begin{equation}
D_{\mathrm{eff}}=D[1+F_{v}(G)+F_{c}(G)],  \label{7.2}
\end{equation}
where $G=%
\frac{4jl_{0}^{2}}{3\sqrt{3}\pi D}\ln \left( \frac{l_{0}}{a_{0}}\right) $,
the term $F_{v}(G)$ is caused by blood flow in the second class vessels and
the capillary system is responsible for the term $F_{c}(G)$. According to (%
\ref{5.16a}) and (\ref{5.16b}) for $G_{0}<G<G_{cc}^{\ast }$ or $%
G_{cc}<G<M^{2/3}$

\begin{equation}
F_{v}(G)=\pi \sqrt{3}\frac{1}{\ln (l_{0}/a_{0})},  \label{7.3a}
\end{equation}
for $G_{cc}^{\ast }<G<G_{cc}$

\begin{equation}
F_{v}(G)=\pi \sqrt{3}\frac{1}{\ln (l_{0}/a_{0})}\left( \frac{G}{G_{cc}^{\ast
}}\right) ^{2},  \label{7.3b}
\end{equation}
and for $G>M^{2/3}$

\begin{equation}
F_{v}(G)=0,  \label{7.3c}
\end{equation}
because for $G>M^{2/3}$ vessels of Class 2 do not exist. By virtue of (\ref
{5.27}) and (\ref{5.3})

\begin{equation}
F_{c}(G)=\gamma M^{4/3}G^{2}.  \label{7.4}
\end{equation}

\section{Effective convective flux}

\label{s7.2}

The results obtained in Section~\ref{s6.3} actually describes the general
convective type effect of blood flow in capillaries of the first subgroup.
Beyond limit (\ref{6.32}) the influence of these capillaries on heat
transfer is practically ignorable. Thus, the corresponding expressions can
be considered to blood for all values of $G$. Therefore, we may assume that
in living tissue there is an effective potential convective flux whose
velocity field $\mathbf{v}_{\mathrm{eff}}(\mathbf{r}\,)$ is specified by
expressions

\begin{equation}
\mathbf{v}_{\text{\textrm{eff}}}(\mathbf{r}\,)=\mathbf{\nabla }\mathcal{P\;}%
\text{.}  \label{7.5}
\end{equation}
Here the velocity potential $\mathcal{P}$ obeys the equation

\begin{equation}
\mathbf{\nabla }^{2}\mathcal{P}=-2j(\mathbf{r}\,)\chi _{v}(\mathbf{r}\,)
\label{7.6}
\end{equation}
and the random field $\chi _{v}(\mathbf{r}\,)$ satisfies the conditions

\begin{equation}
<\chi _{v}(\mathbf{r}\,)>=0,  \label{7.7}
\end{equation}
\begin{equation}
<\chi _{v}(\mathbf{r}\,)\chi _{v}(\mathbf{r}^{\prime })>=g\left( \frac{\mid 
\mathbf{r}-\mathbf{r}^{\prime }\mid }{l_{N}}\right) ,  \label{7.8}
\end{equation}
where the function $g(x)$ is given by expression (\ref{6.10}).

\section{Nonuniformities in heat sink due to the vessel discreteness}

\label{s7.3}

In Chapter~\ref{ch.6} we also have considered the influence of random
nonuniformities in vessel positions on the walker trapping. In limits (\ref
{6.28}) and (\ref{6.32}) where this influence is the strongest, it can be
described by the terms of the same form, viz. $jC\chi _{t}$ and $jC\chi _{v}$%
, respectively. In limits (\ref{6.53}) and (\ref{6.67}) this effect is
suppressed due to $D_{\mathrm{eff}}\gg D$. Therefore, in both these cases we
may describe the walker trapping (heat disappearance) by the term $j(\mathbf{%
r}\,)[1+\chi _{t}(\mathbf{r}\,)]C$, where $\chi _{t}(\mathbf{r}\,)$ is a
random field (denoted by the same symbol) that meets the conditions

\begin{equation}
\left\langle \chi _{t}(\mathbf{r}\,)\right\rangle =0,  \label{7.9}
\end{equation}
\begin{equation}
\left\langle \chi _{t}(\mathbf{r}\,)\chi _{t}(\mathbf{r}^{\prime
})\right\rangle =g\left( \frac{\mid \mathbf{r}-\mathbf{r}^{\prime }\mid }{%
l_{t}}\right) ,  \label{7.10}
\end{equation}
where $g(x)$ is specified by expression (\ref{6.10}) again and

\begin{equation}
l_{t}=l_{n}\mid _{n=n_{t}}=l_{0}2^{-n_{t}(G)}  \label{7.11}
\end{equation}
or by virtue of (\ref{5.4}), for $G\mathbf{\leqslant }G_{cc}$ when the
vessel tree contains countercurrent pairs at level $n_{t}$

\begin{equation}
l_{t}\simeq \lbrack \frac{3\sqrt{3}\pi }{4}\frac{D}{j[\ln
(d_{0}/a_{0})]^{1/2}}]^{1/2},  \label{7.12a}
\end{equation}
and for $G_{cc}<G<M^{2/3}$ when unit veins and arteries belong to level $%
n_{t}$.

\begin{equation}
l_{t}=\left[ \frac{3\sqrt{3}\pi }{4}\frac{D}{j\ln (d_{0}/a_{0})}\right]
^{1/2}.  \label{7.12b}
\end{equation}
For $G>M^{2/3}$ there is no artery or vein of Class 2 and in this case

\begin{equation}
l_{t}=l_{N}.  \label{7.13}
\end{equation}
When $G>M^{2/3}$ the random fields $\chi _{t}(r)$ and $\chi _{v}(r)$ must be
identical, otherwise, $G<M^{2/3}$ they are independent of each other. Such
conditions we may represent in the following form

\begin{equation}
\left\langle \chi _{t}(\mathbf{r}\,)\chi _{v}(\mathbf{r}^{\prime
})\right\rangle =g\left( \frac{\mid \mathbf{r}-\mathbf{r}^{\prime }\mid }{%
l_{N}}\right) \theta (l_{N}-l_{n_{t}})  \label{7.14}
\end{equation}
where $\Theta (x)=0$ for $x<0$ and $\Theta (x)=1$ when $x>0$ and $l_{n_{t}}$
is formally given by expressions (\ref{7.12b}) for $G>M^{2/3}$.

\section{Generalized bioheat transfer equation}

\label{s7.4}

Comparing equations (\ref{6.29}), (\ref{6.50}) and (\ref{6.66}), (\ref{5.30}%
) with each other, taking into account expressions obtained above in this
Section and returning to the variable $T$ (see Chapter~\ref{ch.3}) we may
write the generalized bioheat equation in the following form

\begin{equation*}
c_{t}\rho _{t}\frac{\partial T}{\partial t}=\kappa \lbrack
1+F_{v}(G)+F_{c}(G)]\mathbf{\nabla }^{2}T-c_{t}\rho _{t}\mathbf{\nabla }[%
\mathbf{v}_{\mathrm{eff}}(T-T_{a})]-
\end{equation*}
\begin{equation}
-c_{t}\rho _{t}j\left( \ln \frac{l_{0}}{a_{0}}\right) ^{(\beta
(n_{t})-1)/2}[1+\chi _{t}(\mathbf{r}\,)](T-T_{a})+q_{h}\,.  \label{7.15}
\end{equation}

The system of equation (\ref{7.15}) and expressions (\ref{7.3a}), (\ref{7.3b}%
), (\ref{7.3c}), (\ref{7.4}) forms the proposed generalized model for heat
transfer.

It should be pointed out once again that the proposed model is justified
only when all spatial scales of nonuniformities in the blood flow rate $j$
are well above $l_{t}$. The opposite case can take place, for example, in
hyperthermia of a small tumor, and to investigate the corresponding
influence of the blood flow rate on heat transfer individual analysis is
required.

Concluding the present Chapter we compare the derived equation (\ref{7.15})
with bioheat equations proposed by other authors.

When $G \ll M^{2/3}$ the value $n_t \ll N$ and blood flow in capillaries
practically has no significant effect on heat transfer. In this case, if we,
in addition, ignore random temperature nonuniformities, equation (\ref{7.15}%
) practically coincides with equation (\ref{1.5}) where the unknown
phenomenological parameter

\begin{equation*}
f=\frac{1}{[\ln (l_{0}/a_{0})]^{1/2}}
\end{equation*}
for $G<G_{cc}$. When $G>G_{cc}$ the parameter $f=1$ and equation (\ref{7.15}%
) goes over into the equation similar to equation (\ref{1.3}). We note once
again that it is the countercurrent effect responsible for small values of
the coefficient $f,$ and the effective conductivity model (\ref{1.4}) is
justified for temperature distribution nonuniform on spatial scales of order 
$[\kappa \ln (\l _{0}/a_{0})/(c_{t}\rho _{t}j)]^{1/2}$. For extremely high
blood flow rates, when $G<M^{2/3}$ the effective convective heat transport
can be dominant, which justifies, at least qualitatively, equation (\ref{1.2}%
). \clearpage

\part{Heat transfer in living tissue with extremely nonuniform blood flow
distribution}

\markboth{
{\sc \thepart.{ } Heat transfer in living tissue with extremely \ldots}}{}

Equation (\ref{7.15}) as well as (\ref{1.1}) - (\ref{1.5}) has been obtained
practically by averaging the microscopic equations governing heat transfer
in living tissue on scales of individual vessels. In other words, the
mathematical approach to the heat transfer description associated with
equation (\ref{7.15}) actually considers living tissue in terms of a uniform
(may be anisotropic) continuum. The latter is justified as long as the
tissue temperature and the blood flow distribution over the vascular network
are practically uniform on the scale $l_{v}$, characterizing inhomogeneity
of the living tissue from the heat transfer standpoint. The magnitude of $%
l_{v}$ is approximately equal to the length of vessels, that directly
control heat exchange between the cellular tissue and blood, and for typical
values of $j\sim 3\cdot 10^{-3}s^{-1}$ can be estimated as $l_{v}\sim 0.5cm$
(see Chapter~\ref{ch.2}). When heating or cooling the living tissue is
sufficiently strong the thermoregulation gives rise to substantial local
dependence of the blood flow rate on the tissue temperature \cite{54}.
Therefore, if the size of the tissue domain affected directly is less than $%
l_{v}$ (which typically is the case, at least, in cryosurgery the blood flow
distribution and, consequently, the blood flow rate can became nonuniform
already on scales of order $l_{v}$. In this case to describe heat transfer
in living tissue correctly the bioheat equation should be modified.

In the present part we propose a possible alternative to such a modified
bioheat equation which governs heat transfer in living tissue with extremely
nonuniform blood flow distribution over the vascular network.

\chapter{The bioheat equation and the averaged blood flow rate}

\label{ch.8} 
\markright
{ {\sc \thechapter. The bioheat equation and the averaged blood\ldots }
}

\section{Should the bioheat equation contain the true blood flow rate?}

\label{s8.1}

Equation (\ref{7.15}) has been justified in the previous part for uniform
blood flow distribution over the microcirculatory bed domain. This raises
the question of whether the obtained bioheat equation holds for
substantially nonuniform blood flow distribution and the relative question
how to modify it in order to describe adequately\ heat transfer in living
tissue with nonuniform blood flow rate.

For this purpose we consider the physical effects that are reflected in the
conventional bioheat equation

\begin{equation}
c_{t}\rho _{t}\frac{\partial T}{\partial t}=D\mathbf{\nabla }%
^{2}T-j(T-T_{a})+q_{T}.  \label{8.1}
\end{equation}
The second term on the right-hand side of this equation describes heat
exchange between blood and the cellular tissue as volumetric heat
dissipation. The heat sink term is rigorously justified solely for uniform
blood flow distribution over the vascular network. The matter is that in the
strict sense in tissue there is no local heat dissipation. Blood upon
attaining thermal equilibrium with the surrounding tissue during its motion
in small vessels is withdrawn through large veins practically without heat
loss. The latter is due to the velocity of heat convective transport with
blood flow increasing more quickly than the vessel length as the level
number decreases during the blood motion towards the vein stem. Thus, for
cellular tissue such blood - tissue exchange is local in nature. If blood
flow is substantially nonuniform distributed over the vascular network, then
there can be such a path on the vein tree that the main amount of blood
flows through the veins forming this path. The blood current along this path
may not vary significantly. So, for a blood portion flowing along such a
path to words the stem its balk velocity will increase more slowly than the
vessel length. Therefore, in this case there is a possibility that blood
will lose heat energy during the motion through large veins. The latter
gives rise actually to heat exchange between various tissue regions which
differ significantly in size and such heat exchange can be described either
in terms of the local blood flow rate or by effective heat conductivity.

To demonstrate the validity of such speculations let us consider stationary
solution of equation (\ref{8.1}) where the heat generation rate $q_{T}(%
\mathbf{r}\,)$ differs from zero only inside a certain fundamental domain $%
Q^{\ast }$ of size $l^{\ast }$ and the true blood flow rate $j(\mathbf{r})$
takes the value $j^{\ast }$ inside the domain $Q^{\ast }$ and $j$ at the
exterior points. The value of $j^{\ast }$ is assumed to be large enough so
that $(l^{\ast })^{2}\gg D/j^{\ast }$ and, thus, the formal stationary
solution of equation (\ref{8.1}) must be located in the domain $Q^{\ast }$.

However, the microscopic theory of bioheat transfer can lead to another
result. To show this let us consider blood current $J$ along the path $%
\mathcal{P}_{on}$ the vein tree that originates at the vein $i^{\ast }$
which directly drains the domain $Q^{\ast }$ and terminates at the tree
stem. Due to the conservation of blood at the branching points the blood
current $J(l)$ in a vein of length $l$ belonging to this path is about

\begin{equation}
J(l)\sim \left[ j+j^{\ast }\left( \frac{l^{\ast }}{l}\right) ^{3}\right]
l^{3}.
\end{equation}
Then, according to $\left( \zeta _{n}=\frac{a^{2}v}{2}\right) $ for the
vessels of the path $\mathcal{P}$ the classification parameter%
\index{classification parameter} $\zeta (l)$ treated as a function of $l$
can be represented in the form

\begin{equation}
\zeta (l)\sim 
\frac{l^{2}\ln \left( \frac{l_{0}}{a_{0}}\right) }{D}\left[ j+j^{\ast
}\left( \frac{l^{\ast }}{l}\right) ^{3}\right] .  \label{8.2}
\end{equation}
The minimum $\zeta _{\text{in}}$ of the function $\zeta (l)$ is about

\begin{equation}
\zeta _{\text{in}}\sim \frac{(l^{\ast })^{2}}{D}\ln \left( \frac{l_{0}}{a_{0}%
}\right) [(j^{\ast })^{2}j]^{1/3},  \label{8.3}
\end{equation}
and is attained at

\begin{equation}
l_{\text{in}}\sim \left( \frac{j^{\ast }}{j}\right) ^{1/3}l^{\ast }.
\label{8.4}
\end{equation}
Thus, as follows from (\ref{8.3}), for

\begin{equation}
j\ll j^{\ast }\left[ \frac{1}{(l^{\ast })^{2}j^{\ast }}\right] ^{3}\frac{1}{%
\left( \ln \frac{l_{0}}{a_{0}}\right) ^{3}}  \label{8.5}
\end{equation}
the minimum value of the vessel classification parameter $\zeta _{min}\ll 1$%
. Besides, $\zeta (l^{+})\gg 1$. Therefore under condition (\ref{8.5}) the
path contains heat-conservation veins separated by a collection of
heat-dissipation veins. Thereby, heat energy withdrawn by blood flow through
veins of length $l^{\ast }$ from the domain $Q^{+}$ is lost inevitably at
veins of the length $l_{min}\gg l^{\ast }$ and distributed practically
uniformly over a tissue domain of size $l_{min}$. This conclusion is in
contradiction with the formal solution of equation (\ref{8.1}).

Obviously, these speculations hold true for heat transfer in living tissue
with countercurrent pairs.

Summarizing the aforesaid we can assert that the bioheat transfer equation
should not contain the true blood flow rate. We also note that Chen and
Holmes \cite{13} were the first who proposed an alternative to bioheat
equation which deals with a certain averaged blood flow rate rather than the
true one.

In the following Section we develop certain procedure that enables us to
modify the bioheat equation and conceptually coincides the analyses of heat
transfer in living tissue with the nonuniform blood flow rate formulated
above.

\section{The heat conservation vein tree}

\label{s8.2}

The blood flow rate is typically uniform on scales of capillary length.
Therefore when analyzing properties of heat transfer in living tissue with
the nonuniform blood flow rate we may focus main attention on cases where
the capillary system does not affect significantly heat propagation. Under
such conditions solely a blood flow in the venous bed substantially affects
the walker motion. Therefore we may confine our consideration to the venous
bed and the corresponding blood current pattern $\{J_{i}\}$. The effect of a
blood flow through the arterial bed is practically reduced to
renormalization of diffusion coefficient, which also will be taken into
account in the present Part. Besides, for the sake of simplicity we consider
the case when the vascular network is entirely made up of either unit
vessels $(n_{cc}=0)$ or countercurrent pairs $(n_{cc}=N)$.

When blood flow is nonuniform distributed over the microcirculatory bed, to
classify a given vein $i$ according to its effect on walker motion we should
take into account the whole blood current pattern $\{J_{i}\}$ rather than
the blood current $J_{i}$ in this vessel only. In order to construct the
desired classification, let us specify for each point $\mathbf{r}$ of the
domain $Q_{0}$ a sequence of fundamental domains $\{Q_{nr}\}=\{Q_{0},%
\,Q_{1r},...,Q_{Nr}\}$ where the n-th term $Q_{nr}$ is the fundamental
domain of level $n$ which contains the given point $\mathbf{r}$. Besides, by
the symbol $i_{nr}$ we denote the vein corresponding to the domain $Q_{nr}$,
i.e. the vein of level $n$ that is contained in the domain $Q_{nr}$. On the
venous part of the vascular network the vein collection $\{i_{nr}\}=\{i_{0},%
\,i_{1r},...,i_{Nr}\}$ forms a path leading from the host vein $i_{0}\equiv
i_{0r}$to the vein $i_{Nr}$ corresponding to the elementary domain%
\index{elementary domain} $Q_{Nr}$. In addition, formulae (\ref{4.121}), (%
\ref{4.123}) enables us to match up the number sequence $\{\zeta
_{nr}\}=\{\zeta _{0},\,\zeta _{1r},...,\zeta _{Nr}\}$ for the vein
collection $\{i_{nr}\}$. In what follows we shall assume that the number
sequence $\{\zeta _{nr}\}$ can be interpolated by a function $\zeta (n,r)$
of the continuous variable $n$ being smooth on scales of order unity.

When the blood flow distribution is uniform, according to (\ref{3.5}) and (%
\ref{3.16}) the blood current in a vein of level $n$ is $%
J_{n}=J_{0}/M_{n}=J_{0}2^{-3n}$, where $J_{0}$ is the blood current in the
host vein or, what is the same, the total blood current flowing through the
microcirculatory bed. Whence, taking into account (\ref{3.5}), (\ref{4.121}%
), and (\ref{4.123}) we get $\zeta _{nr}=\zeta _{0}2^{-2n}$ for every point $%
\mathbf{r}$. So, in this case the function $\zeta (n,\mathbf{r})$ is
decreasing with respect to the variable $n$.

When blood flow is nonuniform distributed over the microcirculatory bed the
function $\zeta (n,\mathbf{r})$ for a fixed value of $\mathbf{r}$ does not
have to be monotone. Indeed, let, for example, the resistance to blood flow
along a certain path $\{i_{nr^{\prime }}\}$ be extremely small in comparison
with other possible paths. In this case the blood currents in all the veins $%
\{i_{nr^{\prime }}\}$ are approximately the same and due to $%
l_{n}=l_{0}2^{-n}$ and (\ref{4.121}) and (\ref{4.123}) we get $\zeta
_{n}\approx \zeta _{0}2^{-n}$.

In the framework of the adopted assumptions the host vein corresponds to the
parameter $\zeta _{0}\gg 1,$ whereas the last level veins $\{i_{N}\}$ are
associated with $\{\zeta _{i_{N}}\ll 1\}$. The former inequality follows
from the assumption $l_{0}\gg l_{v}$ because, by virtue of (\ref{4.121}), (%
\ref{4.123}), and (\ref{2.3}) and the estimate $%
\tilde{j}\sim J_{0}/l_{0}^{3}$, we obtain $\zeta _{0}\sim
(l_{0}/l_{v})^{2}\gg 1$. The latter inequalities are justified by that in
the given Section effect of blood flow in capillaries on heat transfer is
assumed to be ignorable. Under these conditions a possible characteristic
behavior of the function $\zeta (n,r)$ for a fixed value of $\mathbf{r}$ is
displayed in Fig.~\ref{Fig17} by the solid and dashed lines for nonuniform
and uniform blood flow distributions, respectively. Due to $\zeta _{0}\gg 1$
and $\{\zeta _{i_{N}}\ll 1\}$ the equation $\zeta (n,\mathbf{r}\,)=1$ has,
at least, one root for every point $\mathbf{r}$ and any root is
substantially greater than one. The integer nearest to the first, i.e. to
the least root of the equation $\zeta (n,\mathbf{r}\,)=1$ will be designated
as $n_{r}$ (Fig.~\ref{Fig17}). In other words, $n_{r}$ is such an integer
that

\begin{equation}
\zeta (n_{r},\mathbf{r}\,)\approx 1  \label{8.6}
\end{equation}
and for all $n<n_{r}$ 
\begin{equation}
\zeta (n_{r},\mathbf{r}\,)>1.  \label{8.7}
\end{equation}

Let $\{i_{nr}\}^{\ast }$ be a part of the vein collection $\{i_{nr}\}$ that
involves all the veins whose level number $0\mathbf{\leqslant }n\mathbf{%
\leqslant }n_{r}$, i.e. $\{i_{nr}\}^{\ast
}=\{i_{0},\,i_{1r},...,i_{n_{r}r}\} $. In this way the collection of
integers $\{n_{r}\}$ allows us to specify the vein system $\mathcal{V}$
consisting of the veins belonging, at least, to one of the possible vein
collections $\{i_{nr}\}^{\ast }$. In other words, the system $\mathcal{V}$
is the greatest connected part of the venous bed that is entirely made up of
the veins for which $\zeta _{i}\geq 1$. The vein collection $\{i\}_{v}$
comprising all the veins whose level number $n=n_{r}$ for some point $%
\mathbf{r}$ may be regarded as a certain ``boundary'' of the tree $\mathcal{V%
}$.

Due to $\zeta _{i}>1$ for all the veins of the tree $\mathcal{V}$ (except
the veins $\{i\}_{v}$ for which $\zeta \simeq 1$) the blood flow in each of
these veins substantially affects the walker motion. Therefore, once a
walker has crossed the boundary of one of these veins, for example, of a
vein belonging to level $n$, in a short time it will be transported by blood
flow into a small neighborhood of a vein belonging to level $(n-1)$ or it
will arrive at the central points of the this vein. The following motion of
the walker will proceed in a similar manner. Thus, as it can be shown
directly replicating practically one-to-one the analysis presented in
Chapter~\ref{ch.5}, if a walker crosses the boundary of the vein system $%
\mathcal{V}$, then in a short time it will arrive at central points of these
veins and leave the tissue domain $Q_{0}$ with blood flow through the host
vein actually without returning into the cellular tissue. This event may be
regarded as walker trapping, thus, we can consider the vein tree $\mathcal{V}
$ in terms of walker traps.

\FRAME{ftbpFU}{5.2368cm}{5.5245cm}{0pt}{\Qcb{Possible characteristic
behaviour of the function $\protect\zeta (n,r)$ for a fixed value of $%
\mathbf{r}$ when the blood flow distribution is nonuniform (the solid line)
and unoform (the thin line).}}{\Qlb{Fig17}}{Fig17}{\special{language
"Scientific Word";type "GRAPHIC";maintain-aspect-ratio TRUE;display
"USEDEF";valid_file "F";width 5.2368cm;height 5.5245cm;depth
0pt;original-width 8.9819in;original-height 9.4818in;cropleft "0";croptop
"1.0008";cropright "1.0012";cropbottom "0";filename
'FIG17.GIF';file-properties "XNPEU";}}

Since the mean distance between the vessels of the tree $\mathcal{V}$ is
mainly determined by the shortest veins, i.e. by the veins belonging to the
``boundary'' of the tree $\mathcal{V}$, the walker trapping is actually
controlled by the veins $\{i\}_{v}$. The more larger vessels of the tree $%
\mathcal{V}$ form the paths of walker fast transport from the veins $%
\{i\}_{v}$ to the host vein. It also should be noted that there can be veins
with the corresponding parameters $\zeta _{i}>1$ which do not belong to the
vessel system $\mathcal{V}$, if blood flow is distributed over the vascular
network extremely nonuniform. The blood flow in these veins may have a
substantial local effect on walker motion. However, once a walker has
crossed the boundary of one of these veins it is rapidly transported by
blood flow only until it reaches a vein for which $\zeta _{i}<1$.
Thereafter, the walker inevitably will leave the latter vein and travel a
distance in the cellular tissue much larger than its length. Thus, blood
flow in veins not belonging to the vessel system $\mathcal{V}$ does not
practically affect walker motion, although the inequality $\zeta _{i}>1$ can
hold for some of these veins (Fig.~\ref{Fig17}). Returning to the terms of
heat transfer the aforesaid allows us to treat the vessel system $\mathcal{V}
$ as a part of the venous bed where blood moves so fast that, first, heat
transport in these vessels is mainly governed by blood convective stream,
and, second, blood has no time to attain thermal equilibrium with the
surrounding cellular tissue.

Since , as it has been mentioned before the walker trapping is mainly
controlled the veins $\{i\}_{v}$ the local life time $\tau (\mathbf{r}\,)$
of walkers being in a small neighborhood of the point $\mathbf{r}$ can be
estimated as (see (5.15a,b))

\begin{equation}
\tau (\mathbf{r}\,)\approx \frac{d_{n_{r}}^{2}}{2\pi D}\ln \frac{d_{n_{r}}}{%
a_{n_{r}}}  \label{8.8}
\end{equation}
where $d_{n_{r}}$ is the mean distance between veins of level $n_{r}$ and in
the given model for the vascular network (\ref{3.11})

\begin{equation}
d_{n_{r}}=\left[ \frac{V_{n_{r}}}{l_{n_{r}}}\right] ^{1/2}=\left( \frac{2}{%
\sqrt{3}}\right) ^{3/2}l_{n_{r}}.  \label{8.9}
\end{equation}

\section{The basic cover of microcirculatory bed domain. The bioheat equation%
}

\label{s8.3}

In order to find the relationship between the characteristics of the vessel
system $\mathcal{V}$ and, consequently, of the walker trapping and the blood
flow distribution described in terms of the blood flow rate $j(\mathbf{r}\,)$
we need the following collections of fundamental domains. Let $Q_{v}$ be the
fundamental domain corresponding to a given vein $i_{vr}$, i.e. the
fundamental domain of the same level as the vein $i_{vr}$ that contains the
given vein as well as the point $\mathbf{r}$ specifying the vein collection $%
\{i_{nr}\}$ which the vein $i_{vr}$ belongs to. In this way considering all
the veins of the collection $\{i_{vr}\}$ we can specify the collection of
fundamental domains $\{Q_{v}\}$ called the basic cover of the tissue domain $%
Q_{0}$. Different domains of $\{Q_{v}\}$ are disjoint and every point $%
\mathbf{r}$ of the domain $Q_{0}$ belongs to one of them and each domain of $%
\{Q_{v}\}$ contains just one vein of the collection $\{i_{v}\}$.

In accordance with the results obtained below on temporal scales about $\tau
(r)$ the walker spreads over a distance of order $l_{n_{r}}$ practically
uniformly. So, considering walker motion in the cellular tissue on spatial
scales of order $l_{n_{r}}$ we can characterize the effect of blood flow in
the vein $i_{vr}$ by the mean rate $r_{w}(\mathbf{r}\,)$ of walker
disappearance in the domain $Q_{vr}$ which is equal to

\begin{equation}
r_{w}(\mathbf{r}\,)=\frac{1}{\tau (\mathbf{r}\,)}C(\mathbf{r}\,).
\label{8.10}
\end{equation}

The basic cover $\{Q_{vr}\}$ allows us to define a new quantity $j_v (r)$
called the averaged blood flow rate according to the formula

\begin{equation}
j_{v}(\mathbf{r}\,)=\frac{1}{V_{nr}}\int\limits_{Q_{vr}}d\mathbf{r}j(\mathbf{%
r}\,)  \label{8.11}
\end{equation}
where $V_{r}$ is the volume of the domain $Q_{vr}\in \{Q_{v}\}$ containing
the given point $\mathbf{r}$. Taking into account (\ref{3.7}) we also may set

\begin{equation}
j_{v}(\mathbf{r}\,)=\frac{1}{V_{nr}}Ji_{vr}.  \label{8.12}
\end{equation}
Expressions (\ref{4.121}), (\ref{4.123}), (\ref{8.6}), (\ref{8.8}) - (\ref
{8.10}), and (\ref{8.12}) lead to the following formula for the walker
disappearance rate

\begin{equation}
r_{w}(\mathbf{r}\,)=j_{v}(\mathbf{r}\,)\left( \ln \frac{l_{0}}{a_{0}}\right)
^{(\beta _{cc}-1)/2}C_{w}(\mathbf{r}\,),  \label{8.13}
\end{equation}
where $\beta _{cc}=0$ for the countercurrent vascular network and $\beta
_{cc}=1$ for the unit vessel network (cf. formula (\ref{5.2})). In addition,
the expressions mentioned above enable us to represent the relationship
between the averaged blood flow rate $j_{v}(r)$ and the length $l_{n_{r}}$
of the corresponding vein $i_{vr}$ of the system $\{i\}_{v}$ in the form

\begin{equation}
l_{n_{r}}^{2}\approx \frac{3\sqrt{3}\pi }{4}\frac{D}{j_{v}(r)}\frac{1}{L},
\label{8.14}
\end{equation}
where $L=\left[ \ln \left( \frac{l_{0}}{a_{0}}\right) \right] ^{(\beta
_{cc}-1)/2}$.

Since, for the veins of the system $\{i\}_{v}$ the classification parameter $%
\zeta (n_{r},\mathbf{r}\,)\newline
=1$ they are not only walker traps but also some exhibit properties heat -
dissipation vessels. In particular vessels with $\zeta _{i}\mathbf{\leqslant 
}1$ give rise to renormalization of the diffusion coefficient $D\rightarrow
D_{\mathrm{eff}}=D(1+F_{v})$. In the case under consideration the
renormalization coefficient $F_{v}$ is constant as it has been shown in
Section~\ref{s5.4}. Therefore, the effect of the blood flow rate in the
veins (or countercurrent pars) of the system $\{i\}_{v}$ is actually reduced
to walker trapping and renormalization of the diffusion coefficient.
Therefore, taking into account expression (\ref{8.3}) we may write the
equation governing evolution of walker distribution in living tissue in
terms of

\begin{equation}
\frac{\partial C}{\partial t}=D_{\mathrm{eff}}\mathbf{\nabla }%
^{2}C-j_{v}\left( \ln \frac{l_{0}}{a_{0}}\right) ^{(\beta _{cc}-1)/2}C+q
\label{8.15}
\end{equation}
where the effective diffusion coefficient $D_{\mathrm{eff}}=D[1+F_{v}]$
where $F_{v}$ is given by expression (\ref{7.3a}).

Equation (\ref{8.15}) is actually the result of averaging microscopic
equations (\ref{3.1}), (\ref{3.2}) over the basic cover $\{Q_{vr}\}$.
Returning to terms of heat transfer, equation (\ref{8.15}) may be rewritten
as

\begin{equation}
c_{t}\rho _{t}\frac{\partial T}{\partial t}=\kappa _{\mathrm{eff}}\mathbf{%
\nabla }^{2}T-c_{t}\rho _{t}j_{v}\left[ \ln \left( \frac{l_{0}}{a_{0}}%
\right) \right] ^{(\beta _{cc}-1)/2}(T-T_{a})+q_{h}  \label{8.16}
\end{equation}
with $\kappa _{\mathrm{eff}}=\kappa \lbrack 1+F_{v}]$. This equation is the
desired bioheat equation governing evolution of the tissue temperature. It
should be noted that in small regions containing vessels with $\zeta _{i}>1$%
, which, however, do not belong to the vessel system $\mathcal{V}$, the
effective diffusion coefficient can exceed the value $D[1+F_{v}]$. This is
the case for a region containing vessels of the sequence $\{i_{nr}\}$ whose
level number is within the interval $[n_{r}^{\ast },n_{r}^{\ast \ast }]$
shown in Fig.~\ref{Fig17}. However influence of this regions on heat
transfer is not significant due to small relative volume of such regions.

Concluding the present Section we would like to point out that according to
equation (\ref{8.15}) or (\ref{8.16}), the characteristic spatial scale $l_D$
of the tissue temperature variations is about

\begin{equation*}
l_{D}=\left[ D_{\mathrm{eff}}/\left( \ln \frac{l_{0}}{a_{0}}\right) ^{(\beta
_{cc}-1)/2}j_{v}\right] ^{1/2}
\end{equation*}
whereas the length of the veins directly controlling heat exchange between
blood and the cellular tissue is $l_{n_{r}}\sim (D/(j_{v}L))^{1/2}$ (see
formula (\ref{8.14})). Thus,

\begin{equation*}
\frac{l_{n_{r}}}{l_{D}}\sim \left[ \frac{D}{D_{\mathrm{eff}}\ln (l_{0}/a_{0})%
}\right] ^{1/2}.
\end{equation*}
In the given approach the value $\ln (l_{0}/a_{0})$ is treated as a large
parameter, which allows us to suppose that $l_{n_{r}}\ll l_{D}$. In
particular, this inequality justifies the adopted basic assumption of
uniformity of the tissue temperature on spatial scales of order $l_{n_{r}}$.

\chapter{Relationship between the averaged and true blood flow rates}

\label{ch.9} 
\markright
{ {\sc  \thechapter. Relationship between the averaged and true\ldots}
}

In order to complete bioheat equation (\ref{8.16}) we need an expression or
equation specifying the relationship between the averaged $(j_v )$ and true $%
(j)$ blood flow rates. In the present Chapter we propose an alternative to
this relation.

\section{The integral form of the relationship between the averaged and true
blood flow rates}

\label{s9.1}

According to the definition of the averaged blood flow rate proposed in the
previous Section, we can write the following formula for the relationship
between the true $(j(\mathbf{r}\,))$ and averaged $(j_{v}(\mathbf{r}\,))$
blood flow rates

\begin{equation}
j_{v}(\mathbf{r}\,)=\int\limits_{Q_{0}}d\mathbf{r}^{\prime }G(\mathbf{r},%
\mathbf{r}^{\prime })j(\mathbf{r}^{\prime }).  \label{9.1}
\end{equation}
Here the kernel%
\index{kernel} of the linear operator%
\index{linear operator} $G$ is given by the expression

\begin{equation}
G(\mathbf{r},\mathbf{r}^{\prime })=\sum_{i\in \{i_{vr}\}}%
\frac{1}{V_{i}}\Theta _{i}(\mathbf{r}\,)\Theta _{i}(\mathbf{r}^{\prime }),
\label{9.2}
\end{equation}
where $\Theta _{i}(\mathbf{r}\,)$ is the characteristic function of the
fundamental domain $\Theta _{i}$ of volume $V_{i}$ corresponding to the vein 
$i$ i.e.

\begin{equation}
\Theta _{i}(\mathbf{r}\,)=\left\{ 
\begin{array}{ccc}
1\,\, & \text{{if}}\,\, & \mathbf{r}\in \Theta _{i} \\ 
0\,\, & \text{{if}}\,\, & \mathbf{r}\not\in \Theta _{i}
\end{array}
\right.  \label{9.3}
\end{equation}
and the sum runs over all the veins of the system $\{i_{vr}\}$. Expression (%
\ref{9.1}) may be regarded as a linear transformation of the field $j(%
\mathbf{r}^{\prime })$. In these terms $G$ is a linear operator which, by
virtue of (\ref{9.2}), is self-adjoint, i.e.

\begin{equation}
G(\mathbf{r},\mathbf{r}^{\prime })=G(\mathbf{r}^{\prime },\mathbf{r}\,)
\label{9.4}
\end{equation}
and meets the condition 
\begin{equation}
\int\limits_{Q_{0}}d\mathbf{r}G(\mathbf{r},\mathbf{r}^{\prime })=1
\label{9.5}
\end{equation}
as it must be due to the conservation of fluid mass.

We note that the averaged blood flow rate $j_{v}$ determined by formula (\ref
{9.1}) is a piecewise constant field with steps at the boundaries of the
domains $\{Q_{vr}\}$. Besides, transformation (\ref{9.1}) leads to some loss
of information due to the existence of the eigenfunctions%
\index{eigenfunction} with the zero eigenvalue%
\index{eigenvalue}.

The partition of the domain $Q_{0}$ into domains of the basic cover is
directly associated with the averaged blood flow rate, (see, e.g. expression
(\ref{8.11})). Therefore, although formula (\ref{9.1}) may be, in principle,
regarded as the desired relationship between the averaged and true blood
flow rate, it is actually unfeasible to use it because the kernel%
\index{kernel} $G(\mathbf{r},\mathbf{r}^{\prime })$ depends on the averaged
blood flow rate rather than the true one. In this case it would be
worthwhile to invert transformation (\ref{9.1}), however, to do it directly
is impossible due to operator (\ref{9.2}) possessing the zero eigenvalue.
Since the particular details in the behavior of the averaged blood flow rate
on spatial scales of order $l_{n_{r}}$ are of little consequence for heat
transfer, we may modify the kernel 
\index{kernel}$G(\mathbf{r},\mathbf{r}^{\prime })$ in order to invert
transformation (\ref{9.1}). This is the subject of the next Section.

\section{The smoothing procedure for the integral operator and the inverse
operator}

\label{s9.2}

The smoothing procedure is actually specified by the replacement of kernel (%
\ref{9.2}) by a certain kernel $%
\tilde{G}(\mathbf{r},\mathbf{r}^{\prime })$ smooth on scales of the basic
cover for which there exists an operator $\tilde{G}^{-1}$ inverse to the
operator $\tilde{G}$%
\index{inverse operator}. In addition the kernel $%
\tilde{G}(\mathbf{r},\mathbf{r}^{\prime })$ also must satisfy conditions (%
\ref{9.4}), (\ref{9.5}) because of these expressions stem from the general
properties of the averaging procedure.

The inverse operator $\tilde{G}^{-1}$ enables us to write the true blood
flow rate $j(\mathbf{r}\,)$ as an explicit functional of $j_{v}(\mathbf{r}%
\,) $.

\begin{equation}
j(\mathbf{r}\,)=\int\limits_{Q_{0}}d\mathbf{r}^{\prime }\tilde{G}^{-1}(%
\mathbf{r},\mathbf{r}^{\prime })j_{v}(\mathbf{r}^{\prime })  \label{9.6}
\end{equation}
where the kernel $\tilde{G}^{-1}(\mathbf{r},\mathbf{r}^{\prime })$ also
meets conditions (\ref{9.4}), (\ref{9.5}). Indeed, from definition,

\begin{equation}
\int\limits_{Q_{0}}d\mathbf{r}\,^{\prime \prime }\tilde{G}^{-1}(\mathbf{r},%
\mathbf{r}\,^{\prime \prime })\tilde{G}(\mathbf{r}\,^{\prime \prime },%
\mathbf{r}^{\prime })=\int\limits_{Q_{0}}d\mathbf{r}\,^{\prime \prime }%
\tilde{G}(\mathbf{r},\mathbf{r}\,^{\prime \prime })\tilde{G}^{-1}(\mathbf{r}%
\,^{\prime \prime },\mathbf{r}^{\prime })=\delta (\mathbf{r}-\mathbf{r}%
^{\prime }).  \label{9.7}
\end{equation}
Then, first, writing expression (\ref{9.7}) for the pairs $(\mathbf{r},%
\mathbf{r}^{\prime })$ and $(\mathbf{r}^{\prime },\mathbf{r}\,)$,
subtracting from one other, and taking into account (\ref{9.4}) we get

\begin{equation}
\tilde{G}^{-1}(\mathbf{r},\mathbf{r}^{\prime })=\tilde{G}^{-1}(\mathbf{r}%
^{\prime },\mathbf{r}\,).  \label{9.8}
\end{equation}
Second, integration of (\ref{9.7}) with respect to $\mathbf{r}^{\prime }$
over $Q_{0}$ and equality (\ref{9.8}) yields

\begin{equation}
\int\limits_{Q_{0}}d\mathbf{r}^{\prime }\tilde{G}^{-1}(\mathbf{r},\mathbf{r}%
^{\prime }).  \label{9.9}
\end{equation}
The linear transformation%
\index{linear transformation} (\ref{9.6}) can be also represented in terms
of the Fourier transforms $j_{F}(\mathbf{k}),j_{Fv}(\mathbf{k})$ of the
fields $j(\mathbf{r}\,),j_{v}(\mathbf{r}\,)$ and the Fourier transform $%
\tilde{G}_{F}^{-1}(\mathbf{k},\mathbf{k}\,^{\prime })$ of the kernel $\tilde{%
G}^{-1}(\mathbf{r},\mathbf{r}^{\prime })$, viz.:

\begin{equation}
j_{F}(\mathbf{k}\,)=\sum_{\mathbf{k}\,^{\prime }}\tilde{G}_{F}^{-1}(\mathbf{k%
},\mathbf{k}\,^{\prime })j_{Fv}(\mathbf{k}\,^{\prime })  \label{9.10}
\end{equation}
where

\begin{equation}
\tilde{G}_{F}^{-1}(\mathbf{k},\mathbf{k}\,^{\prime })=\frac{1}{V_{0}}%
\int\limits_{Q_{0}}\int d\mathbf{r}d\mathbf{r}^{\prime }\tilde{G}^{-1}(%
\mathbf{r},\mathbf{r}^{\prime })\exp [i(\mathbf{kr}-\mathbf{k}\,^{\prime }%
\mathbf{r}^{\prime })].  \label{9.11}
\end{equation}
$V_{0}=\Lambda _{0}^{3}$ is the volume of the microcirculatory bed domain $%
Q_{0}$, and the sum runs over all the vectors $\mathbf{k}=\frac{2\pi }{%
\Lambda _{0}}(n_{x},n_{y},n_{z})$, for $n_{x},n_{y},n_{z}=0,\pm 1,\pm
2\ldots $

Setting $\mathbf{k}\,^{\prime }$ or $\mathbf{k}$ be equal to zero and taking
into account (\ref{9.8}), (\ref{9.9}) we obtain that the function $\tilde{G}%
_{F}^{-1}(\mathbf{k},\mathbf{k}\,^{\prime })$ must obey the following
general properties

\begin{equation}
\tilde{G}_{F}^{-1}(\mathbf{k},0)=\tilde{G}_{F}^{-1}(0,\mathbf{k})=\delta _{%
\mathbf{k},0}.  \label{9.12}
\end{equation}

As it follows from the definition of $j_{v}(\mathbf{r}\,)$ the averaged
blood flow rate $j_{v}(\mathbf{r}\,)$ has to be a smooth function on scales
of the basic cover%
\index{basic cover}. Therefore, the function $j_{v}(\mathbf{r}\,)$ cannot
vary significantly inside any domain $Q_{i_{v}}$ of the basic cover. In
terms of smoothed fields the averaging procedure (\ref{9.6}) is no longer
averaging over the collection of discrete cubes but is specified by a
certain field $l(r)$ whose value at each point $\mathbf{r}$ determines the
characteristic size of neighborhood centered at the point $\mathbf{r}$ over
which the true blood flow rate $j(\mathbf{r}\,)$ should be averaged. Keeping
in mind the Gauss type averaging procedure%
\index{averaging procedure} it is natural to suppose that $[2\pi
l^{2}(r)]^{3/2}$ is equal to the volume $V_{\Gamma
_{v}}=(d_{n_{r}}^{2}l_{n_{r}})$ of the fundamental domain $Q_{n_{r}}$
belonging to the basic cover. Then, from (\ref{8.15}) we obtain

\begin{equation}
l^{2}(r)=%
\frac{\sqrt{3}}{2}\frac{D}{j_{v}(r)L}  \label{9.13}
\end{equation}
where $j_{v}(r)$ is the smoothed averaged blood flow rate. Due to the field $%
j_{v}(r)$ being smooth on scales $j(r)$ the value $l(r)\mathbf{\nabla }%
j_{v}(r)$ cannot substantially exceed $j_{v}(r)$. Therefore, we may seek the
particular expression for the operator $\tilde{G}^{-1}$ in the form of a
certain truncation of the operator $\tilde{G}^{-1}$ expansion into a power
series of $\mathbf{\nabla }$. In order to analyse characteristic features of
such a representation we consider the averaging procedure the value $l(%
\mathbf{r}\,)$ is formally constant, $l(\mathbf{r}\,)=l$. In this case it is
natural to specify the kernel $\tilde{G}_{h}(\mathbf{r},\mathbf{r}^{\prime
}) $ as

\begin{equation}
\tilde{G}_{h}(\mathbf{r},\mathbf{r}^{\prime })=\frac{1}{(2\pi l^{2})^{3/2}}%
\exp \left[ -\frac{(\mathbf{r}-\mathbf{r}^{\prime })}{2l^{2}}\right] .
\label{9.14}
\end{equation}
The integral transformation (\ref{9.1}) with kernel (\ref{9.14}) becomes the
convolution, thus, the Fourier transforms $j_{v}(\mathbf{k}\,),j_{Fv}(%
\mathbf{k}\,)$ of the fields $j(\mathbf{r}\,)$ and $j_{v}(\mathbf{r}\,)$ are
related by the expression for $l\ll \Lambda _{0}$

\begin{equation}
j_{F}(\mathbf{k}\,)=G_{hF}(\mathbf{k}\,)j_{Fv}(\mathbf{k}\,)  \label{9.15}
\end{equation}
where

\begin{equation}
G_{hF}(\mathbf{k}\,)=\frac{1}{(2\pi l^{2})^{3/2}}\int\limits_{\mathcal{R}%
^{3}}d\mathbf{r}\exp \left( -\frac{\mathbf{r}\,2}{2l^{2}}+i\mathbf{kr}%
\right) =\exp \left( -\frac{1}{2}l^{2}k^{2}\right) .  \label{9.16}
\end{equation}
Comparing (\ref{9.10}) and (\ref{9.15}) we find the following expression for
the inverse operator $\tilde{G}^{-1}$:

\begin{equation}
\tilde{G}_{F}^{-1}(\mathbf{k},\mathbf{k}\,^{\prime })=\delta _{\mathbf{k},%
\mathbf{k}\,^{\prime }}\exp \left( \frac{1}{2}l^{2}k^{2}\right) .
\label{9.17}
\end{equation}
Since, we may confine our consideration to the region $kl\mathbf{\leqslant }%
1 $ and the function $\exp (\frac{1}{2}l^{2}k^{2})$ is increasing it is
possible to expand this function in a power series of $k^{2}l^{2}$ and to
truncate the power series at the second term. In this way we get

\begin{equation}
\tilde{G}_{F}^{-1}(\mathbf{k},\mathbf{k}\,^{\prime })=\delta _{\mathbf{k},%
\mathbf{k}\,^{\prime }}\left( 1+\frac{1}{2}l^{2}k^{2}\right) .  \label{9.18}
\end{equation}
Transformation (\ref{9.10}) with kernel (\ref{9.18}) can be rewritten in the
form

\begin{equation}
j(\mathbf{r}\,)=j_{v}(\mathbf{r}\,)-\frac{1}{2}l^{2}\mathbf{\nabla }%
^{2}j_{v}(\mathbf{r}\,).  \label{9.19}
\end{equation}

We point out that such approximation of the function $G_{hF}(\mathbf{k}\,)$
is impossible due to $\exp \left\{ \frac{1}{2}l^{2}k^{2}\right\} $ being a
decreasing function and its truncation similar to (\ref{9.18}) leading to a
wrong result for $lk\sim 1$.

In the general case, i.e. when the field $l^{2}(r$ is nonuniform we may
specify this operator $G^{-1}$ to the second order $\mathbf{\nabla }$ by the
formula for its action on an arbitrary function $\Psi $

\begin{equation}
\tilde{G}^{-1}\Psi =\{1-A_{1}(r)\mathbf{\nabla }^{2}-\mathbf{\nabla }A_{2}(r)%
\mathbf{\nabla }-\mathbf{\nabla }^{2}A_{3}(r)\}\Psi  \label{9.20}
\end{equation}
where $A_{1}(r),A_{2}(r)$ and $A_{3}(r)$ are proportional to $l^{2}(r)$. It
should be noted that form (\ref{9.20}) is the unique expression containing
two vector objects $(\mathbf{\nabla })$ and a scalar field $(l^{2})$. For
the operator $G^{-1}$ determined by expression (\ref{9.20}) the Fourier
transform%
\index{Fourier transform} of its kernel is of the form

\begin{equation}
\tilde{G}_{F}^{-1}(\mathbf{k},\mathbf{k}\,^{\prime })=\delta _{\mathbf{k},%
\mathbf{k}\,^{\prime }}+A_{1F}(\mathbf{k}-\mathbf{k}\,^{\prime })\mathbf{k}%
\,^{\prime }{}^{2}+\mathbf{k}A_{2F}(\mathbf{k}-\mathbf{k}\,^{\prime })%
\mathbf{k}\,^{\prime }+\mathbf{k}\,^{2}A_{3}(\mathbf{k}-\mathbf{k}\,^{\prime
}).  \label{9.21}
\end{equation}
Here $A_{iF}$ is the Fourier transform of the function $A_{i}(\mathbf{r}\,)$ 
$i=1,2,3)$. Condition (\ref{9.12}) immediately gives $A_{1}=A_{2}=0$.
Thereby under the adopted assumptions the general form of the transformation
(\ref{9.6}) is

\begin{equation}
\tilde{G}^{-1}j_{v}=j_{v}-C\mathbf{\nabla }(l^{2}(\mathbf{r}\,)\mathbf{%
\nabla }j_{v})  \label{9.22}
\end{equation}
comparing (9.19) and (9.22) we find $C=\frac{1}{2}$.

Expression (\ref{9.22}) is the desired formula for the smoothed inverse
operator $\tilde G^{-1}$.

\section[Differential form of the relationship between the true and averaged
blood flow rates]{Differential form of the relationship \newline
between the true and averaged blood flow \newline
rates}

\label{s9.3}

Formula (\ref{9.22}) actually specifies the procedure of inverting
relationship (\ref{9.1}). In this way using the identity $\tilde G ^{-1}
\tilde G =1$ we obtain the expression

\begin{equation}
j=\tilde{G}^{-1}j_{v}  \label{9.23}
\end{equation}
{whose right-hand side contains only the averaged blood flow rate due to the
field $l^{2}(\mathbf{r}\,)$ being directly determined by $j_{v}(\mathbf{r}%
\,) $. The substitution of (\ref{9.13}) into (\ref{9.22}) leads equation (%
\ref{9.23}) to the equation determining the desired relationship between the
true and averaged blood flow rates: }

\begin{equation}
j_{v}-D_{v}\mathbf{\nabla }^{2}\ln j_{v}=j  \label{9.24}
\end{equation}
where

\begin{equation}
D_{v}=\frac{\sqrt{3}}{4}\frac{\kappa }{c_{t}\rho _{t}}\frac{1}{L}.
\label{9.25}
\end{equation}
Dealing with a tissue domain bounded by a real physical interface $\sigma $
we need to complete equation (\ref{9.24}) by a certain condition at the
interface $\sigma $. In order to determine this boundary condition let us
consider a domain $Q^{\prime }$ which is practically made up from
fundamental cubes of the basic cover. Then for the averaged blood flow rate $%
j_{v}$ defined by formula (\ref{9.1}) with kernel (\ref{9.2}) we have

\begin{equation}
\int\limits_{Q^{\prime }}d\mathbf{r}j_{v}(\mathbf{r}\,)=\int\limits_{Q^{%
\prime }}d\mathbf{r}j(\mathbf{r}\,).  \label{9.26}
\end{equation}
In terms of smoothed fields relation (\ref{9.26}) leads to the condition

\begin{equation}
\oint\limits_{\partial Q^{\prime }}ds\frac{D_{v}}{j_{v}}\mathbf{\nabla }%
_{n}j_{v}=0  \label{9.27}
\end{equation}
the boundary $\partial Q^{\prime }$ of such a region $Q^{\prime }$.
According to (\ref{9.27}) the term $D_{v}\mathbf{\nabla }\ln j_{v}$ is the
seeming blood flow across the boundary $\partial Q^{\prime }$ due averaging
the blood flow rate in the vicinity of the domain boundary over exterior
points. Since, the averaging procedure never runs over exterior points to
the real interface $\sigma $ we may set

\begin{equation}
\left. \mathbf{\nabla }_{n}j_{v}\right| _{\sigma }=0.  \label{9.28}
\end{equation}
This expression is the desired boundary condition for equation (\ref{9.24}).

\clearpage

\part{Theory of heat transfer in living tissue with temperature
self-regulation}

\markboth{
{\sc \thepart.{ } Theory of heat transfer in living tissue\ldots}}{}

When the tissue temperature $T$ becomes high enough, $T\sim 42\div 44^{o}C$
thermoregulation in living tissue causes significant increase in the blood
flow rate. In particular, under local heating the blood flow rate can
increase by ten-fold. Therefore, to complete bioheat transfer equation
governing evolution of the tissue temperature a theory of living tissue
response to temperature variation in terms of the blood flow rate dependence
on the temperature distribution should be developed. The present Part is
mainly devoted to this subject.

\chapter{Theory of self-regulation under local heating of living tissue}

\label{ch.10} 
\markright
{ {\sc  \thechapter. Theory of self-regulation under local heating\ldots}
}

The microscopic description%
\index{microscopic description} of vessel response to temperature variations
has been stated in Sections~\ref{s2.3} and \ref{s3.4}. There we have
specified the model where the flow resistance $R_{i}$ of vein $i$ is
directly controlled by the mean temperature of blood in this vein and the
flow resistance $R_{i}$ of an artery $i^{\prime }$ is governed by the mean
blood temperature in the corresponding vein $i^{\prime }$. The flow
resistance $\{R_{i}\}$ of all vessels given, we can find the blood
distribution over the vascular network and, thus, the blood flow rate
distribution over the microcirculatory bed domain $Q_{0}$, as depending on
the blood temperature pattern $\{T_{i}^{\ast }\}$ on the vein tree. So,
relating the blood temperature pattern $\{T_{i}^{\ast }\}$ to the tissue
temperature distribution $T(\mathbf{r}\,)$ and, may be, the blood flow rate
distribution $j(\mathbf{r}\,)$ we solve, in principle, the temperature
regulation problem.

\section{Governing equations for blood temperature distribution over the
vein tree}

\label{s10.1}

In this Section we obtain the explicit relationship between the patterns of
blood currents $\{J_{i}\}$ and the blood temperature $\{T_{i}^{\ast }\}$ on
the vascular network on one hand, the blood flow rate $j$ and the tissue
temperature, on the other hand.

As in Part 3 for the sake of simplicity we consider the case when the effect
of blood flow in capillaries on heat transfer is ignorable and the vascular
network either is entirely made up of countercurrent pairs $(n_{cc}=N)$ or
involves unit vessels only $(n_{cc}=0)$. By virtue of the adopted
assumptions, the length $l_{N}$ of the last level vessels is the smallest
spatial scale, thus, in particular, $l_{N}\ll l_{v},l_{D}$ and, thereby, the
blood flow rate $j(\mathbf{r},t)$ can be regarded as a field practically
uniform on the scale $l_{N}$. In this case the relationship between the
blood flow rate $j(r)$ and the blood currents $\{J_{i}\}_{N}$ in the last
level vessels may be given in terms of

\begin{equation*}
J_{ir}=V_{N}j(r),
\end{equation*}
where $J_{ir}$ is, for example, the blood current in the last level vein $%
i_{r}$ that together with the point $\mathbf{r}$ is contained in the same
elementary domain $Q_{Nr}$ of volume $V_{N}$. Then for the adopted vascular
network embedding in the cube $Q_{0}$ from the last expression and equations
(\ref{3.16}) representing the conservation law of blood flow at branching
points we immediately find the following expression for the blood current $%
J_{i}$ in the vein $i$:

\begin{equation}
J_{i}=\int\limits_{Q_{0}}dr\Theta _{i}(r)j(r)\;,  \label{10.1}
\end{equation}
where $\Theta _{i}(r)$ as in Section~\ref{s9.1} is the characteristic
function of the fundamental domain $Q_{i}$ being of the same level as the
vein $i$ and containing this vein inside itself, viz.: 
\begin{equation*}
\Theta _{i}(r)=\left\{ 
\begin{array}{cccc}
1 & , & \,%
\text{if}\, & r\in Q_{i} \\ 
0 & , & \,\text{if}\, & r\notin Q_{i}.
\end{array}
\right. .
\end{equation*}

To describe heat exchange between the cellular tissue and blood, and, thus,
to find the relationship between and the blood temperature pattern $%
\{T_{i}\} $ the temperature field $T(\mathbf{r},t)$ we make use the results
obtained in Section~\ref{s8.2}. According to these results on the vein tree
we can single out a certain connected part (the vessel system $\mathcal{V}$)
which invols vessels for which the classification parameter $\zeta _{i}>1$.
The ``boundary'' of the vessel system $\mathcal{V}$ is made up of veins $%
\{i_{v}\}$ for which $\zeta _{i_{v}}=1$. The system of fundamental domains
corresponding the veins $\{i_{v}\}$ forms the basic cover $\{Q_{v}\}$. In
veins belonging to the vessel system $\mathcal{V}$ except for the vein $%
\{i_{v}\}$ blood moves so fast that heat transport in these vessels is
mainly controlled by blood convective stream and blood has no time to attain
thermal equilibrium with the surrounding cellular tissue as well as for the
countercurrent vascular network with blood in the nearest artery. Therefore,
first, blood in the veins of the system $\mathcal{V}$ should be
characterized by its own temperature $T^{\ast }$. Second, the conservation
of energy at the branching points of the vessel system $\mathcal{V}$, for
example, at a branching point $B_{v}$ can be represented in the form

\begin{equation}
\sum_{B_{v}}T_{\text{in}}^{\ast }J_{\text{in}}=T_{\text{out}}J_{\text{out}}
\label{10.2}
\end{equation}
where $J_{\text{in}},J_{\text{out}},T_{\text{in}}^{\ast },T_{\text{out}}$
are the blood currents and the mean temperatures of blood in the veins going
into this branching point.

The blood flow in veins for which $\zeta _{i}<1$ is in thermodynamic
equilibrium with the cellular tissue and its temperature coincides locally
with the tissue temperature $T$: $T_{i}^{\ast }\simeq T$. The latter
condition is also true for separated veins of Class 1; the veins of the
system $\{i_{v}\}$ exhibit properties of both the heat - conservation and
heat - dissipation veins. The tissue temperature cannot vary significantly
on scales of the basic cover $\{Q_{v}\}$. Therefore, the way how the
temperature of a blood portion varies during the motion from the veins of
level $n>n_{t_{r}}$ to veins of level $n<n_{t_{r}}$ ($n_{t_{r}}$ is the
number level of the vein $i_{vr}$) can be characterized considering the
local tissue temperature $T$ constant. This question has been discussed in
Chapter~\ref{ch.6} where we have shown that the relationship between the
temperature $T^{\ast }$ of blood in veins with $n<n_{t}$ and the tissue
temperature $T$ is different for the unit vein network and the
countercurrent pair network. Since, the effect of blood flow in arteries
with $n\simeq n_{t}$ is not significant for the unit vein network the
temperature of blood does not practically vary during its motion from veins
of level $n>n_{t}$ to veins of level $n<n_{t}$ (see formula (\ref{6.30a}), (%
\ref{6.30b})) for the unit vessel network we may set

\begin{equation}
T_{i_{v}}^{\ast }=T_{i}  \label{10.3a}
\end{equation}
where $T_{i}$ is the tissue temperature in the fundamental domain belonging
to the basic cover $\{Q_{v}\}$ and containing the vein $i_{v}$. Heat
exchange between venous and arterial blood in a countercurrent pair whose
level number $n\approx n_{t}$ is significant. Therefore, venous blood loses
heat energy during its motion from heat - dissipation vein $n>n_{t_{r}}$ to
heat - conservation veins $n<n_{t_{r}}$ at the instant it passes the veins
whose level $n\approx n_{t}$. Formula (\ref{6.31}) allows us to write the
following relationship between $T_{i_{v}}^{\ast }$ and $T_{i}$ for the
countercurrent vascular network:

\begin{equation}
T_{i_{v}}^{\ast }-T_{a}=\frac{1}{[\ln (l_{0}/a_{0})]^{1/2}}(T_{i}-T_{a})
\label{10.3b}
\end{equation}
Expressions (\ref{10.3a}) and (\ref{10.3b}) are actually the ``boundary''
conditions for the system of equations (\ref{10.2}).

Since, the tissue temperature cannot vary significantly on scales of the
basic cover $\{Q_{v}\}$ we may rewrite formulae (\ref{10.3a}), (\ref{10.3b})
as

\begin{equation}
(T_{i_{v}}^{\ast }-T_{a})J_{i_{v}}=\beta _{v}\int\limits_{Q_{v}}d\mathbf{r}%
\,\Theta _{i}(\mathbf{r}\,)(T-T_{a})j  \label{10.3c}
\end{equation}
where $\beta _{v}=1$ for unit veins and $\beta =\frac{1}{\ln (l_{0}/a_{0})}$
for countercurrent pairs. $Q_{iv}$ is the domain of the collection $%
\{Q_{v}\} $ that contains the vein $i_{v}$ and expression (\ref{10.1}) has
been taken into account.

The vascular network embedding allows us to write the following formula for
each branching point, for example, branching point $B$

\begin{equation*}
\Theta _{\text{out}}(\mathbf{r}\,)=\sum_{B}\Theta _{\text{in}}(\mathbf{r}\,)
\end{equation*}
where the sum runs over all the veins going into the branching point $B$.
Then, taken into account (\ref{3.16}) and substituting (\ref{10.3c}) into (%
\ref{10.2}) for, first, for the branching points of the vessel system $%
\mathcal{V}$ that contain the veins $\{i_{v}\}$ and then, successively
performing the summation for any vein $i$ of the vessel system $\mathcal{V}$
, we find

\begin{equation}
(T_{i}^{\ast }-T_{a})J_{i}=\beta _{v}\int\limits_{Q_{0}}d\mathbf{r}\,\Theta
_{i}(\mathbf{r}\,)(T-T_{a})j  \label{10.4}
\end{equation}

For small veins, for example vein $i$, with $n_i > n_{t_r}$ blood is in
thermodynamic equilibrium with the cellular tissue and the tissue
temperature in the fundamental domain $Q_i$ is practically constant.
Therefore, formula (\ref{10.1}) enables us to write an expression similar to
(\ref{10.4}), viz.

\begin{equation}
(T_{i}^{\ast }-T_{a})J_{i}=\int\limits_{Q_{0}}d\mathbf{r}\,\Theta _{i}(%
\mathbf{r}\,)(T-T_{a})j  \label{10.5}
\end{equation}

For $\beta _v =1$ expressions (\ref{10.4}) , (\ref{10.5}) are of the same
form. The latter enables us to regard formula (\ref{10.9}) as the general
relationship between the temperature $T^*_i$ of blood in a vein $i$, the
blood current $J_i$ in this vein, the tissue temperature $T$ and the blood
flow rate $j$ for the unit vessel network.

For the countercurrent vascular network formula (\ref{10.8}) can be regarded
as the general relationship between these quantities if the cofactor $\beta
_{v}$ is treated as a function taking the value $\beta _{v}=1/\sqrt{\ln
\left( \l _{0}/a_{0}\right) }$for the veins of the vessel system $\mathcal{V}
$ and $\beta _{v}=1$ for small veins. In order to complete the given
relationship between $T_{i}^{\ast },J_{i},T$ and $j$ we need to find the
explicit form of the dependence of $\beta _{v}$ on the tissue temperature $T(%
\mathbf{r}\,)$ and the blood flow rate $j(\mathbf{r}\,)$ (or may be the
averaged blood flow rate $j_{v}(\mathbf{r}\,)$). For each of the large veins
belonging to the vessel system $\mathcal{V}$ the corresponding fundamental
domain can be represented as the union of domains belonging to the basic
cover $\{Q_{v}\}$. Therefore, so, using the definition (\ref{9.1}) of the
averaged blood flow rate $j_{v}(\mathbf{r}\,)$, we rewrite the definition (%
\ref{4.123}) of the classification parameter for countercurrent pairs whose
veins belong to the vessel system $\mathcal{V}$ in the form

\begin{equation}
\zeta _{i}=\zeta _{v_{i}}=\frac{\left[ \ln (l_{0}/a_{0})\right] ^{1/2}}{2\pi
Dl_{i}}\int\limits_{Q_{i}}d\mathbf{r}j_{v}(\mathbf{r}\,).  \label{10.6}
\end{equation}
For the veins of the system $\mathcal{V}$ the value $\zeta _{v_{i}}>1$ and $%
\zeta _{v_{i}}\approx 1$ for the ``boundary'' $\{i_{v}\}$ of the system $%
\mathcal{V}$. Since, the field $j_{v}(\mathbf{r}\,)$ is the blood flow rate
averaged over the basic cover $\{Q_{v}\}$ for a vein $i$ not belonging to
the vessel system $\mathcal{V}$ the formal value

\begin{equation*}
\int\limits_{Q_{i}}j_{v}(\mathbf{r}\,)d\mathbf{r}
\end{equation*}
is equal to the blood current in the vein $i$ providing blood flowing
through the vein $i_{v}$ supplying the domain $Q_{i}$ is uniformly
distributed over the corresponding domain $Q_{i_{v}}$. By constitution for
the vein $i_{v}$ the value $\zeta _{v_{i}}=1$. Thus, for every vein not
belonging to the vessel system $\mathcal{V}$ the formal value of $\zeta
_{v_{i}}$ given by expression (\ref{10.6}) must be less than unity; $\zeta
_{v_{i}}<1$ if $i\not\in \mathcal{V}$. These properties of the quantity $%
\zeta _{v_{i}}$ enable us to use $\zeta _{v_{i}}$ as the classification
parameter of belonging to the vessel system $\mathcal{V}$. Therefore, the
dependence of the value $\beta _{v}$ on the vein position on the vessel tree
can be written in the form

\begin{equation}
\beta _{v}=\left\{ 
\begin{array}{cccc}
\lbrack \ln (l_{0}/a_{0})]^{-1/2} & ; & \text{{if}}\, & \zeta _{v}>1, \\ 
1 & ; & \,\text{{if}}\, & \zeta _{v}<1.
\end{array}
\right.  \label{10.7}
\end{equation}

So, the system of expressions (\ref{10.1}), (\ref{10.5}), or the system of
expressions (\ref{10.1}), (\ref{10.4}), (\ref{10.6}), (\ref{10.7}) specifies
the desired relationship between the patterns of the blood currents $%
\{J_{i}\}$ and the blood temperature $\{T_{i}^{\ast }\}$ on one hand, and
the blood flow rates $j(\mathbf{r}\,),j_{v}(\mathbf{r}\,)$ and the tissue
temperature $T(\mathbf{r}\,)$ on the other hand for the unit vessel network
and the countercurrent vascular network, respectively.

The obtained results actually specify the vessel flow resistances $\{R_{i}\}$
as functions of the blood flow rate and the tissue temperature. So, in order
to find the blood flow rate distribution over the microcirculatory bed
domain $Q_{0}$ as a functional of the tissue temperature, which is the
subject of the thermoregulation theory, we need to obtain expressions that
describe the blood current pattern $\{J_{i}\}$ and, thus, the blood flow
rate $j(\mathbf{r}\,)$ in living tissue for given values of the vessel
resistances. The following two Sections are devoted to this problem.

\section{Additional effective pressure sources. The Green matrix for the
Kirchhoff equ\-ations of blood flow redistribution over the vascular network}

\label{s10.2}

To find the relation between the tissue temperature $T(\mathbf{r},t)$ and
the blood flow rate $j(\mathbf{r},t)$ we should solve the system of
Kirchhoff's equations%
\index{Kirchhoff's equations} (\ref{3.16}), (\ref{3.17}). However, when the
tissue temperature is nonuniform in the domain $Q_{0}$, resistances of all
the veins (and the arteries) can be different in magnitude because of the
vascular network response to temperature variations. In this case solving
Kirchhoff's equations%
\index{Kirchhoff's equations} directly is troublesome. In order to avoid
this problem let us introduce quantities $\{\varepsilon \}$ called
additional effective pressure sources (EPSs)%
\index{additional effective pressure sources (EPSs)} which are ascribed to
all the vessels. For the $i$-th artery and vein of level $n_{i}$ the
additional EPS $\varepsilon $ is defined by the expression 
\begin{equation}
\varepsilon =-J_{i}(R_{i}-R_{n_{i}}^{0}).  \label{10.8}
\end{equation}
Formula (\ref{10.8}) allows us to rewrite the second Kirchhoff's equation (%
\ref{3.17}) in terms of: 
\begin{equation}
J_{i}R_{n_{i}}^{0}=\Delta P_{i}+\varepsilon .  \label{10.9}
\end{equation}
The collection of equations (\ref{3.16}) and (\ref{10.9}) for different
branching points and vessels forms the system of Kirchhoff's equations
describing the blood current pattern on a vascular network of the same
geometry where vessels, however, are not sensitive to variations in the
blood temperature and thermoregulation gives rise to the additional EPSs.
The latter vascular network will be referred to\ below as homogeneous one.
In this way, when EPSs have\ the known values the analysis of the blood
current redistribution over the microcirculatory bed with vessels sensitive
to the blood temperature is reduced to solving Kirchhoff's equations for the
corresponding homogeneous vascular network.

Differentiating relation (\ref{10.8}) with respect to the time $t$ and
taking into account equation (\ref{3.27}) we find the following evolution
equation for the quantity $\varepsilon$: 
\begin{equation}
\tau _{n_i} 
\frac{d\varepsilon}{dt}+\varepsilon\left[ f\left( 1-\frac{\varepsilon}{%
J_iR_{n_i}^{\circ }} \right) -\tau _{n_i}\frac{d}{dt}\ln {J_i}\right]
=R_{n_i}^{\circ }\frac 1\Delta \left| (T_i^{*}-T_a)J_i\right|.  \label{10.10}
\end{equation}
In obtaining (\ref{10.10}) we also have assumed that the current $J_i$
cannot change its direction.

The system of equations (\ref{10.10}) and relations (\ref{10.1}), (\ref{10.4}%
) or (\ref{10.5}) specifies the dependence of EPSs on the tissue temperature 
$T(\mathbf{r},t)$ and the blood flow rate $j(\mathbf{r},t)$. If, however, we
find the solution of the system of Kirchhoff's equations (\ref{3.16}) and (%
\ref{10.9}) and, thereby, obtain the blood flow rate as a function of EPSs,
then, we shall be able to get an equation that directly specifies the
relationship between $j$ and $T$. Such a relation with equations (\ref{8.3})
and (\ref{9.26}) forms the desired complete description of heat transfer in
living tissue.

For the venous bed of the corresponding homogeneous vascular network
containing EPSs the solution of the system of Kirchhoff's equations (\ref
{3.16}), (\ref{10.9}) can be written in the form

\begin{equation}
J_{i}\{\varepsilon \}=\sum_{i^{\prime }}\Lambda _{ii^{\prime }}[\varepsilon
_{i^{\prime }}+P\delta _{n_{i^{\prime }}0}]\;.  \label{10.11}
\end{equation}
Here $\Lambda _{ij}$ is the Green matrix%
\index{Green matrix}, i.e. the solution of these equations when $P=0$ and
all $\varepsilon _{i^{\prime }}=0$ except for $\varepsilon |_{i^{\prime
}=j}=1,\,\delta _{n_{i^{\prime }}0}$ is the Kronecker symbol and $n_{j}$ is
the level number of the vein $j$. We note the possibility of representation (%
\ref{10.11}) the results from the linearity of equations (\ref{3.16}) and (%
\ref{10.9}) with respect to the blood currents $\{J_{i}\}$.

Due to the homogeneity of the given vascular network shown sche\-ma\-tically
in Fig.~\ref{Fig18} when the vein $j$ is the host one, the values of $%
\Lambda _{ij}$ are equal for all veins $i$ of the same level. So, from (\ref
{3.16}) we get $2^{3}\Lambda _{ij}=\Lambda _{i^{\prime }j}$ if $n_{j}=0$ and 
$n_{i}=n_{i^{\prime }}+1$ because exactly eight veins go in each branching
point. Whence, it follows that

\FRAME{ftbpFU}{9.0479cm}{11.4576cm}{0pt}{\Qcb{Schematic representation of
the homogeneous venous bed containing EPSs.}}{\Qlb{Fig18}}{Fig18.GIF}{%
\special{language "Scientific Word";type "GRAPHIC";maintain-aspect-ratio
TRUE;display "USEDEF";valid_file "F";width 9.0479cm;height 11.4576cm;depth
0pt;original-width 6.5648in;original-height 8.3333in;cropleft "0";croptop
"1.0006";cropright "1.0007";cropbottom "0";filename
'FIG18.GIF';file-properties "XNPEU";}}

\begin{equation}
\left. \Lambda _{ij}\right| _{n_{j}=0}=\left. 2^{-3n_{i}}\Lambda
_{ij}\right| _{
\begin{array}{l}
\displaystyle\overset{n_{i}=0}{\scriptstyle n_{j}=0}
\end{array}
}\;.  \label{10.12}
\end{equation}
Let us set the total pressure drop across the venous bed $P=1$, all EPSs $%
\varepsilon =0$ and then, sum up equation (\ref{10.9}) along any path $%
\mathcal{P}$ from the host vein to a vein of the last level. In this way
taking into account (\ref{3.20}),(\ref{10.11}) we obtain

\begin{equation*}
P = \sum _{i \in \mathcal{P}} J_i R^0_{n_i} = \sum ^{N}_{n_i=0} \left.
\Lambda_{ij} \right |_{n_j=0} 2^{3n_i} \rho (n_i) R_0
\end{equation*}
and 
\begin{equation}
\left. \Lambda_{ij} \right |_{
\begin{array}{l}
\displaystyle \overset{n_i=0}{\scriptstyle n_j=0}
\end{array}
}\;\cdot R_0\sum_{n=0}^N\rho (n)=1.  \label{10.13}
\end{equation}
Then, introducing quantities $\{Z(n)\}$ defined as 
\begin{equation}
Z(n)=\sum_{n^{\prime }=n}^N\rho (n^{\prime })  \label{10.14}
\end{equation}

\noindent from (\ref{10.12}) and (\ref{10.13}) we get 
\begin{equation}
\Lambda_{ij}|_{n_j=0}=2^{-3n_i}\frac 1{R_0Z(0)}.  \label{10.15}
\end{equation}

To find $\Lambda _{ij}$ for a vein $j$ whose level number $n_{j}\geq 1$ we
consider a path $O_{j}O_{0}$ on the vascular network, shown in Fig.~\ref
{Fig18} by the dashed line which joins the given vein $j$ to the host vein
and directed from higher to lower levels. This path divides the whole venous
bed into the path $O_{j}O_{0}$ and disconnected with each other\ branches
whose initial veins go into branching points on the path $O_{j}O_{0}$. For $%
P=0$ and $\varepsilon =0$ except for $\varepsilon |_{i^{\prime }=j}=1$ in
all veins belonging both to the same level and to the same branch the
corresponding blood currents must be identical. This allows us to transform
the graph shown in Fig.~\ref{Fig18} into the one shown in Fig.~\ref{Fig19},
where the given path $O_{j}O_{0}$ is represented by the sequence of veins
designated by $\{n_{j},n_{j}-1,n_{j}-2,\dots ,1,0\}$ having \ of resistances 
$\{R_{n_{j}}^{0};\,R_{n_{j}-1}^{0};\,R_{n_{j}-2}^{0};\,R_{1}^{0};\,R_{0}^{0}%
\}$. Each vein of this sequence at its terminal points is connected with a
block of identical branches labeled by the level number of the corresponding
vein (Fig.~\ref{Fig19}).

\FRAME{ftbpFU}{11.7959cm}{12.468cm}{0pt}{\Qcb{The block form of the
homogeneous venous bed for $P=0$, $\protect\varepsilon _{i^{\prime }}=0$
except $\protect\varepsilon _{i^{\prime }}|_{i^{\prime }=j}=1$ (a) and the
structure of a subblock (b).}}{\Qlb{Fig19}}{Fig19}{\special{language
"Scientific Word";type "GRAPHIC";maintain-aspect-ratio TRUE;display
"USEDEF";valid_file "F";width 11.7959cm;height 12.468cm;depth
0pt;original-width 10.472in;original-height 11.1111in;cropleft "0";croptop
"0.9985";cropright "1.0017";cropbottom "0";filename
'FIG19.GIF';file-properties "XNPEU";}}

Let $\tilde{J}_{n}$ be the blood current in vein $n$ and $I_{n}$ be the
total blood current in branch block $n$ which are caused by the action of
EPS $\varepsilon _{j}=1$. The total resistance of block $n$ for $1\mathbf{%
\leqslant }n\mathbf{\leqslant }n_{j}$ is

\begin{equation*}
r_{n}=\frac{1}{7}R_{n}^{0}+\frac{1}{7\cdot 8}R_{n+1}^{0}+\frac{1}{7\cdot
8\cdot 8}R_{n+2}^{0}+\ldots
\end{equation*}
i.e. 
\begin{equation}
r_{n}=\frac{1}{(2^{3}-1)}\sum_{p=n}^{N}2^{-3(p-n)}R_{p}^{0}\;,  \label{10.16}
\end{equation}
thereby, from (\ref{3.20}) and (\ref{10.14}) we get for $1\mathbf{\leqslant }%
n\mathbf{\leqslant }n_{j}$ 
\begin{equation}
r_{n}=\frac{1}{(2^{3}-1)}2^{3n}R_{0}Z(n)\;.  \label{10.17}
\end{equation}
In a similar way we obtain the expression for the resistance of the last $%
(n_{j}+1)$-th block: 
\begin{equation}
r_{n_{j}+1}=\frac{1}{2^{3}}2^{3(n_{j}+1)}R_{0}Z(n_{j}+1).  \label{10.18}
\end{equation}

The introduced quantities, i.e. $\tilde{J}_{n},I_{n}$, and $r_{n}$ enable us
to transform the system of equations (\ref{3.16}), (\ref{10.9}) for $P=0$
and $\varepsilon _{i^{\prime }}=0$ except $\varepsilon _{i^{\prime
}}|_{i^{\prime }=j}=1$ into the Kirchhoff's equations associated with the
network shown in Fig.~\ref{Fig19} viz. to the following equations for $1%
\mathbf{\leqslant }n\mathbf{\leqslant }n_{j}$:

\begin{equation}
I_{n}=\tilde{J}_{n}-\tilde{J}_{n-1}\;,  \label{10.19}
\end{equation}
for $2\mathbf{\leqslant }n\mathbf{\leqslant }n_{j}$ 
\begin{equation}
I_{n}r_{n}=\tilde{J}_{n-1}R_{n-1}^{0}+I_{n-1}r_{n-1}\;,  \label{10.20}
\end{equation}
and into two equations that are bound up with the first and the last
elements of the given network: 
\begin{equation}
I_{1}r_{1}=\tilde{J}_{0}R_{0}\;,  \label{10.21}
\end{equation}
\begin{equation}
\tilde{J}_{n_{j}}[r_{n_{j}+1}+R_{n_{j}}^{0}]+I_{n_{j}}r_{n_{j}}=1\;.
\label{10.22}
\end{equation}
It should be noted that the last two equations can be regarded as certain
boundary conditions for the system of equations (\ref{10.19}), (\ref{10.20}%
). Due to $\rho (n)$ being a smooth function of $n$ the ratio $\rho (n)/Z(n)$
may be treated as a small value for $N-n\gg 1$. The latter allows us to find
directly the solution of equations (\ref{10.19}) - (\ref{10.22}) which is
matter of the next Subsection.

\subsection{Continuous solution%
\index{continuous solution} of the Kirchhoff equations}

\label{s10.2.1}

Taking into account (\ref{3.20}), (\ref{10.14}), (\ref{10.17}), and (\ref
{10.18})we can rewrite equation (\ref{10.20})-(\ref{10.22}) in terms of 
\begin{equation}
2^{3}I_{n}Z(n)=(2^{3}-1)%
\tilde{J}_{n-1}\rho (n-1)+I_{n-1}Z(n-1)  \label{10.23}
\end{equation}
for $2\mathbf{\leqslant }n\mathbf{\leqslant }n_{j}$ and 
\begin{equation}
2^{3}I_{1}Z(1)=(2^{3}-1)\tilde{J}_{0},  \label{10.24}
\end{equation}
\begin{equation}
\tilde{J}_{n_{j}}+\frac{1}{(2^{3}-1)}I_{n_{j}}=\frac{1}{R_{0}Z(n_{j})}%
2^{-3n_{j}}.  \label{10.25}
\end{equation}
where, in addition, we have taken into account the identity $z(n+1)+\rho
(n)=z(n)$

First, let us analyse the solution of equations (\ref{10.23}) and (\ref
{10.24}) when $n\sim 1$. Due to smoothness of the function $\rho (n)$ for
such values of $n$ the function $\rho (n)$ as well as $Z(n)$ is supposed to
be constant: $\rho (n)\simeq \rho (0)=1$ and $Z(n)\simeq Z(0)\gg 1$. Then,
from (10.23) and (10.24) we get 
\begin{equation}
2^{3}I_{n}-I_{n-1}=(2^{3}-1)\frac{1}{Z(0)}\tilde{J}_{n-1}  \label{10.26}
\end{equation}
and 
\begin{equation}
2^{3}I_{1}=(2^{3}-1)\frac{1}{Z(0)}\tilde{J}_{0}.  \label{10.27}
\end{equation}
We shall seek the solution of (\ref{10.19}), (\ref{10.26}), and (\ref{10.27}%
) in the form 
\begin{equation}
\tilde{J}_{n}=A_{+}\zeta _{+}^{n}+A_{-}\zeta _{-}^{n}\;\;,  \label{10.28}
\end{equation}
where $A_{+,-};\zeta _{+,-}$ are some constants. Substituting (\ref{10.28})
into (\ref{10.19}) we get

\begin{equation*}
I_{n}=A_{+}\zeta _{+}^{(n-1)}(\zeta _{+}-1)+a_{-}\zeta _{-}^{(n-1)}(\zeta
_{-}-1).
\end{equation*}
Then, the substitution of this expression into (\ref{10.26}) shows us that
the constants $\zeta _{+},\zeta _{-}$ are the roots of the following
equation: 
\begin{equation}
(\zeta -1)(2^{3}\zeta -1)=\frac{2^{3}-1}{Z(0)}\zeta \;.  \label{10.29}
\end{equation}
Whence, it follows that to the first order in the small parameter $1/Z(0)$: 
\begin{equation}
\zeta _{+}=1+\frac{1}{Z(0)};\qquad \zeta _{-}=2^{-3}\left( 1-\frac{1}{Z(0)}%
\right) .  \label{10.30}
\end{equation}
To the same order in $1/Z(0)$ from (\ref{10.19}) and (\ref{10.27}) we find
that the constants $A_{+},A_{-}$ are related by the expression 
\begin{equation}
\frac{A_{-}}{A_{+}}=\frac{1}{(2^{3}-1)Z(0)}\ll 1.  \label{10.31}
\end{equation}

According to (\ref{10.30}) the first term on the right-hand side of (\ref
{10.28}) is an increasing function of $n$, whereas the second one is a
decreasing function. Therefore, by virtue of (\ref{10.31}) the former is
substantially larger than the latter for all $n\sim 1$. So, we may ignore
the second term and regard the first one as a function of the continuous
variable $n$. This allows us to set 
\begin{equation}
\tilde{J}_{n}-\tilde{J}_{n-1}=\frac{\partial \tilde{J}_{n}}{\partial n}
\label{10.32}
\end{equation}
in equation (\ref{10.19}), to assume that $I_{n-1}\approx I_{n}$ in (\ref
{10.26}), and also to rewrite (\ref{10.28}) in terms of 
\begin{equation}
\tilde{J}_{n}\simeq A_{+}\exp \left\{ \frac{1}{Z(0)}n\right\} \;.
\label{10.33}
\end{equation}
In this case the system of equations (\ref{10.19}) and (\ref{10.26}) is
reduced to the equation 
\begin{equation}
\frac{\partial \tilde{J}_{n}}{\partial n}=\frac{1}{Z(0)}\tilde{J}_{n}
\label{10.34}
\end{equation}
and function (\ref{10.33}) is the general solution of the latter equation.
So, ``boundary condition'' (\ref{10.27}) is responsible only for the
existence of the second term in (\ref{10.28}) and, therefore, can be ignored.

As it follows from the form of equations (\ref{10.19}), (\ref{10.20}), their
general solution contains two arbitrary constants, for example, $\tilde{J}%
_{0},\tilde{J}_{1}$, as it takes place for the constants $A_{+},A_{-}$ in
expression (\ref{10.28}). Indeed, all the other values of $\tilde{I}_{n}$
and $\tilde{J}_{n}$ can be determined by iteration. So, there is no solution
of equations (\ref{10.19}), (\ref{10.20}) different from (\ref{10.28}) in
the region $n\sim 1$. Thereby, the influence of ``boundary condition'' (\ref
{10.21}) on the solution of equations (\ref{10.19}) and (\ref{10.20}) is
ignorable and we may seek this solution in the class of smooth functions of
the continuous variable $n$. In particular, in this case the general
solution of equations (\ref{10.19}), (\ref{10.20}) contains only one
arbitrary constant that can be found from ``boundary condition'' (\ref{10.22}%
).

Within the framework of the assumptions adopted above due to (\ref{10.14}),
and (\ref{10.19}) the quantities $\tilde{J}_{n}$ and $I_{n}$ as well as $%
\rho (n)$ and $Z(n)$ are related by the expressions 
\begin{equation}
I_{n}=\frac{\partial \tilde{J}_{n}}{\partial n}\;,  \label{10.35}
\end{equation}
\begin{equation}
\rho _{n}=-\frac{\partial Z(n)}{\partial n}\;,  \label{10.36}
\end{equation}
and the system of equations (\ref{10.19}), (\ref{10.23}), (\ref{10.25})
reduces to the following equation

\begin{equation}
\frac{\partial \tilde{J}_{n}}{\partial n}Z(n)+\tilde{J}_{n}\frac{\partial
Z(n)}{\partial n}=0  \label{10.37}
\end{equation}
subject to the boundary condition 
\begin{equation}
\tilde{J}_{n}\mid _{n=n_{j}}=2^{-3n_{j}}\frac{1}{R_{0}Z(n_{j})}\;.
\label{10.38}
\end{equation}
In obtaining these expressions, we have also taken into account relation (%
\ref{10.32}), supposing that $\mid I_{n+1}-I_{n}\mid \ll I_{n}$ and $\mid
Z(n+1)-Z(n)\mid \ll Z(n)$. Besides, we have ignored the second term on the
left-hand side of (\ref{10.25}) because $I_{n_{j}}=\tilde{J}_{n_{j}-1}-%
\tilde{J}_{n_{j}}\ll \tilde{J}_{n_{j}}$.

The solution of equation (\ref{10.37}) meeting boundary condition (\ref
{10.38}) is of the form

\begin{equation}
\tilde J_n=2^{-3n_j}\frac 1{R_0Z(n)}  \label{10.39}
\end{equation}
and, thus, from (\ref{10.19}) and (\ref{10.32}) we obtain

\begin{equation}
I_n=2^{-3n_j}\frac{\rho (n)}{R_0[Z(n)]^2}\;.  \label{10.40}
\end{equation}

In order to obtain the relation between $\Lambda_{ij}$ and the quantities $%
\tilde{J}_{n},I_{n}$ we, first, classify all possible pairs of veins $%
\{i,i^{\prime }\}$ into two groups. The first group of the pairs designated
by $\{i,i^{\prime }\}_{+}$ comprises all pairs of veins that can be joined
by a path on the vascular network directed either from higher to lower
levels or vice versa, i.e. by a path with a constant direction. Such a path
is depicted in Fig.~\ref{Fig18} by line $O_{j}O_{i_{+}}$. The second group
involves such pairs of veins $\{i,i^{\prime }\}_{-}$ that can be connected
by a path on the venous bed whose direction is changed at a certain
branching point $B$ as shown in Fig.~\ref{Fig18} by the curve $%
O_{j}O_{i_{-}} $. In other words, such a path initially goes, for example,
from the vein $i$ towards the host vein until it reaches a certain branching
point $B_{ii^{\prime }}$ and then, the path runs towards the vein $i$ going
in the opposite direction. In this case we also ascribe to this branching
point the level number $n_{B}$ of the veins going into it and specify the
pair $\{i,i^{\prime }\}_{-}$ also in terms of $\{i,i^{\prime },B_{ii^{\prime
}}\}_{-}$.

If for a vein $i$ whose level number $n_{i}\mathbf{\leqslant }n_{j}$ the
pair $\{ij\}$ belongs to the first group, then, the vein $i$ is one of the
veins of the sequence $\{n_{j},\,n_{j-1},\dots ,1,0\}$, viz. vein $n_{i}$.
In this case as it follows from the definition of the Green matrix $\Lambda
_{ij}=\tilde{J}_{n_{i}}$. Thereby, according to (\ref{10.39}), for a pair $%
\{ij\}_{+}$ where $n_{i}\mathbf{\leqslant }n_{j}$

\begin{equation}
\Lambda_{ij}=2^{-3n_{j}}\frac{1}{R_{0}Z(n_{i})}.  \label{10.41}
\end{equation}
A vein $i$ will belong to the last $(n_{j}+1)$-th block of branches, if its
level number $n_{i}>n_{j}$ and this vein along with the vein $j$ forms a
pair $\{i,j\}_{+}$ of the first group. The total number of such veins for a
fixed value of $n_{i}$ is $2^{3(n_{i}-n_{j})}$ , and the total blood current
in this block is $\tilde{J}_{n_{j}}$. Since, the blood current $\tilde{J}%
_{n_{j}}$ is uniformly distributed among these veins, in anyone of them the
blood current is $2^{-3(n_{i}-n_{j})}\tilde{J}_{n_{j}}$. Therefore,
according to the definition of $\Lambda_{ij}$ and by virtue of (\ref{10.39})
we obtain the following expression for a pair $\{i,j\}_{+}$ where $%
n_{i}>n_{j}$: 
\begin{equation}
\Lambda_{ij}=2^{-3n_{i}}\frac{1}{R_{0}Z(n_{j})}\;.  \label{10.42}
\end{equation}
Now let us consider a pair $\{i,j,B_{ij}\}_{-}$. The given vein $i$ is
located in branch block $n_{B}$ where the total number of veins belonging to
the same level $n_{i}$ is, obviously, $(2^{3}-1)2^{3(n_{i}-n_{B})}$. So, in
this case a blood current in the vein $i$ is $%
(2^{3}-1)^{-1}2^{-3(n_{i}-n_{B})}I_{n_{B}}$. Whence, by virtue of (\ref
{10.17}) for the pair $\{i,j,B_{ij}\}_{-}$ we get 
\begin{equation}
\Lambda_{ij}=-\frac{1}{2^{3}-1}2^{-3(n_{i}+n_{j}-n_{B})}\frac{\rho (n_{B})}{%
R_{0}[Z(n_{B})]^{2}}\;\;.  \label{10.43}
\end{equation}
In formula (\ref{10.43}) the sign '-' means that in the given vein $i$ the
blood current $\Lambda_{ij}$ caused by the additional EPS $\varepsilon
_{j}=1 $ associated with the vein $j$ is directed from lower to higher
levels. In other words, this blood current and the total blood current in
the vein $i$, which is induced by the collective action of all the EPSs and
the real pressure drop $P$ have opposite directions.

We note that expression (\ref{10.15}) is actually contained in formula (\ref
{10.42}) as a special case. So, formulae (\ref{10.41})-(\ref{10.43}) specify
the desired Green matrix $\parallel \Lambda _{ij}\parallel $ completely.
Thereby, these expressions along with definition (\ref{10.8}), equation (\ref
{10.10}), and relation (\ref{10.11}) reduce the model for temperature
self-regulation developed in Section~\ref{s3.4} to one dealing with the
tissue temperature $T(\mathbf{r},t)$, the blood flow rate $j(\mathbf{r},t)$
and the collection of quantities whose evolution is determined by these
fields directly. In order to facilitate perception of the proposed model in
the following Section we shall briefly review its basic grounds, assumption
and relations.

\section{Governing equations of vascular network response}

\label{s10.3}

Within the framework of the proposed model the blood current $J_i$ and the
temperature $T_i^{*}$ of blood in a vessel $i$, the true and averaged blood
flow rate $j, j_v$ and the temperature field $T$ are related by expressions (%
\ref{10.1}) and (\ref{10.4}), (\ref{10.5}) viz.

\begin{equation}
J_{i}=\int\limits_{Q_{0}}d\mathbf{r}\Theta _{i}(\mathbf{r}\,)j(\mathbf{r}%
\,)\;,  \label{10.44}
\end{equation}
\begin{equation}
(T_{i}^{\ast }-T_{a})J_{i}=\int\limits_{Q_{0}}d\mathbf{r}\Theta _{i}(\mathbf{%
r}\,)(T-T_{a})j  \label{10.45}
\end{equation}
for the unit vessel network, and for the countercurrent vascular network

\begin{equation}
(T_{i}^{\ast }-T_{a})J_{i}=\beta _{v}(\zeta _{v})\int\limits_{Q_{0}}d\mathbf{%
r}\Theta _{i}(\mathbf{r}\,)(T-T_{a})j.  \label{10.46}
\end{equation}
Here $\Theta _{i}(\mathbf{r}\,)$ is the characteristic function of the
domain $Q_{i}$, the function $\beta _{v}(\zeta _{v})$ as well the value $%
\zeta _{v}$ are given by expressions (\ref{10.6}), (\ref{10.7}).

The response of the vessels to temperature variations is described in terms
of the additional effective pressure sources (EPSs) specified by (\ref{10.8}%
), viz.:

\begin{equation}
\varepsilon =-J_i(R_i-R_n^0)\;,  \label{10.47}
\end{equation}
which evolve according to equation (\ref{10.10});

\begin{equation}
\tau _n\frac{d\varepsilon }{dt}+\varepsilon \left[ f\left( 1- \frac{%
\varepsilon }{J_iR_n^0}\right) -\tau _n\frac d{dt}\ln J_i\right] =R_n^0\frac
1\Delta \left| (T_i^{*}-T_a)J_i\right| \;.  \label{10.48}
\end{equation}

Here $\tau _n$ is the characteristic time of the n-th level vessel response,
the function $f(x)$ is determined by the quasistationary dependence of the
vessel resistances on the blood temperature, (see formula (\ref{3.28})), $%
R_n^0=R_i|_{T_i^{*}=T_a}$ as a function of $n$ is described by (\ref{3.20}).

The given system of equations is completed by relation (\ref{10.11}) between
EPSs and the blood current pattern $\{J_{i}\}$ on the vascular network

\begin{equation}
J_i=\sum\limits_{i^{\prime }}\Lambda _{ii^{\prime }}[\varepsilon _{i^{\prime
}}+P\delta _{n_{i^{\prime }}0}]\;,  \label{10.49}
\end{equation}
where $P$ is the true pressure drop across the venous bed and the Green
matrix $\Lambda _{ii^{\prime }}$ is specified by formulae (\ref{10.41})-(\ref
{10.43}).

In the following Chapter on the basis of the developed model we shall
consider heat transfer in living tissue characterized by ideal temperature
self-regulation.

\chapter{Theory of ideal temperature self-regulation}

\label{ch.11} 
\markright
{ {\sc  \thechapter. Theory of ideal temperature self-regulation}
}

\section{Ideal response of the vascular network variation}

\label{s11.1}

In the present Section we consider heat transfer in a certain idealized
living tissue where the vascular network response to temperature variations
can control whatever strong heating. By definition (Chapter~\ref{ch.2}) the
thermoregulation process is called ideal when temperature of living tissue
under quasistationary heating cannot go out of the vital temperature
interval $[T_{-},T_{+}]$. This situation will take place if for each vein i
(and artery i) the quasistationary dependence of its resistance $R_{i}^{q}$
on the blood temperature $T_{i}^{\ast }$ is of the form $R_{i}^{q}(T_{i}^{%
\ast })=R_{n_{i}}^{0}\cdot \varphi ^{id}((T_{i}^{\ast }-T_{a})/\Delta )$
where the function $\varphi ^{\text{id}}(x)$ is given by formula (\ref{3.30}%
). Indeed,in this case, if in a certain domain the tissue temperature were
higher than $T_{+}$, the temperature of blood in veins through which blood
is carried away from the given tissue domain should be higher than $T_{+}$
too. Therefore, on the vascular network there would exist a path of zero
resistance. The latter would give rise to an infinitely high blood flow rate
playing the cooling role, owing to which the tissue temperature in the given
domain should drop up to the arterial blood temperature $T_{a}$. We note
that the countercurrent effect leads to a certain renormalization of the
values $T_{+}$ and $T_{-}$ only.

We shall confine ourselves to the following additional assumptions. First,
we assume that the heat generation rate $q(\mathbf{r},t)\geq 0$ at all the
points of the tissue domain $Q_{0}$ and at its boundary $\partial Q_{0}$ the
temperature $T\mid _{\partial Q}$ is maintained at the value $T_{a}(T\mid
_{\partial Q_{0}}=T_{a})$ Chapter~\ref{ch.3}. In this case we may suppose
beforehand that the inequalities $T(r,t)\geq T_{a}$ and $T_{i}^{\ast }\geq
T_{a}$ hold at all the points of the domain $Q_{0}$ and for all the vessels.
Second, the characteristic time $\tau _{n}$ of the vessel response is
assumed to be the same for all the vessels, i.e. $\tau _{n}=\tau $. Third,
we shall ignore the term $\tau _{n}\frac{d}{dt}\ln (J_{i})$ in equation (\ref
{10.48}). This term actually describes dynamic interaction between EPSs and
has no substantial effect on blood flow redistributions over the vascular
network at least in the following two possible limit cases. When $\tau \ll
\tau _{T}$, where $\tau _{T}$ is the characteristic time of tissue
temperature variations, the vascular network response will be
quasistationary, and the term along with the first transient term in (\ref
{10.48}) will be ignorable. When $\tau \gg \tau _{T}$ and, in addition, the
stationary heating of living tissue is not high the value of $\tau _{n}\frac{%
d}{dt}\ln (J_{i})$ should be negligibly small in comparison with $\tau _{n}%
\frac{d}{dt}ln(\varepsilon )$. Indeed, in this case the derivative $\frac{d}{%
dt}ln(J_{i})$ can be estimated as $(\frac{d}{dt}ln(\varepsilon ))\frac{%
\varepsilon }{P}$ and the stationary value of $\varepsilon _{i}/P$ is a
small parameter. Besides, the inequalities $\varepsilon <J_{i}R_{n_{i}}^{0}$
for all the veins are considered to be true in advance because in the given
case these assumptions cannot lead to wrong results as it follows from the
discussions of equation (\ref{11.12}). Dealing with the countercurrent
vascular network we also assume that the vascular network resistance to
blood flow is mainly controlled by large vessels whose level number $n<n_{t}$
for any values of the blood flow rate.

According to (\ref{3.29}) and (\ref{3.30}) for the ideal temperature
self-regulation process $f^{\text{id}}(x)=1$, when $x<1$. Owing to the
adopted assumptions and taking into account expressions (\ref{10.44}), (\ref
{10.46}) we may rewrite equation (\ref{10.48}) in the form

\begin{equation}
\tau \frac{d\varepsilon }{dt}+\varepsilon =R_{n}^{0}\frac{1}{\Delta }\beta
_{cc}\int\limits_{Q_{0}}d\mathbf{r}\Theta _{i}(\mathbf{r}\,)[T(\mathbf{r}%
\,)-T_{a}]j(\mathbf{r}\,)\;.  \label{11.1}
\end{equation}
where $\beta _{cc}=1$ for the unit vessel network and $\beta _{cc}=\left[
\ln (l_{0}/a_{0})\right] ^{-1/2}$ for the countercurrent vascular network.
The latter is justified because dealing with the blood flow redistribution
over the countercurrent vascular network may allows for large vessels with $%
n<n_{t}$ only.

Expression (\ref{10.49}) and equation (\ref{11.1}) enable us to represent $%
J_{i}$ in a vein $i$ of level $n_{i}$ in terms of 
\begin{equation}
J_{i}=J_{ni}^{0}+J_{i}^{a}\;,  \label{11.2}
\end{equation}
where $J_{ni}^{0}$ is the blood current when the tissue temperature $T=T_{a}$
and, thus, $T_{i}^{\ast }=T_{a}$ for all the veins and the additional blood
current $J_{i}^{a}$ caused by the self-regulation process obey the equation 
\begin{equation}
\tau \frac{dJ_{i}^{a}}{dt}+J_{i}^{a}=\frac{\beta _{cc}}{\Delta }%
\int\limits_{Q_{0}}d\mathbf{r}j(\mathbf{r}\,)[T(\mathbf{r}%
\,)-T_{a}]\sum\limits_{i^{\prime }}\Lambda _{ii^{\prime }}R_{n_{i^{\prime
}}}^{0}\Theta _{i^{\prime }}(\mathbf{r}\,)\;.  \label{11.3}
\end{equation}

Expression (\ref{11.3}) can be significantly simplified. For this purpose in
the following subsection we prove a certain identity playing an important
role in the theory of thermoregulation.

\subsection{The ideal thermoregulation identity}

\label{s11.1.1}

Let us show that for a vein $i$ whose level number $1 \ll n_i \ll N$ of
lower order in the small parameter $\rho (n_i)/Z(n_i)$

\begin{equation}
\sum\limits_{i^{\prime }}\Lambda _{ii^{\prime }}R_{n_{i^{\prime
}}}^{0}\Theta _{i^{\prime }}(\mathbf{r}\,)=\Theta _{i}(\mathbf{r}\,)\;.
\label{11.4}
\end{equation}

Every point $\mathbf{r}$ of the tissue domain $Q_{0}$, except points
belonging to a set of zero measure, is located inside just one fundamental
domain $Q_{nr}$ for each level $n$. Thereby, practically for any point $%
\mathbf{r}$ we can specify the collection $\{Q_{nr}\}$ of all the
fundamental domains containing the point $\mathbf{r}$. The domain collection 
$\{Q_{nr}\}$ in its turn specifies the sequence $\{i_{nr}\}$ of connected
veins $\{i_{o},i_{1r},i_{2r},...,i_{Nr}\}$ whose n-th term is the vein $%
i_{nr}$ of level $n$ contained in the domain $Q_{nr}$. The veins $\{i_{nr}\}$
form a continuous path $\mathcal{P}_{r}$ on the venous bed which leads from
the host vein $i_{0}$ to the vein $i_{Nr}$ of the last level.

In formula (\ref{11.4}) the functions $\{\Theta _{i^{\prime }}(r)\}$ are
different from zero for the veins $\{i_{nr}\}$ only, thus,

\begin{equation}
\tilde{\Lambda}_{i}(\mathbf{r}\,)\equiv \sum\limits_{i^{\prime }}\Lambda
_{ii^{\prime }}R_{n_{i^{\prime }}}^{0}\Theta _{i^{\prime }}(\mathbf{r}%
\,)=\sum\limits_{i^{\prime }\in \{i_{nr}\}}\Lambda _{ii^{\prime
}}R_{n_{i^{\prime }}}^{0}\;.  \label{11.5}
\end{equation}
First, let the domain $Q_{i}$ corresponding to the vein $i$ belong\ to the
domain system $\{Q_{nr}\}$ and, thereby, the vein $i$ of level $n_{i}$ be
one of the veins $\{i_{nr}\}$. Then, for this vein, taking into account
expressions (\ref{3.20}), (\ref{10.14}), and (\ref{10.41})-(\ref{10.42})
from (\ref{11.5}) at lower order in $\rho (n_{i})/Z(n_{i})$ we obtain

\begin{equation*}
\tilde{\Lambda}_{i}(\mathbf{r}\,)=\sum\limits_{n_{i^{\prime
}}=0}^{n_{i}-1}\Lambda _{ii^{\prime }}R_{n_{i^{\prime
}}}^{0}+\sum\limits_{n_{i^{\prime }}=n_{i}}^{N}\Lambda _{ii^{\prime
}}R_{n_{i^{\prime }}}^{0}=
\end{equation*}
\begin{equation*}
\sum\limits_{n_{i}^{\prime }=0}^{n_{i}-1}2^{-3(n_{i}-n_{i^{\prime }})}\frac{%
\rho (n_{i^{\prime }})}{Z(n_{i^{\prime }})}+\sum\limits_{n_{i}^{\prime
}=n_{i}}^{N}\frac{\rho (n_{i^{\prime }})}{Z(n_{i})}
\end{equation*}
thus, 
\begin{equation}
\tilde{\Lambda}_{i}(\mathbf{r}\,)=1+\sum\limits_{p=1}^{n_{i}}2^{-3p}\frac{%
\rho (n_{i}-p)}{Z(n_{i}-p)}\simeq 1\;,  \label{11.6}
\end{equation}
because of $\rho (n)$ and $Z(n)$ being smooth functions of $n$.

Now, let the domain $Q_{i}$ not belong to the collection $\{Q_{nr}\}$ and,
thus, the corresponding vein $i$ be none of the veins $\{i_{nr}\}$. In this
case on the path $\mathcal{P}_{r}$ there exists a branching point $B$ of
level $n_{B}$ at which the branch containing the vein $i$ is connected with
the path $\mathcal{P}_{r}$. The branching point $B$ divides the vein
sequence $\{i_{nr}\}$ into two parts according to level number. The veins
whose level number $n_{ir}<n_{B}$ form with the vein $i$ pairs $%
\{i_{nr},i\}_{+}$ of the first group, whereas for the veins whose level
number $n_{ir}\geq n_{B}$ the pairs $\{i_{nr},i\}_{-}$ belong to the second
group and are of type $\{i_{nr},i,B\}_{-}$. Thus, for such a vein $i$
substituting (\ref{3.20}), (\ref{10.42}), and (\ref{10.43}) into (\ref{11.5}%
) and, in addition, taking into account (\ref{10.14}) we obtain

\begin{equation*}
\tilde{\Lambda}_{i}(\mathbf{r}\,)=\sum\limits_{n_{i^{\prime
}}=0}^{n_{B}-1}\Lambda _{ii^{\prime }}R_{n_{i^{\prime
}}}^{0}+\sum\limits_{n_{i^{\prime }}=n_{B}}^{N}\Lambda _{ii^{\prime
}}R_{n_{i^{\prime }}}^{0}=
\end{equation*}
\begin{equation*}
\sum\limits_{n_{i^{\prime }}=0}^{n_{B}-1}2^{-3(n_{i}-n_{i^{\prime }})}\frac{%
\rho (n_{i^{\prime }})}{Z(n_{i^{\prime }})}-\sum\limits_{n_{i^{\prime
}}=n_{B}}^{N}2^{-3(n_{i}-n_{B})}\frac{\rho (n_{B})\rho (n_{i^{\prime }})}{%
Z^{2}(n_{B})}
\end{equation*}
thus,

\begin{equation}
\tilde{\Lambda}_{i}(\mathbf{r}\,)=2^{-3(n_{i}-n_{B})}\left[
\sum\limits_{p=1}^{n_{B}}2^{-3p}\frac{\rho (n_{B}-p)}{Z(n_{B}-p)}-\frac{1}{%
(2^{3}-1)}\frac{\rho (n_{B})}{Z(n_{B})}\right] \;.  \label{11.7}
\end{equation}
Due to $\rho (n)$ and $Z(n)$ being smooth functions of $n$ the main
contribution to the first term in (\ref{11.7}) is associated with $p\sim 1$,
which allows us to expand the function $\rho (n-p)/Z(n-p)$ in a power series
of $p$. In this way accounting for the two first terms from (\ref{11.7}) we
find

\begin{equation*}
\tilde{\Lambda}_{i}(\mathbf{r}\,)=\left\{ 2^{-3n_{i}}\frac{1}{7}\frac{\rho
(n_{B})}{Z(n_{B})}+2^{-3(n_{i}-n_{B})}\right. \cdot
\end{equation*}
\begin{equation}
\cdot \left. \frac{8}{49}\left[ 1-2^{-3n_{B}}(n_{B}+1)+n_{B}2^{-3(n_{B}+1)}%
\right] \frac{d}{dn}\left. \frac{\rho (n)}{Z(n)}\right| _{n=n_{B}}\right\} .
\label{11.8}
\end{equation}

By virtue of the adopted assumptions for the vein $i$ \thinspace\ $n_{i}\geq
n_{B}$, and $n_{i}\gg 1$. So, the vein $i$ in the value $\tilde{\Lambda}%
_{i}(r)$ differ from zero no more than at the second order in the small
parameter $\rho (n)/Z(n)$. Therefore, if the point $\mathbf{r}$ is located
inside the domain $Q_{i}$ corresponding to the vein $i$, then, the domain $%
Q_{i}$ belongs to the collection $\{Q_{n_{r}}\}$ and, thus, $\tilde{\Lambda}%
_{i}(\mathbf{r}\,)=1$. When the domain $Q_{i}$ does not contain the point $%
\mathbf{r}$ it does not belong to $\{Q_{n_{r}}\}$, and, at first order in
the parameter $\rho (n)/Z(n)$ the value $\tilde{\Lambda}_{i}(\mathbf{r}%
\,)=0; $ whence, we immediately obtain formula (\ref{11.4}).

\subsection[Quasilocal equation for the temperature dependence of the blood
flow rate]{Quasilocal equation for the temperature \newline
dependence of the blood flow rate}

\label{s11.1.2}

In what follows we shall confine ourselves to lower order in the small
parameter $\rho (n)/Z(n)$. Then, identity (\ref{11.4}) reduces equation (\ref
{11.3}) to the equation

\begin{equation}
\tau \frac{dJ_{i}^{a}}{dt}+J_{i}^{a}=\frac{\beta _{cc}}{\Delta }%
\int\limits_{Q_{0}}d\mathbf{r}\,\Theta _{i}(\mathbf{r}\,)j(\mathbf{r}\,)[T(%
\mathbf{r}\,)-T_{a}]\;.  \label{11.9}
\end{equation}

Let us choose such a small fundamental domain $Q_{i}$ that its size $l_{i}$
is well below all the characteristic spatial scales of the fields $j(\mathbf{%
r}\,)$ and $T(\mathbf{r}\,)$ (however, as before $l_{i}\gg l_{N}$). In this
case by virtue of (\ref{10.44}) we may set 
\begin{equation}
J_{i}^{a}=\int\limits_{Q_{0}}d\mathbf{r}\,\Theta _{i}(\mathbf{r}\,)[j(%
\mathbf{r}\,)-j_{0}]=V_{i}[j(\mathbf{r}\,)-j_{0}]\;,  \label{11.10}
\end{equation}
and 
\begin{equation}
\int\limits_{Q_{0}}d\mathbf{r}\,\Theta _{i}(\mathbf{r}\,)j(\mathbf{r}\,)[T(%
\mathbf{r}\,)-T_{a}]=V_{i}j(\mathbf{r}\,)[T(\mathbf{r}\,)-T_{a}]\;,
\label{11.11}
\end{equation}
where $\mathbf{r}$ is an arbitrary point of the domain $Q_{i}$, $V_{i}$ is
its volume, and $j_{0}$ is the blood flow rate when the tissue temperature $%
T(r,t)=T_{a}$. Expression (\ref{11.10}) and (\ref{11.11}) allow us to
rewrite equation (\ref{11.9}) in terms of 
\begin{equation}
\tau \frac{\partial j}{\partial t}+j\left[ 1-\beta _{cc}\frac{T-T_{a}}{%
\Delta }\right] =j_{0}\;.  \label{11.12}
\end{equation}
Equation (\ref{11.12}) specifies the vascular network response to
temperature variations and along with the bioheat equation for the
temperature evolution and the relationship between the true and averaged
blood flow rates forms the desired description of heat transfer in living
tissue with ideal thermoregulation which is called the quasi-local model for 
\index{bioheat transfer}.

When deriving equation (\ref{11.12}) we have assumed a priori that within
the framework of the adopted approximations the inequalities $%
\varepsilon_i<J_iR_{n_i}^0$ are true. The validity of this assumption
directly results from (\ref{11.12}). Indeed, otherwise, according to (\ref
{10.8}) the resistance to blood flow at least in one of the veins would be
negative and in this vein the blood current should flow in the opposite
direction and at the corresponding points of the tissue domain the blood
flow rate would be negative. However, as it follows from (\ref{11.12}) the
blood flow rate $j$ is positive for any value of $T$ if at the initial time $%
j\mid _{t=t_0}\geq 0$ for all the points of $Q_0$.

\section[Mathematical modelling of temperature distribution in living tissue
under local strong heating]{Mathematical modelling of temperature \newline
distribution in living tissue under \newline
local strong heating}

\label{s11.2}

In the present section we study some characteristic properties of heat
transfer in living tissue when the size of the region affected directly can
be small or the tissue heating is significantly nonstationary. We show that
under local heating there can be remarkable difference between the averaged
and true blood flow rates. So, for bioheat transfer%
\index{bioheat transfer} in real living tissues the dependence of the
temperature evolution on the averaged blood flow rate rather than the true
one can play a significant role. Besides, evolution of the tissue
temperature in living tissue due to thermoregulation under strong heating
and a similar process in nonbiological media differ markedly in properties.
In particular, delay in vessel response can give rise to anomalous behavior
of a transient process in living tissue.

The properties of the given model has been analyzed numerically. We have
considered the solution of equations (\ref{10.44}), (\ref{10.45}) and (\ref
{11.12}) which in the dimensionless form can be rewritten as \cite{43}:

\begin{equation}
\frac{\partial {\theta }}{\partial t}=\mathbf{\nabla }^{2}\theta -\eta
_{v}\theta +q_{\theta }\;,  \label{11.13}
\end{equation}
\begin{equation}
\eta _{v}-\frac{1}{L}\mathbf{\nabla }^{2}\ln \eta _{v}=\eta \;,
\label{11.14}
\end{equation}
\begin{equation}
\alpha \frac{\partial \eta }{\partial t}+\eta (1-\theta )=1\;.  \label{11.15}
\end{equation}
Here, for the unit vessel network $\theta =\beta _{cc}(T-T_{a})/\Delta $ is
the dimensionless tissue temperature, $\eta =j/j_{0}$ and $\eta
_{v}=j_{v}/j_{0}$ are the dimensionless true and averaged blood flow rates,
all the lengths and the time $t$ are measured in units of $[\kappa
/(c_{t}\rho _{t}j_{0})]^{1/2}$ and $1/j_{0}$ respectively; the constants $%
\alpha =\tau j_{0};$ $L\sim (\kappa _{\mathrm{eff}}/\kappa )\ln
(l_{0}/a_{0}) $; and $q_{\theta }=q/(c_{t}\rho _{t}\Delta )$ are the
dimensionless temperature generation rate.

For the countercurrent vascular network $\theta =\frac{(T-T_{a})}{\Delta }%
\left[ \ln (l_{0}/a_{0})\right] ^{1/2}$; $\eta =j/j_{0};\eta
_{v}=j_{v}/j_{v} $, the unit length and the unit time are $\left\{ \kappa _{%
\mathrm{eff}}\frac{\left[ \ln (l_{0}/a_{0})\right] ^{1/2}}{c_{t}\rho
_{t}j_{v}}\right\} ^{1/2}$, $\left[ \ln (l_{0}/a_{0})^{1/2}\right] /j_{0}$,
respectively, $\alpha =\frac{j_{0}\tau }{\left[ \ln (l_{0}/a_{0})\right]
^{1/2}}$ and $L=(\kappa _{\mathrm{eff}}/\kappa )\ln (l_{0}/a_{0})$. In
numerical calculations we have set $L\sim 4$ because the value $l_{0}/a_{0}$
is typically\ equal to $30-40$ for real microcirculatory beds; $\alpha =0.5$%
; $3.0$ and obtained the solution of equations (\ref{11.13})-(\ref{11.15})
in unbound one-, two-, and three-dimensional spaces for the following form
of the temperature generation rate

\begin{equation}
q_{\theta }(\mathbf{r}\,)=q_{0}\exp \{-\frac{r}{\lambda }_{q}\}\;,
\label{11.16}
\end{equation}
where $\lambda _{q}=0.3;1.0;3.0$.

\FRAME{ftpFU}{8.3604cm}{8.3823cm}{0pt}{\Qcb{The stationary tissue
temperature $\protect\theta $ (a), the true and averaged blood flow rates $%
\protect\eta ,\protect\eta _{v}$ (b) as functions of the radius $r$ in the
three-dimensional space ($\protect\lambda _{q}=0.3$; $L=4$, lines 1 and 2
correspond to $q_{0}=0.5;50;$ respectively).}}{\Qlb{Fig20}}{Fig20}{\special%
{language "Scientific Word";type "GRAPHIC";maintain-aspect-ratio
TRUE;display "USEDEF";valid_file "F";width 8.3604cm;height 8.3823cm;depth
0pt;original-width 8.3057in;original-height 8.3333in;cropleft "0";croptop
"1.0006";cropright "1.0008";cropbottom "0";filename
'FIG20.GIF';file-properties "XNPEU";}}

Fig.~\ref{Fig20}a,b shows the stationary temperature distribution as well as
the stationary distributions of the true and averaged blood flow rates in
the three-dimensional space for $\lambda _{q}=0.3$. As seen from Fig.~\ref
{Fig20}b there is a significant difference between the true blood flow rate
and the averaged one when the temperature field is nonuniform enough and the
temperature maximum attains the boundary $T_{+}(\theta _{+}=1)$ of the vital
interval. Therefore, in this case the nonlocality in the dependence of the
averaged blood flow rate on the tissue temperature is a factor. To
collaborate the last conclusion, Fig.~\ref{Fig21} shows the maxima of the
tissue temperature $\theta _{max}=\theta \mid _{r=0}$ and the

\FRAME{ftbpFU}{4.9907cm}{5.1818cm}{0pt}{\Qcb{ The stationary tissue
temperature $\protect\theta _{max}$ and the blood flow rate $\protect\eta %
_{max}$ at $r=0$ as functions of $q_{0}$ (the three - dimentional space, $%
\protect\lambda _{q}=0.3,L=4$, the thin line corresponds to $\protect\eta %
_{max}(r)$ for $L=\infty $).}}{\Qlb{Fig21}}{Fig21}{\special{language
"Scientific Word";type "GRAPHIC";maintain-aspect-ratio TRUE;display
"USEDEF";valid_file "F";width 4.9907cm;height 5.1818cm;depth
0pt;original-width 8.0739in;original-height 8.361in;cropleft "0";croptop
"1.0014";cropright "0.9992";cropbottom "0";filename
'FIG21.GIF';file-properties "XNPEU";}}

\noindent true blood flow rate $\eta _{max}=\eta \mid _{r=0}$ as functions
of the value $q_{0}$ for the three-dimensional space and $\lambda _{q}=0.3$.
The solid lines display $\theta _{max}$ vs. $q_{0}$ and $\eta _{max}$ vs. $%
q_{0}$ in the given model and the dashed line shows the $\eta _{max}(q_{0})$
dependencies in the case where the averaged blood flow rate has been
formally replaced by the true one in equation (\ref{11.14}), i.e., where the
nonlocality of the $\eta _{v}(\theta )$ dependence has been ignored. In Fig.~%
\ref{Fig12_3} we\FRAME{ftFU}{9.6542cm}{5.1511cm}{0pt}{\Qcb{The stationary
tissue temperature $\protect\theta _{max}$, the true and averaged blood flow
rates $\protect\eta _{max},\protect\eta _{vmax}$ at $r=0$ as functions of $%
q_{0}$ in the three-dimensional space for $\protect\lambda _{q}=0.3$ and $%
1.0 $. (parts (a) and (b), respectively).}}{\Qlb{Fig12_3}}{Fig12_3}{\special%
{language "Scientific Word";type "GRAPHIC";maintain-aspect-ratio
TRUE;display "USEDEF";valid_file "F";width 9.6542cm;height 5.1511cm;depth
0pt;original-width 13.037in;original-height 6.9263in;cropleft "0";croptop
"0.9993";cropright "0.9995";cropbottom "0";filename
'FIG12_3.GIF';file-properties "XNPEU";}} have plotted $\theta _{max},\eta
_{max}$, and $\eta _{vmax}$ versus $q_{0}$ for the different values of the
characteristic size $\lambda _{q}=0.3;1.0$ of the directly heated
three-dimensional region. The same dependencies for two- and one-dimensional
regions and $\lambda _{q}=0.3$ are displayed in \FRAME{ftFU}{3.6054in}{%
1.9493in}{0pt}{\Qcb{The stationary tissue temperature $\protect\theta _{max}$%
, the true and averaged blood flow rates $\protect\eta _{max},\protect\eta
_{vmax}$ at $r=0$ as functions of $q_{0}$ in the two- and one-dimensional
spaces for $\protect\lambda _{q}=0.3$. (parts (a) and (b), respectively).}}{%
\Qlb{Fig12_4}}{Fig12_4}{\special{language "Scientific Word";type
"GRAPHIC";maintain-aspect-ratio TRUE;display "USEDEF";valid_file "F";width
3.6054in;height 1.9493in;depth 0pt;original-width 12.9999in;original-height
6.9998in;cropleft "0";croptop "1";cropright "1";cropbottom "0";filename
'FIG12_4.GIF';file-properties "XNPEU";}}Fig.~\ref{Fig12_4}. As seen from
Fig.~\ref{Fig12_3} and Fig.~\ref{Fig12_4} not only increase in the
characteristic size $\lambda _{q}$ but also decrease in the dimensionality
of the heated region weaken the nonlocality effect of the relation $j_{v}(T)$%
.

\FRAME{ftFU}{8.446cm}{10.5087cm}{0pt}{\Qcb{The maximum tissue temperature $%
\protect\theta _{max}$ (a,c) and the maximum blood flow rates $\protect\eta %
_{max},\protect\eta _{vmax}$ (b,d) vs. the time for the system being
initially at the state $\{\protect\theta =0,\protect\eta =1\}$. (The
three-dimensional space, in the parts (a,b) $\protect\lambda _{q}=3.0,%
\protect\alpha =3.0,$ curves 1,2 correspond to $q_{0}=1.0;5.0$,
respectively; in the parts (c,d) $\protect\lambda _{q}=0.3,$ curves 1,2,3
correspond to $(q_{0}=5.0,\protect\alpha =3.0);(q_{0}=15.0,\protect\alpha
=3.0)$ and $(q_{0}=15.0,\protect\alpha =0.5),$ respectively.}}{\Qlb{Fig12_5}%
}{Fig12_5}{\special{language "Scientific Word";type
"GRAPHIC";maintain-aspect-ratio TRUE;display "USEDEF";valid_file "F";width
8.446cm;height 10.5087cm;depth 0pt;original-width 12.9817in;original-height
16.1573in;cropleft "0";croptop "1.0008";cropright "0.9997";cropbottom
"0";filename 'FIG12_5.GIF';file-properties "XNPEU";}}

Characteristics of the transient process for the system being at the state $%
\{\theta =0,\eta =1\}$ at the initial time are shown in Fig.~\ref{Fig12_5}%
a,b,c,d for $\alpha =0.5;3.0$, and $\lambda _{q}=0.3;3.0$. As it should be
expected, when $\lambda _{q}\gg 1$ and $q_{0}$ is not large enough, such as $%
\theta _{max}\ll 1$ and $\eta \approx 1$, the time increase in the tissue
temperature is monotone (Fig.~\ref{Fig12_5}a). However, if the quantity $%
q_{0}$ is sufficiently large and the tissue temperature $\theta _{max}$
attains values of order one during a time less than the delay time of vessel
response, then the tissue temperature can go out of the vital interval for a
certain time and in the given case the time increase in the tissue
temperature is nonmonotone (Fig.~\ref{Fig12_5}a). Fig.~\ref{Fig12_5}c,d show
that the transient process also possesses, in principle, the same
characteristics when the tissue temperature distribution over living tissue
is substantially nonuniform. As it should be expected, nonmonotony of the
transient process becomes more pronounced as the characteristic time $\tau $
of the vessel response and the size of the region affected directly increase.

\chapter{Heat transfer in living tissue containing a tumor}

\label{ch.12} 
\markright
{ {\sc  \thechapter. Heat transfer in living tissue containing\ldots}
}

In the present Chapter we develop a model for heat transfer in living tissue
containing a tumor whose diameter is substantially less than the
characteristic size of a single microcirculatory bed. Real vascular networks
in normal tissues and in tumors differ in architectonics. However, if a
given idealized vascular network and the real one give rise to the same
pattern of the blood flow rate, then, they will be mathematically equivalent
from the heat transfer standpoint. The latter allows us to make use of the
model developed in the previous Section for a normal microcirculatory bed
after a certain modification of this model.

Typically, for a real tumor response of its vessels to temperature
variations as well as variations in concentration of $O_{2}$, $CO_{2}$, etc.
is depressed. So, under a strong heating in normal tissue the blood flow
rate can increase by tenfold, whereas in tumors it remains practically at
the same level. Under ordinary conditions the blood flow rates in normal
tissue and in a tumor can differ little in magnitude (see e.g.,\cite{54}).

\section{Black spot model of tumor}

\label{s12.1}

Keeping the aforementioned in mind we consider the following model for
living tissue with a tumor \cite{1,ga4}. The vascular network contained in
the tissue domain $Q_{0}$ is assumed to be of the same geometry as one
described in the previous Chapters. The tumor will be treated in terms of a
certain fundamental domain $Q_{t}$ inside which the vessels are
distinguished from other normal vessels by their temperature responses. We
suppose that the temperature response of the normal vessels is ideal whereas
the vessels contained in the tumor domain $Q_{t}$ do not respond to blood
temperature at all. In addition, the resistances of the latter vessels are
assumed to have such values that the distribution of the blood flow rate $%
j_{t}(\mathbf{r}\,)$ over the tumor domain $Q_{t}$ is of a given form when
the tissue temperature coincides with the arterial blood temperature $T_{a}$%
. Besides, in what follows for simplicity we shall study only the case $%
j_{t}(\mathbf{r}\,)\geq j_{0}$. In addition, due to description of heat
transfer in living tissue containing the unit vessel network and in tissue
with the countercurrent vascular network being the same within the
renormalization of the blood flow rate in this Chapter we consider the unit
vessel network only.

Let us introduce a new quantity $T_{s}$ called the seeming temperature that
is specified by the expression 
\begin{equation}
T_{s}(\mathbf{r}\,)=T(\mathbf{r}\,)[1-\Theta _{t}(\mathbf{r}\,)]+\left[
T_{a}+\Delta \frac{j_{t}(\mathbf{r}\,)-j_{0}}{j_{t}(\mathbf{r}\,)}\right]
\Theta _{t}(\mathbf{r}\,)\;,  \label{12.1}
\end{equation}
where $\Theta _{t}(\mathbf{r}\,)$ is the characteristic function of the
tumor domain $Q_{t}$. As seen from (\ref{12.1}) the seeming temperature $%
T_{s}$ coincides with the tissue temperature outside the domain $Q_{t}$ and
is independent of time inside this domain. The seeming temperature $T_{s}$
enables us to specify another system of effective quantities called the
effective temperature of blood. We ascribe to each vein $i$ a value $T_{%
\mathrm{eff}_{i}}^{\ast }$ that, by definition, is related with the seeming
temperature $T_{s}(\mathbf{r}\,)$ by expression similar to (\ref{10.45}),
viz., 
\begin{equation}
J_{i}T_{\mathrm{eff}_{i}}^{\ast }=\int\limits_{Q_{0}}d\mathbf{r}\Theta _{i}(%
\mathbf{r}\,)j(\mathbf{r}\,)T_{s}(\mathbf{r}\,)\;.  \label{12.2}
\end{equation}
In other words, the pattern $\{T_{\mathrm{eff}_{i}}^{\ast }\}$ of the
effective blood temperature, the blood current pattern $\{J_{i}\}$, and the
seeming temperature $T_{s}(\mathbf{r}\,)$ are related with each other in the
same way as do the true blood temperature pattern $\{T_{i}^{\ast }\}$, the
blood current pattern $\{J_{i}\}$, and the true tissue temperature.

Now let us compare the temperature response of the vascular network embedded
in the tissue domain containing the tumor with the response of the
corresponding normal vascular network whose vessels, however, are sensitive
to the effective rather than true blood temperature. First, a small vein
contained inside $Q_{t}$, carries blood away from a certain part of the
domain $Q_{t}$. Therefore, according to (\ref{12.2}), effective temperature
of blood in this vein is totally determined by the seeming temperature $%
T_{s}(r)$ in the tumor domain $Q_{t}$. Thus, for such a small vein sensitive
to the effective blood temperature its resistance will keep a constant value
regardless of variations in the true temperature. Second, it is possible to
ignore the contribution of the tumor domain $Q_{t}$ to the blood current in
a vein whose length $l$ is, at least, twice as large as the mean size $l_{t}$
of the domain $Q_{t}$. Indeed, the total volume of the domain, from which
this large vein carries blood away, is about $V\sim (l/l_{t})^{3}V_{t}$
where $V_{t}$ is the volume of $Q_{t}$. So, for $l\geq 2l_{t}$ we may
suppose that $V\gg V_{t}$. Besides, a self-regulation process has a
considerable effect on heat transfer only when the tissue temperature
approaches near the boundary of the vital temperature interval. In this case
blood flow in vessels being in the normal tissue increases essentially,
whereas blood flow in vessels located in the tumor can vary sufficiently
weaker. The latter also reduces the contribution of the domain $Q_{t}$ to
blood flow in such large veins.

Due to the seeming and true tissue temperature being distinguished in the
tumor domain $Q_{t}$ only, the response of these large veins will be similar
for the two vascular networks. In addition, setting $T(\mathbf{r}\,)=T_{a}$
and substituting (\ref{12.1}) into (\ref{11.12}) we find that the response
of the normal vascular network to the seeming temperature $T_{s}\mid
_{T=T_{a}}$ leads to the required value of the blood flow rate in the domain 
$Q_{t}$.

Therefore, we may describe the response of the vascular network embedded in
the domain containing the tumor in terms of the response of the
corresponding normal vascular network to the effective blood temperature.
The given approach will be called the black spot model. It should be pointed 
\index{black spot model} out that the black spot model can, in principle,
give rise to a wrong result when in the tumor $T-T_{a}\gg T_{+}-T_{a}$.
However, such a high heating of the living tissue is not so interesting. We
also note that the black spot model can describe the vessel response of the
microcirculatory bed when the self-regulation process is not ideal.

Within the framework of the black spot model for the ideal self-regulation
process equation (\ref{11.12}) should be replaced by the equation

\begin{equation}
\tau 
\frac{\partial j}{\partial t}+j\left[ 1-\frac{T_{s}-T_{a}}{\Delta }\right]
=j_{0}  \label{12.3}
\end{equation}
or, in agreement with definition (\ref{12.1}), for $\mathbf{r}\notin Q_{t}$
by the equation 
\begin{equation}
\tau \frac{\partial j}{\partial t}+j\left[ 1-\frac{T-T_{a}}{\Delta }\right]
=j_{0}  \label{12.4}
\end{equation}
and for $\mathbf{r}\in Q_{t}$ by the expression 
\begin{equation}
j=j_{t}(\mathbf{r}\,).  \label{12.5}
\end{equation}
The system of equations (\ref{8.16}), (\ref{9.24}), (\ref{12.3}) or (\ref
{12.4}), (\ref{12.5}) forms the desired description of heat transfer in
living tissue with a tumor.

\section[Mathematical modelling of temperatu\-re distribution in tissue
domain containing a tumour during hyperthermia treatment]{Mathematical
modelling of temperatu\-re \newline
distribution in tissue domain containing \newline
a tumour during hyperthermia treatment}

\label{s12.2}

Within the framework of the given description the characteristics of the
tissue temperature field $T(\mathbf{r},t)$ and the blood flow rate
distribution $j(\mathbf{r},t)$ have been studied numerically for a tumor of
spherical form embedded in the unbound three-dimensional space in which $%
j_{t}=j_{0}$. The model under consideration involves dimensionless equations
corresponding to (\ref{12.4}) and (\ref{12.5}): for $r<\lambda _{t}$

\begin{equation}
\eta =1  \label{12.6}
\end{equation}
and for $r>\lambda _{t}$ 
\begin{equation}
\alpha \frac{\partial \eta }{\partial t}+\eta \lbrack 1-\theta ]=1,
\label{12.7}
\end{equation}
where $\lambda _{t}$ (in units of $[\kappa _{\mathrm{eff}}/(c_{t}\rho
_{t}j_{0})]^{1/2}$) is the radius of the tumor. For the sake of simplicity,
we have considered uniform stationary heating, i. e., in expression (\ref
{11.16}) set $\lambda _{q}=\infty $.

Since, the tissue temperature can vary substantially only on spatial scales
larger than $\sqrt{(D/j_{v})}$ ( $\eta _{v}^{-1/2}$ in the dimensionless
units) in heating the tumor whose radius $l_{t}\gg \sqrt{(D/j_{0})}$ (i. e. $%
\lambda _{t}\gg 1$) the temperature field will become essentially nonuniform
in the vicinity of the tumor already when the tissue temperature attains the
middle points of the vital temperature interval. If $l_{t}\ll \sqrt{(D/j_{0})%
}$ (i. e. $\lambda _{t}\ll 1$), then, for the nonuniform temperature
distribution to occur in the vicinity of such a tumor much stronger heating
of the tissue is required. Indeed in this case the blood flow rate $j$ in
the normal tissue must first increase substantially and become larger than $%
(D/l_{t}^{2})$. Therefore, we have analyzed the dependence of the tissue
temperature field on the heating power for a small $(\lambda _{t}<1)$ and $%
(\lambda _{t}>1)$ large tumor individually.

Fig.~\ref{Fig25} shows obtained distributions of the dimensionless tissue
temperature $\theta (\mathbf{r}\,)$, the true and averaged blood flow rates $%
\eta (\mathbf{r}\,),\eta _{v}(\mathbf{r}\,)$ for the small tumor whose
radius $\lambda _{t}=0.3$. The similar results for the sufficiently large
tumor of radius $\lambda _{t}=3$ are presented in Fig.~\ref{Fig26}. As it
should be expected, under ideal thermoregulation the tissue temperature can
go out of the vital interval in the tumor only, whereas in the normal tissue
the increase in the blood flow rate keeps tissue temperature within this
interval. According to Fig.~\ref{Fig26} the temperature in the large tumor
becomes already noticeably different from the normal tissue temperature when
the latter attains the value $\theta \sim 0.5$ and in the normal tissue the
blood flow rate increases twice. For the small tumor (see Fig.~\ref{Fig25})
the similar difference takes place as the normal tissue temperature comes
near to the boundary of the vital interval $(\theta _{+}=1)$ and in the
normal tissue the blood flow rate increases by tenfold.

We note that once the temperature in the normal tissue has practically
attained the upper boundary of the vital interval $T_{+}=T_{a}+\Delta $ ($%
\theta _{+}=1$) and the temperature in the tumor has become noticeably
higher than $T_{+}$, the stationary temperature distribution in the tumor
domain $Q_{t}$ can be approximately described by the formal equation 
\begin{equation}
\kappa _{\mathrm{eff}}\mathbf{\nabla }^{2}T-c_{t}\rho
_{t}j_{t}(r)(T-T_{a})+q=0  \label{12.8}
\end{equation}
under the boundary condition 
\begin{equation}
T\mid _{\mathbf{r}\in \partial Q_{t}}=T_{+}\;,  \label{12.9}
\end{equation}

\FRAME{ftbpFU}{5.3773cm}{7.0094cm}{0pt}{\Qcb{The distributions of the tissue
temperature $\protect\theta $ (a), the true (thick line) and averaged
(dashed line) blood flow rates $\protect\eta ,\protect\eta _{v}$ (b) in the
vicinity of the tumour whose radius $\protect\lambda _{t}=0.3$ (curves 1,2
correspond to different values of the heat generation rate: $q_{0}=5.0;15.0;$
respectively).}}{\Qlb{Fig25}}{Fig25}{\special{language "Scientific
Word";type "GRAPHIC";display "USEDEF";valid_file "F";width 5.3773cm;height
7.0094cm;depth 0pt;original-width 5.0185in;original-height 8.3333in;cropleft
"0";croptop "1.0006";cropright "1";cropbottom "0";filename
'FIG25.GIF';file-properties "XNPEU";}}

where $\partial Q_{t}$ is the boundary of the tumor domain $Q_{t}$. It also
should be pointed out that equation (\ref{12.8}) contains the true blood
flow rate $j_{t}(\mathbf{r}\,)$ rather than the averaged one $j_{v}(\mathbf{r%
}\,)$.

For a large tumor (whose radius $\lambda _{t}\gg 1$) such replacement is
obviously justified. For a small tumor, $(\lambda _{t}\ll 1)$ it also does
not lead to wrong results. To validate the latter statement with regard to a
small tumor, we analyse the one-dimensional solution of equations (\ref
{11.13}) and (\ref{11.14}) when $\eta =1$ for $x>0$ and $\eta =\eta _{n}$
for $x<0$ where $\eta _{n}\gg 1$ is a constant. By this example we can
demonstrate qualitatively\ the main properties of the temperature
distribution in the vicinity of an interface separating living tissue
regions where the blood flow rates differ significantly in magnitude.
Solving equation (\ref{11.14}) for the given $\eta (x)$ dependence directly
we found that in the region $(\eta _{n}L)^{-1/2}\ll x\ll L^{-1/2}$ the
averaged blood flow rate $\eta _{v}(x)\approx 2(Lx^{2})^{-1}$ and $\eta
_{v}(x)\approx 1$ for $x\geq L^{-1/2}$. Thus, on one hand, for $x\gg (\eta
_{n}L)^{-1/2}$ the averaged blood flow rate practically differs from the
true one in the region $x\mathbf{\leqslant }L^{-1/2}$. On the other hand,
according to (\ref{11.13}) the averaged blood flow rate of value $\eta _{v}$
has a considerable effect on the temperature distribution solely on scale
larger than $L_{\eta }\sim \eta _{v}^{-1/2}$. So, for $\eta _{v}(x)\sim
2(Lx^{2})^{-1}$ the ratio $x/L_{\eta }\sim (2/L)^{1/2}$ can be regarded as a
small parameter when $L\gg 1$ and, thereby, at lower order in $1/L$ we can
ignore the term $\eta _{v}\theta $ in equation (\ref{11.13}) for $0<x\mathbf{%
\leqslant }L^{-1/2}$. Therefore, when in the normal tissue the blood flow
rate $j\gg (D/l_{t}^{2})$ in order to describe temperature distribution in a
small tumor (as well as in a large tumor) we mayreplace in equation (\ref
{11.13}) the averaged blood flow rate by the true one.

\FRAME{ftbpFU}{13.2962cm}{9.3049cm}{0pt}{\Qcb{ The distributions of the
tissue temperature $\protect\theta $ (a), and the true and averaged (dashed
line) blood flow rates $\protect\eta ,\protect\eta _{v}$ (b) in the vicinity
of the tumor whose radius $\protect\lambda _{t}=3$ (curves 1,2 correspond to
different values of the heat generation rate: $q_{0}=1.0;15.0,\;$%
respectively). Fig.a,b show the distribution along the radius, and Fig.c,d
represent the distribution in the plane crossing the tumor domain through
the center.}}{\Qlb{Fig26}}{Fig26}{\special{language "Scientific Word";type
"GRAPHIC";display "PICT";valid_file "F";width 13.2962cm;height
9.3049cm;depth 0pt;original-width 9.2596in;original-height 3.9902in;cropleft
"0";croptop "0.9994";cropright "1.0003";cropbottom "0";filename
'FIG26.GIF';file-properties "XNPEU";}}

\section{Two boundary model%
\index{two boundary model} for freezing of living tissue during
cry\-o\-su\-rgery treatment}

\label{s12.3}

Mathematical analysis of temperature distribution in living tissue during
freezing is useful in the study and optimization of cryosurgical treatment.
For the last few years a number of different approaches to describing a heat
transfer process in living tissue during freezing have been proposed (for a
review see \cite{3,8,15,17,25,51,53}). Within the framework of these
approaches the obtained bioheat equation is of the form (\ref{1.1})
Propagation of the freezing front $\Gamma $ is conventionally described in
terms of the free boundary problem%
\index{free boundary problem} of the Stefan-type:

\begin{equation}
v_{n}\rho _{t}\mathcal{L}=-(\kappa \mathbf{\nabla }_{n}T)\mid _{\Gamma
_{+}}+(\kappa \mathbf{\nabla }_{n}T)\mid _{\Gamma _{-}},  \label{12.10}
\end{equation}
\begin{equation}
T\mid _{\Gamma _{+}}=T\mid _{\Gamma _{-}}=T_{f},  \label{12.11}
\end{equation}
where $\mathcal{L}$ is the latent heat of fusion, $\Gamma _{+}$ and $\Gamma
_{-}$ denote the boundaries of the freezing front on the living and frozen
sides of the tissue, respectively, and $T_{f}$ is the freezing temperature.

A more accurate description of a heat transfer process in living tissue is
obtained if one takes into account the fact that living tissue form an
active, highly heterogeneous medium. Also, when the size of the frozen
region of the tissue is small in comparison with the characteristic length
of the blood vessels that directly control heat exchange between the
cellular tissue and blood, the heterogeneity of living tissue has a
substantial effect on the heat transfer process. Therefore, equation (\ref
{1.1}) which models the blood flow rate in terms of a continuous field $j(%
\mathbf{r}\,)$ has to be modified \cite{3}.

The response of the living tissue to the effects of strong cooling or strong
heating can cause the blood flow rate to vary by an order of magnitude. In
general, if the temperature distribution in the tissue is substantially
nonuniform as, for example, in cryosurgery, then, the temperature dependence
of the blood flow rate is nonlocal and the blood flow rate at a given point
depends on certain characteristics of the temperature distribution rate
other than the tissue temperature at the point only.

In this Section on the developed background we shall formulate a model for
the thermal response in living tissue during the freezing process that will
include the effects of the aforementioned factors \cite{3}.

As it has been shown in Section~\ref{s12.1} heat transfer in living tissue
containing regions where tissue is anomalous in properties can be treated,
at least qualitatively, in terms of heat transfer in living tissue where the
vascular network responds to an effective (seeming) tissue temperature
rather than the real one. Keeping in mind the black spot model we shall
describe living tissue freezing as follows. The frozen region is regarded to
be a certain fundamental domain $Q_{f}$ where resistances of the vessels are
infinitely great, thereby, the blood flow rate in this region is equal to
zero; $j_{f}(\mathbf{r}\,)=0$ if $\mathbf{r}\in Q_{f}$. The frozen domain $%
Q_{f}$ is assumed to be small in comparison with the microcirculatory bed
domain $Q_{0}$. It is natural to treat the freezing temperature $T_{f}$ as
the lower boundary of the vital temperature interval $[T_{-},T_{+}]$, i.e. $%
T_{f}=T_{-}$. Any model for the freezing process should be able to describe
propagation of the freezing front $\Gamma $ where the temperature $T_{f}$ is
attained. Therefore the theory of ideal thermoregulation developed in
Chapter~\ref{ch.11} in this case cannot be directly used because it leads to
infinitely large value of the blood flow rate at $T=T_{f}$.

The matter is that in this model the flow resistance $R_{i}$ of a vessel
becomes zero at $T=T_{f}$. For real vascular network vessels reach the limit
of expanding as temperature decreases and the vessel resistance attains a
certain minimum. This causes the blood flow rate to attain certain large but
finite values in the living tissue domain where the temperature $T\gg T_{f}$%
. Such behavior of the temperature vessel response will be described in
terms of the developed theory of thermoregulation where the blood
temperature dependence $R_{i}(T)=R_{n}^{0}\varphi \left( 
\frac{T^{\ast }-T_{a}}{\Delta }\right) $ of the vessel resistance is shown
in Fig.~\ref{Fig27}a. In the region $T_{vr}<T<T_{a}$ as for the model of
ideal thermoregulation the resistance $R_{i}$ of a vessel $i$ decreases
linearly with the blood temperature $T_{f}^{\ast }$ until its value attains
the temperature $T_{vr}$ near the lower boundary $T_{f}$ of the vital
interval. In the region $T_{f}<T^{\ast }<T_{vr}$ the vessel does not respond
to temperature variations and its flow resistance $R_{i}$ is constant and
equal to $R_{min}=R_{n}^{0}\left( \frac{T_{vr}-T_{f}}{\Delta }\right) $. In
the frozen tissue region, $T<T_{f}$, the vessel resistance formally becomes
infinitely large.

\FRAME{ftbpFU}{8.6283cm}{4.7271cm}{0pt}{\Qcb{Two boundary model for tissue
freezing processes: a - the vessel resistance as a function of the blood
temperature, b - characteristic region of living tissue.}}{\Qlb{Fig27}}{Fig27%
}{\special{language "Scientific Word";type "GRAPHIC";maintain-aspect-ratio
TRUE;display "USEDEF";valid_file "F";width 8.6283cm;height 4.7271cm;depth
0pt;original-width 10.7782in;original-height 5.879in;cropleft "0";croptop
"1.0013";cropright "1.0009";cropbottom "0";filename
'FIG27.GIF';file-properties "XNPEU";}}\bigskip

The black spot model developed in Section~\ref{s12.1} enables us to describe
freezing process of living tissue with vessels responding to temperature
variations as it has been stated above within in the framework of the
following model.

In the frozen region $Q_f$ the tissue temperature evolves according to the
conventional heat conduction equation for solids:

\begin{equation}
c_{t}\rho _{t}\frac{\partial T}{\partial t}=\kappa \mathbf{\nabla }%
^{2}T-q_{f},  \label{12.12}
\end{equation}
where $\kappa $ is the intrinsic tissue conductivity and $q_{f}$ is the rate
of tissue cooling.

Inside the unfrozen living tissue its temperature is governed by the equation

\begin{equation}
c_{t}\rho _{t}\frac{\partial T}{\partial t}=\kappa _{\mathrm{eff}}\mathbf{%
\nabla }^{2}T-j_{v}c_{t}\rho _{t}(T-T_{a}).  \label{12.13}
\end{equation}
Here the effective heat conductivity $\kappa _{\mathrm{eff}}$ is related
with the intrinsic conductivity $\kappa $ by the expression $\kappa _{%
\mathrm{eff}}=\kappa \lbrack 1+F_{v}(G)+F_{c}(G)]$ (see (\ref{7.15}) and, in
addition, we have ignored the metabolic heat generation rate and the
difference between the densities and heat capacities of the cellular tissue
and blood. Following (\ref{12.10}), (\ref{12.11}) we assume that propagation
of the freezing front $\Gamma =\partial Q_{f}$ is controlled, in
mathematical terms, by conditions

\begin{equation}
v_{n}\rho _{t}\mathcal{L}=-\left. (\kappa _{\mathrm{eff}}\mathbf{\nabla }%
_{n}T)\right| _{\Gamma _{+}}+\left. (\kappa \mathbf{\nabla }_{n}T)\right|
_{\Gamma _{-}}  \label{12.14}
\end{equation}
and 
\begin{equation}
T|_{\Gamma _{+}}=T|_{\Gamma _{-}}=T_{f}.  \label{12.15}
\end{equation}

Inside the unfrozen living tissue there are two regions that are different
in thermoregulation properties. In the first one, $Q_{vr}$ (Fig.~\ref{Fig27}%
), adjacent to the frozen domain $Q_f$ the tissue temperature varies from $%
T_f$ to $T_{vr}$, and the blood flow rate is constant and equal to

\begin{equation}
j=j_{0}\frac{T_{a}-T_{+}}{T_{vr}-T_{f}}.  \label{12.16}
\end{equation}
In the second region, where $T<T_{vr}$ the blood flow rate is related to the
local value of the tissue temperature by the equation

\begin{equation}
\tau \frac{\partial j}{\partial t}+j\frac{T-T_{f}}{T_{a}-T_{f}}=j_{0}.
\label{12.17}
\end{equation}
At the interface $\Gamma _{vr}$ of these two domains

\begin{equation}
\left. T\right| _{\Gamma _{vr}^{+}}=\left. T\right| _{\Gamma _{vr}^{-}}
\label{12.18}
\end{equation}
and 
\begin{equation}
\mathbf{\nabla }_{n}\left. T\right| _{\Gamma _{vr}^{+}}=\mathbf{\nabla }%
_{n}\left. T\right| _{\Gamma _{vr}^{-}}  \label{12.19}
\end{equation}
due to the interface $\Gamma _{vr}$ containing no heat sink. The averaged
and true blood flow rates are related, as before, by the equation

\begin{equation}
j_{v}-\frac{\kappa }{c_{t}\rho _{t}L}\mathbf{\nabla }^{2}\ln j_{v}=j.
\label{12.20}
\end{equation}
At the interface $\Gamma _{vr}$ the averaged blood flow rate as well as its
spatial derivatives is continuous, and at the freezing front $\Gamma $,
according to (\ref{9.28})

\begin{equation}
\left. \mathbf{\nabla }_{n}j_{v}\right| _{\Gamma ^{+}}=0.  \label{12.21}
\end{equation}

This model allows \ not only for the phenomena caused by phase transition
during freezing living tissue, but also characteristics of living tissue
response to substantial cooling as well as nonlocality in heat exchange
between the cellular tissue and blood. It should be noted that in spite of
this two boundary model containing the collection of equations (\ref{12.12})
- (\ref{12.21}) within the framework of this model the temperature
distribution can be analyzed not only numerically but also by analytical
methods. The variational principles developed for the Stefan type problems,
(see., e.g., \cite{23,24,ga0,ga01,ga6}) allow one to reduce the system of
equations (\ref{12.12}) - (\ref{12.21}) to interface dynamics of the two
regions in living tissue.

It should be noted that not only cryosurgery problem in living tissue can be
described by free boundary problem. The similar situation we can meet in the
description of the dynamics of local thermal coagulation leading to the
necrosis growth limited by heat diffusion into the surrounding live tissue.
Dealing with this problem we keep in mind the following process. Absorption
of laser light delivered into a small internal region of living tissue
causes the temperature to attain such high values (about $70^{o}C$) that
lead to immediate coagulation in this region. Heat diffusion into the
surrounding live tissue causes its further thermal coagulation, giving rise
to the growth of the necrosis domain. Different mathematical models of the
description of the dynamics of local thermal coagulation in live tissue has
been considered in \cite{lu1}-\cite{lu6} and is outside the problems
considered in this manuscript.

\bigskip

\bigskip

\clearpage

\part{Fluctuations and small scale nonuniformities of the tissue temperature}

\markboth{
{\sc \thepart.{ } Fluctuations and small scale nonuniformities\ldots}}{}

In the previous Chapters we have actually reduced the bioheat transfer%
\index{bioheat transfer} problem to description of heat propagation in a
certain homogeneous continuum with complex properties. However, as for any
heterogeneous medium, the temperature distribution in living tissue can
exhibit spatial nonuniformities and spatiotemporal fluctuations%
\index{spatiotemporal fluctuations} due to the discreteness of vessel
arrangement and random time variations of vessel characteristics leading to
fluctuations in a blood flow. In this Part we study mean characteristics of
spatiotemporal fluctuations and nonuniformities in the tissue temperature
treated as random fields \cite{42}. Such parameters as the mean amplitude
and the correlation length of the temperature nonuniformities can be used in
interpretation of the available experimental data.

\chapter{Characteristics of spatial-temporal fluctuations of the tissue
temperature}

\label{ch.13} 
\markright
{ {\sc  \thechapter. Characteristics of spatial-temporal\ldots}
}

\section{Fluctuations in the tissue temperature due to time variations of
the blood flow rate}

\label{s13.1}

In living tissue blood flow in vessels of a vascular network forms branched
paths of fast heat transfer as well as fast transport of $O_{2}$ and some
other components. Owing to this, heat and mass transfer in living tissue
possesses specific properties, and the blood flow rate treated in terms of a
continuous field $j(\mathbf{r},t)$ is one of the fundamental characteristics
of these processes.

Typically, blood flow in a vessel of length $l$ directly controls the mean
blood flow rate in a tissue domain $Q$ whose size is about $l$, whereas
smaller vessels are responsible for blood flow redistribution over different
parts of this domain. Therefore, fluctuations in vessel resistance to blood
flow in it caused, for example, by time variations in its radius are bound
to give rise to spatiotemporal fluctuations in the blood flow rate $j(%
\mathbf{r},t)$ in the tissue domain $Q$ which are correlated on spatial
scales of order $l$ and on temporal scales determined by the vessel
characteristics. These fluctuations in $j(\mathbf{r},t)$, in their turn,
cause spatiotemporal fluctuations in the tissue temperature. Since, the
vascular network involves vessels of different lengths, both the tissue
temperature and distribution of these components can exhibit fluctuations
characterized by a wide range of spatial and temporal scales.

The purpose of this present Chapter is to investigate the characteristics of
these fluctuations and their dependence on the vascular network
architectonics. Fluctuations in distribution of $O_2, CO_2$ etc. are
expected to have the same properties.

For the sake of simplicity we assume that the heat capacities as well as the
thermal conductivities of the cellular tissue and blood are the same and
independent of temperature, the vascular network involves unit vessels only $%
(n_{cc}=0)$ and consider temperature fluctuations characterized by spatial
scales much larger than the length $l_{n_{t}}$ of vessels directly
controlling heat exchange between blood and the cellular tissue. In this
case the bioheat equation obtained in Chapters ~\ref{ch.6},~\ref{ch.7} can
be rewritten in the form:

\begin{equation}
\frac{\partial }{\partial t}T=D_{\mathrm{eff}}\mathbf{\nabla }%
^{2}T-j(T-T_{a})+q_{h},  \label{13.1}
\end{equation}
where $j_{v}\approx j$ under the adopted assumptions. The heat generation
rate $q_{h}$ that is considered to be constant leads to a uniform
distribution of the tissue temperature with the mean value $%
T_{0}=T_{a}+q_{h}/j_{0}$ where $j_{0}$ is the mean value of the blood flow
rate.

We shall account for temperature fluctuations $\delta T$ caused only by
inherent fluctuations in vessel resistances to blood flow. Therefore,
linearizing equation (\ref{13.1}) with respect to $\delta T$ near $T_{0}$ we
get

\begin{equation}
\frac{\partial }{\partial t}\delta T=D_{\mathrm{eff}}\mathbf{\nabla }%
^{2}\delta T-\left[ j_{0}+\left. \frac{\partial j}{\partial T}\right|
_{T=T_{0}}(T_{0}-T_{a})\right] \delta T-\delta j(T_{0}-T_{a}),  \label{13.2}
\end{equation}
where the derivative $\partial j/\partial T$ is associated with the
temperature dependence of the blood flow rate and $\delta j$ is the blood
flow rate fluctuations inherent to living tissue. In the general case the
derivative $\partial j/\partial T$ is an operator. However, first, when the
difference $(T_{0}-T_{a})$ is substantially less than the width of the vital
temperature interval of living tissue, this term is likely to be small
enough in comparison with $j_{0}$. Second, when the $j(T)$ dependence is a
local function it leads to the renormalization of $j$ only. Therefore, the
term $\frac{\partial j}{\partial T}(T_{0}-T_{a})$ in (\ref{13.2}) will be
ignored.

To analyze the characteristics of temperature fluctuations, first, we shall
find the correlation function

\begin{equation}
G_{\mathbf{r},t}=\left\langle \left\langle \delta T_{\mathbf{r}^{\prime }+%
\mathbf{r},t^{\prime }+t}\delta T_{\mathbf{r}^{\prime },t^{\prime
}}\right\rangle \right\rangle ,  \label{13.3}
\end{equation}
where symbol $\ll ...\gg $ denotes averaging over both the time $t^{\prime }$
and the tissue points $\mathbf{r}^{\prime }$ under the conditions $t=$%
\textrm{constant} and $\mid \mathbf{r}\mid =$\textrm{constant}.

Let us introduce the correlation function%
\index{correlation function} of the blood flow rate fluctuations

\begin{equation}
\Omega _{\mathbf{r},t}=\left\langle \left\langle \delta j_{\mathbf{r}%
^{\prime }+\mathbf{r},t^{\prime }+t}\delta j_{\mathbf{r}^{\prime },t^{\prime
}}\right\rangle \right\rangle .  \label{13.4}
\end{equation}

Then taking into account adopted assumptions from (\ref{13.2}) we obtain the
following relationship between the Fourier transforms of the above two
correlation functions (\ref{13.3}) and (\ref{13.4}) with respect to both the
time $t$ and the coordinates $\mathbf{r}$:

\begin{equation}
G(\mathbf{k},w)=(T_{0}-T_{a})^{2}%
\frac{\Omega (\mathbf{k},w)}{w^{2}+(D_{\mathrm{eff}}k^{2}+j_{0})^{2}}.
\label{13.5}
\end{equation}
Here $w$ and $\mathbf{k}$ are the variables conjugate to $t$ and $\mathbf{r}$%
, respectively. It should be pointed out that when averaging the product $%
(\delta j\delta j)$ in (\ref{13.4}) over the time $t^{\prime }$ only we get
the function

\begin{equation}
\Omega _{\mathbf{r}^{\prime }\mathbf{r},t}=\left\langle \delta j_{\mathbf{r}%
^{\prime },t^{\prime }+t}\delta j_{\mathbf{r}^{\prime },t}\right\rangle ,
\label{13.6}
\end{equation}
which depends on both the variables $\mathbf{r}^{\prime },\mathbf{r}$. This
nonuniformity will be discussed below.

Since, in the case under consideration the mean tissue temperature is
constant in the domain $Q_{0}$ the blood temperature in all the veins is
practically the same. Therefore, implying the mean values all veins and
arteries of one level, for example, level $n$ must be characterized by the
same flow resistance (see (\ref{3.20}))

\begin{equation}
R(n)=R_{0}\varphi \left( \frac{T_{0}-T_{a}}{\Delta }\right) 2^{3n}\rho (n).
\label{13.7}
\end{equation}
Within the framework of the adopted model we may confine our consideration
to the venous bed only the pressure drop $P$ across which is assumed to be
constant.

Random time variations of the blood flow are caused by fluctuations of the
vessel parameters, in particular, vessel radius, biochemical blood
composition, etc. All these factors eventually give rise to random time
variations of the vessel resistances. So, in the present Section time
variations of blood flow rate are considered to occur because of
fluctuations in the vessel resistances. In addition, for the sake of
simplicity, fluctuations in the flow resistance of a vessel are
characterized by a single correlation time.

For a given vein, for example vein $i$, fluctuations $\delta R_{i}(t)$ in
their resistance are described by a correlation time%
\index{correlation time} $1/w(n)$ which can depend on the vein level number $%
n$. In other words, we represent the correlation function of these
fluctuations in the form

\begin{equation}
\left\langle \delta R_{i}(t+t^{\prime })\delta R_{i}(t^{\prime
})\right\rangle =R^{2}(n_{i})\epsilon \Delta (n_{i})\exp (-w(n_{i})\left|
t\right| ),  \label{13.8}
\end{equation}
where $\epsilon $ is a small constant $(\epsilon \ll 1)$, the function $%
\Delta (n)$ accounts for specific details of the correlation function
dependence on $n$ and $\Delta (0)=1$. In particular, $\Delta (n)$ is a
smooth function of $n$ providing that the resistance $R(n)$ is a power
function of $a$, and the ratio $\left\langle (\delta a_{n})^{2}\right\rangle
/a_{n}^{2}$ depends smoothly on $n$. Therefore, in the following for the
sake of simplicity we shall regard both $\Delta (n)$ and $w(n)$ as smooth
functions of $n$. For different vessels fluctuations in their resistances
are assumed to be uncorrelated.

\section{Correlation function of the blood flow rate fluctuations}

\label{s13.2}

In order to find blood flow redistribution over the vascular network caused
by time variations of the vessels resistances and, thereby, to obtain time
variations in the blood flow rate distribution over living tissue, we should
solve the system of the Kirchhoff equations (\ref{3.16}), (\ref{3.17}) to
the first approximation in $\varepsilon $ due to $\varepsilon \ll 1$.

At lower order in $\varepsilon $, i.e. when fluctuations in the resistances $%
\delta R_{i}(t)$ are not accounted for, the solution of the equations (\ref
{3.16}), (\ref{3.17}) describes the uniform distribution of blood flow over
the vascular network and is of the form

\begin{equation}
J_{i}=J_{0}(n_{i})=2^{-3n_{i}}J_{0},  \label{13.9}
\end{equation}
where $n_{i}$ is the level number of the vein $i$ and $J_{0}$ is the total
blood current in the tree stem. Then, actually following the procedure
proposed in Section~\ref{s10.2}. Due to $\varepsilon \ll 1$ to the first
order in $\delta R_{i}(t)$ equation (\ref{3.17}) can be replaced by the
equation

\begin{equation}
J_{i}R(n_{i})=\Delta P_{i}-J_{0}(n_{i})\delta R_{i}(t).  \label{13.10}
\end{equation}
Equations (\ref{13.9}) and (\ref{13.10}) may be regarded as the system of
the Kirchhoff equations describing blood flow distribution over a certain
vascular network of the same architectonics where, however, the vessel
resistances $R(n_{i})$ are constant values and there are some random
effective pressure sources

\begin{equation}
\varepsilon _i = -J_0 (n_i) \delta R_i (t)  \label{13.11}
\end{equation}
associated with these vessels. Being pairwise independent and random
quantities these effective pressure sources $\{\varepsilon _i \}$ cause
fluctuations in the blood currents. As shown in Section~\ref{s10.2} and
Section~\ref{s10.3} linearity of equations (\ref{13.10}), (\ref{13.11}) with
respect to the blood currents allows us to represent the solution of these
equations in terms of

\begin{equation}
J_{i}=J_{0}(n_{i})+\sum_{i^{\prime }}\Lambda _{ii^{\prime }}\varepsilon
_{i^{\prime }},  \label{13.12}
\end{equation}
where the sum runs over all the arteries, and the elements of the matrix$%
\;\left| \left| \Lambda _{ii^{\prime }}\right| \right| $ are specified in
the following way (see Section~\ref{s10.3} expressions (\ref{10.41}) - (\ref
{10.43})). Let us denote by $\{ii^{\prime }\}_{+}$ such a pair of veins $i$
and $i^{\prime }$ that can be joined by a path of constant direction on the
vascular network. This path may be directed either from higher to lower
levels or vice versa (the vein pair $\{i_{1},i_{2}\}_{+}$ in Fig.~\ref{Fig28}%
a). Then, for a vein pair $\{ii^{\prime }\}_{+}$ where $n_{i}<n_{i^{\prime
}} $

\FRAME{ftbpFU}{12.1034cm}{17.9464cm}{0pt}{\Qcb{Schematic representation of
the vein tree.}}{\Qlb{Fig28}}{Fig28}{\special{language "Scientific
Word";type "GRAPHIC";maintain-aspect-ratio TRUE;display "USEDEF";valid_file
"F";width 12.1034cm;height 17.9464cm;depth 0pt;original-width
13.2775in;original-height 19.7497in;cropleft "0";croptop "1.0008";cropright
"1";cropbottom "0";filename 'FIG28.GIF';file-properties "XNPEU";}}

\begin{equation}
\Lambda _{ii^{\prime }}=\frac{1}{R_{\ast }Z(n_{i})}2^{-3n_{i}^{\prime }}
\label{13.13}
\end{equation}
and for $n_{i}>n_{i}^{\prime }$ 
\begin{equation}
\Lambda _{ii^{\prime }}=\frac{1}{R_{\ast }Z(n_{i}^{\prime })}2^{-3n_{i}}.
\label{13.14}
\end{equation}
Here $R_{\ast }=R_{0}\varphi \left( \frac{T^{\ast }-T_{a}}{\Delta }\right) $%
, $Z(n)$ is the function defined by formula (\ref{10.14}). The other pairs
of veins $\{ii^{\prime }\}_{-}$ are characterized by paths with variable
directions, i.e. for a pair $\{ii^{\prime }\}_{-}$ the veins $i,i^{\prime }$
can be joined by a path whose direction becomes opposite at a certain
branching point $B_{ii^{\prime }}$ (the vein pair $\{i_{2},i_{3}\}$ in Fig.~%
\ref{Fig28}a). Let us ascribe to a branching point $B$ the level $n$ of a
vein, that goes in it. Then,

\begin{equation}
\Lambda _{ii^{\prime }}=-\frac{\rho (n_{B_{ii^{\prime }}})}{7R_{\ast
}Z^{2}(n_{B_{ii^{\prime }}})}2^{3(n_{B_{ii^{\prime }}}-n_{i}-n_{j})}.
\label{13.15}
\end{equation}
It should be pointed out that because of $Z(n)$ being a smooth function the
ratio $\rho (n)/Z(n)$ can be treated as a small parameter.

According to (\ref{3.18}) the relationship between the blood current pattern 
$\{J_{i}\}$ and the blood flow rate $j(\mathbf{r}\,)$ is determined by the
formula $J_{i_{r}}=V_{N}j(\mathbf{r}\,)$ where $i_{r}$ is the last level
vein contained among with the point $\mathbf{r}$ in the same fundamental
domain $Q_{Nr}$ of volume $V_{N}$. This formula and expression (\ref{13.12})
enable us to represent the correlation function (\ref{13.6}) in the form:

\begin{equation}
\Omega _{r,r^{\prime },t}=\frac{1}{V_{N}^{2}}\sum_{i}\sum_{i^{\prime
}}\Lambda _{i_{r}i}\Lambda _{i_{r^{\prime }}i^{\prime }}\left\langle
\varepsilon _{i}\varepsilon _{i^{\prime }}\right\rangle .  \label{13.16}
\end{equation}
Then, substituting formula (\ref{13.11}) into (\ref{13.16}) and taking into
account expression (\ref{13.8}) and pair wise independence of the vessel
resistance fluctuations of each other we can rewrite (\ref{13.16}) as

\begin{equation}
\Omega _{\mathbf{r},\mathbf{r}^{\prime },t}=\frac{\epsilon }{V_{N}^{2}}%
\sum_{i}\Lambda _{i_{\mathbf{r}}i}\Lambda _{i_{\mathbf{r}^{\prime
}}i}J_{0}^{2}(n_{i})R^{2}(n_{i})\exp \{-w(n_{i})t\}\Delta (n_{i}).
\label{13.17}
\end{equation}

In order to calculate sum (\ref{13.17}) we divide all the veins into four
groups and consider their contribution into sum (\ref{13.17}) individually.
The first group involves veins that form the first type pairs with both the
venules, i.e. all the veins $\{i\}$ for which $\{i_{r},i\}=\{i_{r},i\}_{+}$
and $\{i_{r},i\}=\{i_{r^{\prime }},i\}_{+}$. These veins make up a path on
the vein tree that originates at the branching point $B_{rr^{\prime }}$
(Fig.~\ref{Fig28}a) where blood streams in the venules $i_{r}$ and $%
i_{r^{\prime }}$ merge into one stream and terminates at the tree stem. In
Fig.~\ref{Fig28}b the given path is displayed by the solid line. For each of
this group, for example, vein $i$ the value $\Lambda _{i_{r}i}$ is
determined by expression (\ref{13.14}), i.e.

\begin{equation}
\Lambda _{i_{r}i}=2^{-3N}\frac{1}{R_{\ast }}\frac{1}{Z(n_{i})}  \label{13.18}
\end{equation}
because $n_{i_{r}}=N$. The same expression gives the value $\Lambda
_{i_{r^{\prime }}i}$. Then, the contribution $\Omega _{r,r^{\prime
},t}^{(1)} $ to expression (\ref{13.17}) is of the form

\begin{equation}
\Omega _{r,r^{\prime },t}^{(1)}=\frac{\epsilon }{V_{N}^{2}}%
2^{-6N}J_{0}^{2}\sum_{n_{i}=0}^{n_{rr^{\prime }}-1}\left[ \frac{\rho (n_{i})%
}{Z(n_{i})}\right] ^{2}\Delta (n_{i})\exp (-w(n_{i})\left| t\right| ),
\label{13.19}
\end{equation}
where $n_{rr^{\prime }}$ is the level number of the branching point $%
B_{rr^{\prime }}$ and we have taken into account the identity

\begin{equation}
\frac{1}{R_{\ast }^{2}}J_{0}^{2}(n_{i})R^{2}(n_{i})=J_{0}^{2}[\rho
(n_{i})]^{2}.  \label{13.20}
\end{equation}

The second group consists of the veins that belong to one of the two paths
on the vein tree which join the venule $i_{r}$ and venule $i_{r}$ with the
branching point $B_{rr^{\prime }}$. All these veins $\{i\}$ form either the
pairs $\{i_{r},i\}_{+}$ and $\{i_{r^{\prime }},iB_{rr^{\prime }}\}_{-}$ or
the pairs $\{i_{r^{\prime }},iB_{rr^{\prime }}\}_{-}$ and $\{i_{r^{\prime
}},i\}_{+}$, which is illustrated in Fig.~\ref{Fig28}c. For this part of the
vein tree formula (\ref{13.18}) gives the values $\Lambda _{i_{r}i}$ and $%
\Lambda _{i_{r^{\prime }}i}$ for $\{i_{r},i\}_{+}$ and $\{i_{r^{\prime
}},i\}_{+}$ and formula (\ref{13.18}) becomes

\begin{equation}
\Lambda _{i_{r}i}=-\frac{1}{7R_{\ast }}2^{-3N}\frac{\rho (n_{rr^{\prime }})}{%
[Z(n_{rr^{\prime }})]^{2}}2^{-3(n_{i}-n_{rr^{\prime }})}  \label{13.21}
\end{equation}
or $\{i_{r},i,B_{rr^{\prime }}\}_{-}$ and the same expression for $%
\{i_{r^{\prime }},i,B_{rr^{\prime }}\}$. The two paths determine an equal
contribution to the sum (\ref{13.17}), thus, the summand in (\ref{13.17})
associated with these veins is

\begin{equation*}
\Omega _{r,r^{\prime },t}^{(2)}=-\frac{2\epsilon }{V_{N}^{2}}%
2^{-6N}J_{0}^{2}\sum_{n_{i}=n_{rr^{\prime }}}^{N}\frac{1}{7}\frac{[\rho
(n_{i})]^{2}\rho (n_{rr^{\prime }})}{Z(n_{i})[Z(n_{rr^{\prime }})]^{2}}%
2^{-3(n_{i}-n_{rr})}\cdot
\end{equation*}
\begin{equation}
\cdot \Delta (n_{i})\exp (-w(n_{i})\left| t\right| )  \label{13.22}
\end{equation}
Due to $\rho (n),Z(n),\Delta (n)$ and $w(n)$ being smooth functions of $n$
the value $\Omega _{r,r^{\prime },t}^{(2)}$ can be estimated as

\begin{equation}
\Omega ^{(2)}_{r,r^{\prime},t} \sim - \frac 27 \frac{2\epsilon}{V^2_N}
2^{-6N}J^2_0 \left [ \frac{\rho (n_{rr^{\prime}})}{Z(n_{rr^{\prime}})}
\right ]^3 \Delta (n_{rr^{\prime}}) \exp (-w(n_{rr^{\prime}}) t )
\label{13.23}
\end{equation}
because in (\ref{13.22}) the terms with $n_i \approx n_{rr^{\prime}}$
determine the main contribution. The third group involves all the veins
belonging to the branch with the node at the branching point $%
B_{rr^{\prime}} $ except for the veins of the second group. (Fig.~\ref{Fig28}%
d). Each vein of the third group, for example, vein $i$ with venules $i_r$
and $i_{r^{\prime}}$ forms either pairs $\{i_r,i,B\}_-$ and $%
\{i_{r^{\prime}},i,B_{rr^{\prime}}\}_-$ or $\{i_r,i,B_{rr^{\prime}}\}$ and $%
\{i_{r^{\prime}},i,B\}$ where $B$ is a certain branching point belonging to
one of the paths made up of the second group veins. For pairs similar to $%
\{i_r,i,B_{rr^{\prime}}\}$ the values $\Lambda _{i_ri}$ and $\Lambda
_{i_{r^{\prime}}i}$ are given by formula (\ref{13.21}) According to (\ref
{3.15}) such pairs as $\{i_r,i,B\}$ correspond to

\begin{equation}
\Lambda _{i_{r}i}=-\frac{1}{7R_{\ast }}2^{-3N}\frac{\rho (n_{B})}{%
[Z(n_{B})]^{2}}2^{-3(n_{i}-n_{B})}.  \label{13.24}
\end{equation}
For fixed level numbers $n_{B}$ and $n_{i}\geq n_{B}$ the total number of
veins is $2(2^{3}-1)2^{3(n_{i}-n_{B})}$. Therefore, the corresponding
summand in (\ref{13.17}) is equal to

\begin{equation*}
\Omega _{r,r^{\prime },t}^{(3)}=-\frac{\epsilon }{V_{N}^{2}}%
2^{-6N}J_{0}^{2}\sum_{n_{B}=n_{rr^{\prime }}}^{N}\sum_{n_{i}=n_{B}}^{N}\frac{%
2}{7}2^{3(n_{i}-n_{B})}\cdot
\end{equation*}
\begin{equation}
\cdot \frac{\lbrack \rho (n_{rr^{\prime }})]^{2}\rho (n_{B})[\rho
(n_{i})]^{2}}{[Z(n_{rr^{\prime }})]^{2}[Z(n_{B})]^{2}}%
2^{-3(2n_{i}-n_{rr}-n_{B})}\Delta (n_{i})\exp (-w(n_{i})\left| t\right| ).
\label{13.25}
\end{equation}
As before, the terms with $n_{i}\sim n_{B}\sim n_{rr^{\prime }}$ give the
main contribution to (\ref{13.25}). So,

\begin{equation}
\Omega _{r,r^{\prime },t}^{(3)}\sim \frac{2}{7}\frac{\epsilon }{V_{N}^{2}}%
2^{-6N}J_{0}^{2}\left[ \frac{\rho (n_{rr^{\prime }})}{Z(n_{rr^{\prime }})}%
\right] ^{4}\Delta (n_{rr^{\prime }})\exp (-w(n_{rr^{\prime }})t).
\label{13.26}
\end{equation}
The forth group of veins consists of the remaining veins. All these veins
form with venules $i_{r},i_{r^{\prime }}$ the pairs of the type $%
\{i_{r},i,B\}_{-},\{i_{r^{\prime }},i,B\}_{+}$ where $B$ is a certain
branching point on the path made up of the first group veins (Fig.~\ref
{Fig28}e). In this case the values $\Lambda _{i_{r}i}$ and $\Lambda
_{i_{r^{\prime }}i}$ are of the form (\ref{13.24}) for\ the fixed number of $%
n_{B}$ and $n_{i}$ the total number of the veins is $%
(2^{3}-1)2^{3(n_{i}-n_{B})}$. Whence we get the following expression for the
corresponding summand

\begin{equation*}
\Omega _{r,r^{\prime },t}^{(4)}=\frac{\epsilon }{V_{N}^{2}}%
2^{-6N}J_{0}^{2}\sum_{n_{B}=0}^{n_{rr^{\prime }}-1}\sum_{n_{i}=n_{B}}^{N}%
\frac{1}{7}2^{3(n_{i}-n_{B})}\cdot
\end{equation*}
\begin{equation}
\cdot \frac{\lbrack \rho (n_{B})]^{2}[\rho (n_{i})]^{2}}{[Z(n_{B})]^{4}}%
2^{-6(n_{i}-n_{B})}\Delta (n_{i})\exp (-w(n_{i})\left| t\right| ).
\label{13.27}
\end{equation}
Since, $\rho (n_{i})$ is a smooth function of $n,$ when summing over $n_{i}$
in (\ref{3.27}) we may set $\rho (n_{i})=\rho (n_{B})$. In this way from (%
\ref{3.27}) we get

\begin{equation}
\Omega _{r,r^{\prime },t}^{(4)}=\frac{\epsilon }{V_{N}^{2}}2^{-6N}J_{0}^{2}%
\frac{8}{49}\sum_{n_{B}=0}^{n_{rr^{\prime }}-1}\left[ \frac{\rho (n_{B})}{%
Z(n_{B})}\right] ^{4}\Delta (n_{B})\exp (-w(n_{B})\left| t\right| ).
\label{13.28}
\end{equation}
Due to $\rho (n)$ being a smooth function of $n$ for $N-n\gg 1$ the ratio $%
\rho (n)/Z(n)\ll 1$ is a small parameter. Therefore, comparing $\Omega
_{r,r^{\prime },t}^{(i)}$ with each other we find that the main contribution
to the value $\Omega _{r,r^{\prime },t}$ is given by the first group veins.
So, formula (\ref{13.19}) enables us to the represent (\ref{13.17}) as

\begin{equation}
\Omega _{r,r^{\prime },t}=\epsilon j_{0}^{2}\int\limits_{0}^{n_{rr^{\prime
}}}dn\left[ \frac{\rho (n)}{z(n)}\right] ^{2}\Delta (n)\exp (-w(n_{B})\left|
t\right| )  \label{13.29}
\end{equation}
Here we have taken into account that the mean total blood current flowing
through the vein tree and the mean value $j_{0}$ of the blood flow rate are
related by the expression $J_{0}=2^{3N}V_{N}j_{0}$ and have converted the
sum with respect to $n$ into the integral over the continuous variable $n$.

Formula (\ref{3.29}) is the desired expression for the correlation function
of blood flow rate fluctuations. The spectral density of these fluctuations
is the Fourier transform of the correlation function with respect to the time

\begin{equation}
\Omega _{\mathbf{r},\mathbf{r}^{\prime }}(w)=2\epsilon
j_{0}^{2}\int\limits_{0}^{n_{\mathbf{r},\mathbf{r}^{\prime }}}dn\left[ \frac{%
\rho (n)}{Z(n)}\right] ^{2}\Delta (n)\frac{w(n)}{w^{2}(n)+w^{2}}.
\label{13.30}
\end{equation}

The mean distance $r$ between the arteries $i_{\mathbf{r}},i_{\mathbf{r}%
^{\prime }}$ of the pairs $\{ii^{\prime }\}$ corresponding to the same
branching point $B_{\mathbf{rr}^{\prime }}$ can be estimated as $r\sim l_{n_{%
\mathbf{rr}^{\prime }}}=l_{0}2^{-n_{\mathbf{rr}^{\prime }}}$, thereby $n_{%
\mathbf{rr}^{\prime }}\sim log_{2}(l_{0}/r)$. Due to the latter estimate and 
$\Omega _{\mathbf{r},\mathbf{r}^{\prime }}(w)$ being a smooth function of $%
n_{\mathbf{rr}^{\prime }}$ on averaging (\ref{13.30}) over the cube $Q_{0}$
for $r\ll l_{0}$ we may set.

\begin{equation}
\left. \Omega _{\mathbf{r}}(w)\approx \Omega _{\mathbf{r}^{\prime },\mathbf{r%
}}(w)\right| _{n_{\mathbf{r}\mathbf{r}^{\prime }}=\log _{2}(l_{0}/r)}.
\label{13.31}
\end{equation}
In the next Section based on the obtained results we shall discuss some
characteristic properties of spatia - temporal fluctuations in the tissue
temperature which are caused by random time variations of the vessel
parameters.

\section{Characteristics of the tissue temperature fluctuations on different
scales}

\label{s13.3}

It is natural to assume that the characteristic time $1/w(n)$ of the vessel
resistance fluctuations decreases with vessel length. So, the value $w(n)$
is likely to be an increasing function of $n$. Typically the resistance of
the vascular network to blood flow is mainly determined by large vessels
rather than small venules and arterioles. Thus, in the given model we must
assume that the function $j(n)$ is an decreasing function of $n$ such that
the formal integral

\begin{equation*}
\int\limits_{0}^{\infty }\rho (n)dn
\end{equation*}
is convergent. The aforementioned allows us to consider in more detail the
special case where $\Delta (n)=1,w(n)=w_{0}\exp (n\nu _{w})$ and $\rho
(n)=\rho (0)\exp (-n\nu _{\rho })$, when $\nu _{w}$ and $\nu _{\rho }$ are
small positive constants but $\nu _{w}N,\;\nu _{\rho }N\gg 1$. In this case
as it follows from (\ref{13.30}),(\ref{13.31}) within the frequency interval 
$w(0)\ll w\ll w(N)$

\begin{equation}
\Omega _{\mathbf{r}}(w)\approx 2\epsilon j_{0}^{2}\frac{\nu _{\rho }^{2}}{%
\nu _{w}}\frac{1}{w}\tan ^{-1}\left[ \frac{w_{0}}{w}\left( \frac{l_{0}}{r}%
\right) ^{\nu _{w}/\ln 2}\right] .  \label{13.32}
\end{equation}
In particular, if $w\ll w_{r}=w_{0}(l_{0}/r)^{(\nu _{w}/\ln 2)}$

\begin{equation}
\Omega _{\mathbf{r}}(w)\approx \pi \epsilon j_{0}^{2}\frac{\nu _{\rho }^{2}}{%
\nu _{w}}\frac{1}{w}  \label{13.33}
\end{equation}
and for $w\gg w_{r}$

\begin{equation}
\Omega _{\mathbf{r}}(w)\approx 2\epsilon j_{0}^{2}\frac{\nu _{\rho }^{2}}{%
\nu _{w}}\frac{w}{w^{2}}\left( \frac{l_{0}}{r}\right) ^{\nu _{w}/\ln 2}.
\label{13.34}
\end{equation}

According to (\ref{13.5}) on spatial scales $r$ where $Dr^{-2}\ll j_{0}$ or $%
Dr^{-2}\ll w$ the Fourier transform $G_{\mathbf{r}}(w)$ of the correlation
function $G_{\mathbf{r},t}$ is directly specified by the function $\Omega _{%
\mathbf{r}}(w)$, viz.

\begin{equation}
G_{\mathbf{r}}(w)\simeq \frac{(T_{0}-T_{a})^{2}}{w^{2}+j_{0}^{2}}\Omega _{%
\mathbf{r}}(w).  \label{13.35}
\end{equation}
In particular, as it follows from (\ref{13.33}) and (\ref{13.35}) if $%
w(0)\ll j_{0}$ then, there is a frequency interval $w(0)\ll w\ll w_{r},j_{0}$
where

\begin{equation}
G_{\mathbf{r}}(w)\sim 1/w,  \label{13.36}
\end{equation}
i.e. in this case fluctuations in the living tissue temperature can exhibit $%
1/f$ behavior.

Concluding the present Section we would like to point out that there is a
certain spatial nonuniformity of the correlation function caused by the
vascular network architectonics. Indeed, in neighborhoods of the points $A$
and $A^{\prime }$ being at a small distance from each other heat transfer
can be controlled by different branches of the arterial bed. Owing to this,
in one direction fluctuations in the blood flow rate and, correspondent, in
the tissue temperature could be correlated, whereas in the opposite
direction such correlations are practically absent. In addition we note,
that because the typical values of the blood flow rate are about $j\sim
10^{-2}\div 10^{-3}s^{-1}$ according to (\ref{13.35}) the fluctuations of
the tissue temperature can exhibit $1/f$ behavior for sufficiently low
frequencies. However, fluctuations in the blood flow rate can exhibit $1/f$
behavior in substantially wider frequency interval (see formula (\ref{13.33}%
)) and can cause similar fluctuations in other physical quantities in living
tissues.

\chapter{Small scale nonuniformities of the tissue temperature}

\label{ch.14} 
\markright
{ {\sc  \thechapter. Small scale nonuniformities of the tissue\ldots}
}

From the standpoint of heat transfer living tissue is a highly heterogeneous
medium. So, the tissue temperature will exhibit spatial nonuniformities even
though a living tissue domain that contains a single microcirculatory bed of
a simple structure is uniformly heated. In this case spatial nonuniformities
of the tissue temperature are determined by intrinsic properties of living
tissue as a heterogeneous medium and may be considered in terms of random
fields.

There are two different type nonuniformities in the temperature. The first
type nonuniformities occur due to influence of blood flow in large vessels
and occur in certain regions whose relative volume is not large. So,
description of these nonuniformities cannot be reduced to continuum
approximation and should be considered individually \cite{36,45,50}. The
second type nonuniformities of the tissue temperature are caused by
relatively small vessels of level $n_{t}$ which directly control heat
exchange between the tissue and blood. The nature of these temperature
nonuniformities is the discreteness of vessel arrangement. We note that a
similar formulation of this problem has been stated in \cite{7}.

In the present Chapter we shall examine the main properties of such random
temperature nonuniformities, in particular, their mean amplitude and
correlation length, which are fundamental characteristics of heat transfer
in living tissue.

\section{The tissue temperature \ non\-uni\-formiti\-es due to random vessel
arrangement}

\label{s14.1}

For the sake of simplicity we confine ourselves to limit (\ref{6.28}) where
the capillary system has no significant effect on heat transfer, the length $%
l_N$ of the last level vessels (arterioles and venules) is well below $%
l_{n_t}$ and spatial nonuniformities in the tissue temperature distribution
are mainly caused by arrangement of the $n_t$-th level arteries and veins.
It should be pointed out that these assumptions are justified for real
microcirculatory beds because for typical values of $j \sim 10^{-3} s^{-1};
\, \kappa \sim 7 \cdot 10^{-3} J/cm \cdot s \cdot K, \, \rho _t \sim 1 g/cm,
\, c_t \sim 3.5 J/g \cdot K$ and setting $l_0 /a_0 \sim 40$ from (\ref{7.12a}%
), (\ref{7.12b}) we get $l_{n_t} \sim 0.5 cm$, which is well above the
characteristic length of real capillaries. In this case, as it follows from (%
\ref{7.15}), the stationary tissue temperature obeys the equation

\begin{equation}
\kappa _{\mathrm{eff}}\mathbf{\nabla }^{2}T-c_{t}\rho _{t}jf[1+\chi _{t}(%
\mathbf{r}\,)](T-T_{a})+q_{h}=0,  \label{14.1}
\end{equation}
where

\begin{equation*}
f=\left[ \ln \frac{l_{0}}{a_{0}}\right] ^{(\beta (n_{t})-1)/2},
\end{equation*}
i.e. $f=1$ for the unit vessel network and $f=[\ln (\l _{0}/a_{0})]^{-1/2}$
for the countercurrent vascular network, the blood flow rate $j$ and the
heat generation rate $q_{h}$ are assumed to be constant over the domain $%
Q_{0}$, the random field $\chi _{t}(\mathbf{r}\,)$ satisfies relations (\ref
{7.9}), (\ref{7.10}) and according to (\ref{7.12a}), (\ref{7.12b}) its
characteristic correlation length

\begin{equation}
l_{t}=\left[ \frac{3\sqrt{3}\pi }{4}\frac{\kappa }{f\rho _{t}c_{t}\ln
(l_{0}/a_{0})j}\right] ^{1/2}.  \label{14.2}
\end{equation}
By virtue of (\ref{6.10}) the maximum of the correlation function $g(x)$ is $%
g(0)\sim 0.16$. Therefore, in equation (\ref{14.1}) the random field $\chi
_{t}(\mathbf{r}\,)$ as well as the nonuniform component $T^{\sim }(\mathbf{r}%
\,)$ of the tissue temperature associated with this field can be treated as
small quantities. Then, linearizing equation (\ref{14.1}) with respect to $%
\chi _{t}(\mathbf{r}\,)$ and $T^{\sim }(\mathbf{r}\,)$ we get

\begin{equation}
\kappa _{\mathrm{eff}}\mathbf{\nabla }^{2}T^{\sim }-c_{t}\rho _{t}jfT^{\sim
}=c_{t}\rho _{t}jf(T^{(0)}-T_{a})\chi _{t}(\mathbf{r}\,),  \label{14.3}
\end{equation}
where the uniform component of the tissue temperature

\begin{equation}
T^{(0)}=\frac{q_{h}}{f\rho _{t}c_{t}j}+T_{a}.  \label{14.4}
\end{equation}
For the Fourier transforms of the functions $T^{\sim }(\mathbf{r}\,)$ and $%
\chi _{t}(\mathbf{r}\,)$:

\begin{equation}
T_{F}(\mathbf{k})=\frac{1}{(2\pi )^{3/2}}\int\limits_{Q_{0}}d\mathbf{r}e^{-i%
\mathbf{kr}}T^{\sim }(\mathbf{r}\,)  \label{14.5}
\end{equation}
and 
\begin{equation}
\chi _{F}(\mathbf{k})=\frac{1}{(2\pi )^{3/2}}\int\limits_{Q_{0}}d\mathbf{r}%
e^{-i\mathbf{kr}}\chi _{t}(\mathbf{r}\,)  \label{14.6}
\end{equation}
equation (\ref{14.3}) takes the form

\begin{equation}
(k^{2}\kappa _{\mathrm{eff}}+c_{t}\rho _{t}jf)T_{F}(\mathbf{k})=-\rho
_{t}c_{t}jf(T^{(0)}-T_{a})\chi _{F}(\mathbf{k}).  \label{14.7}
\end{equation}
Here the domain $Q_{0}$ may be regarded as the unbounded three - dimensional
space $\mathcal{R}^{3}$, and $\mathbf{k}$ is a point of the wave vector
space $\mathcal{R}_{F}^{3}$. Besides, by virtue of (\ref{6.10}), (\ref{7.9})
and (\ref{7.10}) the random field $\chi _{F}(\mathbf{k})$ obeys the
conditions

\begin{equation}
\left\langle \chi _{F}(\mathbf{k})\right\rangle =0  \label{14.8}
\end{equation}
and 
\begin{equation*}
\left\langle \chi _{F}(\mathbf{k})\chi _{F}(\mathbf{k}^{\prime
})\right\rangle =\delta (\mathbf{k}+\mathbf{k}^{\prime })\int d\mathbf{r}%
e^{-i\mathbf{kr}}g\left( \frac{\mid \mathbf{r}\mid }{l_{t}}\right) =
\end{equation*}
\begin{equation}
=\delta (\mathbf{k}+\mathbf{k}^{\prime })\left( \frac{4}{3}l_{t}^{2}\right)
^{3/2}\left\{ \exp \left[ -\frac{2}{3\pi }l_{t}^{2}k^{2}\right] -\exp \left[
-\frac{1}{\pi }l_{t}^{2}k^{2}\right] \right\} ,  \label{14.9}
\end{equation}
where $\delta (\mathbf{k})$ is the $\delta $ - function in the space $%
\mathcal{R}_{F}^{3},$ and also we have taken into account the identity

\begin{equation}
\int\limits_{\mathcal{R}^{3}}d\mathbf{r}\exp (i\mathbf{kr}\,)=(2\pi
)^{3}\delta (\mathbf{k}).  \label{14.10}
\end{equation}
For (\ref{14.5}) the inverse transform is

\begin{equation}
T^{\sim }(\mathbf{r}\,)=\frac{1}{(2\pi )^{3/2}}\int\limits_{\mathcal{R}%
_{F}^{3}}d\mathbf{k}\exp (i\mathbf{kr}\,)T_{F}(\mathbf{k}).  \label{14.11}
\end{equation}
Thereby, due to (\ref{14.2}) and (\ref{14.7}) - (\ref{14.9}) random
nonuniform component $T^{\sim }(\mathbf{r}\,)$ of the tissue temperature
obeys the conditions

\begin{equation}
\left\langle T^{\sim }(\mathbf{r}\,)\right\rangle =0  \label{14.12}
\end{equation}
and 
\begin{equation}
\left\langle T^{\sim }(\mathbf{r}\,)T^{\sim }(\mathbf{r}^{\prime
})\right\rangle =(T^{(0)}-T_{a})^{2}g_{T}\left( \frac{\mid \mathbf{r}-%
\mathbf{r}^{\prime }\mid }{l_{t}}\right) ,  \label{14.13}
\end{equation}
where the correlation function $g_{T}$ of the temperature random
nonuniformities is specified by the expression

\begin{equation*}
g_{T}\left( \frac{\mid \mathbf{r}\mid }{l_{t}}\right) =\frac{1}{(2\pi )^{3}}%
\int\limits_{\mathcal{R}_{F}^{3}}d\mathbf{k}\left( \frac{4}{3}%
l_{t}^{2}\right) ^{3/2}\exp \left[ il_{t}\mathbf{k}\left( \frac{\mathbf{r}}{%
l_{t}}\right) \right] \left\{ \exp \left[ -\frac{2}{3\pi }l_{t}^{2}k^{2}%
\right] -\right.
\end{equation*}
\begin{equation}
\left. -\exp \left[ -\frac{1}{\pi }l_{t}^{2}k^{2}\right] \right\} \left[ 1+%
\frac{4}{3\sqrt{3}\pi }L_{n}l_{t}^{2}k^{2}\right] ^{-2}.  \label{14.14}
\end{equation}
where $L_{n}=(\kappa _{\mathrm{eff}}/\kappa )\ln (l_{0}/a_{0})$.

Expressions (\ref{14.13}) and (\ref{14.14}) specify the desired correlation
function of the random nonuniformities in the tissue temperature. In the
given model the value $L_{n}$ is regarded as a large parameter. The latter
enables us to simplify formula (\ref{14.14}) and, thus, to analyse in more
detail characteristic properties of these temperature fluctuations.

\section[Correlation function of temperature nonuniformities and their
characteristic properties]{Correlation function of temperature \newline
no\-n\-u\-ni\-formities and their \newline
characteristic pro\-pe\-r\-ti\-es}

\label{s14.2}

Converting from the variable $\mathbf{k}$ to the new variable $\mathbf{p}=(1/%
\sqrt{\pi })l_{t}\mathbf{k}$ integral (\ref{14.14}) can be rewritten as

\begin{equation*}
g\left( \frac{\mid \mathbf{r}\mid }{l_{t}}\right) =\frac{1}{(3\pi )^{3/2}}%
\int\limits_{\Re ^{3}}d\mathbf{p}\exp \left[ i\sqrt{\pi }\mathbf{p}\frac{%
\mathbf{r}}{l_{t}}\right] \left( \exp \left[ -\frac{2}{3}p^{2}\right]
-\right.
\end{equation*}
\begin{equation}
\left. -\exp [-p^{2}]\right) \left[ 1+\frac{4}{3\sqrt{3}}L_{n}p^{2}\right]
^{-2}.  \label{14.15}
\end{equation}
By passing to spherical coordinates and integrating over the angles we get

\begin{equation*}
g(x)=\frac{3\sqrt{3}}{4\pi L_{n}^{2}}\frac{1}{x}\int\limits_{0}^{\infty
}dpp\sin (\sqrt{\pi }px)\left( \exp \left[ -\frac{2}{3}p^{2}\right] -\right.
\end{equation*}
\begin{equation}
\left. -\exp [-p^{2}]\right) \left[ p^{2}+\frac{3\sqrt{3}}{4L_{n}}\right]
^{-2},  \label{14.16}
\end{equation}
where $x=\mid \mathbf{r}\mid /l_{t}$. If in (\ref{14.16}) we replace $(p^{2}+%
\frac{3\sqrt{3}}{4}/L_{n})$ by $p^{2}$, the obtained integral will converge.
So at lower order in the small parameter $L_{n}^{-1}$ expression (\ref{14.16}%
) can be represented in terms of

\begin{equation}
g(x) \simeq \frac{3\sqrt{3}}{4\pi L_n^2} \frac{1}{x} \int
\limits_{0}^{\infty} \frac{dp}{p^3} \sin (\sqrt{\pi}xp) \left [ e^{-\frac23
p^2} - e^{-p^2} \right ].  \label{14.17}
\end{equation}
Noting that 
\begin{equation}
\frac{1}{p^2} \left [e^{-\frac23 p^2} - e^{-p^2} \right ] = \int
\limits_{2/3} ^{1} dy e^{-yp^2}  \label{14.18}
\end{equation}
we rewrite (\ref{14.17}) as 
\begin{equation}
g(x) \simeq \frac{3\sqrt{3}}{4\pi L_n^2} \frac{1}{x} \int \limits_{2/3}^{1}
dy \int \limits_{0}^{\infty} dp \frac{1}{p} \sin (\sqrt{\pi}xp) e^{-yp^2}
\label{14.19}
\end{equation}
The formula 
\begin{equation}
\int \limits_{0}^{\infty} dp \frac{1}{p} \sin(\sqrt{\pi}xp) e^{-yp^2} = 
\frac{\pi}{2} erf \left ( \frac{\sqrt{\pi} x}{2 \sqrt{y}} \right )
\label{14.20}
\end{equation}
enables us to transform expression (\ref{14.19}) as

\begin{equation}
g(x) = \frac{3\sqrt{3}}{8 L_n^2} \frac{1}{x} \int \limits_{2/3}^{1} dy \,erf
\left ( \frac{\sqrt{\pi} x}{2 \sqrt{y}} \right ).  \label{14.21}
\end{equation}

Since, for $x\mathbf{\leqslant }1$ in formula (\ref{14.21}) the integrand is
approximately constant, this expression can also be rewritten in the form

\begin{equation}
g_{T}(x)\approx \frac{\sqrt{3}}{8}\frac{1}{L_{n}^{2}}\frac{1}{x}erf\left( 
\frac{\sqrt{\pi }}{2}x\right) .  \label{14.22}
\end{equation}
Therefore, in the case under consideration the characteristic correlation
length of random nonuniformities of the tissue temperature is about $l_{t}$
and, according to (\ref{14.13}) and (\ref{14.22}) their mean amplitude $%
\delta T$ is

\begin{equation}
\delta T\sim \lbrack g_{T}(0)]^{1/2}\mid T^{(0)}-T_{a}\mid \sim \frac{1}{%
L_{n}}\mid T^{(0)}-T_{a}\mid  \label{14.23}
\end{equation}
Thus, the ratio $\delta T/\left| T^{(0)}-T_{a}\right| $ is determined solely
by the characteristic features of the vascular network architectonics and is
independent of such physical parameters of the cellular tissue as $\rho
_{t},\,c_{t},\,\kappa $ and the blood flow rate $j$. In particular, for $%
l_{0}/a_{0}\sim 30$ and $D_{\mathrm{eff}}\sim 2D$ from (\ref{14.23}) we
obtain $\delta T/\left| T^{(0)}-T_{a}\right| \sim 10\%$. In addition, we
would like to point out that these characteristics of heat transfer, among
other possible phenomena mentioned in \cite{3}, can be responsible for
freezing of living tissues within the finite temperature interval rather
than at a fixed value of temperature.

\chapter{Some comments on the bioheat transfer problem and the obtained
results}

\label{ch.15} 
\markright
{ {\sc  \thechapter. Some comments on the bioheat transfer problem\ldots}
}

\section{What the book is about from the standpoint of biophysics and
medicine}

\label{s15.1}

Scientists studying mass and heat propagation in real living tissues as
transport phenomena in certain distributed media, that is at the macroscopic
level, deal with such quantities as the tissue temperature, the
concentration of oxygen or dioxide carbon, etc. averaged on the microscopic
scales. So, they need reliable mathematical models that could be able,
first, to describe adequately experimental data and, second, to predict the
tissue behavior under various conditions. It is especially of great
importance for hyperthermia treatment or cryosurgery when living tissue is
brought closely to boundaries of vital conditions.

However real functioning of living tissues is mainly investigated at the
microscopic level considering tissue elements, including cells and blood
vessels, individually. Therefore, there is a wide gap between macroscopic
modelling and the knowledge obtained by detail analysis of microscopic
processes in living tissues. To build up a bridge between the two levels is
a difficult problem because of complex hierarchical structure of living
tissues and strong interaction of hierarchy levels with each other.

One of the possible ways mostly used at present is the phenomenological
mechanical approach in the framework of which transport phenomena are
described by equations based on certain analogies between living tissue and
various systems of condensed matter. Thus, such equations have to contain a
set of parameters that can be found by the experimental way only. Since, in
this approach specific microscopic properties of living tissue functioning
are not the factor, the results obtained by solving the corresponding
macroscopic equations with their parameters taking certain chosen values can
fit the experimental data formally only. In other words, values of these
parameters chosen in order to explain and predict the tissue behavior under
one conditions can lead to nothing under others. Besides, making such a
theory fit the particular experimental data one pays no attention to the
various specific features of microscopic tissue behavior. So, in this
approach it is very difficult to distinct what microscopic processes are
responsible for the variations in the tissue state as a whole or, moreover,
whether these variations are due to quantitative changes in the intensities
of the present microscopic processes or new ones have come into being.

Another way to build the desired bridge between the microscopic and
macroscopic levels is to develop, as it is usually done in theoretical
physics, a certain relatively rigorous technique converting a microscopic
description of transport phenomena in living tissues into macroscopic one.
The theory of mass and heat transfer developed in this way aggregates all
the characteristic features of the living tissue structure and, so, it is
possible to estimate beforehand values of the corresponding parameters and
to control their variations under different conditions. In this case the
results obtained in modelling enable one to establish the identity of what
processes in living tissue play the main role. In addition, the model
parameters falling beyond the expected intervals indicate that a principally
new process has come into play. It is this problem that has been the main
subject of the present book.

The next Section formally summarizes the main physical results of this book.
Here we go through the book from the beginning schematically singling out
the essence of the adopted model and the clues to understanding what the
main results obtained mean for scientists who do not specialize in the
applied mathematics and theoretical physics.

The proposed model for bioheat transfer%
\index{bioheat transfer} actually allows for the main two fundamental
characteristics of transport phenomena in living tissue, heat diffusion in
cellular tissue and its convective transport with blood flow. Heat
propagation involves the two components on scales of living tissue
structure, however, their relative contribution is different on various
scales.

In contrast to the condensed matter systems heat transfer in living tissue
should be considered simultaneously at all the scales up to the dimensions
of the microcirculation bed region. The matter is that the vascular network
response to local variations in the tissue temperature can cause blood flow
redistribution in all the points of the given microcirculatory bed, whereas
different microcirculatory beds function practically independently of each
other, at least, until the regulation system of the whole organism comes
into play.

On spatial scales $l$ from the capillary length up to lengths of small blood
vessels, $l<0.5\,$cm, the cooperative effect of heat conductivity in
cellular tissue and convective transport with blood flow is the diffusion
type transport with the effective diffusivity $D_{\mathrm{eff}}$. The value
of the effective diffusivity $D_{\mathrm{eff}}=FD$ is determined by the
contribution of all the small vessels which is reflected by the cofactor $F$
relating the diffusivity $D$ of the cellular tissue to the effective
diffusivity $D_{\mathrm{eff}}$. In the general case the coefficient $F$
depends on the mean blood velocity in these small vessels, their
characteristic length, radius, etc., including the maximal length of a
vessel that may be regarded as small. The latter value in turn depends on
the mean blood velocity in these vessels. Due to the microcirculation bed
being organized hierarchically all these values vary with the total blood
flow through the given vascular network and it turns out, as shown in the
present book, the coefficient $F$ becomes constant because variations of the
quantities mentioned above compensate each other. The coefficient $F$ is
actually determined by $\ln (l/a)$, where $l/a$ is the mean ratio between
the length and radius of blood vessels. For real microcirculatory beds $%
l/a\sim 40$ and the value of $F$ is estimated as $F\sim 1\div 3$.
Experimental analysis of heat propagation can only define more exactly this
value. If an experimental value of $F$ is found to be much larger than one
it will means that a certain large artery goes near a heated tissue region
and blood flow in this artery passing a sequence of branching points
transports heat over large distances. The latter process, however, cannot be
characterized by an effective diffusivity, moreover, it is not at all
described by the diffusion type equations and requires an individual
analysis which can be performed using the random walk approach developed in
the present book.

Blood in large vessels of lengths $l>1\,$cm moves so fast that it has
practically no time to attain thermal equilibrium with the surrounding
cellular tissue and, thereby, the blood flow in them carry heat away from
the microcirculatory bed. From the standpoint of heat transfer this effect
is treated as heat dissipation whose intensity is proportional to the
product of the blood flow rate $j$ and the coefficient $f$ of the sink
efficiency. The coefficient $f$ takes into account the heat exchange between
blood flows in large arteries and veins going in the closed vicinity of each
other (counter current vessels). The value of $f$ varies in the interval
from $f\approx 0.5$ for the counter current vascular network and $f=1$ for
vascular networks where the venous and arterial beds are not correlated in
spatial arrangement. In the present book we have shown that the value of $f$
is determined by the microcirculation bed architectonics only for the same
reason as the ratio $D_{\mathrm{eff}}/D$. In no case the heat dissipation
effect can be ignored, that is the coefficient $f$ can be set equal zero.
Experimentally the value of $f$ can be found as the ratio $%
(T_{v}-T_{a})/(T-T_{a})$ where $T_{v}$, $T_{a}$, $T$ are the blood
temperature in large veins, arteries and the tissue temperature.

Besides, in experimental analysis of temperature distribution in living
tissue one should account for that in addition to temperature
nonuniformities due to large vessels there also must be spatial random
nonuniformities in the tissue temperature which are intrinsic in living
tissue and occur at every point of living tissue.

Concerning with bioheat transfer one inevitably meets thermoregulation
problem. In particular, local strong heating can give rise to increase in
the blood flow rate by tenfold due to living tissue response. By now the
thermoregulation problem is far from been well studied even at the
physiological level. In the phenomenological approach this effect is
ordinary described by a local relationship $j(T)$ between the blood flow
rate and the tissue temperature, which is obtained from experimental data of
a heating large living tissue regions. What this relation does describe when
the size of heated tissue region is about or less than $1\div 2\,$cm and the
blood flow rate becomes extremely nonuniform is a question, in particular,
whether the relationship $j(T)$ changes its form or it ceases to exist at
all and the relation between the blood flow rate and the tissue temperature
becomes nonlocal. Especially, variations of blood flow at one point caused
by the temperature increase at another point is a well known effect in
living organisms.

In the present book we tried to develop such a model for thermoregulation
that allows one to attain small scales. We made use of the notion concept of
the effective temperature receptors which take into account the main
characteristics of possible humoral and neurogenic mechanisms of
microcirculation bed self--regulation. It turns out that there are some
conditions under which the local $j(T)$ dependence holds also when the size
of a heated region becomes small. In the general case this relation is
certain to be nonlocal because of a nonlinear response of blood vessels to
temperature variations as well as difference in the delay times of vessels
belonging to various levels of the vascular network. Nevertheless, the local 
$j(T)$ relation may be used as the first rough approximation.

Concerning modelling hyperthermia treatment of tumors, we did not pose the
problem of describing all the processes occurring in tumors under strong
heating. We singled out only that the substantial tumor property from the
standpoint of bioheat transfer which enables the blood flow rate in tumor to
remain at the same level at the same time when the blood flow rate in the
normal tissue surrounding the tumor increases by tenfold. This is one of the
main reasons why the temperature in tumors can remarkably exceed the
temperature in the normal tissue, leading to tumor destruction only.

\section{Brief view on the results obtained for the bioheat transfer problem}

\label{s15.2}

Recapitulating the results obtained above we would like to draw the
following conclusions.

-- Vessels whose level number $n\approx n_{t}$ mainly control heat exchange
between blood and the cellular tissue. These vessels exhibit properties of
both the heat conservation and heat dissipation vessels. Influence of blood
flow in these vessels on heat transfer is actually reduced, first, to
renormalization of the tissue heat conductivity with the renormalization
coefficient actually depending on the characteristic details of the vascular
network only. Second, the vessels of level $n\approx n_{t}$ determine
relationship between the tissue temperature and the temperature of blood in
the ``heat - conservation'' veins. In particular, in the case of a uniform
tissue temperature distribution when level $n_{t}$ consists of unit vessels
the blood temperature in veins of Class~ 1 coincides with the tissue
temperature. When level $n_{t}$ involves countercurrent pairs the blood
temperature in heat - conservation veins is lower that the tissue
temperature. In this case the difference between the venous and arterial
blood temperatures in large vessels is proportional to the difference
between the tissue temperature and the arterial temperature $T_{a}$ with the
constant of proportionality depending on the characteristic details of the
vascular network architectonics.

-- When the blood flow rate is not too high such that the influence of the
capillary system is ignorable and the blood flow rate in practically uniform
on scales of order $l_{n_{t}}$ temperature distribution can be described by
the bioheat transfer%
\index{bioheat transfer} equation with the effective diffusion coefficient
and the heat sink. The countercurrent effect is responsible for the
renormalization of the heat sink term only.

-- Temperature distribution in living tissue is practically independent of
specific details of vessel branching and depends only on characteristic
features of the vascular network architectonics. In particular, it can
depend on the mean distance between vessels of a given level or the mean
number of arteries of the same length which are generated by branching of
one artery whose length is two times as large. These characteristics called
the self-averaging property of heat transfer in living tissue allow one to
consider a model for the vascular network chosen for convenience.

-- The characteristic features of vascular network architectonics, at least,
of microcirculatory beds, can be directly determined by general requirements
on the blood flow rate. For example, equality of the blood flow rate at
different points of the same microcirculatory bed domain (when other
quantities such as the tissue temperature, the concentration of $%
O_{2},CO_{2} $, etc. are constant over the given domain) can specify the
main characteristic details of vessel branching as well as the vascular
network embedding in the corresponding domain.

-- The peculiar properties of heat exchange between blood and the cellular
tissue are the existence of the hierarchical system of branching points
where venous blood streams merge with each other.

-- Depending on the value $G=J_{0}/(2\pi Dl_{0})\ln (l_{0}/a_{0})$ all
arteries and veins of the vascular network can be divided into two classes.
Class 1 involves the vessels called ``heat - conservation'' vessels whose
level number $n<n_{t}(G)$ and the vessels of levels $n>n_{t}$ called ``heat
- dissipation'' vessels forming Class 2. In description of heat transfer in
living tissue the arteries of Class 1 may be considered in terms of the
sources of blood with the temperature $T_{a}$. The basic role of the veins
belonging to Class 1 is reduced to carrying away blood from the tissue
without heat exchange with it. The influence of the vessels, belonging to
Class 2, on heat transfer is ignorable. Blood in these vessels is in thermal
equilibrium with the cellular tissue.

-- When the characteristic total length of capillaries joining given
arterioles to venules is not too long (i.e. when $\gamma \ll 1$ within the
framework of the present model), depending on the value of $G$, the effect
of the capillary system on heat transfer can be of different types.

-- As the blood flow rate increases the capillary system causes convective
heat transport in the tissue and in this case the capillary system may be
considered in terms of porous medium.

-- On further blood flow rate increasing heat transport is controlled again
by diffusive type process in a certain effective medium but with the
effective diffusion coefficient $D_{\mathrm{eff}}$ being a function of the
blood flow rate.

-- When the capillary length is long enough (i.e. when $\gamma \gg 1$) the
range of convectional type transport is absent. In this case heat transfer
is controlled by diffusive type transport and the cellular tissue containing
the capillary system can be treated as an effective uniform medium with the
diffusion coefficient $D_{\mathrm{eff}}$.

-- Temperature distribution, in principle, can be characterized by a wide
range of spatial scales from the length of the smallest vessels (the
arterioles and venules) up to the length of the host artery or vein.
However, the basic scale which should be taken into account is of order $%
l_{n_{t}}$ for the diffusive type transport and $l_{N}$ (the venules length)
for the convectional type.

-- In the case when heat transfer is controlled by heat conservative vessels
which can be regarded as a typical case for real living tissue, the ratio
between the mean amplitude of temperature random nonuniformities and the
mean value of $(T-T_a)$ is mainly determined by the characteristic features
of the vascular network architectonics and is about 10\%.

-- Spatial temporal fluctuations in the tissue temperature and the blood
flow rate caused by random time variations in the parameters of vessels
forming a hierarchical vascular network can exhibit $1/f$ behavior.

-- When the blood flow rate becomes substantially nonuniform on spatial
scales of order $l_{n_t}$ the tissue temperature evolution is controlled by
the averaged blood flow rate rather than true one. The relationship between
the true and averaged blood flow rates can be written in the differential
form.

-- The vascular network response to temperature variations in living tissue
is represented in terms of variations in vessel resistances to blood flow
with the total resistance of vascular network being determined by a large
number of vessels belonging to different hierarchy levels. The flow
resistance of each artery - vein pairs is considered to be governed by the
temperature of blood in the corresponding vein. This blood temperature
dependence of vessel resistance forms a cooperative mechanism of living
tissue self-regulation which is based on individual response of each vessel
to the corresponding hierarchical piece of information and leads to
thermoregulation due to self - processing of information. With blood flow
resistance being a linear function of blood temperature this temperature
self-regulation becomes ideal, i.e. local variations in the tissue
temperature give rise to variations in the blood flow rate at the same
points only. Within the framework of ideal self-regulation time variations
of the true blood flow rate are locally governed by the tissue temperature.

-- Nonideality of the vessel response not only alters the form of local
relationship between the true blood flow rate and the tissue temperature but
also gives rise to a nonlocal dependence of the blood flow rate on the
tissue temperature distribution over living tissue. Inequality of the delay
times for vessels belonging to different levels can cause a nonlocal
relationship between the blood flow rate and the tissue temperature during
transient processes.

-- When the capillary system has no significant effect on heat transfer and
thermoregulation can be treated as ideal, bioheat transfer is governed by
the system of three equations, namely: the parabolic equation for the tissue
temperature evolutions with the heat sink term proportional to the averaged
blood flow rate; the elliptic equation determining the relationship between
the averaged and true blood flow rate; and the ordinary differential
equation which relates time variations in the true blood flow rate to local
values of the tissue temperature.

-- When the living tissue region affected directly is small, the difference
between the true and averaged blood flow rates can become substantial which
has a remarkable effect on the temperature distribution in living tissue.

-- Time delay in vessel response can give rise to nonmonotone growth of the
tissue temperature under rapid heating of living tissue. As a result, the
tissue temperature can go beyond the vital temperature interval for a time
of the same order as the time delay.

-- Due to depression of temperature response of tumor vessels the
temperature in a tumor exceeds significantly the mean tissue temperature
under to uniform sufficiently strong heating of living tissue. The value of
heat generation rate needed for such heating depends on the tumor size.

-- Freezing processes in living tissue can be described by the two boundary
model which deals with propagation of freezing front separating the frozen
and living regions of the tissue and the boundary of a certain living tissue
domain where the blood flow rate has attained it maximum.

\clearpage


\end{document}